\documentclass[a4paper]{report}

\usepackage{setspace}
\usepackage[english]{babel}
\usepackage{amsmath}             % Loads the amsmath package for math extras.
\usepackage{graphicx}            % This allows uploading of graphics.
\usepackage{amssymb}
\usepackage{appendix}
\usepackage{slashed}
\usepackage{tensor}
\usepackage{extarrows}
\usepackage{bm}

\makeatletter

\newcommand{\shorteq}{%
  \settowidth{\@tempdima}{-}% Width of hyphen
  \resizebox{\@tempdima}{\height}{=}%
}
\makeatother

\begin{document}

\begin{titlepage}
\begin{center}
\vspace*{1in}
{\LARGE Inter-Theory Relations in Physics: Case Studies from Quantum Mechanics and Quantum Field Theory}
\par
\vspace{1.5in}
{\large Joshua Rosaler}
\par
\vfill
A Thesis submitted for the degree of Doctor of Philosophy (DPhil)
\par
\vspace{0.5in}
Faculty of Philosophy
\par
\vspace{0.5in}
University of Oxford
\par
\vspace{0.5in}
January 11, 2013
\end{center}
\end{titlepage}

\title{Inter-Theory Relations in Physics: Case Studies from Quantum Mechanics and Quantum Field Theory}
\author{Joshua Rosaler}
\date{Submitted: January 11, 2013}

%\maketitle

\pagenumbering{roman}

\begin{singlespace}
\tableofcontents
\end{singlespace}

\chapter*{Acknowledgements}

\begin{singlespace}

This thesis has grown out of a long-standing desire to understand how the highly abstract and beautiful mathematical descriptions furnished by our most fundamental theories of physics connect both with each other and with everyday experience. This research has been fueled in large part by the belief that inquiry into the relations among current physical theories can be facilitated greatly by the application of insights that have arisen primarily in the philosophy of science literature on reduction and emergence. Conversely, it also has been motivated by the belief that the philosophical discourse on reduction and emergence would benefit greatly from, and can be made more precise through, deeper and more extensive engagement with the finer details of scientific theory and practice. 

As will no doubt become apparent to the reader in the pages to follow, my intellectual debt to David Wallace in this thesis is hard to overstate. There are very few parts that have not been greatly improved, or even spawned, by some crucial piece of insight that I gained in conversation with David or by reading his work.  I have done my best to credit him fully, though it's possible if not likely that there are places where I've failed to do so. It goes without saying that I am very lucky to have had him has my dissertation supervisor. 

I also owe a great deal of thanks to Simon Saunders, who provided a tremendous amount of assistance with early drafts of the material in Chapter 2, and pushed me to pursue questions concerning generality of theory reductions that gave rise to the material presented in the latter portions of Chapter 1, as well as providing many helpful insights on that material. Harvey Brown also provided many helpful comments on drafts of Chapter 2, and my understanding of a wide range of issues has benefited greatly from our many hours spent discussing philosophy of physics over lunch. Chris Timpson gave very helpful feedback on the material on limits and on generality in reduction in Chapter 1. Exchanges with Ward Struyve, Samuel Colin, Gary Bowman, Owen Maroney, James Anglin, Nazim Bouatta,  Jeremy Butterfield, Cian Dorr and I'm sure others whose names, for no good reason, escape me at the moment, also contributed to my thinking on many of the issues covered in the thesis.  

I am especially grateful to Chris Timpson and Jeremy Butterfield for their invaluable comments on an earlier draft of this thesis. 
 
In addition, I would like to  Oxford University Press, Pembroke College and the University of Oxford Clarendon fund for their generosity in funding this research. 

Finally, and most of all, I would like to thank my parents and our German shepherd Percy, to the three of whom this work of the last four years of my life is dedicated. 
  
\end{singlespace}

\begin{abstract}
\begin{singlespace}
\small

The relationship that is widely presumed to hold between physical theories and their successors, in which the successors in some sense explain the success of the theories they replace, is known commonly as 'reduction.' I argue that one traditional approach to theory reduction in physics, founded on the notion that a superseded theory should simply be a mathematical limit of the theory that supersedes it, is misleading as a general picture of the relationship whereby one theory encompasses the domain of empirical validity of another. I defend an alternative account that builds upon a certain general type of relationship between dynamical systems models describing the same physical system, further suggesting how this relationship can be generalized in its core features to cases where neither model is a dynamical system. I demonstrate how this particular relationship resembles the methodological prescriptions set out by Ernest Nagel's more general approach to reduction across the sciences.

  After clarifying these points of general methodology, I go on to apply this approach to a number of particular inter-theory reductions in physics involving quantum theory. I consider three reductions: first, connecting classical mechanics and non-relativistic quantum mechanics; second, connecting classical electrodynamics and quantum electrodynamics; and third, connecting non-relativistic quantum mechanics and quantum electrodynamics. In all cases, a certain core set of mechanisms, employing decoherence together with variations of Ehrenfest's Theorem, serves to underwrite the occurrence of approximately classical behavior. For concreteness, I consider two particular realist interpretations of quantum theory - the Everett and Bohm theories - as potential bases for these reductions. However, many of the technical results concerning these reductions pertain also more generally to the bare, uninterpreted formalism of quantum theory.

\end{singlespace}
\end{abstract}

\pagenumbering{arabic}

\chapter{The Methodology of Theory Reduction in Physics}
\label{ch1-corrections}

The progress of physics since the era of Kepler and Newton suggests that new fundamental theories should be required to bear a special relationship to their predecessors called `reduction,' which is supposed to ensure that newer theories encompass all of the genuine successes of their predecessors. The relationships between special relativity and Newtonian mechanics and between statistical mechanics and thermodynamics are often taken as paradigmatic examples of theory reduction in physics. However, this relationship is also usually taken to hold between general relativity and special relativity, between general relativity and the theory of Newtonian gravitation, between quantum theories and classical theories, between relativistic and non-relativistic quantum theories, and between quantum field theories and quantum mechanics. However, the possibility of reductions involving quantum theory continue to be somewhat more controversial because of the measurement problem and widespread disagreement about the possible nature of wave function collapse.

Broadly, the goal of this thesis is to elaborate a philosophical picture of reduction that I develop in general terms in this chapter, and then to use this picture to clarify how certain instances of reduction involving quantum theory come about. The approach of this thesis will be to take as potential bases for these reductions two competing versions of quantum theory which offer unambiguous, realist accounts of measurement processes and which explain, rather than merely postulating (as positivist and empiricist versions of the theory do), the appearance of wave function collapse: namely, Everett's `Many Worlds' theory and Bohm's `hidden variables' theory. Because these theories share an essential piece of mathematical structure - a wave function that always evolves according to the Schrodinger equation - it is natural to consider these theories in parallel. I emphasize that my primary goal is not to weigh the relative merits of these two interpretations of quantum theory, but to consider how various reductions would work differently between them. However, if one theory succeeds with a particular reduction where the other does not, we should take this as a basis for preferring one to the other; on the other hand, it may turn out that both succeed in underpinning these reductions on their own terms, in which case the position of this thesis will remain neutral with respect to the two. 

My analysis of reduction throughout this thesis rests upon a broadly realist (rather than a positivist or empiricist) view of scientific theories, as well as on the view that physical theories and science more generally should be expected to progress toward a state of greater unification over time, and that this progress reflects an underlying unity in nature itself. I do not attempt to defend realism or the unity of science as a whole, except insofar as my analyses potentially lend greater credibility to the realist interpretations of quantum theory that I consider by demonstrating their success at underpinning certain reductions, and except insofar as the specific reductions that I examine serve to illustrate the possibility of subsuming (and thereby unifying) the successes of different theories under a single framework. Rather, I take these philosophical doctrines as the starting point and general setting for my analysis of reduction. In my treatment of the particular reductions that I consider, I proceed on the expectation that such reductions can be performed (thus assuming the possibility of unification from the outset), addressing the question of \textit{how}, exactly, to perform them by considering what particular results must hold in order for these reductions to go through.

\section{Introduction: Outline and Structure of the Thesis}

First, a very brief summary of the thesis: Chapter 1 spells out the methodology of reduction in physics  - or `physical reduction,' as I call it - that I employ throughout the thesis. Chapters 2, 4 and 5 elaborate a number of particular reductions according to this methodology, while Chapter 3 provides an introduction to prerequisites of quantum field theory needed for Chapters 4 and 5. Chapter 6 considers how the account of physical reduction developed in the present chapter serves to resolve certain difficulties with another closely related account of reduction.  

To be more specific, the present chapter, Chapter 1, is devoted to setting out the methodology for physical reduction that I apply to particular reductions in later chapters, and to placing this methodology within the context of existing accounts of reduction in science. In it, I make three central arguments concerning the methodology of physical reduction (which I label in boldface for the sake of emphasis).\ 

\begin{enumerate}
\item \textbf{ Limits and Their Limitations:} First, the conventional notion that superseded theories in physics are, generally speaking, limiting cases of the theories that supersede them, is simplistic. In fact, this notion of reduction fails to characterise many of the inter-theory relations that it is most often purported to characterise. 
\item \textbf{ Dynamical Systems Reduction:} Within a realist, semantic view of physical theories (where the semantic view identifies a theory with the collection of its models), a more appropriate characterisation of the reduction relation in physics can be given in cases where the models of the theories concerned can be given in terms of some dynamical map on some state space. This approach to reduction, which I call the dynamical systems approach, is built around an insight extracted from reductions in statistical mechanics but that I argue can be extended more generally to reduction in physics as a whole; in the philosophical literature, it has been discussed independently in publications by Marco Giunti, Jeremy Butterfield and Jeffrey Yoshimi and was introduced to me in different forms by David Wallace and David Albert, who as far as I know also came to it independently in the context of research into the foundations of statistical mechanics. By comparison with the more popular Nagelian and limit-based approaches, this way of approaching reduction in physics has received relatively little attention in the literature on reduction; however, I argue here that it is central to an accurate, general account of theory reduction in physics. Moreover, I argue that dynamical systems reduction exhibits a number of important parallels with Ernest Nagel's classic account of theory reduction - or rather, with a particular refinement of Nagel's account - but also exhibits a number of disanalogies with Nagel's approach as well.
\item \textbf{ Template-Based Reduction:} I argue that the project of effecting a reduction between two theories within the dynamical systems framework is most perspicuously approached through the use of what I call `reduction templates,' which, briefly, are incomplete proofs of reduction. In reducing the theoretical description of a particular system provided by a high level theory to the theoretical description of that same system provided by a lower level theory, the generality of the reduction  - that is, the range of systems within the domain of the higher level theory to which the reduction applies - runs inversely to the degree of completeness of the proof. Complete proofs of reduction often require reference to the specific details governing the system in question, while an understanding of the general mechanisms that apply across a wide range of systems may require abstraction away from system-specific details. Through the use of templates, the process of  reduction can be compartmentalised into those components of the reduction which apply across of wide range of instances of the superseded theory's success, and those which are more specialised to particular systems or sets of systems. For a clear sense both of the general mechanisms that underpin the reduction, as well as the more system-specific details that are required to make the reduction complete, I claim that one should provide multiple templates: first, at a high level of generality and possibly a low level of completeness, and then successively customising the template to progressively narrower specifications and smaller sets of systems, thereby filling the gaps in the more general templates. 
\end{enumerate}

In addition to the general claims about reduction that I make in Chapter 1, in Chapters 2,4, and 5 I apply a dynamical sytems, template-based approach to a number of particular reductions: 1) in Chapter 2, the reduction between classical Newtonian mechanics (NM) and quantum mechanics (QM) \footnote{Note that I use `classical' here to describe theories which are non-quantum, and `Newtonian' to describe theories which are non-quantum \textit{and} nonrelativistic. }, 2) in Chapter 4, the reduction between quantum mechanics (QM) and relativistic quantum electrodynamics (QED) 3) in Chapter 5, the reduction between classical electrodynamics (CED) and relativistic quantum electrodynamics (QED).

In providing the templates for these reductions, I endeavour to make clear which results remain in need of proof, either because of the need to consider systems at a level of detail that goes beneath the level of generality that I seek to achieve in my analysis, or because of a technical conjecture which, though plausible, I have left unproven. In developing these templates, my approach is, in some respects, to work backward from the result to be proven, assuming that the reduction can be performed, and to consider what results need to hold for the reduction to be effected in the manner I have suggested.

Given the realist background of this project, the quantum theories involved in the reductions I consider - QM and QED - must be given realist interpretations, which, in contrast to operationalist or positivist interpretations, associate a relatively concrete metaphysical picture to the theory. I adopt as bases for my analysis the Everett and Bohm versions of these theories, which both possess the  property, essential to my analysis, that they are non-collapse interpretations of quantum theory. For reasons that I elaborate in Chapter 2, this fact makes it especially natural and convenient to consider the Everett and Bohm interpretations in parallel.

Where the Bohm theory is concerned, I employ Bohm's original formulation of the nonrelativistic quantum mechanics of a spinless particle, and Bell's formulation of the nonrelativistic quantum mechanics of a spin-1/2 particle \cite{bohm1952suggested}, \cite{bell2004speakable}. In the case of QED, I consider Struyve and Westman's minimalist model \cite{struyve2006new}. Where the Everett theory is concerned, I hope that the discussion will be of interest even to readers skeptical of Everett's theory, or of realist interpretations of quantum theory more generally; the technical discussions pertaining to Everett's theory apply equally well to the bare, uninterpreted formalism of quantum mechanics without collapse (for a defense of the view that Everett's theory simply is a reification of the bare QM formalism without collapse, see Wallace's \cite{wallace2012emergent}). Insofar as any interpretation of quantum theory is likely to `piggyback' on the empirical success of the bare formalism - as all the leading realist interpretations do -  results pertaining to the bare formalism, whether with regard to reduction or other considerations, are likely to have strong relevance to any realist interpretation of quantum theory. For this reason, the results concerning particular reductions that I present in chapters 2, 4 and 5 carry a significance that is interpretation-neutral.
 
In non-collapse interpretations of quantum theory, the phenomenon of decoherence is responsible for \textit{effective} wave function collapse. However, as I explain in chapter 2, the particular decoherence condition that ensures effective collapse in Everettian theories does not suffice to ensure effective collapse in their Bohmian counterparts. In Bohmian theories, the requirement for effective collapse, namely that branches of the quantum state be disjoint with respect to the configuration space of the additional variables, or `beables,' as John Bell called them (to contrast them with `observables'), amounts to a special kind of decoherence condition whose mathematical specification depends on the choice of beable. With regard to the reduction of classical theories to Bohmian quantum theories, I argue that the vanishing of the `quantum potential,' which is often cited as a sufficient condition for classical behaviour, is in some important respects a red herring with regard to the explanation of classical behavior within any Bohmian theory. I demonstrate that an approach which first considers the detailed structure of the wave function resulting from decoherence, and only then the determines effect of the wave function on the Bohmian configuration, is more transparent and better reflects the nature of realistic classical systems.

In the reductions relating to quantum field theory that I consider in Chapters 4 and 5, the Schrodinger picture of quantum field theory proves to be an extremely useful tool for analysing the nonrelativistic and classical domains of QED. I provide a detailed review of the Schrodinger picture of quantum field theory, as well as of Bohmian QFT, before presenting a template-based dynamical systems analysis of the classical and nonrelativistic domains of QED. 

In chapter 6, I  consider how a dynamical systems approach to reduction, despite its deep parallels with Nagelian reduction, addresses some of the major concerns about Nagelian reduction, at least within the context of physical reductions.

%In the remainder of chapter 1, I begin by discussing two broad classes of approach to reduction in physics: limit-based and Nagelian. I discuss potential refinements of both, and argue that limit-based reductions do not accurately represent the character of the relationship between successive theories in physics. I then develop the dynamical systems approach to reduction, highlighting a number of parallels with Nagelian reduction and its refinements. Finally, I consider the degree of generality that can be retained in the reduction of theories, and argue that this matter is best addressed through the use of reduction templates.

%In chapter 2, I provide a template for the NM/QM reduction within the dynamical systems framework. Most of my analysis in the Everettian context summarises much of the existing literature on the emergence of classical behavior in quantum theory, although the certain portions in which I specialise this analysis to the dynamical systems framework are original. My analysis in the context of Bohm theory is largely original. The beginning of the chapter contains a preliminary discussion of decoherence and effective wave function collapse in order to lay the groundwork for discussions of macroscopic Newtonian behavior. 

%In chapter 3, I provide a template for the QM/QED reduction within the dynamical systems framework, for both the Everett and Bohm interpretations.   

%In chapter 4, I provide a template for the CED/QED reduction within the dynamical systems framework, for both the Everett and Bohm interpretations. 

\section{Two Views of Physical Reduction}

In \cite{nickles1973two}, Nickles observes that the term `reduction' is typically used in opposite senses in the philosophy and physics literatures. Given a high-level (i.e. less encompassing, less fundamental) theory $T_{h}$ and a low-level (i.e., more encompassing, more fundamental) theory $T_{l}$ , the physics literature typically speaks of $T_{l}$ reducing to $T_{h}$, while the philosophical literature speaks of $T_{h}$ reducing to $T_{l}$. This is, to some extent, a matter of convention. The physicist's sense of reduction calls to mind uses of `reduction' that designate simplification, as when a complex fraction is reduced to a simpler one in arithmetic, while the philosopher's sense calls to mind uses of `reduction' that signify some sort of subsumption or inclusion into a broader framework, as in `the reduction of chemistry to physics,' or the `reduction of mathematics to logic.' Both senses are true to different uses of the word. In this thesis, I have chosen to employ the philosopher's sense of the term. 

Thomas Nickles is often credited with being the first to underscore the distinction between reduction in the philosopher's sense and reduction in the physicist's sense \cite{nickles1973two}. Yet the distinction, as Nickles draws it, is not solely a matter of convention as to whether the high level theory is said to reduce to the low level theory or the low level theory to reduce to the high level theory. Once the conventions are made to agree, there remains a substantive difference between the meaning of the term `reduction' as it is most often employed in the physics literature and the meaning of the term as it is most often employed in the philosophy literature. The philosopher's notion is based on an account of reduction given by Ernest Nagel while the physicist's views reduction essentially as a matter of taking mathematical limits \cite {NagelSS} 
\footnote{Of course, one may question whether it is entirely appropriate or fair to identify one sense of reduction as the physicist's and the other as the philosopher's. There are, after all, instances of physicists employing what is effectively reduction in the philosopher's sense (arguably, textbook proofs of the Ideal Gas Law on the basis of statistical mechanics are examples of this \cite {kittel1969thermal}) and of philosophers employing reduction in the physicist's sense (see, for instance, \cite{BattermanDD}).}. 

Having reversed the physicist's convention to agree with the philosopher's, these two notions of reduction as Nickles defines them, and which he designates $reduction_{1}$ for the philosopher's sense and $reduction_{2}$ for the physicist's sense, can be defined as follows:

\begin{quote}
\textbf{$Reduction_{1}$:} (Nagelian Reduction) $T_{h}$ $reduces_{1}$ to $T_{l}$ if the laws of $T_{h}$ can be derived from those of $T_{l}$, possibly along with some auxiliary assumptions, either exactly or (more often) as approximations, in all cases in which $T_{h}$ is approximately accurate.
\end{quote}

\noindent

\begin{quote}
\textbf{$Reduction_{2}$:} (Limit-Based Reduction)  $T_{h}$ $reduces_{2}$ to $T_{l}$ if there exists some set of parameters $\{ \epsilon_{i} \}$ defined within $T_{l}$ such that $\lim_{\{ \epsilon_{i} \rightarrow 0\}} T_{l} = T_{h}$. \cite{nickles1973two},\cite {batterman2007intertheory}
\footnote{Note that if one has  $\lim_{\{ \epsilon_{i} \rightarrow \infty \}} T_{l} = T_{h}$, or  $\lim_{\{ \epsilon_{i} \rightarrow a \}} T_{l} = T_{h}$ where $0<a<\infty$, one can always redefine the parameters $\{ \epsilon_{i} \}$ so that the limit approaches 0.} 
\footnote{In Nickles' original definition of $reduction_{2}$, the sense of reduction is the inverse of the one I give here, in that on Nickle's definition the superseding theory $T_{1}$ $reduces_{2}$ to the superseded theory $T_{2}$, rather than vice versa as in the definition that I provide. This inversion merely reflects an arbitrary choice of convention, and I choose the opposite convention to the one that Nickles chooses. The reason for this is so that $reduction_{2}$ and $reduction_{1}$ are defined according to the same choice of convention (that is, with the superseded theory reducing to the superseding theory), thereby facilitating an analysis of the truly substantive differences between these concepts of reduction.}

\end{quote}

\noindent Both of these definitions of reduction as they stand are still quite vague. In Nagelian reduction, what does it mean to `derive' the laws of one theory from those of another, given that many of the theories listed above have radically different ontologies, and are often formulated in drastically different mathematical and conceptual frameworks? Concerning limit-based reduction, what does it mean to take the limit of a $\textit{theory}$, given that the notion of a mathematical limit is usually defined for functions? 

Moreover, one may doubt that these two concepts of reduction are wholly mutually exclusive: for example, perhaps the limit-based notion of reduction can be subsumed into the Nagelian one, since taking limits might be construed as a form of deduction. However, as we will see when we discuss reduction in the Nagelian sense in more detail, in cases where $T_{h}$ employs terms not used in $T_{l}$, Nagelian reduction requires the use of additional assumptions often referred to as `bridge laws' to translate these terms into the terms of $T_{l}$. Limit-based reduction, insofar as it constitutes a well-defined framework for reduction at all, typically makes no mention or use of such assumptions. Thus, in such cases of `heterogeneous reduction,' as Nagel calls it, there is indeed a clear distinction to be made between the two approaches, on the basis of whether or not bridge rules are employed. 

%A further difference, apparent already at the present level of specification of the two concepts of reduction, is that the first notion requires only that in cases where $T_{h}$ succeeds, these successes can be deduced on the basis of $T_{l}$. The second, on the other hand, implies that there is a wholesale convergence of $T_{l}$'s laws and predictions to those of $T_{h}$ in a particular limit, implying that the domain of $T_{h}$'s successful applications can be characterized uniformly in terms of certain parameters within $T_{l}$ approaching certain values. As I argue below, this latter sense of reduction is unrealistic, and does not accurately characterize many of the theory relations that it is purported to characterise.  

Finally, apart from these two notions of reduction, there are several other approaches to reduction that have been proposed in the philosophy literature: notably, Kim's functional model of reduction and Hooker's `New Wave' model  \cite {kim2000mind}, \cite {hooker1981towards}. Both of these have been presented as alternatives to Nagelian reduction, primarily in the context of discussions of reduction and emergence in philosophy of mind. Marras, however, has argued that Kim's account is only superficially distinct from Nagel's \cite {marras2006emergence}. Likewise, Fazekas has argued that the main purported difference between New Wave and Nagelian reduction, that New Wave reduction succeeds without employing bridge laws, is obviated by the tacit, surreptitious invocation of assumptions that are effectively equivalent to bridge laws. Given that discussions of Kim's account are largely specialised to philosophy of mind, it would take the analysis too far afield to see how, if at all, his account can be translated to the context of physical reductions. Insofar as Hooker's account does hold some natural application to physical reduction, Fazekas has argued, convincingly in my view, that it, too, collapses into a particular refinement of Nagelian reduction \cite {fazekas2009reconsidering}. 

For these reasons - the fact that these other approaches to reduction have been formulated primarily in the context of philosophy of mind, and the possibility that, when applied to physical reduction, they collapse into Nagelian reduction - I will retain a focus on the two approaches to physical reduction that Nickles discusses: limit-based and Nagelian. In sections \ref{Limit} and \ref{Nagel}, I suggest how these accounts can be made more precise. In section \ref{Limit}, I argue that even on the most plausible clarification of the limit-based view of reduction, this account fails to accurately characterise the general nature of the relation between high- and low- level physical theories. In section \ref{Nagel}, I discuss a particular refinement of Nagelian reduction and list some of the most common concerns with it. In section \ref{DSReduction}, I describe an alternative account of physical reduction, which I designate dynamical systems (or DS) reduction, and highlight its similarities to Nagelian reduction; crucially, though, this approach concerns the reduction of individual \textit{models} of the high-level theory to individual \textit{models} of the low-level theory, rather than the wholesale reduction of entire theories.

\subsection{Limits and Their Limitations} \label{Limit}

The literature on limit-based approaches to reduction is extensive. Batterman, Butterfield, Rohrlich, Schiebe, Redhead and Post, among others, all have explored different applications and implications of this approach \cite{butterfield2011emergence}, \cite{BattermanDD}, \cite{Rohrlich}, \cite{Scheibe1}, \cite{Scheibe2}, \cite{Scheibe3},  \cite{Redhead}, \cite{Post} . A detailed review of the literature on this topic is beyond the scope of this thesis, and I limit myself to considering its two most commonly cited applications: the NM/SR reduction and the CM/QM reduction. 

The physics and philosophy of physics literatures are replete with claims that Newtonian mechanics is a `limiting case' of special relativity as $c \rightarrow \infty$, or as $\frac{1}{c} \rightarrow 0$ (abbreviated here as $NM = \lim_{c \rightarrow \infty}SR$) and that classical mechanics is a limiting case of quantum mechanics as $\hbar \rightarrow 0$ (abbreviated here as $CM = \lim_{\hbar \rightarrow 0}QM$) (see, e.g., \cite {inonu1953contraction}, \cite {sakurai1995modern}). These claims entail that, as some parameter or set of parameters $\{ \epsilon_{i} \}$ approach zero, the theory $T_{l}$ somehow `goes into' $T_{h}$. However, in the case of Newtonian mechanics and special relativity, $\frac{1}{c}$ never approaches zero for any system, nor does it $\textit{approach}$ anything since $c$ is always a constant. In the case of classical and quantum mechanics, analogous considerations apply: $\hbar$ never approaches zero for any system since it is a constant with a definite value. Nevertheless, from a purely mathematical perspective it is often the case that if one $\textit{does}$ allow the values of these parameters to vary, then one often does retrieve equations that are approximately Newtonian or classical in form. However, it would be obviously incorrect to say that physical systems which behave in Newtonian fashion are those with very large values of $c$ and that those which behave classically are those with small values of $\hbar$, since all systems have the same values for these quantities. For this reason, the physical significance of results concerning the $\hbar \rightarrow 0$ and  $\frac{1}{c} \rightarrow 0$ limits  is obscure.

A more sophisticated formulation of $reduction_{2}$, then, should explain why results derived by taking $\frac{1}{c}$ and $\hbar$ to zero should be physically significant given that in reality the values of these quantities are fixed. A natural answer to this concern would be to consider the possibility that what is $\textit{really}$ meant by $\frac{1}{c}$ approaching zero or $c$ approaching infinity is that $c$ is very large in comparison to a certain relevant, variable set of velocities characterizing the system in question, and that what is really meant by $\hbar$ approaching zero is that $\hbar$ is very small in comparison to some relevant measure of action (the units of $\hbar$) $S_{cl}$ characterizing the system. On this formulation of $reduction_{2}$, the appropriate limit to take in the NM-SR case is the limit as some $\textit{dimensionless}$ parameter $\epsilon \equiv \frac{v}{c}$ approaches zero, while the appropriate limit to take in the CM-QM case is the limit as the (again) \textit{dimensionless} parameter $\epsilon \equiv \frac{\hbar}{S_{cl}}$ approaches zero.

A question that immediately presents itself about this refinement of $reduction_{2}$, in which the parameters $\epsilon_{i}$ are required to be dimensionless, is how to identify appropriate definitions for the quantities $v$ and $S_{cl}$: for a given system, which velocities, exactly, are relevant to the limit, and how precisely does one compute the `typical action'? If one were able to identify such a set of velocities or such a definition of the typical action of a system, it might then be possible to formulate derivations that extract Newtonian or classical equations from relativistic or quantum theories, respectively, without taking $\frac{1}{c}$ and $\hbar$ to zero, but rather by taking the value of some corresponding dimensionless quantity such as $\epsilon \equiv \frac{v}{c}$ or $\epsilon \equiv \frac{\hbar}{S_{cl}}$, to zero - specifically, by varying $v$ or $S_{cl}$, each of which may assume distinct values from system to system. In this more sophisticated formulation, Newtonian systems could conceivably be characterised as those with velocities much smaller than $c$, and classical systems as those with typical actions large in comparison with $\hbar$. 

Once we have specified that the relevant parameters characterising the limit in $reduction_{2}$ should be dimensionless, we are still left with another worry, relevant specifically to the NM/SR case. Strictly speaking, the limit of SR as all velocities go to zero is a theory in which nothing moves, and in which there is only one reference frame. What is really meant, then, when NM is identified as a limiting case of SR is not that models of NM are strictly speaking the \textit{limits} of some models of SR as the relevant velocities go to zero, but rather that they provide some first- (or potentially higher-but-finite-) order approximation to these models of SR. More generally, what must be meant by $reduction_{2}$, if it is to include the NM/SR case as an example, is that for systems in which the relevant $\epsilon_{i}$ are sufficiently small, $T_{h}$ provides a good approximation to $T_{l}$ (whether zeroth, first, second, or higher order in the $\epsilon_{i}$) - not necessarily, as Nickles has suggested, that $T_{h}$ is a limit of $T_{l}$ as $\epsilon_{i} \rightarrow 0$. 

On this further refinement of $reduction_{2}$, though, there remain yet other questions about the limit-based concept of reduction. The $\epsilon_{i}$ being small are supposed to be a sufficient condition for a given model of $T_{l}$'s being approximated by some model of $T_{h}$; is this also a necessary condition? Purported instances of $reduction_{2}$, particularly the NM/SR and CM/QM cases, seem to suggest that it may also be necessary; however, to lend the definition of $reduction_{2}$ the greatest potential viability, I will refrain from encumbering it with this additional constraint, and assume that $reduction_{2}$ takes it only as a sufficient condition that the $\epsilon_{i}$ be small.   

A final worry about $reduction_{2}$, as specified thus far, pertains to cases where $T_{h}$ and $T_{l}$ are formulated within radically different mathematical and conceptual frameworks, such as the CM/QM reduction. In these cases, taking the relevant $\epsilon_{i}$ to be small may not be sufficient to establish any uniquely obvious correspondence between models of the two theories, as the mathematical concepts involved in the two models still will be radically different even after taking the limit. That is, beyond taking $\epsilon_{i}$ to be small, some clear correspondence between the frameworks of the two theories is typically still needed for the reduction to be effected, as will become apparent when we further consider the CM/QM case. 

Putting this last worry aside, a less ambiguous and more refined version of $reduction_{2}$ might be formulated as follows:

\begin{quote} \label{Reduction2}
$Reduction_{2} (refined)$:  $T_{h}$ $reduces_{2}$ to $T_{l}$ if there exists some set of dimensionless parameters 
$\{ \epsilon_{i} \}$ defined within $T_{l}$ such that when $\{ \epsilon_{i} \}$ are sufficiently small, $T_{h}$ approximates $T_{l}$. 
\end{quote}

\noindent Without some clear correspondence between the concepts populating the mathematical frameworks of the two theories, it may not be possible to decide in general whether a given model of $T_{l}$ is approximated by some model of $T_{h}$. As we will see, in the case of NM/SR, the correspondence between the variables in the two theories presents itself fairly immediately: position in SR corresponds to position in NM, time in SR (within a narrow range of inertial reference frames) corresponds to time in NM, etc.. In the case of CM/QM, this is less the case. Nevertheless, when I refer to $reduction_{2}$ below, the reader should understand it in terms of the refined version I have elaborated here.

What theories might satisfy this updated construal of $reduction_{2}$? I argue below that neither the NM-SR reduction nor the CM-QM reduction does. (Throughout, I employ the term classical to mean non-quantum, and Newtonian to mean non-quantum and non-relativistic; so, relativistic systems may be classical, but not Newtonian.) As I argue shortly, in the relativistic case there does not appear to be any set of velocities such that whenever $c$ is large in comparison to these, Newtonian behavior is always approximately retrieved. Similarly, in the quantum mechanical case there does not appear to be any precise definition of the typical action of a system such that whenever $\hbar$ is small in comparison to it, this suffices to ensure classical behavior of the system in question (a claim I defend below as well as in Chapter 2 when I consider the emergence of classical behavior within the Everettian and Bohmian formulations of quantum theory). Moreover, in the CM-QM case, there are quantum systems such as spin-1/2 particles or (near-) plane waves, which simply do not yield any counterpart in the classical theory in this limit (since taking $\frac{\hbar}{S_{cl}}$ to be small does not destroy interference or superposition effects). The notion that every model of a quantum system has a `classical limit,' in the sense that it approximates some model of classical mechanics, is incorrect. 

To take an example other than the NM/SR and CM/QM cases, classical optics is often said to be the limit of wave optics as $\frac{\lambda}{d} \rightarrow 0$, where $\lambda$ is the wavelength of light and $d$ the typical dimensions of the object on which it is impinging. Yet, applying the refined version of $reduction_{2}$ described above, one can identify systems in wave optics which, as one takes $\frac{\lambda}{d}$ progressively smaller, do not return any system in the less encompassing theory of geometric optics. For example, a system of standing waves in a metallic cavity does not produce the ray-like behavior of waves in geometric optics, since standing waves, which are essentially wave-like and do not propagate in any single direction, will continue to occupy the cavity even as one shortens their wavelength (while maintaining the relationship between the wavelength and the relevant cavity dimension necessary to sustain a standing wave). From this example and the two discussed below, it follows that one cannot generally expect a given system in a low level theory to give rise to a system contained within the high level theory in the manner prescribed by $reduction_{2}$.

%Moreover, even putting to the side these other worries, Newtonian mechanics should be construed as a first-order approximation to SR at low velocities. Typically, though the limit as $\epsilon \rightarrow 0$ of $f(x+\epsilon)$ is not the first order approximation to $f(x+\epsilon)$, but the zeroth order approximation $f(x)$. Thus, even in this respect it is incorrect to classify Newtonian mechanics as a limiting case of special relativity as velocities go to zero. The limit of SR as all velocities go to zero is a theory in which nothing moves. If anything, NM is a first-order approximation to SR, not the limit of SR as relevant velocities go to zero. Thus, if the reduction NM/SR is to account as an example of limiting-case reduction, we must adjust our definition of limiting-case reduction to incorporate first-order as well as zeroth order approximations to the low level theory. 

%In addition, for theory relations in which $T_{h}$ contains terms that do not occur in $T_{l}$, it may be the case that 

\subsubsection{Problems with $Reduction_{2}$ in the NM/SR Case}

I'll begin by considering the possibility of a type-2 reduction in the case of the reduction of Newtonian mechanics to special relativity. With regard to the choice of dimensionless parameters characterising the reduction, it is possible to interpret $v$ in at least two different ways: we may interpret $v$ as the velocity of a body as measured from a given frame, or we may interpret it as the relative speed of two frames of reference used to describe the motion of the same system.

One account of the connection between relativistic and Newtonian mechanics that adopts the latter interpretation is \cite{inonu1953contraction}, in which Wigner and Inonu demonstrate that the group of transformations relating inertial reference frames in Newton's theory, the Galilean group, is a `contraction' of the group of transformations relating inertial frames in special relativity, the Lorentz group - where a group is a contraction of another group, roughly, if it is a limit of the other group as some parameter $\epsilon$ is taken to zero. They do so by taking $\epsilon = \frac{1}{c}$ and $c \rightarrow \infty$.

While from a mathematical point of view Wigner and Inonu's results are unassailable, as an account of reduction between Newtonian and special relativistic theories they certainly do not suffice.The variable $c$ is constant, so the ``counterfactual" limit $c \rightarrow \infty$ is not physically realistic. Strictly speaking, it is the variable $\frac{c}{v}$ where $v$ is the relative speed of the two frames, and not $c$ itself, which must be taken to $\infty$ (or, rather, very large). Such a limit amounts to the physical requirement the relative velocities of the two frames be small in comparison to $c$.

However, reinterpreted as the claim that the Galilean group approximates the Lorentz group for small $\frac{v}{c}$, Wigner and Inonu's result still does not suffice to demonstrate the claim stated in the refined definition of $reduction_{2}$, namely that the NM and SR agree to arbitrary precision when $v/c$ is sufficiently small. When $\frac{v}{c}$ is small, the Lorentz transformation equations do not converge \textit{uniformly} to the Galilean transformation equations: the Lorentz transformation equation for the time coordinate contains a residual dependence on the $x$ coordinate, even when $\frac{v}{c}$ is taken to be small: $t' \approx t-\frac{v}{c^{2}}x$, so that for arbtirarily small $\frac{v}{c}$, there will be an $x$ such that the difference between the time coordinates of the two frames at $x$ - a relativistic effect - can be made arbitrarily large. This x-dependence of the time coordinate can only be eliminated if one additionally insists that $x$ is such that $\frac{\frac{v}{c^{2}}x}{t} << 1$. If one does not impose this additional constraint, then for small but non-zero $v$, say 5 miles per hour, there will continue to be significant difference in standards of simultaneity between the two frames for sufficiently large $x$.  
%(This does not undermine Wigner and Inonu's claim that the Lorentz group converges to the Galiliean group when $\frac{v}{c} \rightarrow 0$, but only highlights the fact that the convergence is non-uniform: it is true that for fixed (x,t), for any $\delta>0$ there is an $\eta$ such that $\frac{v}{c^{2}}x<\delta$ $\forall \frac{v}{c}>\eta$; it is not true that for any $\delta>0$, there exists a value of $\frac{v}{c}$ such that $\frac{v}{c^{2}}x < \delta$ $\forall x$.)

%This limit can be interpreted as the claim that when $\frac{v}{c} \rightarrow 0$, all conceivable predictions of SR are approximately the same as those of NM.

%When the condition $v/c<<1$ is imposed, the Lorentz transformation equations do not return the Galilean transformation equations. They $\textit{almost}$ do, but the equation for the transformation of the time coordinate contains a residual dependence on the $x$ coordinate: $t' \approx t-\frac{v}{c^{2}}x$. This x-dependence of the time coordinate can only be eliminated if one additionally insists that $x$ is such that $\frac{\frac{v}{c^{2}}x}{t} << 1$. If one does not impose this additional constraint, then for small but non-zero $v$, say 5 miles per hour, there will continue to be significant difference in standards of simultaneity between the two frames for sufficiently large $x$.

But even if one imposes this additional restriction on $x$, while the Galiliean transformations are guaranteed to agree with Lorentz transformations to arbitrary precision, the pair of restrictions $\frac{v}{c}<<1$, $\frac{\frac{v}{c^{2}}x}{t} << 1$ places no limit on the speeds of bodies measured from these two frames (as distinguished from the relative speed of the two frames). The fact that these bodies cannot travel faster than light with respect to either frame, or the that internal energy contributes to inertia, for example, is a non-Newtonian effect, one that may obtain even in the presence of the condition $v/c<<1$, since this is a condition on the relative velocities of the frames and not on the motions of bodies measured from these frames. In order to surmount this difficulty one must impose the additional restriction that $\frac{v'}{c}<<1$, where $v'$ is the upper bound on the speeds of bodies in the system, measured from either frame.

Yet even if one imposes this third condition, one is still confronted with cases in which relativistic, non-Newtonian effects arise. For example, in a conducting wire, electrons move relatively slowly, at a drift velocity of fractions of a meter per second. If one has two wires, then from the drift frame of the electrons in one of the wires, there is a disparity between positive and negative charge densities in the other wire caused by relativistic length contraction, since the positive and negative charges in the other wire are moving at different speeds. Thus, from the point of view of this test charge, there is an electrostatic force from the other wire that either attracts or repels it depending on the direction of the current. While for any single electron this relativistic effect is miniscule since its velocity is so small, for $10^{23}$ electrons, it is perfectly detectable \cite {purcell2011electricity}. Thus, even given the above constraints, miniscule relativistic effects can become detectable if they are compounded over many degrees of freedom. What constraint ought one to impose in order to preclude non-Newtonian effects like these? One could try to limit the number of particles in the system, but the advisability of doing so seems doubtful, especially since most of the reliably non-relativistic classical systems we know of contain large numbers of particles. 

To be sure, the attraction of the wires can be modelled nonrelativistically as a magnetostatic effect from the original reference frame in which the wires are stationary. However, given that the Principle of Relativity applies in Newtonian mechanics as in Special Relativity, it is equally legitimate to describe the system from the frame of the moving electrons, and to regard the force as electrostatic. The presence of an electrostatic force in the electron frame as compared with the absence of any such force in the rest frame of the wires is undeniably a relativistic effect, since the use of Galiliean as opposed to Lorentz transformations between the frames would require the electrostatic force to be zero in the electron frame if it is zero in the wire frame, which is not the case.

\subsubsection{Problems with $Reduction_{2}$ in the CM/QM Case}

Like the NM/SR reduction, the CM/QM reduction is frequently cited as a case of $reduction_{2}$ - that is, CM is frequently characterised as a `limiting case' of QM. I have already argued that naive formulations of $reduction_{2}$, which do not require that relevant parameters be dimensionless, fail to characterise any inter-theory relations for the simple reason that constants of nature don't vary. Might the more refined formulation of $reduction_{2}$ that I have proposed succeed in characterising the CM/QM case? I have offered evidence to the effect that this refinement does not characterise the NM/SR case, and will likewise argue now that it fails to characterise the CM/QM case. 

In the CM/QM case, $reduction_{2}$, on my refinement, amounts to the claim that models of QM for which the characteristic classical action is large in comparison with $\hbar$ - that is, for which $\frac{\hbar}{S_{c}}<<1$ - each approximate some model of classical mechanics. In attempting to assess the validity of this claim, a difficulty immediately arises: without some way of knowing which elements of a classical model are supposed to correspond to a given element of the quantum model under consideration, how are we to determine whether the quantum model approximates some classical model? This was not a problem in the NM/SR case, as a correspondence between the models of the two theories naturally presented itself there: spatial positions as measured from a particular reference frame in SR corresponded to spatial positions as measured from the some reference frame in NM, the coordinate time and standard of simultaneity in a particular SR reference frame correspond to the measure of time and standard of simultaneity in some model of NM. But the case of CM/QM is different. Models of CM describe point particles moving in space under the influence of some force laws (though these models may be given various formulations, such as those associated with the Lagrangian and Hamiltonian formalisms). Models of QM describe the time evolution of some matrix elements either of state vectors or operators on a Hilbert space; position and momentum in these models are typically associated with non-commuting operators rather than c-numbers as in CM, and even in the limit where the typical classical action of a quantum system is large, the position and momentum operators do not become c-numbers; whatever the value of the typical classical action, operators on Hilbert space and c-numbers defined, say, on some phase space, are different sorts of mathematical objects, so the behavior of one could never mimic or approximate that of the other - it's not clear what it would even mean for an infinite-dimensional matrix of a position operator, consisting of a continuous infinity of numerical entries, to approximate a classical position, which consists of only three numbers. 

Given the drastically different mathematical frameworks employed in models of CM and of QM, some correspondence must be established between the two before the CM model can be regarded as an approximation to the QM model. Indeed, this assertion simply reflects the need for the `bridging principles' that play such a central role in Nagelian accounts of reduction, which I describe in the next section. One natural correspondence between the frameworks of quantum and classical mechanics, however, does seem fairly obvious: classical position and momentum correspond to the \textit{expectation values} of the quantum mechanical position and momentum operators, rather than to the operators themselves. Let us take for granted this correspondence, even though $reduction_{2}$ makes no reference to the need for such bridging assumptions between the two theories. Does $reduction_{2}$ successfully account for the relationship between CM and QM if we allow that expectation values of quantum operators correspond the values of corresponding classical quantities? It does not, because a quantum system's typical classical action being large in comparison to $\hbar$ does not ensure that expectation values, say, of position and momentum, follow the same dynamics as their classical counterparts. Indeed, the dynamics of these expectation values depends essentially on the choice of quantum state, and an arbitrary choice of quantum state certainly will not yield classical evolutions for these expectation values, even in cases where the typical classical action is large relative to $\hbar$ (say, as a result of the system's mass being large). Any state that is not a narrow wave packet state in nearly any system other than a simple harmonic oscillar, will yield non-classical evolutions for these expectation values. 

In the preceding examination of the $reduction_{2}$ approach to physical reduction, I have attempted to make every reasonable allowance for this account (and some allowances one might fairly consider \textit{un}reasonable) in an effort to clarify its methodology and to seek out instances in which it succeeds. I have shown that in the two cases that are most often cited cited as examples of this kind of reduction, it in fact does not apply. This, I claim, is strong enough reason for abandoning $reduction_{2}$ as a general characterisation of inter-theory relations in physics. 

\vspace{5mm}

Before moving on from $reduction_{2}$ for good, though, it is worth noting in connection with limit-based reduction the work of Robert Batterman on this subject. Batterman has argued in \cite {BattermanDD} that the singular nature of the limits that are purported to characterise certain inter-theory relations - such as the relation between quantum and classical mechanics - precludes the reduction of the high-level theory to the low-level theory and signals the existence of a new, emergent (in the sense of being irreducible to the lower-level model) theory characterising what Batterman calls the `asymptotic borderland' between the theories. While I do not attempt to address Batterman's claims in this thesis, the view taken here is that the asymptotic nature of the limits in question does \textit{not} signal the failure of reduction between the theories, because limit-based reduction is not the proper way to characterise reduction in physics to begin with. Batterman's analysis of asymptotic relations between theories simply shows that $\textit{if}$ reduction between physical theories is defined along the lines of $reduction_{2}$, then some of the cases that we thought were instances of reduction aren't.

Jeremy Butterfield has also written extensively on the role of limits in theory reduction in physics. In particular, he has argued that limits offer a way of reconciling the categories of reduction and emergence, widely regarded in the philosophical literature as mutually exclusive \cite{butterfield2011emergence}. Again, though, because I claim that a limit-based approach does properly characterise the general nature  is not the right way of construing reduction between physical theories, these results will not enter the rest of my analysis.  

I've argued that the concept of inter-theory reduction in physics that takes a high-level theory be a limiting case of a low-level theory is either too vague to be useful or, upon further elaboration, wrong. But there is no denying that the limit-based results that appear throughout physics are strongly suggestive of some deep connection between the theories in question. Whatever the nature of this connection, their physical significance lies not in providing an overarching explanation of the success of the high-level theory. What the physical significance of these limit-based results is, precisely, I leave as an open question.

\subsection{Nagelian Reduction and its Critics} \label{Nagel}

According to the account of reduction set out in Ernest Nagel's \textit{The Structure of Science}, reductions can be broadly classified into two categories: homogeneous and inhomogeneous. In the former, the theory to be reduced contains no terms which are not contained in the reducing theory, while in the latter it does. An example of a homogeneous reduction is the reduction of Kepler's theory of planetary motion to Newton's Theory of Gravitation \cite{dizadji2010s}. An example of a heterogeneous reduction is the reduction of thermodynamics, which employs the concept of temperature, to the Newtonian mechanics of microscopic particles, which contains no reference to temperature. 

\subsubsection{Nagel's Formal Criteria for Reduction: `Connectability' and `Derivability'}

Recognizing that `no term can appear in the conclusion of a formal demonstration unless the term also appears in the premises,'  Nagel asserts that in the case of an inhomogeneous reduction, something beyond the low-level theory (which Nagel calls the `primary science') is necessary to perform a derivation of the laws of the higher-level theory (which Nagel refers to as the `secondary science'). He claims that there are two formal conditions that must be satisfied in order to effect an inhomogeneous reduction of the higher- to the lower- level theory, criteria which Nagel designates the condition of `connectability' and the condition of `derivability':

\begin{quote}
\begin{singlespace}
(1) Assumptions of some kind must be introduced which postulate suitable relations between whatever is signified by `A' [a term in the secondary science] and traits represented by theoretical terms already present in the primary science. The nature of such assumptions remains to be examined; but without prejudging the outcome of further discussion, it will be convenient to refer to this condition as the `condition of connectability.' (2) With the help of these additional assumptions, all the laws of the secondary science, including those containing the term `A,' must be logically derivable from the theoretical premises and their associated coordinating definitions in the primary discipline. Let us call this the `condition of derivability.'
\end{singlespace}
\end{quote}

\noindent While Nagel takes reduction essentially to be deduction, and thus to center on criterion (2), the condition of connectability provides the additional element required to effect a deduction of a higher-level theory containing terms that do not appear in the lower-level theory, in that it provides for a lexicon of sorts to translate the terminology of the higher-level theory into that of the lower-level theory. 

\vspace{5mm}

The central example that Nagel employs to illustrate his model of reduction is the reduction of the Ideal Gas Law ($pV=nRT$), as understood in the context of classical thermodynamics, to the laws of Newtonian mechanics, which were assumed during the period when statistical mechanics was formulated to govern the fundamental microscopic constituents of gases. He notes that while the term `temperature' had an accepted meaning in the context of thermodynamics, given in terms of experimental measurements using thermometers and other devices, as well as in terms of the theoretical role that it played in thermodynamical laws, the term made no appearance in the low-level theory, Newtonian mechanics. Pointing to the usual derivation of the Ideal Gas Law from classical statistical mechanics that one finds in most introductory textbooks on statistical mechanics, Nagel takes note of the strategy used to resolve this difficulty: namely, to associate the thermodynamical term temperature with a quantity understandable within the framework Newtonian mechanics, namely average molecular kinetic energy. More precisely, the derivation of the Ideal Gas Law from the assumptions of Newtonian physics postulates the relation $\langle K.E. \rangle = \frac{3}{2}kT$, thereby satisfying Nagel's connectability condition and permitting the derivation of the Ideal Gas Law from Newtonian physics (combined with some assumptions about Newtonian initial conditions). The connection between temperature and molecular kinetic energy has come to serve as the paradigmatic example of what is sometimes referred to as a `bridge principle,' `bridge law' or `bridge rule' or `reduction function' - that is, one of Nagel's connecting assumptions.

\subsubsection{Nagel's Model, Refined}

%Schaffner, Hooker, Fazekas, GNS

Since Nagel put forward his original model of reduction, a number of modifications and refinements of this model have been proposed. There isn't space here to review all of them, so I will only discuss the particular modifications which are employed in the refinement of Nagel's views that I consider in my analysis.  

Schaffner, one of the early commentators on and developers of Nagel's work on reduction, observed that Nagel's account of reduction is, strictly speaking, too stringent, since reductions in practice rarely if ever yield derivations of the higher level theory $T_{2}$ from the lower level theory $T_{l}$, but rather yield derivations of some modified or corrected version $T_{h}^{'}$ of $T_{h}$ that employs the same vocabulary as $T_{h}$; $T_{h}^{'}$ is sometimes referred to as the `analogue theory' of $T_{h}$.  According to Schaffner, bridge laws can then be understood as those relations that link all terms in $T_{h}^{'}$ that do not appear in $T_{l}$ with terms in $T_{l}$. One then derives $T_{h}^{'}$ from a combination of $T_{l}$ and the set of bridge laws \cite{schaffner1967approaches}. 

Like Schaffner, Hooker proposed a revision of Nagel's model that accommodates the fact that in practice it is often not, strictly speaking, $T_{h}$ that gets derived from $T_{l}$. However, unlike Schaffner, Hooker claims that the theory that one does derive from $T_{l}$, which Hooker calls the `image theory,' $T_{h}^{*}$, of $T_{h}$, should be formulated in the vocabulary of $T_{l}$ rather than that of $T_{h}$. Thus, no bridge laws are required to derive $T_{h}^{*}$ from $T_{l}$. Once an image theory has been derived, says Hooker, one can regard $T_{h}$ as having been reduced by virtue of an `analogue relation' that $T_{h}$ bears to $T_{h}^{*}$ \cite {hooker1981towards}. While it is not clear precisely what Hooker's analogue relation consists in, Hooker claims that because what one really derives in a reduction is an image theory and not the theory $T_{h}$ itself, even inhomogeneous reductions do not require the use of bridge laws. 

The refinement of Nagel's account that I consider here, dubbed the Generalized Nagel-Schaffner (GNS) model by Dizadji-Bahmani, Frigg and Hartmann in \cite {dizadji2010s}, consolidates both Schaffner's and Hooker's
\footnote{The GNS model does not explicitly draw on Hooker's work, although it does independently of this work, recognise the need for what Hooker calls an image theory in the process of reduction.} 
insights, and is based largely on Schaffner's Generalized-Replacement-Reduction (GRR) model 
\footnote{the distinction largely being that the former does not adopt Schaffner's view of bridge laws (see \cite{schaffner1994discovery}) }. 
Specifically, the GNS model incorporates both Hooker's image and Schaffner's analogue theories. On this model, reduction can be understood as a three-step process, starting with the basic ingredients of a low-level theory $T_{l}$, a high-level theory $T_{h}$, and a set of bridge laws:

\begin{enumerate}
\item Derive the image theory $T_{h}^{*}$ for some restricted boundary or initial conditions within the low level theory $T_{l}$. This step refines Nagel's derivability condition; it is the image theory $T_{h}^{*}$, not the high-level theory  $T_{h}$, that is derived from $T_{l}$ on this account of reduction. 
\item Use bridge laws to replace terms in $T_{h}^{*}$, which belong to the vocabulary of the low level theory, with corresponding terms belonging to the high level theory. This yields the analogue theory $T'_{h}$ (as is made clear in step 3, the sense of `analogue' here is different from the sense in which Hooker uses it, in that it refers to the relation between $T'_{h}$ and $T_{h}$, not between  $T_{h}^{*}$ and $T_{h}$ as in Hooker's account). This, along with step 3, refines Nagel's connectability condition. 
\item  If the modified theory $T'_{h}$ is `strongly analogous' to the high level theory $T_{h}$, the high level theory has been reduced to $T_{l}$.  The `strong analogy' relation is sometimes also characterised as Ôapproximate equalityÕ, Ôclose agreementÕ, or Ôgood approximationÕ and can be understood in any of these senses.  This step contributes an additional component to Nagel's connectability condition. 
\end{enumerate}

\noindent Henceforth, when I speak of Nagelian reduction, I will construe it according to the GNS model, unless explicitly stated otherwise. Moreover, note that Nagel's connectability condition on this refinement consists of two `connections': first, the bridge laws that link the image theory $T_{h}^{*}$ and the analogue theory $T_{h}^{'}$, and second, the rather vaguely defined `analogue relation' that connects the analogue theory $T_{h}^{'}$ to the high level theory $T_{h}$.

\subsubsection{Problems for Nagelian Reduction}

In \cite{dizadji2010s}, the authors provide a comprehensive survey of common criticisms of Nagelian approaches to reduction. Here, I restrict my focus to a few of these, quoting the authors directly:

\begin{itemize}
\begin{singlespace}
\item \textit{The syntactic view of theories}. Nagel formulated his theory in the framework of the so-called syntactic view of theories, which regards the- ories as axiomatic systems formulated in first-order logic whose non-logical vocabulary is bifurcated into observational and theoretical terms. This view is deemed untenable for many reasons, one of them being that first-order logic is too weak to adequately formalise theories and that the distinction between observational and theoretical terms is unsustainable. This, so one often hears, renders Nagelian reduction untenable.
\item  \textit{The content of bridge laws}. There is a question about what kind of statements bridge laws are. Nagel considers three options: they can be claims of meaning equivalence, conventional stipulations, or assertions about matters of fact. The third option can be broken down further, since a statement connecting two quantities could assert the identity of two properties, the presence of a (merely) \textit{de facto} correlation between them, or the existence of a nomic connection. Although the issue of the content bridge laws is not per se an objection, it is a question that has often been discussed in ways that gave rise to various objections, in particular in connection with multiple realisability, to which we turn now.
\item  \textit{Bridge laws and multiple realisability}. The issue of multiple realisability (MR) is omnipresent in discussions of reduction. A $T_{P}$-property [in my notation, $T_{h}$-property ] is multiply realisable if it corresponds to more than one different $T_{F}$ -properties [in my notation, $T_{l}$-properties ]. The standard example of a multiply realisable property is that of pain: Pain can be realised by different physical states, for instance in a humanÕs and in a dogÕs brain. The issue also seems to arise in SM because, as Sklar points out (Sklar 1993, 352), temperature is multiply realisable. MR is commonly considered to undermine reduction. ... [One] argument from MR is that, in order to reduce $T_{P}$ -phenomena to $T_{F}$ -phenomena, $T_{P}$ -properties must be shown to be Ônothing over and aboveÕ $T_{F}$ -properties. That is, it must be shown that $T_{P}$ -properties do not exist as something extra or in addition to $T_{F}$ -properties: There is only one group of entities, $T_{F}$ -properties. Showing this requires the identification of $T_{P}$ - properties with $T_{F}$ -properties. But a multiply realisable $T_{P}$ -property is not identifiable with a $T_{F}$-property. This undercuts reduction.
\item \textit{Strong analogy}. Strong analogy is essential to GNS. This raises three issues. The first is that the notion of strong analogy is too vague and hard to pin down to do serious work in a reduction. It is a commonplace that everything is similar to everything else, and hence saying that one theory is analogous to another one is a vacuous claim.
\item \textit{The Epistemology of Bridge Laws}. How are bridge laws established? Nagel points out that this is a difficult issue since we cannot test bridge laws independently. The kinetic theory of gases can be put to test only after we have adopted Equation 5 as a bridge law, but then we can only test the entire ÔpackageÕ of the kinetic theory and the bridge law, while it is impossible to subject the bridge law to independent tests. While this is not a problem if one sees bridge laws as analytical statements or mere conventions, it is an issue for those who see bridge laws as making factual claims.   \cite{dizadji2010s}
\end{singlespace}
\end{itemize}

\noindent As we will see, the alternative account of reduction that I propose, dynamical systems, or DS, reduction, adopts certain elements of the GNS account, but considers the reduction of individual models of a high-level theory to models of a low-level theory, rather than the reduction of whole theories. In the concluding chapter, I explain how DS reduction eliminates much of the vagueness and ambiguity that lies at the root of most of these criticisms of Nagelian reduction, and thereby addresses these concerns in the cases where it applies.

\section{Dynamical Systems (DS) Reduction} \label{DSReduction}

While Nagel and Schaffner were proponents of the (now widely repudiated) syntactic view of theories, the approach adopted in this dissertation is the semantic view. Van Fraassen, famously, has characterised the distinction between the semantic and the syntactic views of scientific theories as follows: 

\begin{quotation}
\begin{singlespace}
The syntactic picture of a theory identifies it with a body of theorems, stated in one particular language chosen for the expression of that theory. This should be contrasted with the alternative of presenting a theory in the first instance by identifying a class of structures as its models. In this second, semantic, approach the language used to express the theory is neither basic nor unique; the same class of structures could well be described in radically different ways, each with its own limitations. The models occupy centre stage. \cite{van1980scientific}
\end{singlespace}
\end{quotation}

\noindent Although there is, of course, more to the difference between the syntactic and semantic views than this now-popular slogan communicates, the subtleties of the distinction will not concern us as the remainder of the analysis will take place within the semantic view, in which, as Van Fraassen puts it `the models occupy centre stage.' In section \ref{DSGNSParallels}, I argue that, despite Nagel's own disposition toward the syntactic view, Nagelian (read: GNS) reduction applies equally well within the semantic view, since nothing about the central requirements of GNS reduction, apart from the reference to the reduction of theories rather than to the reduction of models of those theories, uniquely requires a syntactic intepretation of the theories concerned: as we will see, the concepts of image theory, bridge rule, analogue theory and strong analogy all can be carried over into the semantic framework, with the only major difference being that in this context the three steps of GNS reduction involve an image \textit{model}, bridge rules, an analogue \textit{model} and strong analogy. In this dissertation, the particular models that I consider are all of a particular sort: namely, they can be formulated as \textit{dynamical systems}, a notion I define shortly. The basic elements of Nagelian reduction, as applied to the reduction of dynamical systems models, constitute what I refer to as `dynamical systems reduction,' or DS reduction, and furnish the methodological groundwork for the particular reductions considered in Chapters 2, 4 and 5.

The methodology elaborated here seeks both to generalise and develop an approach that has been applied to certain reductions in statistical mechanics, demonstrating how it may also be used more widely to describe relations between theories outside of this context. In section \ref{OriginsPrecursors}, I discuss work by a number of authors that either anticipates or paves the way for the DS approach.

\subsection{Models of Dynamical Systems}

A dynamical systems model $M$ of a theory $T$ consists of a state space $S$ and a dynamical map $D$ on $S$; formally, we can write $M=(S,D)$.  In all models that I consider, the state space $S$ is endowed with the minimum structure of a differentiable manifold with a norm. The dynamical map is a differentiable function of time and of the state $x$ in $S$ such that for fixed $t$, $D$ specifies a bijection of $S$ onto itself, and such that $D$ is the identity map on $S$ when $t=0$:

\begin{equation}
D: \mathbb{R} \times S \ \longrightarrow \ S,
\end{equation}

\begin{equation}
D: (t,x_{0})   \ \longmapsto \  x(t), 
\end{equation}

\begin{equation}
D_{t}: S \ \longrightarrow \ S,
\end{equation}

\begin{equation}
D: (0;x_{0}) \ \longmapsto \  x_{0}.
\end{equation}

\noindent The dynamical map specifies the time evolution of points in $S$, so that $x(t)=D(t;x_{0})$, where $x_{0}$ is the state of the system at time $t=0$. The requirement that the dynamical map at fixed time be one-to-one ensures that the dynamics are deterministic.

For example, the model of a single massive, spinless particle in CM is given by $M=(\Gamma,D_{CM})$, where $\Gamma$ is the corresponding classical phase space and 

\begin{equation} \label{CMDynamicalMap}
D_{CM}\big[t;(x_{0},p_{0}) \big] = \bigg(e^{ \{ \circ, H  \} t } x \big|_{x_{0},p_{0}}, e^{ \{ \circ, H  \} t } p \big|_{x_{0},p_{0}}\bigg), 
\end{equation}

\noindent where \footnotesize $ e^{ \{ \circ, H  \} t } f(x,p) \equiv f(x,p) + \{ f(x,p), H \} t  + \frac{1}{2!}  \{ \{ f(x,p), H \}, H \} t^{2} +  \frac{1}{3!}  \{ \{ \{ f(x,p), H \}, H \} , H \} t^{3} + ...  $ \normalsize, and $\{ , \}$ denotes the Poisson bracket, defined by $\{ f,g \} \equiv \partial_{x} f \partial_{p} g - \partial_{x} g \partial_{p} f$, with $f$ and $g$ some arbitrary differentiable functions on phase space.  

The model of a single massive, spinless particle in QM is given by $M_{QM} =(\mathcal{H},D_{QM})$, where $\mathcal{H}$ is the Hilbert space of a single massive spinless particle and 

\begin{equation} \label{QMDynamicalMap}
D_{QM} \big[t; | \psi_{0} \rangle \big] = e^{-i \hat{H} t}  | \psi_{0} \rangle.
\end{equation}

\noindent where \scriptsize $e^{-i \hat{H} t}  | \psi_{0} \rangle \equiv \big( \hat{I} + (-i \hat{H} t) + \frac{1}{2!} (-i \hat{H} t)^{2} + \frac{1}{3!} (-i \hat{H} t)^{3} + ... \big) | \psi_{0} \rangle $ \normalsize.

All models of the theories considered in this dissertation, including the quantum theories, employ dynamical maps that are deterministic. I leave the question as to how to extend the dynamical systems approach to indeterministic systems for future work. The interpretations of quantum theory that I examine - the Everett and Bohm theories - both treat wave function collapse as an effective process induced by decoherence, and the stochastic aspects of quantum theory as merely apparent. The underlying dynamics in both cases is fully deterministic.

\subsection{Laws of Motion in DS Models} \label{Laws}

GNS reduction requires that the laws of $T_{h}$ be derivable, in an approximate sense, from the laws of $T_{l}$, along with some auxiliary assumptions that include bridge rules. In a semantic, dynamical systems context, the laws of a theory often correspond to the dynamical maps of its models; typically, the dynamical maps associated with the different models of a theory will all have some common form. 

Conventionally, though, the dynamical laws of a theory are not specified in the form of dynamical maps, but equivalently in terms of differential equations. In the case of theories whose maps are deterministic, it will be possible to model the system's dynamics in terms of a first order differential equation, or some set of first order differential equations. Starting from the solution $x(t) = D(t; x_{0})$ and differentiating both sides with respect to time, we have the first order differential equation of motion

\begin{equation}
\frac{dx}{dt}   = f(x, t)
\end{equation}

\noindent where $f(x, t) = \frac{\partial}{\partial t} D(t; x_{0}) \big|_{t=0,x_{0}=x} $. 

For example, in the case of CM, the first order differential equations corresponding to the dynamical map provided in eq. (\ref{CMDynamicalMap}) are

\begin{equation}
\begin{split}
& \frac{dx}{dt} = \{x, H\} \\
& \frac{dp}{dt} = \{p, H\},
\end{split}
\end{equation}

\noindent which are simply Hamilton's equations. 

In the case of QM, the first order differential equation corresponding to the dynamical map provided in eq. (\ref{QMDynamicalMap}) is

\begin{equation}
i \frac{\partial}{\partial t} | \psi \rangle = \hat{H} | \psi \rangle,
\end{equation}

\noindent the standard form of Schrodinger's equation.

\subsection{Symmetries of Dynamical Systems}

A function $T:S \rightarrow S$ is a dynamical symmetry of the dynamical systems model $(S,D)$ if it is an automorphism of $S$ satisfying the condition

\begin{equation}
D(t;T(x_{0})) = T(D(t;x_{0})) \ \text{for all} \ x_{0} \in S \ \text{for all} \ t\in \mathbb{R}.
\end{equation}

\noindent That is, $T$ carries solutions of the equations of motion into other solutions of the equations of motion. Or, equivalently, if one takes the trajectory/solution associated with the function $x(t)$, with $x(0) = x_{0}$, then the function $T(x(t))$ also consitutes a solution to the equation of motion with initial condition $T(x_{0})$.

For example, in classical Hamiltonian mechanics, consider the model of a single particle in 3 dimensions moving in a spherically symmetric potential, so that the Hamiltonian takes the form

\begin{equation}
H = \frac{p^{2}}{2m} + V(r)
\end{equation}

\noindent where $r$ is the distance of the particle from some fixed origin. Then the map $T$ given by

\begin{equation}
T(\vec{x},\vec{p}) = \big(e^{ \{ \circ, \vec{L}\cdot \hat{n}\theta  \} } \vec{x}, e^{ \{\circ,\vec{L}\cdot \hat{n}\theta \} } \vec{p} \big)
\end{equation}

\noindent where $\vec{L} = \vec{x} \times \vec{p}$ is the angular momentum, and which constitutes a rotation of the position and momentum about the axis $\hat{n}$ and the angle $\theta$, is a dynamical symmetry (as well as what is known in classical Hamiltonian dynamics as a canonical transformation generated by the function $\vec{L}\cdot \hat{n}$). This follows straightforwardly from the fact that for the above Hamiltonian 

\begin{equation}
\{ L_{i}, H \} = 0 \ \text{for all} \ i, \ \text{where} \ i=x,y,z
\end{equation}

\noindent (see for instance, \cite{GoldsteinCM} Ch. 9). Other dynamical symmetries for the above Hamiltonian include spatial translations (which are generated by the momentum function $p$ on phase space) and and time translations (which are generated by the Hamiltonian $H$ itself).

\subsection{Reduction of Dynamical Systems}

In this subsection, I set out formal criteria for DS reduction, first offering some remarks to motivate these criteria. I also provide a somewhat simplified example to illustrate this approach. More complicated examples are discussed in later chapters of the thesis.

\subsubsection{Bridge Maps and Bridge Rules}

As we will see, the appropriate DS counterpart to Nagelian bridge laws is a differentiable, time-independent function $B$ from the low level state space $S_{l}$ into the high level state space $S_{h}$:

\begin{equation}
B: S_{l}  \longrightarrow S_{h}
\end{equation}

\begin{equation}
B: x^{l} \longmapsto B(x^{l}),
\end{equation}

\noindent where $x^{l} \in S_{l}$. The function $B$ will typically be many-one, and satisfies certain added conditions to be discussed below. Its mathematical domain may be the whole of $S_{l}$ or a subset of $S_{l}$, and its image the whole of $S_{h}$ or a subset of $S_{h}$. As we will see, the bridge map will serve to identify those structures in the low-level model that approximately emulate the behavior of states in the high-level model. 

If $x^{h} \in S_{h}$ and $x^{h} =B(x^{l})$, denote its inverse image under the bridge map $E_{x_{h}}$, so that 
 
\begin{equation}
E_{x_{h}} \equiv \big\{ x_{l} \in S_{l} \big|  B(x^{l})=x^{h}  \big\}.
 \end{equation}
 
 \noindent Thus, the set  $E_{x^{h}}$ is the set of states that correspond to $x^{h}$ under the bridge map. As we will see, though, there is a dynamical constraint which further restricts which $x^{l}$ physically instantiate a given $x_{h}$ under the bridge map. 

For example, let us consider the CM and QM models of a single spinless particle of mass $m$ in an external potential. One possible bridge map between $S_{QM}=\mathcal{H}_{1p}$, the Hilbert space of a single spinless particle, and $S_{CM}=\Gamma_{1p}$, the phase space of a single massive particle with no internal degrees of freedom, is given by taking expectation values of the quantum mechanical position and momentum operators:

\begin{equation}
B_{QM}^{CM}: \mathcal{H}_{1p}  \longrightarrow \Gamma_{1p}
\end{equation}

\begin{equation}
B_{QM}^{CM}: | \psi \rangle \longmapsto \bigg( \langle \psi | \hat{x}| \psi \rangle , \langle \psi | \hat{p}| \psi \rangle \bigg) 
\end{equation}

\noindent The bridge map thus associates with each element $|\psi \rangle$ of $\mathcal{H}_{1p}$ an element $(x',p')$ of $\Gamma_{1p}$, such that $(x',p')= \big( \langle \psi | \hat{x}| \psi \rangle , \langle \psi | \hat{p}| \psi \rangle \big) $. Note that this map is many-one, since there will be many $|\psi \rangle$ which map to the same $(x',p')$. Its domain is the whole of $S_{QM}=\mathcal{H}_{1p}$ and its image the whole of $S_{CM}=\Gamma_{1p}$. Furthermore, note that 

\begin{equation}
E_{(x',p')} \equiv \bigg\{ | \psi \rangle \in \mathcal{H}_{1p}  \ \big|  \ \big( \langle \psi | \hat{x}| \psi \rangle , \langle \psi | \hat{p}| \psi \rangle \big) \  =  (x',p')  \bigg\}.
 \end{equation}

\noindent In other words, $E_{(x,p)} $ is the set of quantum states whose expectation values for position and momentum are, respectively, $(x,p)$.

In highlighting the parallels between the DS account and the GNS account of reduction later on, it will be important to distinguish between the function that carries elements of the low-level space $S_{l}$ to elements of the high level space $S_{h}$, which I call the bridge \textit{map}, and the assignment of variable names to the images under the bridge map of elements in $S_{l}$,  which I call a bridge \textit{rule}. For example, the bridge map $B_{QM}^{CM}$ carries $| \psi \rangle$ into $( \langle \psi | \hat{x}| \psi \rangle , \langle \psi | \hat{p}| \psi \rangle )$; the bridge rule simply makes the assignment of the variables $ (x',p')$ to $( \langle \psi | \hat{x}| \psi \rangle , \langle \psi | \hat{p}| \psi \rangle )$. At this stage, such a distinction may seem trivial, and the GNS account does not bother to distinguish between these two steps. However, as we will see, the distinction turns out to play an important role in distinguishing between image and analogue models in the DS account. 

Note how the bridge map and bridge rule have gone some of the way toward satisfying Nagel's connectability criterion (as interpreted within a semantic, dynamical systems approach): they associate with each $x_{h}$ a set of $x_{l}$, thereby providing a clear correspondence between a certain portion of the mathematical formalism of $T_{h}$, namely $S_{h}$, and a certain portion of the mathematical formalism of $T_{l}$, namely $S_{l}$. However, the full analogy between bridge maps as I have defined them here and bridge laws as they are envisaged in the GNS account will become apparent only once the DS account of reduction has been fully laid out.

\subsubsection{Induced Dynamics}

Given a model $M_{l}=(S_{l},D_{l})$ of $T_{l}$, a model $M_{h}=(S_{h},D_{h})$ of $T_{h}$, and a bridge map $B: S_{l}  \longrightarrow S_{h}$, the dynamical map $D_{l}: S_{l} \rightarrow S_{l}$  induces, through the bridge map, a dynamics $D^{x^{l}_{0}}_{h}(t;x_{0}^{h})$ on $S_{h}$. Specfically, every dynamical trajectory $x^{l}(t) = D_{l}(t;x_{0}^{l})$ on $S_{l}$ that remains in the domain of $B$ has an image $x'^{h}(t) = B\big( D_{l}(t;x_{0}^{l}) \big)$ in $S_{h}$. Generally, the trajectory $x'^{h}(t)$ may depend on the particular choice of initial condition $x_{0}^{l}$, not just on the image $x_{0}^{h}\equiv B(x_{0}^{l})$ to which  $x_{0}^{l}$ maps under $B$.  

For example, in the CM/QM case we have been considering, the dynamics on Hilbert space induces through the bridge map/rule a dynamics on phase space:

\begin{equation}
B_{QM}^{CM} \bigg( D_{QM} \big(t; | \psi_{0} \rangle \big) \bigg) = \big(x'(t), p'(t) \big) =   \big( \langle \psi_{0} | e^{i\hat{H} t}  \hat{x} \ e^{-i\hat{H} t} | \psi_{0} \rangle, \langle \psi_{0} | e^{i\hat{H} t}  \hat{p} \ e^{-i\hat{H} t} | \psi_{0} \rangle \big)
\end{equation}

\noindent However, it is important to note that the induced trajectory on $S_{h}$ is sensitive to the choice of initial condition $| \psi_{0} \rangle$ in the Hilbert space $S_{l}$. In this sense, the dynamical map induced by the low-level dynamics via the bridge map does not in general prescribe an autonomous (in sense of being determined only by high-level states and not depending on the specific low-level states that instantiate them) dynamics on $S_{h}$.

\subsubsection{Reducing Dynamics} \label{DSRDynamics}

In order for a DS reduction to take place between a model $M_{h} = (S_{h}, D_{h})$ of $T_{h}$ and a model $M_{l} = (S_{l}, D_{l})$ of $T_{l}$, it is necessary that the induced dynamics $B \circ D_{l}$ on $S_{h}$ approximate, in the sense of $S_{h}$'s norm, the dynamics $D_{h}$. This requirement, which I denote the DSR (Dynamical Systems Reduction) condition, can we written

\begin{equation} \label{DSRApprox}
B (D_{l}(t;x^{l}_{0})) \approx D_{h}(t; B(x^{l}_{0}))
\end{equation} 

\noindent or, more precisely, 

\begin{equation} \label{DSR}
\bigg|  B \big(D_{l}(t;x^{l}_{0})\big) - D_{h}\big(t;B(x^{l}_{0})\big) \bigg|_{h} < \delta,
\end{equation} 

\noindent for some domain $d$ of states in $S_{l}, $where $| \ |_{h}$ designates the norm on $S_{h}$ and $\delta$ is a prescribed margin of error characterising the accuracy of the approximation. Note that the left hand side of (\ref{DSRApprox}) corresponds to the dynamics induced on $S_{h}$ by $D_{l}$ through $B$, with initial condition $x_{0}^{l}$, while the right hand side corresponds to the dynamics of $T_{h}$ applied to $x^{h}_{0}\equiv B(x^{l}_{0})$, the image of $x^{l}_{0}$ under $B$.

The DSR condition guarantees that some element of the low level-model, prescribed by some function of the low-level state determined by the bridge map $B(x^{l})$, behaves approximately in the same manner as a state in the high-level model. As we will see shortly, the requirement for DS reduction will be formulated in terms of the existence of a bridge map satisfying the DSR condition; however, in order to avoid trivialising counterexamples, the DSR condition must be supplemented with certain additional constraints on the bridge map $B(x^{l})$. The first of these constraints is that the bridge map not depend explicitly on the time $t$. Without this constraint, the DSR condition would be trivial insofar as it would be satisfied between any two models for which the cardinality of the low-level model was greater than or the same as that of the high-level theory; it is straightforward to see that one could simply absorb any differences of dynamical structure into the bridge map itself. As I explain now, a further constraint, pertaining to the symmetries of the two models, also must be imposed on the bridge map.

\subsubsection{Reducing Symmetries} \label{DSRSymmetries}

Beyond time-independence of the bridge map, it is also necessary that the bridge map be compatible with the dynamical symmetries of the high-level model in a sense that I now elaborate. For any dynamical symmetry $T_{h}$ of the high-level model and any $x^{l}_{0} \in d \subset S_{l}$ such that $T_{h}(B(x^{l}_{0})) \in B(d)$ (where recall that $d$ is the domain for which the DSR condition holds approximately, and $B(d)$ is its image under $B$), there should exist some symmetry $T_{l}$ of the low-level model such that

\begin{equation} \label{DSRSymm}
T_{h}(B(x^{l}_{0}) )\approx B(T_{l}(x^{l}_{0})).
\end{equation}

\noindent The rationale for imposing this condition is that it serves to ensure that not only the trajectory $D_{h}(t;x_{0}^{h})$, but also its image  $T_{h}(D_{h}(t;x_{0}^{h}))$ under the symmetry $T_{h}$, is reduced by some solution to the low-level model - which is to say, there exists some solution of the low-level model whose image under $B$ approximates the transformed high-level trajectory - so long as the transformed high-level trajectory remains in the image domain $B(d)$. Note that if the DSR condition is satisfied for some high-level solution, so that  $B(D_{l}(t;x^{l}_{0})) \approx D_{h}(t;B(x_{0}^{l}))$, and the condition (\ref{DSRSymm}) holds, then the solution $T_{h}(D_{h}(t;x^{'h}_{0})) = D_{h}(t;T_{h}(x^{'h}_{0}))$ of the high-level dynamics also satisfies the DSR condition in the form $B(D_{l}(t;T_{l}(x_{0}^{l}))) \approx D_{h}(t;B(T_{l}(x_{0}^{l})))$. This can be seen as follows:

\begin{align}
& D_{h}(t;B(x_{0}^{l})) \approx B(D_{l}(t;x^{l}_{0})) \\
& T_{h}(D_{h}(t; B(x_{0}^{l})) \approx T_{h}(B(D_{l}(t;x^{l}_{0}))) \  \ [\textit{Apply $T_{h}$ to both sides}]   \\
& D_{h}(t;T_{h}(B(x_{0}^{l})) \approx   B(T_{l}(D_{l}(t;x^{l}_{0}))) \ \  [\textit{Use $T_{h} ( D_{h}(t;x_{0}^{h}) = D_{h}(T_{h}(x_{0}^{h}))$ and (\ref{DSRSymm})}]   \\
& D_{h}(t;B(T_{l}(x_{0}^{l})) \approx   B(D_{l}(t;T_{l}(x^{l}_{0})))  \ \ [\textit{Use (\ref{DSRSymm}) and $T_{h} ( D_{h}(t;x_{0}^{h}) = D_{h}(T_{h}(x_{0}^{h}))$}].   
\end{align}

\noindent Thus, if condition (\ref{DSRSymm}) is satisfied, then if a high-level trajectory is approximated by the image under $B$ of some trajectory in the low-level model, it follows the transformation of that high-level trajectory under a symmetry of the high-level model will be approximated by the image under $B$ of some other trajectory in the low-level model.  

However, we should only demand that the transformed trajectory $T_{h}(D_{h}(t;x^{'h}_{0}))$ be approximated by the image of some trajectory in the low-level model if the symmetry tranformation $T_{h}$ does not carry the trajectory - or some chosen segment of it - outside of the image domain $B(d)$, or outside the domain of applicability of the high-level model. For instance, Galilean symmetries of Newtonian models include boosts by velocities with magnitude greater than the speed of light; we should not insist that the high-level trajectories obtained under this symmetry transformation be approximated by the image of some trajectory in a given low-level model (e.g. a model of SR) for the simple reason that these transformed trajectories are unphysical - that is, they do not describe real physical systems since they are outside the high-level model's domain of applicability. In our reductions, we should only insist that those parts of a high-level model that serve to describe real physical systems be approximated by some low-level model (though it may sometimes be the case that a low-level model does nevertheless reduce even the unphysical parts of the high-level model, as is the case between nonrelativistic CM and nonrelativistic QM, which both incorporate Galilean boosts of arbitrarily high velocity). 

I assume here that, for some sufficiently constrained class $C$ of physical systems, the domain of applicability of $M_{h}$ to $C$ is circumscribed by the image $B(d)$ of $M_{h}$'s domain $d$ in $M_{l}$, where $M_{l}$ is a low-level model that also serves to describe $C$. As discussed in the Introduction, I presume throughout this thesis a convergence of successive models to the truth, proceeding on the expectation that lower-level models will indeed turn out to represent a strictly more accurate approximation to reality - i.e. to the class $C$ of physical systems in question - than do their high-level counterparts and that, as a consequence, the domain of applicability of some $M_{h}$ to some $C$ is circumscribed by any $M_{l}$ that also describes $C$.

In addition to criterion $(\ref{DSRSymm})$, we should require that the group structure characterising the action of the symmetries of the high-level model be approximated, within $B(d)$, by the group structure induced through the bridge map by the group structure characterising the symmetries of the low-level model. That is, we should require that if

\begin{equation}
T_{h}^{1}(B(x^{l}) )\approx B(T_{l}^{1}(x^{l})) \ \text{for all} \ x^{l} \in d \  \text{such that} \ T_{l}^{1}(x^{l}) \in d
\end{equation}

\noindent and 

\begin{equation}
T_{h}^{2}(B(x^{l}) )\approx B(T_{l}^{2}(x^{l})) \ \text{for all} \ x^{l} \in d \ \text{such that} \ T_{l}^{2}(x^{l}) \in d, 
\end{equation}

\noindent and $T_{l}^{1} \circ T_{l}^{2} (x^{l}) \in d$, and $T_{h}^{1} \circ T_{h}^{2} (B(x^{l})) \in B(d)$ , then

\begin{equation}
T_{h}^{1} \circ T_{h}^{2} (B(x^{l})) \approx B(T_{l}^{1} \circ T_{l}^{2} (x^{l})).
\end{equation}

\noindent Thus, the bridge map can be regarded as an approximately structure-preserving function between the state spaces of the two models, where the preserved structure is associated not only with the dynamics of the models but also with their dynamical symmetries.

\subsubsection{Formal Criteria for DS Reduction} \label{DSRReduction}

Having made these motivating remarks, we are now in a position to state formal conditions for dynamical systems reduction:

\

\noindent \underbar{\textbf{DS Reduction:}}
\begin{quotation}
A model $M_{h}$=($S_{h}$, $D_{h}$) of $T_{h}$ describing some class $C$ of physical systems reduces over time scale $\tau$ and to within margin of error $\delta$ to a model $M_{l}$ =($S_{l}$, $D_{l}$) of $T_{l}$ also describing $C$ only if there exists differentiable function $B: S_{l} \rightarrow S_{h}$ that does not depend explicitly on time, and a nonempty subset $d \subset S_{l}$, such that   

\begin{enumerate}

\item for any $x^{l}_{0} \in d$
\begin{equation}
\bigg|  B \big(D_{l}(t;x^{l}_{0})\big) - D_{h}\big(t;B(x^{l}_{0})\big) \bigg|_{h} < \delta,
\end{equation}
for all $0\leq t \leq \tau$;

\item 
\begin{enumerate}
\item for every dynamical symmetry $T_{h}$ of $M_{h}$ and for every $x^{h} \in B(d)$ such that $T_{h}(x^{h}) \in B(d)$, there exists a dynamical symmetry $T_{l}$ of $M_{l}$ and an $x^{l} \in d$, such that $x^{h}=B(x^{l})$  and

\begin{equation}
T_{h}(B(x^{l}) )\approx B(T_{l}(x^{l}));
\end{equation}

\item if

\begin{equation}
T_{h}^{1}(B(x^{l}) )\approx B(T_{l}^{1}(x^{l})) \ \text{for all} \ x^{l} \in d \  \text{such that} \ T_{l}^{1}(x^{l}) \in d,
\end{equation}

\noindent and

\begin{equation}
T_{h}^{2}(B(x^{l}) )\approx B(T_{l}^{2}(x^{l})) \ \text{for all} \ x^{l} \in d \ \text{such that} \ T_{l}^{2}(x^{l}) \in d, 
\end{equation}

\noindent and $T_{l}^{1} \circ T_{l}^{2} (x^{l}) \in d$, and $T_{h}^{1} \circ T_{h}^{2} (B(x^{l})) \in B(d)$, then

\begin{equation}
T_{h}^{1} \circ T_{h}^{2} (B(x^{l})) \approx B(T_{l}^{1} \circ T_{l}^{2} (x^{l})).
\end{equation}

\end{enumerate}

\end{enumerate} 

\end{quotation}
 
\noindent These conditions should be understood as necessary conditions for one dynamical system to reduce to another. Whether they are sufficient depends on the possibility of finding trivialising counterexamples - i.e., examples such that for any two DS models for which the cardinality of $S_{l}$ is higher than that of $S_{h}$, one can find a bridge map $B$ satisfying the specified conditions. If such examples can be found, then further conditions must be imposed on the bridge map $B$. What the above conditions are meant to capture are two of the most salient requirements that must be satisfied for a mathematical structure defined in a low-level model - specified by the bridge map - to emulate, or approximately instantiate, the dynamical behavior and other physically salient aspects of the state in the high-level model. I leave it to future work to ascertain whether any further conditions need be placed on the bridge map, and if so, what these conditions are.  

In the simple example that I consider later in this chapter, I demonstrate that condition 1) is satisfied and show that 2) is satisfied for two particular symmetries of the high-level model, leaving it to the reader to extrapolate how the other symmetries of the high-level model are to be reduced. In the more involved reductions considered in the body chapters of this thesis, I focus on demonstrating condition 1), leaving it to future work to demonstrate the validity of condition 2); nevertheless, it is not difficult to surmise in many of these cases how the demonstration of condition 2) for the various symmetries of the high-level model should go.

%Intuitively, then, DS reduction offers a number of criteria for a low-level model of a physical system to \textit{instantiate} some high-level model of that same system - that is to say, for elements of the low-level model to fill the role of various elements in the high-level model.   For instance, the image domain $B(d)$ instantiates the high-level state space $S_{h}$, or rather that subset corresponding to states of the high-level model that can realistically be used to describe the system in question. The induced dynamics $B(D_{l}(t;x_{0}^{l})$ instantiates the high level dynamics $D_{h}(t; B(x_{0}^{l})$. For appropriate low-level symmetries $T_{l}$, the induced transformation $B(T_{l}(x_{0}^{l})$ instantiates some high-level symmetry $T_{h}(B(x_{0}^{l}))$

\subsubsection{Reduction v. Mathematical Analogy}

The requirement of DS reduction that both models describe the same physical system - or, more generally, class $C$ of physical systems - is included in the formal requirements for DS reduction so as to rule out pairs of models that are mathematically similar in structure but in which the success of one at describing some physical system cannot be reasonably regarded as accounting for the success of the other, since the two models are used to describe completely different physical systems. 
For example, the Schrodinger equation for a free massive particle takes the form of a diffusion equation with imaginary coefficient; as a result there exists a direct mapping between between models of, say, heat diffusion in three dimensions and the quantum mechanical model of a single free particle. Yet it is clear that we would not want to say that the theory of heat diffusion in three dimensions serves in any respect to explain the success of the Schrodinger equation in modelling the behavior of free (low-energy) particles; the parallels between the models simply provide a case in which similar mathematical structures happen to be applicable in distinct physical contexts. 
%For example, there exist strong mathematical analogies between certain models of classical statistical mechanics and certain models of quantum field theory (see, for instance, \cite{peskin1996introduction}, chs. 8 and 9). While it may be possible to find a map between the state spaces of the two theories satisfying requirements 1) and 2), it would not be reasonable to conclude on these grounds that the success of, say, classical statistical mechanics \textit{accounts} in any way the success of the quantum field theory (in the way that, say, that the success of SR can account for the success of NM), especially since the relation between them involves a so-called `Wick Rotation' from a real-valued to an imaginary-valued time parameter. 
Thus, one must distinguish between mere mathematical analogy, which occurs when the same or related mathematical structures happen to be applicable in different physical contexts, and reduction, in which similarities of mathematical structure serve to account for the fact that two distinct models can be successfully employed in describing the \text{same} physical system.

\subsubsection{A Note on the Question of Relativistic Covariance}

Note that the DSR Condition assumes a common time parameter for the high- and low- level models. Moreover, in making this choice of time parameter, the DS account of reduction requires that the manifest covariance of any relativistic models involved in the reduction be sacrificed. Note, however, that the reference to a common time parameter in the DSR condition is incompatible only with models formulated in a \textit{manifestly} relativistically covariant fashion, but not with models that are covariant. Thus, the reference to a particular time parameter in the DSR condition does \textit{not} preclude the inclusion of Lorentz-invariant models within this frameowork. When I turn to relativistic theories in Chapters 4 and 5, my focus will be on Hamiltonian and Schrodinger picture formulations of the models that I consider, which, although not manifestly covariant, are covariant.

\subsubsection{DS Reduction and Laws of Motion}

It is often more convenient to specify the dynamics of a DS model in the form of first-order differential equations, rather than in the form of a dynamical map. Let us examine how the DSR condition should be formulated when the dynamics of the high- and low- level models are prescribed in this way. As discussed in section \ref{Laws}, the dynamical map of $M_{h}$ specifies the solutions $x^{h}(t) = D_{h}(t; x^{h}_{0})$ to the differential equation

\begin{equation} \label{HighLevelDynamics}
\frac{d x^{h}}{dt} = f_{h}(x^{h},t)
\end{equation}

\noindent where $f_{h}(x^{h},t) = \frac{\partial }{\partial t} D_{h}(t; x_{0}^{h}) \big|_{t=0, x_{0}^{h} = x^{h} }$,  and likewise the dynamical map of $M_{l}$ specifies the solutions $x^{l}(t) = D_{l}(t; x^{l}_{0})$ to the differential equation

\begin{equation}
\frac{d x^{l}}{dt} = f_{l}(x^{l},t),
\end{equation}

\noindent  where $f_{l}(x^{l},t) = \frac{\partial }{\partial t} D_{l}(t; x_{0}^{l}) \big|_{t=0, x_{0}^{l} = x^{l} }$.

At the level of differential equations, DSR condition will be satisfied if the induced trajectory $x'^{h}(t) \equiv B_{l}^{h}(x^{l}(t))$ approximately satisfies the differential equations of $T_{h}$:

\begin{equation} \label{dDSRAnalogue}
\frac{d x'^{h}}{dt} \approx f_{h}(x'^{h},t)
\end{equation}

\noindent or, more explicitly, if 

\begin{equation} \label{dDSR1}
\frac{d}{dt} B_{l}^{h}\big(x^{l}(t)\big) \approx f_{h}\bigg(B_{l}^{h}\big(x^{l}(t)\big), t \bigg).
\end{equation}

\noindent Note that while this relation is a sufficient condition for the DSR condition to hold, it is not a necessary condition. We can see that it is a sufficient condition by integrating both sides of (\ref{dDSR1}) with respect to time:  

\begin{equation}
\begin{split}
& \int_{0}^{t} dt' \  \frac{d}{dt'} B_{l}^{h}(x^{l}(t')) \approx   \int_{0}^{t} dt'  \ f_{h}\bigg(B_{l}^{h}\big(x_{l}(t')\big), t' \bigg) \\
& B_{l}^{h}(x^{l}(t)) - B_{l}^{h}(x^{l}_{0})\approx   D_{h}(t; B_{l}^{h}(x^{l}_{0})) -  D_{h}(0; B_{l}^{h}(x^{l}_{0}))   \\
& B_{l}^{h}(D_{l}(t;x^{l}_{0})) \approx  D_{h}(t;B_{l}^{h}(x^{l}_{0})) 
\end{split}
\end{equation}

\noindent where in going from the first line to the second line I have used that $f_{h}(x^{h},t) = \frac{\partial }{\partial t} D_{h}(t; x_{0}^{h}) \big|_{t=0, x_{0}^{h} = x^{h} }$, and in going from the second to the third I have used that $ B_{l}^{h}(x^{l}_{0})=D_{h}(0; B_{l}^{h}(x^{l}_{0}))$. Note that for the condition $(\ref{dDSR1})$ to be sustained over some time period $\tau$, the domain $d$ should be such that the image dynamics roughly preserve the set $d$; that is, that they map states in $d$ to other states in $d$, at least on the timescale $\tau$.

While  (\ref{dDSR1}) is sufficient for the DSR condition to hold, it is not necessary insofar as there may exist induced trajectories on the high-level state space that remain close (in the sense of the $S_{h}$'s norm) to the trajectory prescribed by the high-level model but such that the time derivative of these trajectories does not remain close in value to the derivatives prescribed by ($\ref{HighLevelDynamics}$). For example, consider a trajectory rapidly oscillating with small amplitude around the trajectory prescribed by the high-level dynamics; the values of the states will be close, so that the DSR condition is satisfied, but the time derivatives will differ drastically so that  (\ref{dDSR1}), or alternatively, (\ref{dDSRAnalogue}), is not. In all reductions considered in later chapters, the stronger condition (\ref{dDSR1}) will be proven, rather than the condition (\ref{DSRApprox}).

\vspace{5mm}

%As with the condition 1., in cases where the relevant symmetry $T_{h}$ is continuous and parametrised by some parameter $\alpha$, to prove the conditions 2. and 3. it suffices to prove the approximate validity of differential equations, namely,

%\begin{equation}
%\frac{d}{d \alpha} B(x^{l}(\alpha)) \approx g_{h}(x_{h},t)
%\end{equation}

%\noindent where x^

%\begin{equation}

\subsubsection{A Simple Example of DS Reduction: Classical Mechanics and Quantum Mechanics (w/o Environmental Decoherence)}

\

\noindent \underbar{\textit{Condition 1: Dynamics}}

\

To continue with the models of a single spinless particle in the CM/QM reduction, note that the DSR condition in differential form requires, in the case $H = \frac{p^{2}}{2m} + V(x)$, $\hat{H} = \frac{\hat{p}^{2}}{2m} + V(\hat{x})$, that

\begin{equation}
\begin{split}
& \frac{d}{dt} \langle \hat{x} \rangle \approx  \big\{ \langle x \rangle, H \big( \langle \hat{x} \rangle, \langle \hat{p} \rangle \big) \big\}_{\langle \hat{x} \rangle,\langle \hat{p} \rangle} = \frac{1}{m} \langle \hat{p} \rangle \\
& \frac{d}{dt} \langle \hat{p} \rangle \approx  \big\{ \langle p \rangle, H\big( \langle \hat{x} \rangle, \langle \hat{p} \rangle \big) \big\}_{\langle \hat{x} \rangle,\langle \hat{p} \rangle} = - \frac{\partial V(\langle \hat{x}) \rangle }{\partial \langle \hat{x} \rangle},
\end{split}
\end{equation}

\noindent where the subscript $\langle \hat{x} \rangle$, $\langle \hat{p} \rangle$ on the Poisson brackets indicates differentiation with respect to $ \langle \hat{x} \rangle$ and $\langle \hat{p} \rangle$, rather than with respect to $x$ and $p$. Employing the bridge rule substitutions $x' \equiv \langle \hat{x} \rangle$, $p' \equiv \langle \hat{p} \rangle$, these can be written in a form more reminiscent of the original classical equations that they serve to reduce: 

\begin{equation}
\begin{split}
& \frac{d x'}{dt} \approx  \big\{x', H \big( x', p' )\big\}_{x',p'} = \frac{1}{m} p' \\
& \frac{d p' }{dt} \approx  \big\{ p' , H\big(x', p' \big) \big\}_{x',p'} = - \frac{\partial V(x') }{\partial x'},
\end{split}
\end{equation}

\noindent It can be shown using Ehrenfest's Theorem that these approximate equations, representing the differential DSR condition as applied this pair of models, hold within the domain $d_{CM}$ of states consisting of wave packets that are simultaneously narrow in both position and momentum (to within the constraints of the Uncertainty Principle). Ehrenfest's Theorem states that, for any state of a quantum system with the above-specified Hamiltonian, the following relation holds:

\begin{equation}
\frac{d\langle \hat{p} \rangle}{dt} = -\left\langle \hat{ \frac{\partial V}{\partial x }}  \right\rangle 
\end{equation}

\noindent (see, for instance, \cite{merzbacherQM}, p. 41, or almost any other graduate level quantum mechanics text, for a proof of this theorem). Note that this does not suffice to ensure that expectation values of position and momentum evolve approximately according to Newtonian equations. For this, it is necessary that the stronger condition, 

\begin{equation} \label{EhrenfestNarrow}
\frac{d\langle \hat{p} \rangle}{dt} \approx  - \frac{\partial V(\langle \hat{x} \rangle) }{\partial \langle \hat{x} \rangle  }
\end{equation}

\noindent hold. It is well-known that, as a result of Ehrenfest's Theorem, this condition holds approximately for states that are narrowly peaked in position and momentum. Thus, for the domain of states consisting of narrow wave packets, the DSR condition between the high- and low-level models is approximately satisfied. However, the timescale on which this is so will typically be restricted by the timescale on which wave packets tend to spread under the dynamics, as they tend to do in generic situations.

The relation (\ref{EhrenfestNarrow}) suffices to ensure the validity of condition 1) for DSR reduction, which in this particular case takes the form, 

\begin{equation}
\begin{split}
& \bigg| \langle \psi_{0} | e^{i\hat{H} t}  \hat{x} \ e^{-i\hat{H} t} | \psi_{0} \rangle -  e^{ \{ \circ, H  \} t } x \big|_{\langle \hat{x} \rangle_{0} , \langle \hat{p} \rangle_{0}} \bigg| < \delta_{x}, \\
& \\
& \text{and} \\
& \\
& \bigg| \langle \psi_{0} | e^{i\hat{H} t}  \hat{p} \ e^{-i\hat{H} t} | \psi_{0} \rangle -  e^{ \{ \circ, H  \} t } p \big|_{\langle \hat{x} \rangle_{0} , \langle \hat{p} \rangle_{0}} \bigg| < \delta_{p},
\end{split}
\end{equation}

\noindent where $\langle \hat{x} \rangle_{0} \equiv \langle \psi_{0} |  \hat{x} | \psi_{0} \rangle$ and  $\langle \hat{p} \rangle_{0} \equiv \langle \psi_{0} |  \hat{p} | \psi_{0} \rangle$, for $0 \leq t \leq \tau$, where $\tau$ is timescale on which wave packets become widely spread out on spatial dimensions characteristic of the variation of the potential $V(x)$ (for a more precise characterisation of this length scale, see for instance \cite{allori2002seven}). The norm employed on phase space is simply the difference of the positions and of the momenta. Less formally, we can write this condition as 

\begin{equation} 
\begin{split}
& \langle \psi_{0} | e^{i\hat{H} t}  \hat{x} \ e^{-i\hat{H} t} | \psi_{0} \rangle \approx   e^{ \{ \circ, H  \} t } x \big|_{\langle \hat{x} \rangle_{0} , \langle \hat{p} \rangle_{0}}  \\
& \\
& \text{and} \\
& \\
& \langle \psi_{0} | e^{i\hat{H} t}  \hat{p} \ e^{-i\hat{H} t} | \psi_{0} \rangle \approx   e^{ \{ \circ, H  \} t } p \big|_{\langle \hat{x} \rangle_{0} , \langle \hat{p} \rangle_{0}},  
\end{split}
\end{equation}

\noindent where, again, the approximation should be understood as being relative to some specified margins of error $\delta_{x}$ and $\delta_{p}$.

\vspace{5mm}

\

\noindent \underbar{\textit{Condition 2: Symmetries}}

\

I demonstrate the validity of condition 2), concerning the relation between the symmetries of the models, with regard to rotations and Galilean boosts in classical mechanics. In principle, these conditions should be shown to hold for all symmetries and states of the high-level model such that both the states and their mappings under the symmetry are in the image domain $B(d)$, which here consists of the entire classical phase space $\Gamma$. While I limit myself here to considering these two symmetries, following these examples it should be straightforward for the reader to demonstrate these conditions for other symmetries of the given classical model.  

\vspace{5mm}

\noindent \textit{Symmetry 1: Rotation}

In the case of a Hamiltonian system with spherically symmetric potential $V(r)$, the rotations about the origin constitute a group of dynamical symmetries.  Condition 2a) for rotations is ensured by the fact that

\begin{equation}
\left( \langle \psi |e^{i \vec{\hat{L}}\cdot \hat{n} \theta } \    \hat{x}  \     e^{-i \vec{\hat{L}}\cdot \hat{n} \theta }     | \psi  \rangle,  \langle \psi |e^{i \vec{\hat{L}}\cdot \hat{n} \theta } \    \hat{p}  \     e^{-i \vec{\hat{L}}\cdot \hat{n} \theta }     | \psi  \rangle     \right)  \approx  e^{ \{ \circ, \vec{L}\cdot \hat{n}  \theta \} }   \left(  \langle \psi | \hat{x} | \psi \rangle, \langle \psi|  \hat{p} | \psi \rangle \right)
\end{equation}

\noindent for $| \psi \rangle \in d$. This condition is satisfied as a consequence both of the Baker-Hausdorff Lemma, which states that

\begin{align} \label{BakerHausdorff}
 e^{i\lambda \hat{B}} \hat{A} e^{-i \lambda \hat{B}} &= \hat{A} + i\lambda  \left[ \hat{B}, \hat{A} \right] + \frac{( i\lambda )^{2}}{2!}  \left[ \hat{B},  \left[ \hat{B}, \hat{A} \right] \right] + \frac{( i\lambda )^{3}}{3!}  \left[ \hat{B}, \left[ \hat{B},  \left[ \hat{B}, \hat{A} \right] \right] \right] + ... \\
& \equiv e^{\left[  i \lambda \hat{B} , \circ \right]} \hat{A}
\end{align} 

\noindent (see, for instance \cite{sakurai1995modern}, p.96) and of the result that for narrow wave packet states $|  \psi_{q',p'} \rangle$

\begin{equation} \label{CommutatorPB}
\langle \psi_{q',p'} |  \left[ f(\hat{x},\hat{p})    ,  g(\hat{x},\hat{p})   \right]  |  \psi_{q',p'} \rangle \approx i \left\{ f(x,p) , g(x,p)  \right\} \big|_{q',p'}
\end{equation}

\noindent where $f(x,p)$ and $g(x,p)$ are the unique classical functions associated with the quantum operators $f(\hat{x},\hat{p})$ and $g(\hat{x},\hat{p})$ and do not vary significantly on scales of action equal to $\hbar$; this can be derived through fairly extensive manipulation of the canonical commutation relation $[\hat{x},\hat{p}] = i$. Note that this relation makes explicit the physical correspondence between Poisson brackets and commutators. Whereas Dirac originally postulated the correspondence on the basis of the algebraic similarities between the two brackets, rather than on the asummption that one structure physically underwrites the other, the DSR condition serves to illustrate the physical basis for this formal correspondence.

Condition 2 b) for rotations takes the form

\begin{equation}
\begin{split}
& \left( \langle  \psi |       e^{i \vec{\hat{L}}\cdot \hat{m} \phi }  e^{i \vec{\hat{L}}\cdot \hat{n} \theta }  \   \hat{x}  \ e^{-i \vec{\hat{L}}\cdot \hat{n} \theta }  e^{-i \vec{\hat{L}}\cdot \hat{m} \phi }  | \psi \rangle, \langle  \psi |       e^{i \vec{\hat{L}}\cdot \hat{m} \phi }  e^{i \vec{\hat{L}}\cdot \hat{n} \theta }  \   \hat{p}  \ e^{-i \vec{\hat{L}}\cdot \hat{n} \theta }  e^{-i \vec{\hat{L}}\cdot \hat{m} \phi }  | \psi \rangle \right)    \\
& \approx   e^{ \{ \circ, \vec{L}\cdot \hat{n}  \theta \} }   e^{ \{ \circ, \vec{L}\cdot \hat{m}  \phi \} } (\langle \psi | \hat{x} | \psi \rangle, \langle \psi|  \hat{p} | \psi \rangle)
\end{split}
\end{equation}

\noindent  and is likewise satisfied for $ | \psi \rangle $ a narrow wave packet state. Again, this result follows as a consequence of (\ref{BakerHausdorff}) and (\ref{CommutatorPB})

\vspace{5mm}

\noindent \textit{Symmetry 2: Galilean Boosts}

The dynamical map associated with a one-particle classical Hamiltonian $H = \frac{p^{2}}{2m} + V(x)$ above will not generally commute with a boost by some velocity $v$, which therefore will not serve as a dynamical symmetry of the model. However, if we consider the two-particle case in which the potential depends only on the spatial distance between the particles, so that $H = \frac{p_{1}^{2}}{2m_{1}} + \frac{p_{1}^{2}}{2m_{1}} + V(|x_{1}-x_{2}|)$, then a boost of both particles by the same velocity $v$ will commute with the dynamical map associated with this Hamiltonian. Thus, a Galilean boost in this case will count as a symmetry of the model. A Galilean boost by velocity $v$ takes the form 

\begin{align}
& x'_{1}=x_{1}-vt \\
& x'_{2}=x_{2}-vt \\
& p'_{1}=p_{1}-m_{1}v \\
& p'_{2}=p_{2}-m_{2}v.
\end{align}

\noindent In the quantum mechanical model, there is likewise a symmetry of the dynamics that typically also goes under the name of a Galilean boost. As in the classical model, these transformations are parametrised by a velocity $v$; under such a transformation, the wave function $\psi(x_{1},x_{2},t)$ transforms to $\psi'(x_{1}',x'_{2},t')$, given by

\begin{equation}
\psi'(x_{1}',x'_{2},t') = e^{-i(m_{1} v \cdot x_{1}+m_{2} v \cdot x_{2} -  \frac{1}{2}m_{1} v^{2} t-\frac{1}{2}m_{2} v^{2} t)} \psi(x_{1},x_{2},t)
\end{equation}

\noindent with $x'_{1}=x_{1}-vt$, $x'_{2}=x_{2}-vt$ and $t'=t$ (see, for instance \cite{merzbacherQM}, p.75). It is straightforward to see that under the bridge map given by the expectation value, the quantum mechanical Galiliean boost induces a classical Galilean boost: 

\begin{align}
&\left( \langle \psi' |  \hat{x}_{1}  | \psi' \rangle,   \langle \psi' |  \hat{x}_{2}  | \psi' \rangle;  \langle \psi' |  \hat{p}_{1}  | \psi' \rangle, \langle \psi' |  \hat{p}_{2}  | \psi' \rangle  \right) \\
&= \left( \langle \psi |  \hat{x}_{1}  | \psi \rangle - v t , \langle \psi |  \hat{x}_{2}  | \psi \rangle - v t ;   \langle \psi |  \hat{p}_{1}  | \psi  \rangle - m_{1}v,  \langle \psi |  \hat{p}_{2}  | \psi  \rangle - m_{2}v  \right),
\end{align}

\noindent thereby satisfying condition 2a). Thus, for any Galilean boost on phase space, there exists a corresponding transformation on Hilbert space that induces it via the expectation value. To satisfy condition 2b), though, it is necessary that the composition of two Galilean boosts on phase space , by $v$ and then by $v'$, agree approximately with the transformation induced under the bridge map by the composition of the corresponding boosts on Hilbert space. The composition of two boosts on Hilbert space gives

\begin{equation}
\psi''(x''_{1}, x''_{2},t'') = e^{-i\left[m_{1} \left(v+v'\right)  \cdot x_{1} + m_{2} \left(v+v'\right)  \cdot x_{2} - \frac{1}{2} m_{1}\left( v^{2} + v'^{2}\right)t - \frac{1}{2} m_{2}\left( v^{2} + v'^{2} \right) t \right]}\psi(x_{1},x_{2},t)
\end{equation}

\noindent with $x''_{i}=x_{i}- (v+v')t$ for $i=1,2$ and $t''=t$. Note that this is equal to a single boost by $v+v'$ up to a global time-dependent phase factor $( m_{1} + m_{2} ) (v \cdot v')  t$, which does not make a difference to any of the amplitudes of the theory, or to rays in projective Hilbert space. Under the composed boosts, it is straightforward to see that

\begin{align}
&  \left( \langle \psi'' |  \hat{x}_{1}  | \psi'' \rangle,   \langle \psi'' |  \hat{x}_{1}  | \psi'' \rangle; \langle \psi'' |  \hat{p}_{1}  | \psi''  \rangle,  \langle \psi'' |  \hat{p}_{2}  | \psi''  \rangle \right) \\
& =    \left( \langle \psi |  \hat{x}_{1}  | \psi \rangle - \left( v+v' \right) t ,  \langle \psi |  \hat{x}_{2}  | \psi \rangle - \left( v+v' \right) t  ; \langle \psi |  \hat{p}_{1}  | \psi  \rangle - m_{1} \left(v+v' \right), \langle \psi |  \hat{p}_{2}  | \psi  \rangle - m_{2} \left(v+v' \right)  \right)
\end{align}

\noindent thereby ensuring the validity of condition 2b) with respect to classical Galilean symmetry.

\vspace{10mm}

\noindent \underbar{\textit{Limitations of the Simple Quantum Model}}

Note that the quantum models to which the classical models considered so far have been reduced make no mention of environmental decoherence, and thus allow for arbitrary coherent superpositions of the degrees of freedom in question. Moreover, in chaotic systems, the quantum models predict that initially narrow wave packets will spread on fairly short time scales beyond the coherence lengths that typically characterise the macroscopic or mesoscopic systems that exhibit approximately Newtonian behavior (see \cite{wallace2012emergent} Ch.3 for detailed discussion of this point). Thus, although the classical model considered here may serve as an effective (if only approximate) description of such systems, the quantum model does not insofar as it will, on relatively short timescales, predict coherence lengths that disagree dramatically with those observed in these systems. Thus, it is necessary to replace the quantum model considered here with a more sophisticated one that takes account of environmental degrees of freedom and thereby continually suppresses the coherence length of the system in question; this is the goal of the next chapter. Nevertheless, the reduction involving the quantum model without environment helps to provide a simplified illustration the basic components of DS reduction, if we momentarily allow ourselves to overlook its shortcomings as a description of real, approximately Newtonian systems.  

%\noindent \underbar{\textit{Limitations of the QM Model w/o Environment}}

%While the classical model discussed here serves as an effective description of many macroscopic classical systems, the quantum model that I have been discussing in the context of DS reduction does not suffice as a description of these same systems. The primary reason is that any realistic quantum description of such macroscopic systems must incorporate the influence of the environmental degrees of freedom that invariably interact with such systems; while these interactions can be ignored in the classical description of these systems in cases where friction/dissipation is negligible, the prevalence of entanglement in quantum models entails that even a single microscopic degree of freedom in the environment (for example, a high frequency photon) will have a dramatic effect on the amplitudes associated with the macroscopic system when the two interact. So, although the example above does provide an instance of DS reduction between models, the low-level quantum model in this case is limited in its capacity to describe general macroscopic systems insofar as it does not incorporate the potential effects of environmental degrees of freedom. 

\subsection{Transitivity of DS reduction}

Under certain conditions, if a model $M_{1}$ reduces to another model $M_{2}$ and $M_{2}$ reduces to $M_{3}$, then it will be true that $M_{1}$ reduces to $M_{3}$. Specifically, the domain $d_{1} \subset S_{2}$ associated with the bridge map $B_{2}^{1}: S_{2} \rightarrow S_{1}$ must be in the image of the domain $d_{2} \subset S_{3}$ under the bridge map $B_{3}^{2}: S_{3} \rightarrow S_{2}$. If this is the case, then $M_{1}$ and $M_{3}$ satisfy the DSR conditions with bridge map $B_{3}^{1} : S_{3} \rightarrow S_{1}$ equal to the composition of the bridge maps $B_{2}^{1}$ and $B_{3}^{2}$, so that $B_{3}^{1} \equiv B_{2}^{1} \circ B_{3}^{2} $. The domain in $S_{3}$ associated with the bridge map  $B_{3}^{1}$ is equal to $ d_{2} \cap (B_{3}^{2})^{-1}(d_{1})$, the intersection of $d_{2}$ and the inverse image under $B_{3}^{2}$ of $d_{1}$. The timescale associated with the reduction will be the smaller of the timescales associated with the two component reductions. 

I will prove transitivity for DSR condition 1), in differential form. The proof of transitivty of condition 2) can be carried out in similar fashion and so is omitted. Given that the two component reductions hold, we have

\begin{equation} \label{TopReduction}
\frac{d}{dt} B_{2}^{1}\big(x^{2}(t)\big) \approx f^{1}\big(B_{2}^{1}\big(x^{2}(t)\big), t \big),
\end{equation}

\noindent for some timescale $\tau_{1}$ and

\begin{equation}
\frac{d}{dt} B_{3}^{2}\big(x^{3}(t)\big) \approx f^{2}\big(B_{3}^{2}\big(x^{3}(t)\big), t \big)
\end{equation}

\noindent for some timescale $\tau_{2}$. Also, $x^{1}(t) = B_{2}^{1}(x^{2}(t))$, and $x^{2}(t) = B_{3}^{2}(x^{3}(t))$. We want to show that  

\begin{equation}  \label{Trans}
\frac{d}{dt} B_{3}^{1}\big(x^{3}(t)\big) \approx f^{1}\big(B_{3}^{1}\big(x^{3}(t)\big), t \big),
\end{equation}

\noindent for $x^{3}(t) \in d$ with $d \equiv  d_{2} \cap (B_{3}^{2})^{-1}(d_{1}) $, and $0 \leq t \leq \tau$ with $\tau = \min \{ \tau_{1}, \tau_{2} \}$, where $\tau_{1}$ is the timescale for the 1-to-2 reduction and $\tau_{2} $ the timescale for the 2-to-3 reduction. By the Chain Rule, we can expand the left-hand side of (\ref{Trans}) as follows

\begin{equation} \label{TransChain}
\frac{d}{dt} x^{1}\left(x^{2}\left(x^{3}(t)\right)\right) = \frac{d x^{1}}{d x^{2}} \frac{d x^{2}}{d x^{3}} \frac{d x^{3}}{dt} = \frac{d x^{1}}{d x^{2}} \frac{d x^{2}}{d x^{3}} f^{3}\left(x^{3}(t),t \right)
\end{equation}

\noindent where I have employed the dynamical equation for $M_{3}$, $\frac{d x^{3}}{dt} = f^{3}(x_{3},t)$. Separately, and also by the Chain Rule, the reduction of $M_{1}$ to $M_{2}$ entails

\begin{equation}
\frac{dx_{1}}{dx_{2}} \frac{ d x_{2}}{dt} \approx f^{1}\left(x^{1}\left(x^{2}(t)\right),t \right),
\end{equation}

\noindent which in turn entails

\begin{equation}
\frac{dx_{1}}{dx_{2}} f^{2}(x^{2},t) \approx f^{1}\left(x^{1}\left(x^{2}(t)\right),t\right).
\end{equation}

\noindent Likewise, the reduction of $M_{1}$ to $M_{2}$ entails

\begin{equation}
\frac{dx_{2}}{dx_{3}} f^{3}(x^{2},t) \approx f^{2}\left(x^{1}\left(x^{2}(t)\right),t\right).
\end{equation}

\noindent Putting these results together, eqn ( \ref{TransChain}) becomes

\begin{equation}
\frac{d}{dt} x^{1}\left(x^{2}\left(x^{3}(t)\right)\right) =  \frac{d x^{1}}{d x^{2}} \frac{d x^{2}}{d x^{3}} f^{3}\left(x^{3}(t),t \right) \approx \frac{d x^{1}}{d x^{2}}  f^{2}\left(x^{2}(t),t \right) \approx f^{1}\left(x^{1}\left(x^{2}(x^{3}(t)\right)\right),t).
\end{equation}

\noindent Alternatively, we can write this as 

\begin{equation}  \label{Trans}
\frac{d}{dt} B_{3}^{1}\big(x^{3}(t)\big) \approx f^{1}\big(B_{3}^{1}\big(x^{3}(t)\big), t \big),
\end{equation}

\noindent which is the result we wanted. Again, this result will only hold if \textit{both} of the component reductions hold, so the reduction timescale here is the minimum of $\tau_{1}$ and $\tau_{2}$. Moreover, it will also only hold if the point $x^{3}(t)$ remains in the domain $d_{2}$, so that the $2-3$ reduction holds and its image under $B_{3}^{2}$ is in the domain $d_{1}$, so that the $1-2$ reduction holds; thus, the $1-3$ reduction is only guaranteed to hold for $x_{3}(t) \in d_{2} \cap (B_{3}^{2})^{-1}(d_{1}) $.

\subsection{Generalising DS Reduction}

Dynamical systems reduction only applies in cases where $T_{l}$ and $T_{h}$ can be modelled as dynamical systems with a common time parameter. However, for theories that probe successively more fundamental levels of physical reality, it is likely this will not continue to be the case. A more general framework for reduction that encompasses DS reduction as a special case will need to be developed to accommodate these reductions. 

Wallace has suggested a concept of reduction in which theories are modelled in terms of whole histories rather than in terms of state space evolutions in dynamical systems. The latter sort of model can be easily subsumed into the former by treating the trajectories of the dynamical system as the histories, and the dynamics of the theory as a constraint on allowable histories. However, not all conceivable models which are formulated in terms of histories need be formulable as a dynamical system. For example, solutions to the Einstein field equations which are not globally hyperbolic may not possess a global description in terms of the time evolution of some state on some state space, but rather only in terms of histories with no globally definable time parameter. Wallace defines his concept of reduction as follows: 

\begin{quote}
\begin{singlespace}
Given two theories A and B, and some subset $D$ of the histories of A, we say the A instantiates B over domain $D$ iff there is some (relatively simple) map $\rho$ from the possible histories of A to those of B such that if some history h in $D$ satisfies the constraints of A, the $\rho(h)$ (approximately speaking) satisfies the constraints of B. (It will often be convenient to speak of the history h as instantiating $\rho(h)$, but this should be understood as shorthand for the more detailed definition here.) ...

This instantiation relation (I claim) is the right way of understanding the relation between different scientific theories - the sense in which one theory may be said to ``reduce" to another \cite{wallace2012emergent}.
\end{singlespace}
\end{quote}

\noindent While such a histories-based approach seems a promising generalisation of DS reduction, Wallace's definition does not specify any precise constraints on the map $\rho(h)$ (other than perhaps the implicit constraint that $\rho$ be sufficiently `simple'). Future elaborations of Wallace's notion of reduction should specify the necessary constraints on $\rho$.

\subsection{Precursors of the DS Approach} \label{OriginsPrecursors}
 
The central idea of DS reduction, that the high-level dynamics composed with some bridge map should yield approximately the same state as does the bridge map composed with the low-level dynamics over the same time-period - in short, that the dynamics should commute with the bridge map - is an old one. While its applications to physics have thus far been restricted primarily to the context of reductions in statistical mechanics - where the bridge map consists of some sort of coarse-graining function - I have argued above, and will continue to argue throughout the remainder of the thesis, that this insight applies much more broadly to reductions between any two theories whose models can be formulated as dynamical systems, as is the case with most current physical theories. As we have seen above, the DS approach develops this basic insight into a more formal and more general approach to reduction, supplementing it with further constraints on the bridge map, including time-independence and compatibility with the symmetries of the models in question. Thus, it represents a full-fledged alternative to the limit-based and Nagelian approaches that have dominated the literature on reduction in physics. In the present sub-section, I discuss the work of a number of authors that also addresses, with some variations, the core insight on which the DS approach is based, highlighting differences from the DS approach where they occur; it is worth noting here that none of these approaches imposes the additional condition requiring compatibility of the bridge map with the symmetries of the models, nor do any explicitly require the bridge map - or rather their counterpart to the bridge map - to be time-independent. To distinguish the general idea that dynamics should commute with some function between state spaces of the high- and low- level models from its formulation specifically within the context of DS reduction, I will refer to the general idea as the `dynamical commutation' condition, and to my own formulation of it as DSR condition 1) (I may occasionally refer to it also simply as the DSR condition). 

I was led to the basic idea of DS reduction in the context of my own research through discussions with with my doctoral thesis supervisor, David Wallace, who has for some time been advocating the dynamical commutation approach informally in discussion, in particular as regards the derivation in quantum mechanics of effective subsystem dynamics for density matrices from the dynamics a larger system; moreover, Wallace's \cite{wallace2012emergent}, Ch. 3 briefly proposes a generalisation, which I discuss in a later section, of the dynamical commutation condition to reductions of models formulated in the mathematical language of histories rather than of states evolving in time; this approach thus involves a map not between state spaces but rather between the history spaces of the models; however, Wallace does not impose any precise constraints on this map, as is necessary to avoid the condition being satisfied trivially. Finally, I also encountered a variant of the dynamical commutation condition in David Albert's Columbia University course on the foundations of statistical mechanics.

\vspace{5mm}

Both Giunti and Yoshimi have suggested their own variants of the dynamical commutation condition with regard to the reduction of dynamical systems generally, though the potential applications that concern them lie within philosophy of mind and in discussions of reduction in philosophy of science generally; they do not specifically discuss applications of this approach to reductions in physics, where (I claim) it is especially salient \cite {giunti2006emulation}, \cite {yoshimi2012supervenience}. Moreover, Giunti requires that his bridge map counterpart, which he calls an `emulation,' be an injective, or one-to one, function. As we have seen, the bridge map of my DS approach impose no such requirement, and may be many-one. Yoshimi, on the other hand, requires that his counterpart to the bridge map, which he calls a `supervenience function,' be an onto function between state space. Again, the bridge map of DS reduciton imposes no such requirement. Moreover, neither Giunti nor Yoshimi demand compatibility of their bridge map counterparts with the symmetries of the models, nor do they explicitly insist that it not be explicitly dependent on time (though perhaps this may be regarded as an assumption implicit in their analyses). 

\vspace{5mm}

While much of his work on reduction and emergence focuses on limit-based and Nagelian approaches, Butterfield also discusses inter-level relations in physics in terms of dynamical systems. Like DSR condition 1), the core condition for reduction that Butterfield's analysis draws on, which he calls `meshing' of `macro-' and `micro-' level dynamics, involves the commutation of some `coarse-graining' function between micro- and macro-level state spaces with the time-evolution prescribed on those spaces. The macro-level state space is identified with a partition of the micro-level state space, and the coarse-graining function simply maps an element of the micro-level space into the cell of the partition to which it belongs. On Butterfield's account, the closest analogue to what I call the high-level model is the macro-level model; to what I call call the low-level model, the micro-level model; and to what I call the bridge map, the coarse-graining function \cite{ButterfieldDynSys}. Note that Butterfield's terminology draws heavily on examples of reduction in statistical mechanics.  

Butterfield characterises the dynamics of a macro- and micro- models as `meshing' relative to a particular partitioning $\mathcal{P} = \{ C_{i} \}$ of the micro-level state space $\mathbb{S}$ when the set obtained by applying the micro-evolution law $T: \mathbb{S} \rightarrow \mathbb{S}$ to an element of $\mathcal{P}$ is itelf an element of  $\mathcal{P}$, so that for any $i$, $T(C_{i}) = C_{j}$ for some $j$. Thus, the micro-level dynamics $T: \mathbb{S} \rightarrow \mathbb{S}$ induces, via the coarse-graining, some macro-level dynamics $\bar{T}: \mathcal{P} \rightarrow  \mathcal{P}$. This will not be the case for an arbitrary partition of $\mathbb{S}$ since two microstates in the same partition may evolve under the microdynamics into separate elements of the partition, so that micro-level determinism gives rise to macro-level indeterminism (where the macro state space corresponds to the partitioning of the micro state space). 

However, Butterfield acknowledges that this concept of meshing may not apply to many realistic cases in which one dynamical system purportedly reduces to another - such as the reduction of models involving the Boltzmann, Navier-Stokes and diffusion equations to some micro-physical mechanical model - and so suggests that the following modifications and allowances to his notion of meshing may be required before these realistic examples can be counted as instances of it(I quote directly from Butterfield here):

\begin{itemize}
\item `the meshing may not last for all times;
\item the meshing may apply, not for all micro-states $s$, but only for all except a ``small'' class;
\item the coarse-graining may not be so simple as paritioning $\mathbb{S}$; and indeed
\item the definition of the micro-state space $\mathbb{S}$ may require approximation and-or idealisation, especially by taking a limit of a parameter: in particular, by letting the number of microscopic contitutents tend to infinity, while demanding of course that other quantities, such as mass and density, remain constant or scale appropriately.' \cite{ButterfieldDynSys}
\end{itemize}

\noindent Indeed, all of the first three of these considerations are already built into the definition of DS reduction. DS reduction is defined only relative to a particular timescale and margin of error and for a particular, potentially limited, domain $d$ of states in the low-level state space; moreover, the bridge-map of DS reduction need not yield a partitioning of the low-level space (that is, the inverse images under $B$ of points in $S_{h}$ need not form a partition of $S_{l}$; indeed, it will not necessarily be the case that every point in $S_{h}$ even \textit{has} an inverse image). Butterfield's fourth concern only comes into play in certain special cases, for example in reductions where quantum field theory or statistical mechanics furnishes the reducing model, since both of these theories typically involve taking limits as the number of degrees of freedom in the theory goes to infinity. In the case of quantum field theory, which I do consider in this thesis, this fourth concern of Butterfield's is averted by taking a `cut-off' approach to quantum field theory and thereby treating the QFT model in question as a model of a large-but-finite, rather than an infinite, number of degrees of freedom. 

While the modifications to his meshing condition that Butterfield suggests anticipate a number of differences between meshing and DS reduction, it will be worthwhile to explore these differences in a bit more detail. One essential difference, just noted, is that while Butterfield's meshing condition requires that the coarse-graining function (his counterpart of the bridge map) be associated with some partition of the low-level state space, the bridge map need not take as its domain the whole of $S_{l}$, and therefore need not prescribe a partitioning of $S_{l}$; moreover, the bridge map need not take the whole of $S_{h}$ as its image, providing still another reason why the high-level state space cannot in general be regarded on DS reduction as a partition of the low-level space. 

Furthermore, if a micro-level system obeys Butterfield's meshing condition with respect to some partition, then for any macro-level initial condition - i.e., some partition cell - it must be the case that the deterministic dynamics induced on the partition by the micro-level dynamics yield the same result irrespective of the microcondition that instantiates that initial macrocondition. Since any element of the partition can serve as the macro- initial condition, and since every element of the micro-level state space belongs to some element of the partition, Butterfield's meshing condition requires that the \textit{whole} micro-level state space (or at least all but a very small subset of this space) serve as the domain that approximates the macro-level dynamics under coarse-graining; by contrast, in DS reduction, the domain of $S_{l}$ whose induced dynamics under the bridge map approximates the high-level dynamics is not required to be the entirety of the low-level space.

Finally, on Butterfield's approach, the coarse-graining function associated with a partition that respects the meshing condition is not required to respect the symmetries of the low-level model insofar as it does not require that for any symmetry of the deterministic macrodynamics, there will be some symmetry of the micro-level dynamics that induces it under coarse-graining - nor does not entail that the group structure of the micro-level symmetries induce the group structure of the macro-level symmetries on the partition. 

A final, though potentially less substantive, difference between Butterfield's account of dynamical commutation and the DS approach is that while the inspiration for the DS approach comes from examples of reduction in statistical mechanics, on the DS approach the reduced and reducing models need not correspond, respectively, to models of macroscopic and microscopic phenomena, nor does the bridge map need to correspond to a `coarse-graining' in any sense other than its often being a many-one function (certainly, it is not required to furnish a partition of $S_{l}$, nor is it required to map onto the whole of $S_{h}$). Of course, if Butterfield is using the terms `macro-' and `micro-' merely to suggest some analogy with statistical mechanical reductions, and not by way of restricting this approach to reductions in which high- and low- level descriptions correspond respectively to `macro-' and `micro-' level phenomena, then this distinction collapses to some extent into one of terminology.

\vspace{5mm}

Within statistical mechanics, Lanford's Theorem provides an explicit instance of the dynamical commutation \footnote{Thanks to Jeremy Butterfield for pointing me to this example.} (see, for instance, \cite{Lanford1},  \cite{Lanford2},  \cite{Lanford3}, \cite{UffinkValente}). Lanford's Theorem shows that the Boltzmann equation, which describes the behavior of the distribution $f_{t}(\vec{x},\vec{p})$ in 6-dimensional $\mu$-space of particles in a dilute gas (and assumes the molecules in the gas are modelled as solid spheres), can be derived from the formalism of classical Hamiltonian mechanics, which prescribes via the Liouville equation the time evolution of a probability distribution $\rho_{t}(\vec{x}_{1},\vec{p}_{1},...,\vec{x}_{N},\vec{p}_{N})$. The theorem establishes a particular bridge or correspondence between probability distributions $\rho$ on phase space and distributions $f$ on $\mu$-space, such that to any probability distribution $\rho$ there corresponds a unique $f$, but such that there are in general many $\rho$ that may yield the same $f$ under this correspondence. The theorem then shows that provided certain constraints are imposed on the initial phase space probability distribution $\rho_{0}$ at some time $t=0$, the evolution of $f$ induced by the evolution of $\rho$ via this correspondence approximately satisfies Boltzmann's equation for some time scale $\tau$ (what this time scale turns out to be depends on the strength of the assumptions made about the evolution of $\rho$). Thus, Lanford's Theorem shows that, applied to some domain of possible initial probability distributions $\rho_{0}$, the low-level dynamics (The Liouville Equation) for some time $t$ followed by an application of the bridge map or coarse-graining yields approximately the same final distribution $f_{t}$ as does the bridge map followed by an application of the high-level dynamics (The Boltzmann equation) for the same time $t$, thus satisying the dynamical commutation condition. 

\vspace{5mm}

Werndl has shown that for every deterministic dynamical system, there is an indeterministic model that reproduces the same empirical predictions to within some given margin of error, and also that for every indeterministic dynamical model, there is a deterministic one that is observationally indistinguishable from it, again to within some margin of error  \cite{Werndl1}, \cite{Werndl2}, \cite{Werndl3}, \cite{Werndl4}. All of the models considered in this thesis are deterministic, although it is possible (particularly in the case of the quantum theories I consider) that observationally equivalent stochastic models can be chosen in place of these; in such a case, it would be necessary to extend the account of reduction among deterministic models that I provide to reductions among indeterministic models, as well as to reductions of deterministic to indeterministic models, and reductions of indeterministic to deterministic models. 

\vspace{5mm}

Finally, Peter Smith offers an account of approximate truth of models of dynamical systems that closely resembles in certain respects the account of strong analogy between analogue and high-level models that I discuss here (see \cite{SmithDynApprox} and \cite{SmithChaos}, Ch.5). To be sure, the notion of strong analogy that I consider here concerns a notion of closeness between distinct dynamical systems models, while Smith is concerned with the notion of closeness between these models and the behavior of the physical systems that the models describe. While from a conceptual point of view the question of what it means for one model to approximate another and the question of what it means for a model to approximate the behavior some actual physical system are clearly distinct, according to Smith the latter sense of approximation can be understood as a similarity of geometrical structures associated respectively with the model and the physical system; likewise, in the account of strong analogy between dynamical systems models that I give here, what it means for one model to approximate another can also be understood as similarity of geometrical structure. For example, a classical phase space model of a simple pendulum can provide an approximately true description of a real physical pendulum insofar as it is possible to plot, on the same phase space, both the trajectory predicted by the model and the trajectory of the real pendulum and to show that these trajectories are `close' within some margin of error $\epsilon$ for some time period; crucially, this sense of closeness is specified by the geometrical structure - usually some norm - of the state space in question. In the sense of strong analogy that I discuss here, a (say) quantum model of the simple pendulum will approximate a classical phase space model if the phase space trajectory induced by the quantum model through the relevant bridge map approximates, in the geometrical sense furnished by a norm on phase space, the trajectory prescribed by the classical model.  

\vspace{5mm}

My primary goal in this thesis is to demonstrate that the way of thinking about reduction in terms of dynamical commutation can be applied widely beyond the realm of statistical mechanics, and beyond the realm of macro- to micro- reductions, and that it can be extended and refined into a general account of intertheory relations in physics that is distinct from the limit-based and Nagelian approaches. Nevertheless, it may incorporate elements of both of these approaches; in particular, in the next section I will be discussing some of the parallels between DS and Nagelian reduction.

%\underbar{Newtonian Mechanics and Special Relativity}

%Consider a model of Newtonian mechanics in Hamiltonian form, with dynamics determined by the equations

%\begin{equation}
%\begin{split}
%& \frac{dx}{dt} = \{x, H\} \\
%& \frac{dp}{dt} = \{p, H\},
%\end{split}
%\end{equation}

%\noindent with  

%\begin{equation}
%H = \frac{p^{2}}{2m} +V(x).
%\end{equation}

%\noindent Now consider a model of special relativity with 

\section{DS Reduction and Nagelian Reduction: Parallels} \label{DSGNSParallels}

Perhaps the most salient parallel between DS and Nagelian reduction is that both make use of special correspondences between the elements of the high- and low- level descriptions of a particular system. More specifically, the bridge maps and bridge rules of DS reduction serve much the same purpose as the bridge laws of GNS reduction, insofar as they identify those elements of the low-level description that approximately mimic the behavior of particular elements in the high-level description. 

However, the analogy between the two approaches extends further than this. Recall that the GNS account of theory reduction distinguishes four `theories': the low level theory $T_{l}$, the high level theory $T_{h}$, the image theory $T_{h}^{*}$, and the analogue theory $T'_{h}$. Recall that on the GNS approach, $T_{h}^{*}$ is formulated in the language of $T_{l}$ and deduced from $T_{l}$ without the use of bridge laws; $T'_{h}$ is then obtained from $T_{h}^{*}$ by straightforward bridge law substitution, and is formulated in the language of $T_{h}$; if the reduction is successful $T'_{h}$ will be `strongly analogous' to $T_{h}$. It is in this sense that a high level theory $T_{h}$ may be reduced to a low level theory $T_{l}$ on the GNS account. On the semantic, DS approach to physical reduction, I claim, the portion of the reduction that involves demonstrating that DSR condition 1) is satisfied proceeds much according to this same basic outline, with a major revision being that it is \textit{models} rather than whole theories that are reduced. Let us make the parallels between GNS and the dynamical component of DS reduction more explicit. 

\subsubsection{Nagel's Homogeneous/Inhomogeneous Distinction}

Nagel introduces the distinction between homogeneous and inhomogeneous reductions as the motivation for introducing bridge laws into his account of reduction. Yet it is worth noting here that this distinction can be quite vague once we begin to probe it further. 

For example, Nagel uses the reduction of the Galilean theory of terrestrial gravitation to Newton's Universal Theory of Gravitation as an example of a homogeneous reduction, explaining that all of the terms employed in the former - position, time, mass, force - are all also contained in the latter. However, there is room to doubt this classification when one considers the nature of the constant $g=9.8 m/s^{2}$. In Galileo's theory, $g$ appears as a primitive constant whose value is unexplained, whereas in Newton's it is equated to $\frac{G M_{E}}{R^{2}_{E}}$ (where $M_{E}$ is the mass of the earth and $R_{E}$ is the radius of the earth). Likewise, in the reduction of the thermodynamic Ideal Gas Law to statistical mechanics, temperature $T$ is a basic quantity in the context of thermodynamics, while in the context of statistical mechanics it is equated to $\frac{2}{3 k_{B}} \langle K.E.\rangle$. Given this parallel, it seems rather arbitrary of Nagel to claim the term $T$ occurs only in the reduced but not the reducing theory, making the relation $T = \frac{2}{3 k_{B}} \langle K.E.\rangle$ a bridge law, while claiming that $g$ occurs in both reduced and reducing theories, making the relation $g = \frac{G M_{E}}{R^{2}_{E}} $ something other than a bridge law (say, a definition of a non-basic quantity - namely, $g$ - in Newton's theory).

%In the context of the gravitational case, Nagel would presumably not want to identify $g$ as a term that occurs only in the high-level theory but not in the low-level theory, since he characterises the reduction as homogeneous. Instead, he would likely characterise $g$ as a quantity that occurs both in Galileo's theory and in Newton's theory, where in Newton's theory this term is simply a non-basic term that can be defined in terms of more fundamental quantities. Yet it is not clear why he should not then choose to likewise regard temperature as a term occuring in both thermodynamics and statistical mechanics, where in statistical mechanics it is simply a non-basic term that can be defined in terms of more fundamental quantities.   

Yet we can still see the motivation for Nagel's homogeneous/inhomegneous distinction: in the Galileo/Newton reduction, the theoretical frameworks of the two theories arguably share a lot more in common than do the frameworks of the two theories involved in the thermo./stat.-mech. reduction. Yet, as we have seen in the former example, even in cases where the theoretical frameworks are quite similar, some links between them - albeit ones that may seem quite obvious and natural, sometimes so much so that they are left implicit - will be required to effect a reduction of one theory to the other (in the Galileo/Newton case, the definition $g = \frac{G M_{E}}{R^{2}_{E}} $ is indeed required before one can derive Galileo's $F=mg$ as an approximation). Whether one regards the theoretical frameworks of the reduced and reducing theories as sufficiently different to characterise the links as bridge laws (rather than, say, as definitions of non-fundamental quantities in the reducing theory) is to a significant extent arbitrary.  

Many of these considerations carry over to reduction in the DS approach, where the high- and low- level models may involve very similar, or very clearly distinct, mathematical structures, but in which there is no clear division between models whose mathematical structures are `the same' - there will always be some differences, assuming the models are not identical - and those in which they are clearly distinct. One might be inclined to call reductions between models whose state spaces take the same general form homogeneous and those in which their forms are different inhomogeneous. Yet, in the end it is a matter of arbitrary choice what particular kinds of similary of mathematical structure one relies on in the classification of reductions as homogeneous or inhomogeneous. 

As an example, one might be inclined to regard as homogeneous the DS reduction of a Newtonian phase space model of the motion of some centers-of-mass to the Newtonian phase space model of the motion of the system's smaller constituents. The classification as homogeneous would be motivated by the fact that both state spaces are symplectic manifolds (albeit of different dimension), and in addition may have Hamiltonians of the same general $H=\sum_{i} \frac{p_{i}^{2}}{2m_{i}} + \sum_{i \neq j} V(|x_{i}-x_{j}|)$ form. However, even if one classifies such a reduction as homogeneous, a bridge map, given by the center of mass function and the corresponding formula for momentum, is still required. As an example of an inhomogeneous reduction, one might take the example considered repeatedly in this chapter, in which some classical phase space model is reduced to some quantum model; in this case, one state space is a symplectic manifold and the other a Hilbert space; the bridge map is given by the expectation values of the position and momentum operators. Again, though, the question of which particular differences of mathematical structure are to determine the classification as homogeneous or inhomogenous is a matter of arbitrary choice. 

Moreover, whether the state spaces are more or less similar in their mathematical form, all DS reductions require bridge maps that are constrained to fulfill the same set of requirements. Unlike in Nagel's approach, where some reductions are said to require bridge laws and some are not, the counterparts of bridge laws in the DS approach, bridge maps, are required of all reducitons (though, again, in some cases these bridge maps may be particularly natural or obvious). Consequently, it does not appear that in the context of DS reduction very much hangs on the homogeneous/inhomogeneous distinction, insofar as it can be drawn.

\subsubsection{Image Models, Bridge Rules, Analogue Models and `Strong Analogy'}

On the DS account of the reduction of a high-level model $M_{h}$ to a low level model $M_{l}$, one can, by analogy with the GNS approach, identify an image model $M_{h}^{*}$ and an analogue model $M'_{h}$. It is the analogue model that approximates the high-level model $M_{h}$.

The image model $M_{h}^{*}$ is formulated using elements of the model $M_{l}$ - that is, in terms of the mathematical structures defined on $M_{l}$'s state space - and can be deduced from $M_{l}$ solely on the basis of a restriction to a particular domain of states in $S_{l}$. Its dynamics consist of the composition of the bridge map $B$ and $D_{l}$:

\

\noindent \underbar{\textit{Image Model Dynamics:}}

\

\begin{equation} \label{dDSR}
\frac{d}{dt} B\big(x^{l}(t)\big) \approx f_{h}\big(B \big(x^{l}(t)\big), t \big)
\end{equation}

\noindent for $x^{l} \in d$, where $d$ is some domain of states in $S_{l}$ roughly preserved under the dynamics for some limited timescale $\tau$. Recall that satisfaction of image model dynamics for some such domain $d$ suffices to ensure satisfaction of the DSR condition. 

By further analogy with the GNS account, the analogue model is obtained from the image model through bridge rule substitutions,

\

\noindent \underbar{\textit{Bridge Rules:}}

\

\begin{equation}
x'^{h}\equiv B(x^{l})
\end{equation}

\

\noindent and its dynamics are specified by the equation of motion:

\

\noindent \underbar{\textit{Analogue Model Dynamics:}}

\

\begin{equation}
\frac{d x'^{h}}{dt} \approx f_{h}(x'^{h},t).
\end{equation}

\noindent Note that the expression $B(x^{l})$, which occurs in the image model, is an expression built from structures within the low level model $M_{l}$ - in this sense it is formulated in the mathematical `language' of the low level model. On the other hand, the bridge rule equivalent of this expression, $x'^{h}$, is to be understood as an object of the sort defined within the high level model $M_{h}$: specifically, an element of the high level state space $S_{h}$. In this sense the analogue model $M'_{h}$ is formulated in the mathematical `language' of the high level model.
% In the state-space/dynamics notion I've been using to specify models, one could write $M_{h}^{*} = (d_{h}, B_{l}^{h} \circ D_{l})$, and $M_{h}^{'} = (B_{l}^{h}(d_{h}), B_{l}^{h} \circ D_{l})$, where $B_{l}^{h}(d_{h}) \subset S_{h}$ is the image of $d_{h}$ in $S_{h}$.  

From this we can see that the image model is typically formulated in notation which lays bare the detailed construction of the $M_{l}$ structures that approximately instantiate the dynamics of $M_{h}$, while the analogue model neatly packages and conceals the internal makeup of these structures by assigning to them simpler variable expressions which are a mathematical form familiar to $M_{h}$. Figuratively speaking, the image model enables us to `look under the hood' of the analogue model $M'_{h}$ to see in detail how its structures are assembled from the mathematical components of $M_{l}$.

For a reduction to take place in the GNS account, the analogue model $M'_{h}$ must be `strongly analogous' to the high level model $M_{h}$. Within the context of the GNS model the condition of strong analogy is highly ambiguous, though is intended to include some requirement of approximate agreement between $M'_{h}$ and  $M_{h}$. On the DS approach, the relation of strong analogy is unambiguous, and requires that 

\

\noindent \underbar{\textit{`Strong Analogy'}}

\

\begin{equation}
\big| x'^{h}(t) - x^{h}(t) \big| < \delta \ \forall \ 0 \leq t \leq \tau,
\end{equation}
 
\noindent where $\tau$ again is the reduction timescale. Note that this `strong analogy' claim is just the DSR condition rewritten using bridge rule substitution $x'^{h}(t) \equiv B \big(D_{l}(t;x^{l}_{0})\big)$ and the definition $x^{h}(t) \equiv D_{h}\big(t; B(x^{l}_{0}) \big)$. \footnote{One could object that there is still an ambiguity as to what the appropriate norm to take on the high level space $S_{h}$ is. In all the examples I consider, however, the appropriate choice of norm is always obvious.}

\vspace{5mm}

The basic elements of DS reduction can be consolidated in a manner that directly parallels the three steps of GNS reduction, with an additional step to show that the dynamical symmetries of the high-level model are approximately replicated via the bridge map by those of the low-level model :

\

\noindent \underbar{\textbf{DS Reduction in Four Steps:}}

\

\begin{enumerate}
\item Derive the image model $M_{h}^{*}$ as an approximation to the dynamics of $M_{l}$ within some restricted domain of states within the low level state space $S_{l}$. This amounts to deducing a relation of the form $
\frac{d}{dt} B\big(x^{l}(t)\big) \approx f_{h}\big(B\big(x^{l}(t)\big), t \big)$ from the dynamics of $M_{l}$ and a restriction to a particular domain of states in $S_{l}$. Note the derivation of an image model amounts to a proof of the DSR condition. This step refines Nagel's derivability condition; note that it is the image model $M_{h}^{*}$, not the high-level model  $M_{h}$, that is derived from $M_{l}$ on this account of reduction. 
\item Use bridge rules $x'^{h} \equiv B(x^{l})$ to replace the terms $ B(x^{l})$ occurring  in $M_{h}^{*}$, and which are constructed using the mathematical structures of $M_{l}$, with corresponding terms belonging to the high level model. This yields the analogue model $M'_{h}$, whose dynamics are specified by the dynamical equation $\frac{d x'^{h}}{dt} \approx f_{h}(x'^{h},t)$. This, along with step 3, refines Nagel's connectability condition. 
\item  On the reduction timescale $\tau$ on which the image model holds, $M'_{h}$ is `strongly analogous' to the high level model $M_{h}$, in the precisely defined sense that $\big| x'^{h}(t) - x^{h}(t) \big| < \delta \ \forall \ 0 \leq t \leq \tau$. This step contributes an additional component to Nagel's connectability condition. 
\item (no analogue in GNS) Prove that the bridge map respects the symmetries of the high-level model by demonstrating that conditions 2 a) and b) for DS reduction are satisfied.
\end{enumerate}

\noindent The parallels with the GNS approach, and the manner in which the essential elements of this account are all paralleled within a semantic, dynamical systems approach should at this point be clear. Note also that there is no reason that dynamical systems reduction, as spelled out here, need be restricted to dynamical systems in physics, though all of the examples I consider here are reductions of this sort. On a final note, it is in the sense specified by DS reduction that I will refer to a model $M_{l}$ of $T_{l}$ \textit{instantiating} some model $M_{h}$ of $T_{h}$ - that is, $M_{l}$ instantiates $M_{l}$ if the DSR condition between the two is satisfied. I also may refer more specifically to $B(d)$ as approximately instantiating some subset of the high-level state space $S_{h}$, and to the image model as approximately instantiating the dynamics $D_{h}$ of the high-level model.

\section{DS Reduction and Nagelian Reduction: Disanalogies}

The first and most general distinction between DS and Nagelian reduction is that the former concerns the reduction of individual models while the latter concerns the reduction of theories. Nagelian reduction, moreover, specifically requires the derivation of the \textit{laws} of the high-level theory from those of the low-level theory. In the case of DS reduction, to be sure, it is also necessary that the laws of the high-level model - which I take it are most naturally associated in the DS picture with the equations of motion of the model - be derivable from those of the low-level model in the sense that it is possible to derive some image laws from the low-level model, which serve to approximate the laws of the high-level theory via bridge rules and the strong analogy relation. 

Yet models of physical theories involve much more mathematical structure than simply their dynamics - for example, the structures associated with their state spaces and the dynamical symmetries on those state spaces. In models of classical Hamiltonian mechanics, for example, the dynamical equations, as expressed in terms of Poisson brackets with the Hamiltonian, are but a portion of the larger symplectic structure of the phase space manifold, which serves as a unified geometrical framework in which to understand not only the dynamics but the symmetries of the theory as well as the whole fomalism of canonical transformations. In models of non-relativistic quantum mechanics, likewise, the dynamical law specified by the Schrodinger equation is but a portion of the larger mathematical apparatus associated with Hermitian operators, unitary transformations, and the like. Unlike Nagelian reduction, which focuses on the derivation of the high-level theory's \textit{laws}, DS reduction more generally seeks to identify substructures of the low-level model that approximately instantiate the structures of the high-level model in some domain. While the dynamical laws of the high-level model certainly represent one crucial piece of the high-level model's structure that must be instantiated by the low-level model (and the fact of this instantiation is one that must be derived from the low-level model), they do not exhaust it. As we have seen, the image domain $B(d)$ instantiates that part of the high-level theory's state space $S_{h}$ that can be used to accurately model the physical system in question, while $B(T_{l})$ approximately instantiates some high-level dynamical symmetry $T_{h}$ of the theory for some $T_{l}$; likewise, the group composition structure of the high-level symmetries $T_{h}^{1} \circ T_{h}^{2}$ is approximately instantiated by the group composition structure $B(T_{l}^{1} \circ T_{l}^{2})$ of the low-level symmetries induced via the bridge map. 

%Of course, the fact that this is so for a given pair of models, for a given bridge map, is something that must be derived, and it is with respect to the derivation of the image of the high-level equations of motion in the low-level model that the parallels between DS and Nagelian reduction occur. 

One final difference between DS and Nagelian reduction that bears discussion is that, while the bridge maps of DS reduction and the bridge laws of Nagelian reduction do fulfill similar roles, DS reduction is framed in terms of the $\textit{existence}$ of a mathematical function (the bridge map) satisfying certain criteria, while the bridge laws of Nagelian reduction are understood as separate assumptions made independently of the high-and low- level theories, which are necessary to derive the appropriate analogue to the high-level laws. That is, the DS approach takes the high- and low- level models of a system as given, and a reduction is said to occur only if a certain mathematical relationship obtains between these models - namely, the existence of a function between the state space satisfying the necessary mathematical conditions given above. The Nagelian approach, on the other hand, treats bridge laws as independent auxiliary assumptions that supplement the low-level theory to facilitate the derivation of an analogue to the laws of the high-level theory.

\section{Generality in Reduction} \label{Templates}

Authors on the subject of theory reduction, particularly in physics, have tended toward one of two approaches to this subject, which I call the `systematic' and `piecemeal' approaches to reduction. There is no clear-cut division between these two categories, but rather a spectrum of possible approaches between these extremes which aspire to varying degrees of generality.

Advocates of more systematic approaches to reduction tend to assume that among systems which exhibit $T_{h}$ regularities, the explanation in terms of $T_{l}$ as to why they do so can be given on the basis of general results that connect the formalisms of the two theories and which are presumed to apply to all such systems. That is, they tend to view the problem of reduction exclusively in terms of generalities which can then be applied to particular systems, much in the way an algebraic identity applies generally for all numbers in a particular set. The details that are most salient to the explanation of the reduction, however, are always the same, independently of the system being investigated.

Advocates of more piecemeal approaches implicitly deny the possibility of such generality in explaining why certain $T_{l}$ systems exhibit $T_{h}$ regularities. Ultimately, they deny that it is possible to abstract away so completely from the details of the system in question in explaining why it obeys regularities characteristic of $T_{h}$. The most extreme version of this view insists that reductions must be accounted for on a case-by-case basis - that is, for each system that exhibits $T_{h}$ regularities, one must provide a separate and distinct $T_{l}$-based explanation of why it does so that is tailored specifically to that system.

The systematic and piecemeal approaches are illustrated, respectively, by work that tries to explain reduction entirely on the basis of formal mathematical results, on the one hand, and by work which tries to explain, for a system defined to within narrow parameters, how behavior characteristic of $T_{h}$ can be accounted for by $T_{l}$, on the other. Landsman's well-known treatment of the quantum-classical correspondence (detailed in \cite{landsman1998mathematical}), for instance, falls more toward the systematic end of the spectrum in that it rests primarily on formal mathematical correspondences between quantum and classical mechanics without considering the details of particular systems, while Joos and Zeh's calculations of decoherence rates for different environments consisting of air molecules, dust particles, and photons (see \cite{caldeira1981influence} and \cite{joos1985emergence}) fall closer to the piecemeal end of the spectrum.

\vspace{5mm}

Let us take a moment to consider how the distinction between systematic and piecemeal approaches to reduction relates to the distinction between $reduction_{1}$ and $reduction_{2}$. The type-1 approach to reduction, to which, I have argued, the dynamical systems approach belongs, can be systematic since there is nothing inherent in the definition of DS reduction, or more broadly of Nagelian reduction, that automatically precludes the existence of some general result that accounts for all successful applications of the theory $T_{h}$. On the other hand, a type-1 reduction also can be piecemeal, since it is also possible that the set of $T_{l}$ systems that instantiate the models of $T_{h}$ are highly distinct in the sense that their underlying $T_{l}$ descriptions differ widely.

A type-2 reduction, if any such reductions exist, is necessarily systematic since it requires that the behavior of any approximately $T_{h}$ system be retrieved from some $T_{l}$ system as some set of parameters $\{ \epsilon_{i} \}$  in $T_{l}$ approach zero. (Note that my analysis in section \ref{Limit} does not preclude the occurence of instances of $reduction_{2}$; it only precludes regarding $reduction_{2}$ as a general characterisation of inter-theory relations in physics). Thus, if any system exhibits $T_{h}$ regularities, one need look no further than the values of $\{ \epsilon_{i} \}$ and the associated limits for an explanation of this fact; all other particularities of the system are irrelevant to the reduction.

It is worth clarifying here the difference between a systematic type-1 reduction and a type-2 reduction (which is necessarily systematic) since it is reasonable to ask whether there are any strongly systematic reductions which are of type-1 but not of type-2. To take one example, circuit theory is instantiated by Maxwell's equations and its success can be explained quite systematically, and with a great deal of generality, on the basis of Maxwell's theory: at a certain intermediate level detail, Ohm's law, and rules governing governing capacitors, inductors and various other circuit elements all can be derived in similar fashion from Maxwell's equations (see, for example, \cite{purcell2011electricity}) although at the most fine-grained level of explanation, results from solid state physics pertaining the specific materials used to make these devices must be invoked. Yet, despite the generality of these derivations at an intermediate level of detail, it does not seem that there is any simple limit of the form given in the definition of $reduction_{2}$ that neatly encapsulates all such derivations. Therefore, this serves as an example of systematic-ness within reduction that is not associated with type-2 reduction.

%As already noted, $T_{1}$ explanations of $T_{2}$ behavior are systematic or piecemeal as a matter of degree. The derivation of the laws of circuit theory from Maxwell's equations is to a large extent systematic in the sense of applying to a wide range of $T_{1}$ systems meeting a certain set of broad criteria (see, for example, ~\cite{PurcellEM}), though at a certain level of detail, fine-grained specifics regarding the microscopic constitution of the resistor, capacitor, inductor or other circuit element in question may become salient.  

Moreover, while formal limits tend to feature most centrally in type-2 reductions, this does not mean that they may not also play some partial role in either systematic or piecemeal type-1 reductions. 
%For example, Butterfield's approach to reduction is most accurately classified as systematic type-1, though it employs limit-based results. Piecemeal type-1 reductions, such as particular decoherence models, may consider limits in which parameters such as mass or number of environmental degrees of freedom, or some combination of parameters, approach infinity, though the relevance of such limits may be limited to the particular system or set of systems in question, rather than being of universal relevance to all $T_{1}$ systems, as in the case of type-2' reductions. 
The essential difference between the manner in which limits are employed in type-2 reductions and the manner in which they may be employed in type-1 reductions is this: in type-2 reductions, for any model of \textit{any $T_{l}$ system at all}, we expect to retrieve some model $M_{h}$ of $T_{h}$ describing the same system when we dial the values of the parameters $\{ \epsilon_{i} \}$ in that model sufficiently close to zero. In type-1 reductions, this is not necessarily the case since there is no requirement in type-1 reductions that every $T_{l}$ system correspond to some $T_{h}$ system, but only that every model of  $T_{h}$ that successfully (if approximately) describes a real physical system be instantiated by some domain of \textit{some} $T_{l}$ model; the explanation as to how this instantiation occurs may rely partially upon some particular limit-based result, combined with a number of other non-limit-based results and assumptions.

\subsection{Reduction and Explanation}

The task of any type of reduction considered here, whether type-1 or type-2, systematic or piecemeal, can be regarded as a task of $\textit{explanation}$ - specifically, of explaining why some model of $T_{h}$ succeeds at describing some system or set of systems on the basis of some model of $T_{l}$. The criteria for scientific explanation are notoriously controversial and seemingly variable across different parts of science, so it is worthwhile taking a moment to place my analysis within this broader discussion. 

It is straightforward to see that both $reduction_{1}$ and $reduction_{2}$ fit squarely into the traditional deductive-nomological model of explanation, which takes explanation to be \textit{deduction} of the explanandum  from premises that include at least one universal law and typically also some auxiliary premises  \cite{hempe1965aspects}.  In the context specifically of DS reduction, the explanandum is the dynamics of the image model $M_{h}^{*}$, while the universal law is the dynamics of the low-level $M_{l}$, and the auxiliary assumptions are given by a domain restriction within the state space $S_{l}$. The bridge rules, analogue model and strong analogy relationship all complement the derivation of the image model dynamics for purposes of clarifying the connection between the image model $M_{h}^{*}$ and the high-level model $M_{h}$. 

While $reduction_{1}$ and $reduction_{2}$ are well-accomodated by the DN model of explanation, this model faces a number of well-known difficulties \cite{salmon1984scientific}. Nevertheless, the particular nuances which make trouble for the DN model, such as the need to account for assymmetries of explanation and to ensure that premises are relevant to the conclusion, need not concern us here since they do not arise in the cases I consider. Moreover, while the DN model was formulated by Hempel within the framework of logical empiricism, nothing in the DN model's central elements suggests that it cannot also be applied within a semantic, realist framework. A number of more current models of explanation, such as Salmon's causal-mechanical model and Kitcher's unificationist model, have endeavoured to refine and update the deductive-nomological model so as to avoid the above-mentioned difficulties, and do so within a realist framework. Others, such as van Fraassen's constructive empiricist account, have sought to replace it wholesale \cite{salmon1984scientific}, \cite{kitcher1981explanatory},\cite{van1980scientific}. I leave it as an open question, which, if either, of the main realist refinements of the DN model best accommodates the examples I consider.

\vspace{10mm}

In section \ref{Systematic}, I explain why wholly systematic approaches to reduction do not generally succeed, illustrating my point with the by-now familiar examples of the NM/SR and CM/QM intertheory relations. I then consider the completely piecemeal approach in section \ref{Piecemeal}, but argue that it is too weak a position. I contend, by reference to particular examples, that it is usually possible to retain a degree of generality and systematic-ness in deriving the image model dynamics from the base model $M_{l}$, without having to produce a totally new derivation for each system separately. I go on in section \ref{Templates} to elaborate and defend a `semi-systematic,' or template-based, approach to theory reduction that reconciles the role of system-specific details in reductions with the desire to understand the general mechanisms and principles at work in reductions across a wide range of systems. Not surprisingly, the degree of generality that can be retained in accounts of theory reduction will depend on the particular pair of theories, and models, in question.

\subsection{Completely Systematic Approaches to Theory Reduction} \label{Systematic}

%The claim that a wholly systematic approach to theory reduction in physics is generally appropriate, or the claim that it is not, is ultimately an empirical assertion which must be assessed on the basis of particular examples.
%This is because the question of whether - and if so, how - one can explain on the basis of a superseding theory why a superseded theory succeeds in its domain is one that must be answered separately for each pair of theories in question, since the details of the explanation will depend crucially on what the theories happen to be and what the data is that supports them. For each pair of theories that does admit an explanation for every successful application of the superseded theory, either the explanation will be the same for every such application or it won't. If in such cases we find that successes of the superseded theory do not typically admit of a single over-arching explanation in terms of the superseding theory, but rather require separate explanations for different successes, then it would be demonstrably false to claim that systematic approaches to reduction are generally effective at providing such explanations. On the other hand, if in all cases where the superseded theory's successes could be explained in terms of the superseding theory, we were to find that the same overarching explanation applied in every case, then this would provide strong evidence for the assertion that systematic approaches to reduction are generally effective.

Below, I argue that a completely systematic approach to reduction does not generally succeed in physics, and support this view with two examples: the reduction of Newtonian mechanics to special relativity (by `special relativity,' I mean Lorentz covariant classical dynamics) and the reduction of classical mechanics to quantum mechanics.

\subsubsection{Completely Systematic Approaches to the NM-SR Reduction}

If a completely systematic type-2 reduction of NM to SR is probably not possible, as argued in section \ref{Limit}, might a completely systematic type-1 reduction be? That is, for all physical systems whose behavior can be approximated by some Newtonian model, is it possible to provide a single general derivation of the corresponding image laws of NM that applies to all such systems? I claim that one cannot, and offer an example to illustrate my point.

Consider an SR model of an idealised system consisting of two masses connected by a spring that is sufficiently stiff and sufficiently compressed that the total system contains enough potential energy to make the rest mass of the total system substantially exceed the sum of the individual constituent masses; the spring itself is assumed to have effectively zero rest mass. (While perhaps not very realistic, one can at least in principle model such a system in SR.) Assume that the center of mass the combined system behaves in Newtonian fashion and that the centers of mass of each of the masses do as well. An explanation as to why this is so will have to include some explanation as to why the energy stored in the spring is not released (for instance, because it is held together by powerful clamp). For, if the potential energy (which is comparable to the rest energies of the two masses) were converted to kinetic energy, the bodies would fly off at some substantial fraction of the speed of light and non-Newtonian effects would become apparent.    %\footnote{It is possible to object that this example is unrealistic and therefore ineffective. However, I claim that it is no less realistic than, say, the example of the Twin Paradox. The effects of time dilation illustrated by the twin paradox have been tested, but not by sending a spaceship off at close to the speed of light and then comparing the ages of the twins some time later; such an execution of the experiment would be unrealistic, since we currently do not possess the capacity to accelerate bodies as massive as people to substantial fractions of the speed of light. Rather, the effects illustrated by this paradox have been tested through our ability to measure time differences to extraordinary precision using atomic clocks; the famous experiment by Hafele and Keating, in which an atomic clock was sent around the world on an airplane while leaving an identical clock on earth, confirmed this effect. Similarly, in the case of the masses on the spring, our ability to detect differences in inertia between cases where the spring is compressed and uncompressed will depend on our ability to measure differences in inertia very finely - as finely as time was measured on the atomic clocks in Hafele and Keating's experiment. I claim that if such a minute difference in inertia were to be detected between the two cases, the explanation of the object in the compressed case}.
Consider also a system whose total rest mass is the same as the total mass of the earlier system, but this time in which the mass is due entirely to the rest masses of its constituents, and not to the contributions of internal energy resulting from their interactions or motion. In such a case, the explanation as to why such a system behaves in Newtonian fashion will be of a different nature, since there is no need to explain why potential energy is not being released. In this case, the fact that the body isn't travelling too close to the speed of light (relative to some fixed inertial frame) and that there aren't excessively strong external forces acting on it should suffice to account for its Newtonian behavior. Thus, in each case, system-specific details concerning the nature of the internal binding forces (or lack thereof) within the bodies will play some role in the explanation of Newtonian behavior.

To take a more realistic example, consider a heavy atomic nucleus that follows an approximately Newtonian trajectory in a cyclotron (say, the circular trajectory of a moving charge in a constant magnetic field). The explanation as to why the trajectory of the nucleus is approximately Newtonian will depend in part on an explanation as to why the binding energy of the nucleus is not released, causing fragments of the nucleus to fly off at some significant portion of the speed of light. This explanation will depend on details of the internal constitution and binding of the nucleus in question and, in particular, on its half-life. Again, system-specific details play some role in the explanation of Newtonian behavior, precluding a single reduction that encompasses all Newtonian systems. 

By demonstrating the need to invoke system-specific details when attempting provide a complete explanation of Newtonian behavior in different systems, these examples illustrate that a reduction of NM to SR that is both complete and completely general - or, to keep to my terminology, completely systematic -  does not exist. However, given that the NM-SR reduction ought to have been at the outset the most likely example of completely systematic reduction, this particular case does not bode well for the widspread applicability of completely systematic approaches to reduction. 

Having demonstrated that no completely systematic account of the reduction of NM to SR, either of type 1 or type 2, is available, I will now turn to my second example, which also will be the focus of Chapter 2: the reduction of classical to quantum mechanics.

\subsubsection{Completely Systematic Approaches to the CM-QM Reduction}

If a completely systematic type-2 reduction of CM to QM is probably not possible, as argued in section \ref{Limit}, might a completely systematic type-1 reduction be? That is, for all physical systems whose behavior can be approximated by some classical model, is it possible to provide some overarching demonstration of the approximate accuracy of classical models that encompasses all of such systems? The answer again is patently no, but the arguments to this effect depend to some degree on what one thinks counts as a successful application of classical mechanics.

There is potentially a wide range of things one could mean by the term `classical.' From the point of view of quantum theory, the least stringent notion of classicality that we can adopt is apparent definiteness of the values of variables like position and momentum, absent any dynamical constraints on their evolution. As I argue later on, decoherence with respect to an appropriate pointer basis, combined with a solution to the measurement problem, will suffice to reproduce this attribute of classicality. However, people usually mean more than just apparent definiteness when they speak of the empirical success of classical mechanics; they also mean that certain dynamical constraints on the evolution of position and momentum variables are satisfied by the systems in question. Such dynamical constraints, in turn, come in different varieties. For example, we might require that the relevant variables obey Newton's Second Law of Motion, without placing any constraints on the allowable force laws that may appear in this law. Such behavior will include, among other things, behavior involving contact forces and friction such as the simple pendulum, mass on a spring, or normal force systems with and without friction. Also involving contact forces are the equations of classical fluid dynamics, such as the Navier-Stokes equation, which are derived from Newton's Second Law (combined with additional assumptions affecting the form of the contact forces involved). On the other hand, one could restrict one's notion of classical behavior further to systems described in terms of conservative forces, which can be characterised in terms of a simple classical potential - for example, the mass on a spring and the simple pendulum. And one could even further restrict one's attention to classical behavior which consists only of behavior that can be described in terms of fundamental force laws, such as electromagnetism, in which the conservative forces arise from fields rather than contact forces.

These different classes of systems, which encompass different models of classical behavior, will, of course, have different quantum mechanical underpinnings. For classical systems involving a fundamental force law, the potential that appears in the classical equation of motion will be the same as the one appearing in the quantum equation of motion. On the other hand, for classical systems which involve contact forces or friction, the classical potential, if one exists, the potential appearing the Schrodinger equation of the underlying quantum model will be extremely complicated and different from the potential that appears in the underlying microscopic quantum equations (it will likely only match the potential employed in the classical model in some average sense). To be more specific, consider two distinct systems described by the classical model of the harmonic oscillator: the first a mass on a spring and the second an electric charge moving in a tube bored through an axis of a uniform spherical charge distribution (in which case the electric field will vary linearly with distance from the center of the sphere) \footnote{I assume that the charge is sufficiently massive that energy losses through radiation can be neglected.}. In the second case, the classical potential generated by the electric field will be the same potential that appears in the underlying quantum model of the charge'es behavior. In the first case, the fact the that one can employ a harmonic oscillator potential to describe the motion of the block is something that needs to be explained in terms of the complex microscopic constitution of the spring - at the microscopic level, this potential will be wildly fluctuating on the length scale of the atoms making up the spring.

These two applications of the same classical model must be reduced separately, and no complete reduction can be given that encompasses both. Thus, a completely general, completely systematic, type-1 reduction of classical models of macroscopic systems cannot be given, especially if one adopts a relatively inclusive construal of what counts as classical behavior. Consideration of details specific to the system in question, or to the class of systems into which it falls, will be required for a totally comprehensive reduction of the classical to the quantum model of that system.

\subsection{Completely Piecemeal Approaches to Reduction} \label{Piecemeal}

The failure of completely systematic approaches to reduction, either of type 1 or of type 2,  might lead one to take the view that reductions must be performed in type-1, piecemeal fashion. A quote, again from David Wallace, suggests such a piecemeal approach:

\begin{quote}
\begin{singlespace}
Crucially: this `reduction,' on the instantiation model, is a local affair: it is not that one theory is a limiting case of another \textit{per se}, but that, \textit{in a particular situation}, the `reducing' theory instantiates the `reduced'  one. Consider the first example above, for instance. The reason that classical mechanics is applicable to the planets of the Solar System is not because of some \textit{general} [italics mine] result that classical mechanics is a limiting case of quantum mechanics. Rather, the \textit{particular system} [italics mine] under consideration - the solar system - is such that some of its properties approximately instantiate a classical-mechanical dynamical system. Others do not, of course: it is not that the solar system is approximately classical, it is that it (or a certain subset of its degrees of freedom) instantiate an approximately classical system.

... The real story of the relations between scientific theories is not a story about a tower of theories, with particle physics at the bottom and macroeconomics at the top: rather, it is a patchwork of domain-relative instantiations \cite{wallace2012emergent}.
\end{singlespace}
\end{quote}

\noindent According to Wallace, the fact that certain quantum systems approximately instantiate classical Newtonian systems - or rather, that, for certain systems, quantum models of those systems instantiate certain Newtonian models of those same systems - is not something that can be accounted for systematically by some general mathematical result, either involving a limit or, he seems also to suggest, of any other form, but something that must be explained on a piecemeal basis.

However, as I now argue, it is still possible to retain some measure of generality in our accounts of reduction, and we can do better than to provide reductions in a totally piecemeal, case-by-case fashion. Indeed, it would be surprising if the rather striking formal results relating the dynamical and kinematical structures of superseded and superseding theories in physics did not have $\textit{some}$ fairly widespread relevance to actual instances of reduction; to say, for instance, that the result that $\lim_{\frac{v}{c} \rightarrow 0}\frac{1}{\sqrt{1-\frac{v^{2}}{c^{2}}}} =1$, and the fact that plenty of relativistic equations return Newtonian ones as a result, has no widespread relevance to the emergence of NM behavior from SR is (for reasons which I take to be self-evident) utterly implausible. Limits or other general formal results, while they do not in themselves constitute complete reductions, often do lend significant insight into the mechanisms and principles that relate the high- and low-level theory models of many or all such systems. 

\vspace{5mm}

\subsubsection{Piecemeal Reduction vs. the Pluralism of Cartwright and Dupre}

Wallace's piecemeal approach to reduction may call to mind the pluralism of Cartwright and Dupre, who, like Wallace, take pains to underscore the patchwork nature of regularities described by scientific theories and to characterise these regularities as islands of order in a much vaster sea of irregularity. Yet the patchwork of Wallace's view differs dramatically in certain crucial respects from that of Cartwright and Dupre. Wallace's patchwork respects the hierarchical distinction between high- and low- level theories, with theories in physics at the bottom and those in chemistry, biology, and psychology successively further up; moreover, it respects the reducibility, in the sense specified by Wallace's concept of instantiation, of high- to low- level theories. Cartwright and Dupre's pluralist view, on the other hand, is strongly anti-reductionist in that it denies the reducibility of higher- to lower- level theories, and moreover opposes the very distinction between high- and low- level theories \cite{cartwright1999dappled}, \cite{DupreDisorder}. Thus, while both views commonly acknowledge the patchwork nature of scientific regularities, this is simply a reflection of the fact that both strive to grapple - in very different ways - with the same fact about about the way in which science describes nature.

As a consequence of the differences just cited, the reductionist and pluralist accounts may be further distinguished in terms of the way they characterise the relationship between the \textit{domains} of different theories.  In Wallace's patchwork, the domains of higher-level theories are contained in those of low-level theories; that is, systems that instantiate the laws or models of a higher-level theory also instantiate those of a lower-level theory (though the reverse is not generally true). 
%\footnote{In fact, one might regard the distinction between high- and low-level theories as . When we characterise a particular theory as high- or low- level, then, this may be taken simply as a reflection of our \textit{expectation}, pending detailed demonstation, that the domain of the high-level theory is contained in that of the low-level theory.}. 
Cartwright and Dupre's patchwork imposes no requirement that the domain of a high-level theory, say in biology, be contained in that of some low-level theory in physics. In this sense, the various patches making up the patchwork on the pluralist view are on more level footing than they are on the reductionist view; the domains of theories in physics and in biology are simply different on the pluralist view and the former are not required to contain the latter; in fact, the pluralist view explicitly requires that this kind of containment does not occur. Thus, Wallace's view will typically ascribe much larger domains to low-level theories than will the pluralist view of Cartwright and Dupre, since Wallace's view requires these domains to subsume those of higher-level theories while Cartwright and Dupre's denies this subsumption. 

This difference between the reductionist and pluralistic accounts of the patchwork nature of scientific regularities can in part be traced back to a difference in the degree to which they condone extrapolation from the observed success of scientific theories in the carefully controlled contexts where they are often tested, to their applicability in the much vaster, and typically much more complex, world outside of these contexts. Wallace, along with most of the scientific and philosophical communities, accepts the legitimacy of such extrapolations, implicitly taking them as a natural induction on the base of data extracted from experiments. 
%from those experimentally tractable contexts in which we can use our theories to make reliable predictions to those contexts where the complexity of the systems involved prohibits us at least in practice from making precise predictions about their behavior. 
For instance, while most of what is regarded as the confirming evidence for the Standard Model of particle physics is drawn from scattering experiments performed in particle accelerators, physicists typically assume that these laws also apply in contexts highly remote from particle accelerator experiments, such as occur in efforts to describe the evolution of the early universe.

Cartwright and Dupre, on the other hand, regard such inferences as far too cavalier and argue that we ought to be more reserved in our extrapolations. Cartwright, for instance, claims that the theories of physics and the other sciences apply only \textit{ceteris paribus}, under the carefully tuned conditions under which they are typically tested. Thus her view goes beyond mere \textit{skepticism} about claims, for instance, that biological systems fall within the domain of the Standard Model or any potential successor to the Standard Model; it centers on an explicit \textit{denial} of such claims. Thus, she argues that the patchwork of scientific practice most strongly supports a metaphysical picture in which it is not just the contexts in which we can apply our theories to make predictions that are disjointed, but nature itself.

In summary, one must not overlook the fact that the piecemeal approach to reduction is just that - an approach to \textit{reduction} - and thus, unlike Cartwright's view, will generally support the idea of higher-level regularities being reduced to lower-level ones. The kind of diversity that is supported by the piecemeal approach to reduction is only diversity in the sense of a single underlying theory providing different \textit{explanations} of different high-level phenomena, though in contrast to Cartwright and Dupre's pluralism, these different explantions may be given within the same underlying theoretical framework; Wallace's piecemeal approach to reduction is thus entirely compatible with the idea that there is a single underlying, universal theory of fundamental physics, and that the many disparate patches of higher-level regularity all fall within the domain of this one theory. Cartwright's view is explicitly anti-reductionist and anti-unfication, and the diversity that is suggested by Cartwright's view is diversity at a much deeper, metaphysical level, since it not only suggests different explanations for different higher-level regularities, but ultimately that the need for different explanations reflects a world that is dappled not only in terms of our ability to discern patterns in high-level phenomena, but fundamentally.

%Nancy Cartwright's view of scientific laws, according to which the regularities captured by different scientific theories, including the theories of fundamental physics, apply only within isolated patches, where the conditions are tuned just so as to bring these regularities about - that is, the laws of the special sciences and of fundamental physics in particular hold only \textit{ceteris paribus}, but not generally.  

%In summary, while both views emphasize the disparate, patch-like nature of higher-level regularities, the resemblance between the two views ends here. According to Cartwright, the most natural explanation for the disparate variety of concepts and laws that science uses to describe nature is that nature itself is inherently pluralistic or `dappled' \cite{cartwright1999dappled}, \cite{cartwright1983laws}. 

\subsection{Reduction Templates} \label{Templates}

%On the other hand, I should note that while the domain of any superseded physical theory can only be characterized as a patchwork, my point of view differs dramatically from Cartwright's in that it still supports the universality or ``imperialism" of physics at the fundamental level - or rather, supports the view that as physical theories proceed to progressively more fundamental scales, they become more universal, and the patches constituting their patchwork of their applicability both grow and merge. The world is not fundamentally dappled, as Cartwright suggests; it is only dappled at levels of regularity which are emergent.

%However, even those who acknowledge the universality of fundamental physics and reducibility of superseded theories to their successors nevertheless might take the position that each instance of reduction to the more fundamental theory must be explained on a case-by-case basis. That each success of the emergent theory must be reduced separately from others.

Although it is not possible to give a completely general, systematic account of how a superseded theory $T_{h}$ reduces to a superseding theory $T_{l}$, it is often possible to retain a significant degree of generality in explanations of why certain  systems modelled in $T_{l}$ also may be approximately modelled in $T_{h}$, and therefore to do better than to approach theory reduction on a completely piecemeal basis. The degree of generality that can be retained in such explanations depends strongly on the two theories in question, though in most if not all of the cases mentioned in the first paragraph of the Introduction to this chapter, is substantial. 

Given that any system in the class is described both by a high-level model $M_{h}$ and a low level model $M_{l}$, the deductive portion of the reduction consists of deriving the image model $M_{h}^{*}$ from $M_{l}$. Systems with models in $T_{h}$ can be paritioned into classes, such that reductions to $T_{l}$ of the $T_{h}$ models of systems in the same class follow the same `template,' and reductions of systems in different classes follow different templates. A template is an incomplete proof, or outline, of the basic steps and principles and mechanisms that are involved in deriving the image model $M_{h}^{*}$ from $M_{l}$. A given template may take for granted assumptions that can only be proven by considering the full particularities of the individual system in question, or that are merely plausible conjectures awaiting proof. Systems within a given reduction template may differ substantially in their $T_{l}$ description, though at a certain intermediate level of detail, the basic outlines of the reasoning that explains their approximate $T_{h}$ behavior are the same.

To a degree, the separation into reduction classes of systems in $T_{h}$'s domain is arbitrary, and depends in particular on how detailed the template associated with the class is. Ultimately, the most detailed possible explanation of why a particular $T_{l}$ system exhibits $T_{h}$ regularities will take account of things like the exact state of the system and its exact microscopic constitution; such an approach amounts essentially to cranking the relevant initial conditions through the appropriate equations of motion and then reading the information relevant to the $T_{h}$ level of description off from it. At this level, each template amounts to a completely detailed, rigorous proof of reduction, and each separate system has its own reduction template and is the sole member of its reduction class. However, such explanations in practice are never given because we are not capable of gaining this kind of detailed information about the systems we care about. Moreover, they tend to obscure the general principles and mechanisms at work across a wide range of instances of reduction. If, on the other hand, one is willing to sacrifice some detail in the template by making certain general, plausible assumptions about the system in question (which one can go back and try to prove rigorously later if one likes), then one may gain some insight into these mechanisms and principles and where they fit in to the overall scheme of the full reduction. Templates significantly reduce the labor involved in explaining particular instances of reduction by compartmentalising the derivation of the image model on the one hand into those parts that require specific reference to the system in question, and, on the other hand, those parts of the derivation that are uniform across the reduction class.

Thus, a template-based approach to reduction is one in which a variety of explanations of the approximate $T_{h}$  behavior of a particular system, varying according to level of detail, are possible. Explanations provided by more general, less detailed templates will apply across a correspondingly larger reduction class of systems; however, the derivations of the model of the systems in the reduction class may take a number of claims for granted, since proving these claims would require considering the particular details of different systems within the class. On the other hand, less general, more detailed templates will tend to take the particularities of individual systems into account, and therefore only apply over smaller reduction classes; the reduction classes of these more detailed templates should be subsets of the reduction classes corresponding to less detailed templates for the same system. 

Both kinds of templates, general and detailed, are necessary to a full understanding of reduction: the former because it illustrates the general principles and mechanisms at work in the reduction of the $T_{h}$ to the $T_{l}$ models of a wide range of systems, and the latter because it `fills in' or completes the more general template with a detailed demonstration of the assumptions taken for granted in the more general template. Thus, accounts of reduction provide the most insight not when they are given exclusively at the finest level of detail, nor when they are given exclusively at the greatest level of generality, but rather when they are given in stages, with earlier stages corresponding to  a template at the greatest level of generality and later stages corresponding to progressively more detailed templates, whose reduction classes narrow at each step to account for more details specific to the system under investigation. Each stage can be seen as combing over the same deductive path between $T_{l}$ and $T_{h}$, but each time in progressively finer detail, so that both the broad outlines and the fine details of the explanation can be understood.

%I claim offer the most insight when they are given in stages, starting from the most general, least rigorous explanation, involving the largest reduction classes, and then proceeding in progressively more detail to more fine-tuned explanations, which apply over progressively smaller reduction classes.

%and in which the detail and rigor of the explanation will tend to vary inversely with its generality. That is, if one wishes to provide an explanation of the mechanisms at work across a wide variety of similar reductions, one must sacrifice some rigor and detail to the account. On the other hand, a completely rigorous account of reduction requires that one take full account of the details of the system in question, and therefore such an account will lose any generality, applying only to the particular system in question.

Below, I return briefly to the two examples of the CM-QM reduction and the NM-SR reduction, and suggest in broad terms how a template-based approach might be applied to them.

\subsubsection{Reduction Templates for the NM-SR Reduction}

I argued above that the reduction of Newtonian mechanics to special relativity cannot be given a systematic formulation, either of type 1 or type 2. On the other hand, I claim that we can do better than to explain reductions of Newtonian behavior to SR on a totally piecemeal, case-by-case basis. 

Earlier, I argued that Newtonian systems with significant internal energy (e.g., binding energy comparable to their rest energy) will require a different explanation for their Newtonian behavior than Newtonian systems that do not have significant internal energy; on a template-based approach, these two sets of systems would belong to distinct reduction classes, where the reduction template for the former class will include some account of why this energy is not released. However, at a less detailed level of explanation, in which the rest of mass of composite bodies is taken as an assumption rather than as something to be explained, and the internal structure of these bodies not considered, these two sets of systems might belong to the same reduction class and the reduction of their NM models to their SR models follow the same reduction template. 

%I explained that reductions of Newtonian behavior to the Lorentz covariant description that underpins it would be qualitatively different depending on whether the force involved was a fundamental force such as electromagnetism, or a contact force. I claim simply that these two kinds of systems belong to distinct reduction classes, and that their behavior therefore is explained according to distinct reduction templates. 

\subsubsection{Reduction Templates in the CM-QM Reduction}

As in the case of the NM-SR reduction, I argued above that in the case of the CM-QM reduction, systematic reductions of type 1 and 2 are not available.

As in the NM-SR case, the failure of totally systematic accounts does not require us to consider all classical systems on a case-by-case basis. The largest, most general reduction class for the quantum-classical reduction might be simply those systems which appear to have definite values for properties such as position and momentum. As I argue below, decoherence combined with a solution to the measurement problem should suffice to explain the classicality - construed in this broad sense - of these systems.

However, we might identify further subclasses of this reduction class: in particular, the subclass of systems which approximately obey Newtonian equations of motion. Then we might identify further subclasses of this class corresponding to Newtonian systems involving contact forces and those involving fundamental forces. In the latter case the potential appearing in classical equations is the same as the potential appearing in the underlying quantum equations, while in the latter the two potentials are different (assuming these contact forces even admit description in terms of a classical potential). Proceeding further, we can distinguish subclasses of the `contact force' reduction class corresponding to fluid systems exhibiting regularities like the Navier-Stokes equation (which is derived from Newton's second law and thus counts as an instance of it application) and to systems like the simple pendulum. The microscopic origins of the phenomenological force laws employed in these two sets of cases will be substantially different. On the other hand, we can also identify subclasses of the `fundamental force' reduction class corresponding, say, to gravity and electromagnetism.

At the most fine-grained level, there will be a reduction class for every system and every state of every system (for some states may yield classical behavior while others don't). 
%The reduction template for such a class will amount to a completely detailed explanation of the system's classical behavior, and will entail evolving the initial condition of the system according to Schrodinger's equation and using this solution to identify its macroscopic regularities. 
Yet, as was suggested above, if one proceeds immediately to this level of description without first providing the more general reduction templates, the general principles and mechanisms which underlie the emergence of many instances of classical behavior, such as decoherence, Ehrenfest's theorem, and the compounding of micro into macro degrees of freedom, will be completely obscured.

In the following chapter, I will illustrate the application of a template-based approach to the CM-QM reduction, in particular by focusing on the reduction class consisting of classical systems involving only basic force laws like the gravitational and electrostatic forces.

\subsubsection{Template-based Reduction and the Patchwork Nature of Higher-Level Regularities}

Earlier, when discussing the piecemeal approach to reduction, I pointed to a superficial similarity between this view and Nancy Cartwright's `dappled' picture of physical laws: namely, that both view higher-level regularities as islands of order in a much vaster sea of irregularity. Yet, on further analysis, the fact they share this should come as no surprise, for both views are simply making an effort to come to terms with what is, after all, a fact about science that \textit{any} account of higher level regularities in science must come to grips with. Scientific theories \textit{do} often tend to operate in isolated patches; the laws of genetics do apply to living organisms, but not to the inorganic matter and energy and space and time that make up most of the cosmos; the Standard Model works well at predicting the results of scattering experiments, but none of us can use it to predict the population dynamics of badgers in the UK; the laws of circuit theory work well at predicting the currents and voltages in a circuit, but not for predicting the scattering cross sections for hadron collisions. No understanding of the relation between different sets of scientific regularities, whether they are characterized by different theories within a single science or theories between different sciences, can be plausible without somehow accommodating this fact.    

The template-based approach to reduction, I believe, accommodates the diversity and patchwork nature of higher-level regularities more effectively than does the piecemeal approach to reduction. Like the piecemeal approach, the template-based view accommodates the diversity of higher-level regularities by providing different explanations where necessary for different patches of regularity, but all within the framework of the same fundamental theory; on the template-based view the different explanations correspond to different reduction templates.  
Where the template-based approach surpasses the piecemeal approach is in the fact that the template-based approach provides a framework that facilitates the identification and emphasis of the general principles and mechanisms that apply across different patches of higher-level regularity, thereby enabling a more general, systematic understanding of the particular reductions under consideration than can be achieved by following the purely piecemeal approach. 
%Ultimately, a completely rigorous account of the reduction of a particular classical system will require that one take.

%However, we learn a great deal by considering the commonalities among different reductions, and the general principles which allow these reductions to take place. Thus, the subdivision into templates is extremely illuminating from a conceptual point of view.

\subsubsection{Reduction Templates and Multiple Realisation}

Template-based reduction is ideally suited to accommodate the fact of multiple realisation. On a DS approach, multiple realisability corresponds to the fact that more than one model of a theory $T_{l}$ can serve to reduce to the same model of $T_{h}$, or that there is more than one bridge map through which reduction can occur between a single pair of models $M_{l}$ and $M_{h}$, or that more than one low-level state $x^{l}$ may instantiate the same high-level state $x^{h}$ (I discuss multiple realisation in DS reduction further in the concluding chapter). The image models $M_{h}^{*}$ corresponding to these distinct realisations may be different, and so the templates for deriving these image models will be distinct at some fine-grained level of detail. However, it may happen that these distinct, more fine-grained templates, may have a sufficient number of steps in common that they can be seen as refinements of the same more general template. 

%\subsubsection{Reduction Templates and Kitcher's Schematic Arguments}

%Philip Kitcher's well-known unificationist account of explanation introduces the concept of an `argument pattern,' and because of certain superficial similarities, it may appear \textit{prima facie} that the notion of a reduction template  is simply an iteration of this idea. After briefly explaining the concept of an argument pattern, I argue that the two concepts are essentially different. 

\section{Summary}

Ultimately, the approach to physical reduction that I advocate in this dissertation can be classified as a dynamical systems approach, which is neither wholly systematic nor wholly piecemeal, but based instead on the use of reduction templates. In deducing an image model $M_{h}^{*}$ for some model $M_{h}$ of $T_{h}$ from a model $M_{l}$ of $T_{l}$, a complete proof may be difficult or practically impossible to come by in the case of highly complicated systems. As we will see in the next chapter, in such a case plausible but unproven assumptions about the system in question need to be made, and the deduction of $M_{h}^{*}$ within a particular domain of $M_{l}$ should be given in the form of a template rather than a complete proof. For simpler systems, it may be possible to provide templates that are sufficiently complete that they constitute complete proofs of the image model. 

In summary, the goal of template-based reduction as much to tell a story about the relationship between the alternative descriptions of the physical world offered by different physical theories as it is to give a proof of the approximate accuracy of the high-level theory's models on the basis of the low-level theory's models (the story, if told in sufficient detail, should amount to such a proof). Necessarily, given the semantic, dynamical systems approach taken here, as well as the specialisation to theories in physics, it will be a story told in the language of mathematics.

%Through a general discussion of two reductions, I have suggested a way of approaching reductions in physics that enables one to retain a degree of generality, so that one illustrates the general principles and mechanisms at work in these reductions without overstating the bounds of their applicability. Ultimately, the validity of this approach can only be supported by demonstrating that it fits with particular concrete cases of reduction, which I do in Chapters 2, 3 and 4. Specifically, in chapters 3 and 4, I hope to show that in explaining the emergence of classical field theoretic behavior from quantum field theory, and also in explaining the emergence of non-relativistic quantum theory from relativistic quantum field theory, one can employ different reduction templates to gain an understanding of the principles and mechanisms that underpin the reduction of a wide range of different classical or non-relativistic-quantum systems to their quantum and relativistic-quantum quantum counterparts, respectively. In both chapters, as in chapter 2, both the Everettian and Bohmian formulations of the relevant quantum theories will be considered. 

\chapter{The Classical Domain of Non-Relativistic Quantum Mechanics}
\label{ch2}

The first two theories that I consider as an application of dynamical systems, template-based reduction are Newtonian mechanics and nonrelativistic quantum mechanics. Specifically, I consider the reduction of a class of models of Newtonian mechanics describing $N$ macroscopic centers of mass interacting through a time-independent potential, to a class of corresponding models within Everettian and Bohmian quantum mechanics. Throughout, I refer to the Everett theory as the Bare/Everett theory as a reminder that, from a mathematical point of view, Everett's theory just is the bare formalism of quantum theory, prescribing unitary dynamics for a vector in a Hilbert space without collapse. I do not address more abstract algebraic approaches to quantum theory in this thesis; the interested reader can consult, for example, Landsman's \cite{landsman1998mathematical}, or \cite{landsmanClassicalQuantum}. 

In section \ref{MeasProb}, I discuss the measurement problem as it relates to the reduction of classical to quantum mechanics, and my reasons for considering the Everett and Bohm theories in parallel.  In section \ref{Models}, I present the model of Newtonian mechanics to be reduced, and the models of Everettian and Bohmian quantum mechanics to which I reduce it. In section \ref{Decoherence}, I describe the basic mechanisms of decoherence, measurement and effective wave function collapse in Bohmian and Everettian quantum mechanics, including a review of the decoherent histories framework in the Schrodinger picture. In section \ref{CMEverett}, I provide a template for the DS reduction of the model of Newtonian mechanics to the corresponding model in the Bare/Everett theory. Finally, in section \ref{CMBohm}, I provide a template for the DS reduction of the model of Newtonian mechanics to the corresponding model in the Bohm theory.

\section{The Measurement Problem} \label{MeasProb}

Given the realist background of this thesis, any attempt to reduce classical to quantum theory must take account of the quantum measurement problem. For, it is only through some resolution to this problem that the connection between the microscopic indeterminacy (relative to familiar variables such as position and momentum) of quantum theory and the apparent macroscopic determinacy, and determinism, of classical theory can be elucidated according to the demands of the realist. It is for this reason that, in considering the reductions that I do, I have adopted two proposed resolutions to the measurement problem, the Everett and Bohm theories. 

However, I hope that my analysis will contain points of interest for readers skeptical of the Everett and Bohm theories, or of realist approaches to quantum theory more generally. As Wallace has argued at length, Everett's theory, from a mathematical point of view, \textit{is} just the bare formalism of quantum theory without collapse \cite{wallace2012emergent}. The distinction between the two, as far as usage goes, comes from the added points of metaphysical and epistemological interpretation that the Everett theory attaches to the bare formalism. The most controversial claim of Everett's theory, that each branch of the total wave function, as defined by the requirement of decoherence, corresponds to an independent `world' with its own determinate reality, emerges as a consequence of taking the theory's mathematics seriously as a guide to the structure of the physical world; `taking the math seriously,' on this view,  entails not inserting \textit{ad hoc} exceptions to the theory's laws for purposes of agreeing with the experimental data, as advocates of more traditional positivist or operationalist approaches, such as the famous Copenhagen Interpretation, are often accused of doing. Since much of the thesis concerns the bare formalism of quantum theory, and any successful interpretation of quantum theory is likely to incorporate this formalism in some fashion or other, the skeptic about Everett and Bohm may still find a few pieces of pertinent material in the pages to follow. This material will be concentrated primarily in the sections pertaining to the Bare/Everett theory.

\subsection{Motivation for Considering the Bare/Everett and Bohm Theories in Parallel}

One motivation for considering the Everett and Bohm theories together is that, if one is going to attempt to effect these reductions within the context of the Bohm theory, it is necessary anyway first to effect them within the context of the Bare/Everett theory (indeed, the project of this thesis grew out of my initial investigations into the classical domain of Bohm's theory). The reason for this is that in the Bohm theory, the dynamics associated with the added structure of the theory - namely, the guidance equation for the added variables, designated `beables'  by John Bell (of the famous Bell Inequality), one the Bohm theory's foremost proponents - depends on the value of the quantum state, but the value of the quantum state does not depend on the dynamics or configuration of the additional variables. Thus, in order to assess the behavior of the beables it is necessary, at least in a formal mathematical sense, to go through the Bare/Everett theory in determining the unitary evolution of the wave function. Where Everettians and Bohmians disagree, primarily, is on whether the structure associated with the wave function is sufficient to save the appearances, and, if not, on whether the additional structure of Bohm's theory does enable the theory to save the appearances. 

Brown and Wallace have argued that Bohm's theory is Everett's theory `in denial,' in the sense that Everett's theory already contains all of the necessary mathematical structure to save the appearances, and so the additional configurations of Bohm's theory are therefore merely epiphenomenal `idle wheels.' However, their argument relies on the presumption that Everett's theory does indeed save the appearances - by no means a consensus opinion. Accepting Brown and Wallace's point that, if Everett's theory does indeed save the appearances, Bohm's additional configurations are superfluous, I nevertheless maintain a consideration of Bohm's theory out of recognition of the possibility that Everett's theory may, for one reason or other, fail to save the appearances, and that Bohm's additional configurations may offer the mechanism needed to address its shortcomings. The criticism that is currently the source of most informed skepticism about Everett's theory is that it cannot adequately explain the role of probability - specifically, the success of the Born, or $|\psi|^{2}$, Rule - in ordinary quantum mechanics. Deutsch and Wallace have proposed a derivation of the Born Rule from the principles of rational decision theory, which has been notably defended by Greaves \cite{deutsch1999quantum}, \cite{wallace2007quantum}, \cite{greaves2004understanding}. For further discussion of the `Everett in denial' charge against the Bohm theory, the reader should consult, in particular, the exchange between Brown/Wallace and Valentini \cite{r2005solving}, \cite{valentini2008broglie}.

In section \ref{Decoherence}, I provide a brief summary of the accounts of measurement offered by the Bare/Everett and Bohm theories. 

\section{The Models} \label{Models}

In this section, I describe the detailed structure of the models that I consider in the context of the CM/QM reduction, specifying the state space $S$ and dynamical map $D$ of each.  

Consider a system consisting of $N$ macroscopic centers of mass interacting through a simple potential. Newtonian mechanics, Everettian quantum mechanics and Bohmian quantum mechanics all provide different descriptions of this system. 

\subsection{The Newtonian Model}

Consider a classical system modelled in the classical Hamiltonian framework, consisting of $N$ centers of mass with positions and momenta $\{(X_{1},P_{1}, ...,X_{N}, P_{N}) \}$ and masses $\{ M_{1},...,M_{N}\}$, with the standard Hamiltonian $H= \sum_{i} \frac{P_{i}^{2}}{2M_{i}} + V(X_{1},...,X_{N})$. The state space is classical N-particle phase space, 

\

\noindent \underbar{\textit{State Space}}

\

\begin{equation}
S = \Gamma_{N}
\end{equation}

\noindent To condense the notation, I will write $(X,P) \equiv (X_{1},P_{1}, ...,X_{N}, P_{N})$, and $F(X,P)\equiv F(X_{1},P_{1}, ...,X_{N}, P_{N}) $. The first order dynamical equations of evolution are Hamilton's equations:

\

\noindent \underbar{\textit{Dynamics}}

\

\begin{equation}
\begin{split}
& \frac{d X_{i}}{dt} = \frac{\partial H}{\partial P_{i}} = \frac{P_{i}}{M_{i}}  \\ 
& \frac{d P_{i}}{dt} = - \frac{\partial H}{\partial X_{i}} = - \frac{\partial V}{\partial X_{i}}. 
\end{split}
\end{equation}

\noindent Together, these reproduce Newton's Second Law:

\begin{equation}
M_{i} \frac{d^{2}X_{i}}{dt^{2}} = - \frac{\partial V}{\partial X_{i}} 
\end{equation}

\noindent Alternatively, as we have seen, the phase space dynamics can be prescribed in terms of the dynamical map $\big( X(t),P(t) \big) = D \big[ t; (X_{0},P_{0}) \big] = \big(e^{\{\circ,H\}t}X \big|_{X_{0},P_{0}},e^{\{\circ,H\}t}P \big|_{X_{0},P_{0}}\big)$.

\subsection{The Everett/Bare-QM Model}

In the context of a bare-QM/ Everettian picture, the quantum mechanical model of $N$ particles in isolation from their environment has as its state space the Hilbert space of $N$ spinless particles:

\

\noindent \underbar{\textit{State Space w/o Environment}}

\

\begin{equation}
S = \mathcal{H}_{S},  
\end{equation}

\noindent where the subscript $S$ denotes the system consisting of the central macroscopic degrees of freedom under consideration. The first-order dynamics of the model are specified by the Schrodinger equation:

\

\noindent \underbar{\textit{Dynamics w/o Environment}}

\

\begin{equation}
i \frac{\partial }{\partial t} | \psi \rangle = \hat{H}_{S} | \psi \rangle,  
\end{equation} 

\noindent where $\hat{H}_{S} = \sum_{i} \frac{\hat{P}_{i}^{2}}{2M_{i}} + V(\hat{X}_{1},...,\hat{X}_{N})$ and $| \psi \rangle \in \mathcal{H}_{S}$. 

\vspace{5mm}

More realistic models of macroscopic systems incorporate environmental degrees of freedom and their interaction with the macroscopic degrees of freedom whose classicality we wish to explain. The state space of this model is

\

\noindent \underbar{\textit{State Space w/ Environment}}

\

\begin{equation}
S = \mathcal{H}_{S} \otimes \mathcal{H}_{E},  
\end{equation}

\noindent where the subscript $E$ denotes the environmental degrees of freedom, consisting of any degrees of freedom external to the centers of mass (which may include internal degrees of freedom of the bodies in question in addition to degrees of freedom not included in these bodies). The dynamics on the combined Hilbert space of the macro- and micro- degrees of freedom is determined by the Schrodinger equation,

\

\noindent \underbar{\textit{Dynamics w/ Environment}}

\

\begin{equation}
i \frac{\partial }{\partial t} | \chi  \rangle = \big( \hat{H}_{S} + \hat{H}_{E} + \hat{H}_{I} \big) | \chi \rangle,
\end{equation}

\noindent where again $\hat{H}_{S} = \sum_{i} \frac{\hat{P}_{i}^{2}}{2M_{i}} + V(\hat{X}_{1},...,\hat{X}_{N})$, $\hat{H}_{E}$ denotes the environmental Hamiltonian, $\hat{H}_{I}$ the Hamiltonian governing the interaction between the macroscopic degrees of freedom and the environment, and $| \chi \rangle \in \mathcal{H}_{S} \otimes \mathcal{H}_{E}$. 

In the following discussion, I leave the forms of $\hat{H}_{I}$ and $\hat{H}_{E}$ unspecified, though more detailed models, such as the well-known Caldeira-Legett model, do specify the forms of these. My choice to leave the forms of $\hat{H}_{I}$ and $\hat{H}_{E}$ reflects the generality of the template that I seek to provide for the reduction of $CM$ to $QM$. Accounts of macroscopic classical behavior within the context of more specific models, like the Caldeira-Legett model, correspond to templates which, though more complete as demonstrations of reduction, are correspondingly less general.

\subsection{The Bohm Model}

The Bohmian model of a system of $N$ particles incorporates all of the mathematical structure of the Bare/Everett model, but adds an additional component to the state space and to the dynamics. In addition to a quantum state residing in a Hilbert space and the corresponding dynamics, the Bohm theory posits the existence of a spatial configuration for each of the $N$ particles in the system, along with an accompanying dynamics for this configuration. The state space thus consists of two spaces: a Hilbert space, and a configuration space. For purposes of concision, the state space can be regarded as the Cartesian product of the two:

\

\noindent \underbar{\textit{State Space w/o Environment}}

\

\begin{equation}
S =  \mathcal{H}_{S} \times \mathbb{Q}_{S}
\end{equation}

\noindent where, again, the subscript $S$ denotes the system consisting of the central macroscopic degrees of freedom under consideration. The first-order dynamics of the model are given by the first order (in time) equations:

\

\noindent \underbar{\textit{Dynamics w/o Environment}}

\

\begin{equation}
\begin{split}
&i \frac{\partial }{\partial t} | \psi \rangle = \hat{H}_{S} | \psi \rangle, \\
& \frac{dQ_{i}}{dt}= \frac{1}{M_{i}} \nabla_{i} S(X) \ \big|_{Q}
\end{split}
\end{equation}

\noindent where, again, $\hat{H}_{S} = \sum_{i} \frac{\hat{P}_{i}^{2}}{2M_{i}} + V(\hat{X}_{1},...,\hat{X}_{N})$, $| \psi \rangle \in \mathcal{H}_{S}$,  $Q_{i} \in \mathcal{Q}_{S}$ and $\psi(X,t) \equiv \langle X | \psi (t) \rangle \equiv R(X,t) e^{iS(X,t)}$.

In addition, the Bohm theory as originally formulated by Bohm was stipulated to include an additional constraint, that the epistemic probability distribution over possible initial configurations $Q_{0}$ is the Born Rule distribution  $|\langle X | \psi_{0} \rangle|^{2}$. The combined dynamics of the configuration and the quantum state possess a property known as equivariance that ensures that if this is the case, the probability over configurations at any later time $t$ is $|\langle X| \psi(t) \rangle|^{2}$. However, work by Valentini and Westman, and Valentini, Russell and Towler has argued that it may not be necessary to postulate the Born Rule distribution, but that instead this distribution - or rather an arbitrarily close approximation to it - can be explained as a consequence of the dynamics of the Bohm theory, in that these dynamics carry an arbitrary initial distribution, after coarse-graining, into a distribution that very closely approximates the Born Rule distribution \cite{valentini2005dynamical}, \cite{towler2012time}. 

\vspace{5mm}

More realistic models of such systems incorporate environmental degrees of freedom and their interaction with the macroscopic degrees of freedom whose classicality we wish to explain. The state space of this model is

\

\noindent \underbar{\textit{State Space w/ Environment}}

\

\begin{equation}
S = \big( \mathcal{H}_{S} \otimes \mathcal{H}_{E} \big) \times \big( \mathbb{Q}_{S} \oplus  \mathbb{Q}_{E}  \big),  
\end{equation}

\noindent where the subscript $E$ denotes the environmental degrees of freedom, consisting of any degrees of freedom external to the centers of mass (which may include internal degrees of freedom of the bodies in question in addition to degrees of freedom not included in these bodies). The dynamics of the quantum state and Bohmian configuration are determined by the first-order equations,

\

\noindent \underbar{\textit{Dynamics w/ Environment}}

\

\begin{equation}
\begin{split}
& i \frac{\partial }{\partial t} | \chi  \rangle = \big( \hat{H}_{S} + \hat{H}_{E} + \hat{H}_{I} \big) | \chi \rangle, \\
& \frac{dQ_{i}}{dt}= \frac{1}{M_{i}} \nabla_{X_{i}} S(X,y) \ \big|_{Q,q} \\
& \frac{dq_{j}}{dt}=\frac{1}{m_{j}} \nabla_{y_{j}} S(X,y) \ \big|_{Q,q},
\end{split}
\end{equation}

\noindent where $| y \rangle $ is a position eigenstate of $\mathcal{H}_{E}$, $\langle X,y| \chi \rangle = R(X,y) e^{iS(X,y)}$, $Q_{i} \in  \mathbb{Q}_{S}$ and $q_{j} \in  \mathbb{Q}_{E}$. Moreover, the center of mass Bohmian configuration $Q_{s}$ is defined by the relation $Q_{S,i} = \frac{\sum_{j} m_{j} q_{j}}{\sum_{j}m_{j}}$, where the sum is over all microscopic fundamental (from the perspective of Bohmian NRQM) particles contained in the macroscopic body in question. Rimini and Peruzzi discuss the conditions under which the center of mass Bohmian configuration $Q_{S}$ obeys a guidance equation of the usual form prescribed above \cite{peruzzi2000compoundation}.

\subsection{A Few Comments About the Models}

For a reduction of macroscopic Newtonian behavior to either Everettian or Bohmian quantum theory to be complete, the class of system-environment models for the quantum theories described above, in which the degrees of freedom of the central system $S$ (also known as the `relevant' degrees of freedom) are centers of mass of macroscopic bodies,  must be derived from a more fundamental model in which all of the degrees of freedom are microscopic - since, after all, the center of mass degrees of freedom are simply weighted averages over the microscopic degrees of freedom of the bodies in question. Deriving these models from the microscopic models is an important part of a complete reduction of macroscopic classical behavior to microscopic classical behavior, but one that is likely to involve system-specific details regarding the specific material consitution of the body in question the particular binding interactions that join the microscopic particules into a single macroscopic body. Again, more detailed templates should be provided to fill in the gaps in the template provided here. 

Moreover, given that no environment is included in the Newtonian model, some explanation as to why it has been incorporated into the quantum model of the same system should be provided. Briefly, the answer is that, as a result of quantum entanglement, external degrees of freedom which exert a negligible effect on the degrees of freedom in a Newtonian model - such as very tiny particles, or electromagnetic radiation - and therefore can be left out of the model, can have a profound effect on the dynamics of these degrees of freedom in the context of the quantum models. We shall see this in more detail in the coming sections.

\section{Decoherence, Measurement and Effective Wave Function Collapse} \label{Decoherence}

A quantum measurement on a subsystem $S$ of closed system $SAE$ (where $A$ constitutes any measuring apparatus that may be present and $E$ the microscopic environment) in a pure state is a unitary, dynamical process on $SAE$'s Hilbert space that establishes a particular kind of correlation between the degrees of freedom of $S$ and the degrees of freedom external to $S$. The presence or absence within $E$ of human observers is, on the following exposition, immaterial to the physical description of the measurement process, which, like any other physical process, is modelled throughout as a unitary evolution on a Hilbert space. The process whereby such correlations are established in measurements is an example of the more general phenomenon of quantum decoherence. In this section, I review the concepts of quantum measurement and decoherence, explaining how they are manifested in the particular contexts of the Everett and Bohm theories, and how they give rise to the appearance of effective wave function collapse in these theories. In the process, I emphasise, following Maroney and Hiley, that the sort of decoherence that suffices to produce effective collapse in Everett's theory does not suffice to do so in Bohm's; a more specific kind of decoherence is required to induce effective collapse in Bohm's theory. 

I begin with a brief, preliminary mathematical review of projection valued measures (PVM's) and positive operator valued measures (POVM's), since these notions underpin much of modern quantum measurement theory, as well as the decoherent histories framework. 

\subsection{PVM's and POVM's} \label{PVM}

Given a Hilbert space $\mathcal{H}$, a projection valued measure (PVM) on that Hilbert space is a set of operators $\{\hat{P}_{i} \}$ on $\mathcal{H}$ satisfying the following criteria:

\begin{itemize} 
\item $\sum_{i} \hat{P}_{i} = \hat{I}$ 
\item $\hat{P}_{i} \hat{P}_{j} = \delta_{ij} \hat{P}_{j} $ (no sum over like indices). 
\end{itemize}

\noindent Any Hermitian operator $\hat{A}$ on $\mathcal{H}$ can be decomposed in some PVM, according to the operator's spectral decomposition:

\begin{equation}
\hat{A} = \sum_{i} a_{i} \hat{P}_{i}.
\end{equation}

\noindent Different possible measurement outcomes correspond in the bare quantum formalism to operators $\hat{P}_{i} $, and, if the state of the system is $| \psi \rangle$, the probability of the outcome $a_{i}$ on measurement is $\langle \psi | \hat{P}_{i}  |  \psi \rangle$. Of course, on a realist interpretation of quantum theory such as the Everett or Bohm theory, the sense in which the $\hat{P}_{i}$ correspond to determinate outcomes must be elaborated, and the fact that the probability of the outcome $a_{i}$ is $\langle \psi | \hat{P}_{i}  |  \psi \rangle$, must be demonstrated. 

An example of a PVM is the set  $\{ \int_{\Delta_{i}} dX | X \rangle \langle X | \}$, where $\Delta_{i}$ form a partition of the configuration space, and $| X \rangle$ are position eigenstates. Another is the set $\{ \int_{\Omega_{i}} dP | P \rangle \langle P | \}$, where $\Omega_{i}$ form a partition of the momentum space, and $| P \rangle$ are momentum eigenstates.

\vspace{5mm}

Given a Hilbert space $\mathcal{H}$, a positive operator valued measure (POVM) on that Hilbert space is a set of operators $\{\hat{\Pi}_{i} \}$ on $\mathcal{H}$ satisfying the following criteria:

\begin{itemize} 
\item $\sum_{i} \hat{\Pi}_{i} = \hat{I}$ 
\item $\langle \psi| \hat{\Pi}_{i} | \psi \rangle \geq 0$ for all $| \psi \rangle \in \mathcal{H}$, for all $i$. 
\end{itemize}

\noindent As in the case of PVM's, each operator $\hat{\Pi}_{i}$ corresponds to a measurement outcome or to a range of such outcomes, the probability of which is $ \langle \psi |\Pi_{i}  |\psi \rangle$  when the system is in state $|\psi \rangle$. As Wallace notes in \cite{wallace2012emergent}, the first criterion ensures that the probabilities of all the outcomes sum to $1$; the second ensures that the probability of each outcome is a positive number. Note that any PVM is a POVM, since $\langle \psi | \hat{P}_{i}  |  \psi \rangle =  \langle \psi | \hat{P}_{i}^{2}  |  \psi \rangle$, and all of the eigenvalues of  $\hat{P}_{i}^{2}$, which are the squares of the eigenvalues of $\hat{P}_{i}$, must be non-negative; in fact, for elements of a PVM, all eigenvalues of a $\hat{P}_{i}$ are either $0$ or $1$.

An example of a POVM is the set $\{ \int_{\Sigma_{i}} dX dP \ | X,P \rangle \langle X,P | \}$, where $\Sigma_{i}$ form a parition of the phase space and $ | X,P \rangle$ is a coherent state of fixed width centered about the configuration $X$ and the momentum $P$ \cite{wallace2012emergent}. For our purposes, it will suffice to identify a coherent state $ | X,P \rangle$ with a Gaussian wave packet of the form

\begin{equation}
\langle x | X,P \rangle = \frac{1}{L^{1/2} \pi^{1/4}} e^{-iPx} e^{\frac{-(x-X)^{2}}{2L^{2}}}
\end{equation}

\noindent that is narrowly peaked in both position and momentum and such that the uncertainties in position and momentum $\Delta x$ and $\Delta p$ satisfy the minimum uncertainty condition $\Delta x \Delta p=\frac{1}{2}$ (the constant $L$ enforces this condition). The coherent state $ | X,P \rangle$ is sometimes defined as an eigenstate of an annihilation operator $\hat{a} = \frac{1}{\sqrt{2}}(\sqrt{m \omega} \hat{X} + \frac{1}{\sqrt{m \omega}} \hat{P} ) $ with some complex eigenvalue $\alpha$ and for some $m$ and $\omega$ (one can check that the wave packet state given is indeed an eigenstate of this operator), though it will suffice here to understand by a coherent state simply a minimum uncertainty wave packet narrowly peaked both in position and in momentum.

In general, the elements of this POVM will not be mutually orthogonal and so will not constitute a PVM. However, if the cells $\Sigma_{i}$ of the phase space partition have dimensions in position and momentum that are significantly larger than the position and momentum widths of the coherent states $ | X,P \rangle$, the POVM elements $\hat{\Pi}_{i} \equiv \int_{\Sigma_{i}} dX dP \ | X,P \rangle \langle X,P |$ form an approximate PVM, since

\begin{equation} \label{PhaseSpacePVM}
\hat{\Pi}_{i} \hat{\Pi}_{j} \approx \delta_{ij}\hat{\Pi}_{i},
\end{equation}

\noindent thereby ensuring approximate orthogonality of the projectors in the POVM. 

For a much more thorough account of PVM's and POVM's, and of quantum measurement more generally, the reader should consult Busch, Lahti and Mittelstaedt's excellent monograph, `The Quantum Theory of Measurement' \cite{BuschQTM}.

\subsection{Quantum Measurement} \label{Measurement}

Let $S$ be a subsystem consisting of the degrees of freedom we wish to measure, $A$ a subsystem consisting of a measuring apparatus, and $E$ the degrees of freedom external to $S$ and $A$ (e.g., air molecules, photons, etc.) which I shall call the `environment.' Ignore for the moment any potential interaction or entanglement with the environment and let us simply consider the interaction between $S$ and $A$, which I assume for the moment to be unitary. If $A$ is to be an effective measuring apparatus, then it should be the case that if the states $\{ | s_{i} \rangle \}$ constitute a basis for $S$ associated with some observable of $S$, and $| a_{r} \rangle$ is some initial `ready' state of the apparatus then over the time of the measurement interaction between $A$ and $S$

\begin{equation}
| s_{i} \rangle \otimes | a_{r }\rangle \rightarrow | \theta_{i} \rangle \otimes | a_{i} \rangle,
\end{equation}

\noindent where $\langle a_{j}| a_{i} \rangle \approx 0$ for $i \neq j$ (if i is a discrete index) or for $i$ sufficiently different from $j$ (if $i$ is a continuous index) and $| \theta_{i} \rangle$ is an arbitrary state of $A$. If $| \theta_{i} \rangle=| s_{i} \rangle$, then the measurement is classified as a `nondisturbing' measurement and the states $| s_{i} \rangle$ are called `pointer states'; otherwise, it is classified as a `disturbing' measurement (the paradigmatic example of a disturbing measurement is an ideal photon measurement, in which the $ |\theta_{i} \rangle$ would be the vacuum state of the electromagnetic field for every $i$ \cite{de2002foundations}). The first analysis of measurement in the nondisturbing case, in which the apparatus was itself treated quantum mechanically, was famously given by von Neumann in his seminal work \cite{vonNeumann1932}. \footnote{Note that the term `pointer state' is sometimes used to refer to states of the measuring device - typically some `pointer' on the measuring device - rather than of the measured system itself.}

It follows from the linearity of the evolution that if the system $S$ starts out in a coherent superposition $\sum_{i} c_{i} | s_{i} \rangle$, then the measurement interaction induces the evolution

\begin{equation}
(\sum_{i} c_{i} | s_{i} \rangle) \otimes | a_{r }\rangle \rightarrow \sum_{i} c_{i} | \theta_{i} \rangle \otimes | a_{i} \rangle,
\end{equation}

\noindent where again the states $| a_{i} \rangle$ of $A$ that become correlated to the different basis states $| s_{i} \rangle$ are mutually orthogonal. 

Let us now incorporate the environment into the analysis, allowing for interaction between the environment $E$ and the system $SA$ and assuming that the total system $SAE$ is in a pure state that always evolves unitarily according to the Schrodinger equation \footnote{In typical cases of quantum measurement, the system $S$ will be microsopic and $A$ macroscopic, so it will often only be $A$ that interacts directly with the environment.}. If the states $| \theta_{i} \rangle \otimes | a_{i} \rangle$ are such that they suffer minimal entanglement with the environment under interaction with the environment, then this interaction will induce the evolution

\begin{equation}
| \theta_{i} \rangle \otimes | a_{i} \rangle \otimes | E_{0} \rangle  \rightarrow | \theta_{i} \rangle \otimes | a_{i} \rangle \otimes | E_{i} \rangle,
\end{equation}

 \noindent where, for reasons that I discuss further below in the section on environmental scattering,  $ \langle E_{j} | E_{i} \rangle \approx 0$ for $i \neq j$ (if i is a discrete index) or for $i$ sufficiently different from $j$ (if $i$ is a continuous index). Thus, again by linearity, if $S$ begins in the initial pure state $\sum_{i} c_{i} | s_{i} \rangle$ we can expect the measurement interaction to induce the following evolution:
 
 \begin{equation}
 (\sum_{i} c_{i} | s_{i} \rangle) \otimes | a_{r }\rangle  \otimes  | E_{0} \rangle  \rightarrow \sum_{i} c_{i} | \theta_{i} \rangle \otimes | a_{i} \rangle \otimes | E_{i} \rangle,
 \end{equation}
 
 \noindent where again $\langle a_{j}| a_{i} \rangle \approx 0$ and  $ \langle E_{j} | E_{i} \rangle \approx 0$ for $i \neq j$ (if i is a discrete index) or for $i$ sufficiently different from $j$ (if $i$ is a continuous index). Typically, because $E$ contains many (often on the order of $10^{23}$ or more) degrees of freedom, this process whereby the states $| E_{i} \rangle$ correlated to the states $| s_{i} \rangle$ become mutually orthogonal - which is widely referred to as `decoherence' - will be effectively irreversible, at least on timescales short of quantum-mechanical Poincare recurrence times (see, for example, \cite{bocchieri1957quantum} for a discussion of Poincare recurrence in quantum mechanics).
 
In cases where it is only the system $A$ that interacts directly with $E$, the states $| a_{i} \rangle$ that undergo minimal entanglement with environment are sometimes also referred to as pointer states, and as we shall see are typically constrained to be localised in both position and momentum as a result of environmental scattering. 
 
 \vspace{5mm}
 
On the Everett interpretation, a measurement requires both orthogonality of apparatus states ($\langle a_{j}| a_{i} \rangle \approx 0$)  and decoherence:
 
 \begin{equation} \label{DecoherenceEverett}
\textbf{Decoherence:} \  \langle E_{j}|E_{i} \rangle \approx 0 \  \text{for} \ i \neq j. 
\end{equation} 
 
 \noindent The irreversible process of decoherence is conventionally regarded by Everettians as inducing effective wave function collapse.

Measurement, and efective collapse of the quantum state, in Bohm's theory occurs only if the sets of states and $\{ |a_{i} \rangle \}$ and $\{ |E_{i} \rangle \}$ are not just orthogonal but specifically non-overlapping in configuration space - that is, only if $ \langle a_{j}|y \rangle \langle y|a_{i} \rangle \approx 0 \  \ \ \forall y \ \in \mathbb{Q}_{A} \ \text{for} \ i \neq j$ and 

\begin{equation} \label{DecoherenceBohm}
\textbf{Configuration Space Decoherence:} \langle E_{j}|y \rangle \langle y|E_{i} \rangle \approx 0 \  \ \ \forall y \ \in \mathbb{Q}_{E} \ \text{for} \ i \neq j.
\end{equation}

\noindent Note that this condition entails $\ref{DecoherenceEverett}$, but is not entailed by it. The condition that $\{ |a_{i} \rangle \}$ be non-overlapping in $A$'s configuration space $\mathbb{Q}_{A}$ serves to ensure that the Bohmian configuration of $A$ becomes appropriately correlated to the state of $S$, while the condition that $\{ |E_{i} \rangle \}$ be non-overlapping in $E$'s configuration space $\mathbb{Q}_{E}$ serves to ensure not only that the Bohmian configuration of the environment $E$ becomes appropriately correlated to the states of $A$ and $S$, but also that this process is efffectively irreversible. Note that what I have called `configuration space decoherence,' Maroney and Hiley elsewhere has called `superorthogonality,' also identifying it as the necessary condition for effective collapse in Bohm's theory (see \cite{MaroneyDensity} and \cite{bohm1995undivided}); in addition, Bohm, in his original  account of quantum measurement in pilot wave theory, emphasises the need for wave packets to be disjoint in the beable configuration space in order to induce effective collapse (though he does not employ the term `beable,' which was later coined by Bell) \cite{bohm1952suggestedii}. If (\ref{DecoherenceBohm}) is satisfied, the branches of the total quantum state will have disjoint configuration space supports and only one branch $|s_{i} \rangle \otimes |a_{i} \rangle \otimes | E_{i}\rangle $ will govern the evolution of the total Bohmian configuration of $SAE$; all other branches can be disregarded, although they are still present, and the state has in this sense effectively collapsed onto a single branch.

Thus, in bare QM, a quantum measurement generally establishes a correlation between the quantum state of system $AE$ at some time after the measurement interaction and the state of system $S$ at some time \textit{prior} to the measurement interaction (in measurements of the first kind, the states of $B$ will in addition be correlated with the states of $A$ after the measurement). In the Bohm theory, a quantum measurement also does this, but additionally, and more importantly, establishes a correlation between the \textit{configuration} of system $AE$ after the measurement interaction and the quantum state of system $S$ at some time prior to the measurement interaction.

%If the number of microscopic degrees of freedom in $B$ is macroscopically large $(\sim 10^{23})$, the effectivecollapse process, whereby either condition (\ref{DecoherenceEverett}) or condition (\ref{DecoherenceBohm}) comes to be satisfied, will often be irreversible (up to times short of the quantum mechanical Poincare recurrence time of the system; see \cite{bocchieri1957quantum} for details). The timescales on which either relation \ref{DecoherenceEverett} or \ref{DecoherenceBohm} comes to be satisfied are, in realistic models where the number of degrees of freedom in $B$ is macroscopically large, extremely short. 

\subsubsection{A Few Points About Measurement in Bohm's Theory}

In the Bohmian model of quantum measurement, $AE$'s configuration $q_{AE}$, which lies in the region of $AE$'s configuration space where the value of $\phi_{i}(y) \equiv \langle y | a_{i},E_{i} \rangle$ ($| a_{i},E_{i} \rangle \equiv | a_{i}\rangle \otimes |E_{i} \rangle$ and $y$ denotes position in the total configuration space of $AE$ ) is non-negligible, for some $i$, becomes correlated to the state $s_{i}(x) \equiv \langle x | s_{i} \rangle$ - not, in general, to $S$'s configuration $q_{S}$. The `outcome' of a measurement corresponds to a particular index $i$; the Bohmian configuration $q_{AE}$ of system $AE$ `registers' the outcome $i$ if it lies in a region where $\phi_{i}(y)$ is non-negligible. This aspect of effective collapse in pilot wave theory is consistent with our usual notion of collapse from conventional quantum theory. The `outcome' of the measurement is identified by the index $i$ corresponding to the packet that the total system point enters. Thus, a measurement occurs in Bohm's theory when the Bohmian configuration specifically of the external degrees of freedom succeed, irreversibly, in picking out a branch of the total system's wave function. In the case where the states $| s_{i} \rangle$ are eigenstates of an operator $\hat{A}_{S}$  that is degenerate, the dynamics will associate $q_{AE}$ with some subspace of states in $S$'s Hilbert space, rather than with an individual state.

The configuration $q_{S}$, on the other hand, does not play a major role in the measurement process. By equivariance, it is constrained to lie in a region where one of the $s_{i}(x)$ is non-negligible. However, unlike the states $\phi_{i}(y)$, the different $s_{i}(x)$ may and often do have substantial overlap in $S$'s configuration space (for example, in a measurement of angular momentum), so that the configuration  $q_{S}$ does little to distinguish among them. It is the fact that $\phi_{i}(y)$ do not have substantial configuration space overlap that is crucial to ensuring a well-defined measurement outcome in this example.

For convenience, we define the set $AE_{i}(t) \subset \mathbb{Q}_{AE}$, where $ \mathbb{Q}_{AE}$ is the configuration space of $AE$, as the the subset of $AE$'s configurations that `register' the outcome $i$:

\begin{equation} \label{Region}
AE_{i}(T) \equiv  \text{supp}_{\epsilon}[\phi_{i}(y)],
\end{equation}

\noindent where I call $\text{supp}_{\epsilon}[f(y)] \equiv \{ y \in \mathbb{Q}_{B} | \ |f(y)| > \epsilon , \text{where} \ \epsilon > 0 \}$ the `$\epsilon$-support' of the function $f$ (though occasionally I may abuse terminology and refer to it simply as the support). Note that as long as the configuration space decoherence condition is satisfied, and one does not choose $\epsilon$ too small, $AE_{i}(t)  \cap AE_{j}(t) = \varnothing $ for $i \neq j$ - that is, the sets $AE_{i}$ are disjoint. Thus, the measurement has outcome $i$ if the regions defined by (\ref{Region}) are disjoint, and if $q_{AE}(t) \in AE_{i}(t)$. Note that on this definition, the measurement can only have an outcome if the wave packets  $\phi_{i}(y)$ have disjoint $\epsilon$-support, for some sufficiently small value of $\epsilon$. Also note that the smallest value $\epsilon$ such that the $AE_{i}(t)$ are disjoint serves to characterize the strength of the correlations between $q_{AE}$ and $s_{i}(x)$ established by the measurement; measurements can be characterized as `strong' or `weak' depending on the minimum value of $\epsilon$ that leads to disjoint $AE_{i}(t)$. Unless stated otherwise, I will assume that the minimum value of $\epsilon$ that yields disjoint $AE_{i}$ in a measurement in Bohm's theory is extremely small - in other words, that configuration space decoherence has taken effect - making the measurement strong. 

\vspace{5mm}

To address the measurement problem, an interpretation of quantum theory must both explain the appearance of determinate measurement outcomes and the fact that measurement outcomes conform to Born Rule probabilities. In the Bohm theory, there is no indeterminacy about the systemÕs configuration - the Bohmian configuration is always well-defined at every moment in time. Why then the need for any kind of decoherence at all? Because, even in spite of the determinacy of the Bohmian configuration, the measurement \textit{outcome} itself will not be determinate if the configuration space decoherence condition is not satisfied. Moreover, regarding the second matter of the Born Rule probabilities, if measurement outcomes are not clearly defined or determinate to begin with, it will not be possible to assign probabilities to them.  

Addressing the matter of determinacy of outcomes first, the configuration $q_{AE}$ only registers a measurement outcome if the regions $AE_{i}$ are disjoint; this, in turn, requires that the configuration space decoherence condition is satisfied. In the absence of configuration space decoherence, the regions $AE_{i}$ may overlap; if $q_{AE}$ lies in the overlap of more than one of the regions $AE_{i}$, the measurement will be indeterminate, since in this case $q_{AE}$ fails to single out a unique branch of the quantum state. Even though the configuration $q_{AE}$ itself is determinate, the outcome is not. Thus, on Bohm's theory, a measurement process is indeterminate for the simple and fairly banal reason that it fails dynamically to establish a correlation between the configuration of $AE$ and the state of $S$, much as a measurement in classical mechanics would be indeterminate if the configuration of the pointer on a measuring device were not dynamically correlated in the appropriate manner with the state of the system being measured.  

\vspace{5mm}

On a separate matter, it is worth noting that in cases where there is a continuous infinity of branches in the total quantum state, the disjointness of the branches' configuration space supports does not entail effective collapse as straightforwardly as it does in cases where the number of branches is discrete. In the discrete case, the configuration space decoherence condition guarantees the existence of regions of configuration space between any two adjacent branches such that the wave function is effectively zero there; by equivariance, if the configuration is in the support of one branch, it will not be able to transition to another branch because the Bohmian dynamics do not permit it to traverse regions where the wave function is effectively zero. Thus, for as long as the branches remain disjoint, the configuration is guaranteed to remain in its branch. The same reasoning does not apply in the case where the expansion of the quantum state consists of a continuous infinity of disjoint branches. For, although two branches sufficiently separated in their indices may have disjoint supports, there may be a continuum of intermediate branches between them, such that the magnitude of the quantum state between the branches never becomes negligible \footnote{Thanks to David Wallace for pointing this out to me.}. The suppression of drift between disjoint branches of the Bohmian configuration in this case takes more effort to see. In the remainder of my analysis, I  assume without proof that disjointness of branches precludes drift between those branches even in the case of continuously indexed branches; the reader should take this as a conjecture - which I henceforth refer to as the `No Drift Conjecture' - awaiting proof.

\subsection{Environmental Decoherence: Localisation as a Result of Environmental Scattering}

Generally speaking, `decoherence' is the process whereby the degrees of freedom external to some system - whether these degrees of freedom consist of some purpose-constructed measuring apparatus or particles of dust, air, radiation, etc. in the environment - interact and become entangled with that system. Assume that the system $S$ consisting of the degrees of freedom of interest and the system $E$ consisting of any degrees of freedom external to $S$ are iniitially in some product state:

\begin{equation}
| \psi \rangle  \otimes | E_{0} \rangle = \left( \alpha |\psi_{1} \rangle +  \beta |\psi_{2} \rangle  \right) \otimes  |E_{0} \rangle. 
\end{equation}

\noindent Assume further that the interaction between $S$ and $E$ is such that over time scales where the internal dynamics of $S$ can be ignored,

\begin{align}
& |\psi_{1} \rangle \otimes | E_{0} \rangle \rightarrow  |\psi_{1} \rangle \otimes | E_{1} \rangle    \\
&  |\psi_{2} \rangle \otimes | E_{0} \rangle \rightarrow  |\psi_{1} \rangle \otimes | E_{2} \rangle,
\end{align}

\noindent as will be the case when $|\psi_{1} \rangle$ and $|\psi_{2} \rangle $ belong to the set of pointer states of $S$, where `pointer states' here are defined as those states of $S$ that suffer minimal entanglement with $E$ as a result of the interaction between $S$ and $E$ (much of the analysis in this subsection draws on \cite{schlosshauer2008decoherence} and \cite{JoosZehBook}). From these assumptions, it follows that

\begin{equation}
\left( \alpha |\psi_{1} \rangle +  \beta |\psi_{2} \rangle  \right) \otimes  |E_{0} \rangle \rightarrow  \alpha |\psi_{1} \rangle \otimes  | E_{1} \rangle  +  \beta |\psi_{2} \rangle \otimes | E_{2} \rangle 
\end{equation}

\noindent over timescales where the internal dynamics of $S$ can be ignored. The reduced density matrix of the system $S$ is 

\begin{equation}
\hat{\rho}_{S} = \frac{1}{2}\left\{  |\psi_{1} \rangle  \langle \psi_{1} | + |\psi_{2} \rangle  \langle \psi_{2} |   +   |\psi_{1} \rangle  \langle \psi_{2} |   \langle E_{2} | E_{1} \rangle  + |\psi_{2} \rangle  \langle \psi_{1} |   \langle E_{1} | E_{2} \rangle   \right \}
\end{equation}

\noindent When $\langle E_{2} | E_{1} \rangle \approx 1$, the state of the combined systems is a product state and they are completely unentangled (as is the case at the beginning of the interaction). If $\langle E_{2} | E_{1} \rangle \approx 0$, then 

\begin{equation}
\hat{\rho}_{S} \approx \frac{1}{2}\left\{  |\psi_{1} \rangle  \langle \psi_{1} | + |\psi_{2} \rangle  \langle \psi_{2} |    \right \}.
\end{equation}

The macroscopic objects of our everyday experience interact perpetually with their environments, which include degrees of freedom associated, for example, with dust particles, air molecules, neutrinos and photons, via the scattering of these particles off of the object in question. It is well-known that the pointer states of typical macroscopic systems under such scattering interactions are states that are localised both in position and in momentum; more specifically, they are the coherent states $| q,p \rangle$, where $q$ and $p$ denote the position or spatial configuration $q$ and the momentum $p$ about which the coherent state is peaked  (see, for instance, \cite{zurek1993coherent} for arguments to this effect). 

Thus, if $|\psi_{1} \rangle = | q_{1},p_{1} \rangle $ and $|\psi_{2} \rangle = | q_{2},p_{2} \rangle $, then

\begin{align}
& |q_{1},p_{1} \rangle \otimes | E_{0} \rangle \rightarrow  |q_{1},p_{1} \rangle \otimes | E_{1} \rangle    \\
& |q_{2},p_{2} \rangle \otimes | E_{0} \rangle \rightarrow  |q_{2},p_{2} \rangle \otimes | E_{2} \rangle,
\end{align}

\noindent and 

\begin{equation} \label{TwoPackets}
\left( \alpha |q_{1},p_{1} \rangle +  \beta |q_{2},p_{2} \rangle  \right) \otimes  |E_{0} \rangle \rightarrow  \alpha |q_{1},p_{1} \rangle \otimes  | E_{1} \rangle  +  \beta |q_{2},p_{2} \rangle \otimes | E_{2} \rangle 
\end{equation} 

\noindent  over timescales on which the internal dynamics of $S$ can be ignored. The reduced density matrix of the system $S$ is 

\begin{equation}
\hat{\rho}_{S} = \frac{1}{2}\left\{  |q_{1},p_{1} \rangle  \langle q_{1}, p_{1} | +  |q_{2},p_{2} \rangle  \langle q_{2}, p_{2} |   +    |q_{1},p_{1} \rangle  \langle q_{2}, p_{2} |   \langle E_{2} | E_{1} \rangle  +|q_{2},p_{2} \rangle  \langle q_{1}, p_{1} |   \langle E_{1} | E_{2} \rangle   \right \}
\end{equation}

\noindent If $\langle E_{2} | E_{1} \rangle \approx 0$, then 

\begin{equation}
\hat{\rho}_{S} \approx \frac{1}{2}\left\{  |q_{1},p_{1} \rangle  \langle q_{1}, p_{1} | +  |q_{2},p_{2} \rangle  \langle q_{2}, p_{2} |    \right \}.
\end{equation}

\noindent If the environment just consists of a single particle, then if the wavelength of the environmental particle is sufficiently small, the single particle will be sufficient to result in orthogonality of environmental states: $\langle E_{2} | E_{1} \rangle \approx 0$. If, on the other hand, the wavelength is large in comparison with $q_{1}-q_{2}$ and the scattering sufficiently weak, then we will have $\langle E_{2} | E_{1} \rangle \approx 1$ as the result of a single scattering event. By contrast, if the environment consists of sufficiently many long-wavelength particles all of which scatter off of $S$, then this will also cause orthogonality of environmental states $\langle E_{2} | E_{1} \rangle \approx 0$ . For models in which there are many individual scattering events, but in which each on its own is insufficient to induce a strong measure of entanglement,  it is possible to show for a variety of models of the environment that

\begin{equation} \label{EnvDecay1}
\langle E_{2} | E_{1} \rangle \propto  e^{- \Lambda  (q_{1} - q_{2})^{2} t},
\end{equation}

\noindent where the constant $\Lambda$ is determined by the details of the scattering process in question. In the limit where $|q_{1} - q_{2}|$  becomes sufficiently large that each individual scattering event is sufficient to resolve the difference between the two wave packets, 

\begin{equation} \label{EnvDecay2}
\langle E_{2} | E_{1} \rangle \propto  e^{- \Gamma  t}
\end{equation}

\noindent where $\Gamma$ is the total scattering rate, again a constant determined by the details of the scattering process (see \cite{schlosshauer2008decoherence}, Chs. 2 and 3 for a more detailed explanation of these effects).

More generally, if one considers the density matrix $\hat{\rho}_{S}$ of $S$ obtained by tracing over the degrees of freedom in $E$ in the total density matrix for the combined system $SE$, then in cases of many weakly entangling scattering events it  can be shown that on timescales for which the internal dynamics of $S$ can be ignored (as a result of the large of mass of $S$), $\hat{\rho}_{S}$ evolves so that

\begin{equation}
\langle q' | \hat{\rho}_{S}(t)| q \rangle  =  \langle q' | \hat{\rho}_{S}(0)| q \rangle e^{- \Lambda  (q-q')^{2} t},
\end{equation}

\noindent where again, $\Lambda$ is a constant determined by the details of the model in question. That is, the off-diagonal elements of the position-space density matrix will decay exponentially in time, and with the square of the position-space width $|q-q'|$ of the initial coherent superposition. 

Typically, the time scales on which coherence of superpositions of macroscopically different positions of $S$ is lost in such models will be extremely short - much shorter than timescales on which the $S$'s internal dynamics, associated with the Hamiltonian $\hat{H}_{S}$, will induce significant changes in the state of $S$, or of $SE$ as a whole. On the basis of such models, calculations of decoherence timescales for various kinds of macro-(or meso-)scopic systems have been performed (see, for instance,  \cite{schlosshauer2008decoherence}, p. 135). For example, a dust grain (diameter about $10^{-3}$cm) in a coherent superposition of different positions typically will lose coherence within $10^{-31}s$ as a result of its interaction with the atmosphere around it. In the best laboratory vacuum that we are able to create, decoherence due to whatever few air molecules remain, as well as due to things like background radiation and neutrinos which we are unable to screen out, will typically occur in a dust grain in about $10^{-18}s$. As one would expect, the larger the object, the more it interacts with its environment and the harder it is to screen off from interaction with the environment.

Moreover, although the fundamental equations of our quantum models are all time-reversible, the process of environmental decoherence is in practice irreversible (at least on timescales short of Poincare recurrence timescales). When a system becomes entangled with its environment, the coherence of the superposition that is initially localised in the system itself becomes a property of the total system $SE$. In practice, this process cannot be reversed since we do not control the behavior of all degrees of freedom in the environment. For discussion of how the effective irreversibility of decoherence process is to be reconciled with the reversibility of the fundamental quantum mechanical equations of motion, the reader can consult $\cite{wallace2012emergent}$, Ch.9.

To summarise, for typical macroscopic systems, environmental decoherence will very rapidly destroy the coherence of any superpositions of macroscopically differing positions - typically on timescales much shorter than the characteristic timescales of the system's own internal dynamics. However, if the macroscopic system is in a spatially narrowly localised state - specifically, a coherent state that is also narrowly peaked in momentum - then entanglement with the environment will not occur for as long as this continues to be the case. The combined system $SE$ will evolve as a product state, and the system $S$ will evolve unaffected by $E$, solely according to the internal dynamics prescribed by $\hat{H}_{S}$.

\subsubsection{Decoherence Master Equations}

Much of the study of environmental decoherence focuses on effective equations of motion for the reduced density matrix $\hat{\rho}_{S}$. Equations derived for a wide variety of environments all have the same general form, known as the Caldeira-Leggett equation:

\begin{equation} \label{CaldLegg}
i \frac{d \hat{\rho}_{S}}{dt} = [ \hat{H}_{S} + \frac{1}{2}M \Omega^{2} \hat{X}^{2}, \hat{\rho}_{S} ]  - i \Lambda \eta k_{B} T \left[ \hat{X},\left[ \hat{X}  , \hat{\rho}_{S} \right] \right] + \frac{\eta}{2M} [\hat{X},\{ \hat{P},\hat{\rho}_{S} \}].
\end{equation}

\noindent Caldeira and Leggett first derived this equation in the context of a model, known as the Caldeira-Leggett model. in which the environment consists of many independent harmonic oscillators whose positions couple linearly with that of $S$ \cite{caldeira1983path}; since this model was originally proposed, the general form of the Caldeira-Leggett equation has found much broader applications beyond this particular model (see \cite{schlosshauer2008decoherence} for a more detailed discussion of this equation and of the Caldeira-Leggett model).The right-hand side of this equation consists of four components: 1) The system $S$'s unitary dynamics, prescibed by $\hat{H}_{S}$, 2) a renormalisation term (the term proportional to $\Omega^{2}$), 3) A decoherence term (the term proportional to $\Lambda$), 4) a dissipation term (the last term), which accounts for effects of classical friction. The constants $\Omega$, $\eta$ and $\Lambda$ are all determined by the particular model in question, while $M$ is simply the mass that enters into the Hamiltonian $\hat{H}_{S}$. In the systems that I consider here, the renormalisation and dissipation terms can be neglected by comparison with the other two terms. It is the decoherence term that is ultimately responsible for supression of macroscopic coherence of superpositions on the very rapid timescales characteristic of decoherence.

%because the coherent states constitute a basis for $S$'s Hilbert space, one generally can express the overall state of the combined system $SE$ as superposition over an arbitrary number (in fact, a continuum) of coherent states, rather than just of the two as  in (\ref{TwoPackets}):

%\begin{equation}
%\int dq \ dp \ \alpha(q,p) | q,p \rangle \otimes | \phi(q,p) \rangle.
%\end{equation}

%\noindent In such a case, one can write down the density matrix of such a system

%If we consider the density matrix of the subsystem $S$,

%\begin{equation}
%\hat{\rho} =
% \begin{pmatrix}
%  |\alpha|^{2} & \alpha \beta^{*}  \langle E_{2} | E_{1} \rangle\\
%   \alpha^{*} \beta  \langle E_{1} | E_{2} \rangle &  |\beta|^{2},
% \end{pmatrix}
%\end{equation}

\subsection{The Decoherent Histories Framework}

The decoherent histories framework, originally developed by Griffiths and Gell-Mann and Hartle, is an effective tool for analysing the branching structure of the quantum state under the unitary evolution prescribed by Schrodinger's equation (see, for example, \cite{griffiths1984consistent}, \cite{gell1993classical}, \cite{DowkerKentConstHist}, \cite{halliwell1995review}; for a briefer introduction, see for instance \cite{landsmanClassicalQuantum}). It is especially useful when considering the behavior of closed, macroscopic quantum systems, and for this reason has become a cornerstone of quantum cosmology. It also provides an illuminating perspective from which to examine the appearance of classical behavior in quantum systems. 

The decoherent histories formalism makes essential use of PVM's and POVM's reviewed in section \ref{PVM}. While many presentations of the decoherent histories framework present it within the context of the Heisenberg picture (see, for example, \cite{halliwell1995review}), so that the pertinent PVM and POVM operators are time-dependent, as in most (though not all) of the thesis I remain in the Schrodinger picture, where PVM and POVM operators are time independent and it is the quantum state, rather than the operators, that evolves. I adopt the Schrodinger picture partly because it facilitates the extension of the analysis to the Bohm theory, but also because the picture that it offers, in terms of states evolving in a Hilbert space, is arguably more intuitive than that offered by the Heisenberg picture (though this latter point is likely to be a matter of personal preference to some extent). 

Consider the usual time evolution equation for a closed quantum system in a pure state:

\begin{equation}
| \chi(t) \rangle = e^{-i \hat{H}t} | \chi_{0} \rangle.
\end{equation}

\noindent Given an arbitrary PVM, we may divide the time interval $t$ into $N$ equal slices, inserting a factor of the identity $\sum_{i_{k}} \hat{P}_{i_{k}}$ at each time interval:

\begin{align} \label{Wfcn}
| \chi(t) \rangle &=   \left(\sum_{i_{N}} \hat{P}_{i_{N}}\right) e^{-i \hat{H} \frac{t}{N}}  ...    e^{-i \hat{H} \frac{t}{N}}   \left(\sum_{i_{1}} \hat{P}_{i_{1}}\right) e^{-i \hat{H} \frac{t}{N}} | \chi_{0} \rangle \\
& =  \sum_{i_{1},...,i_{N}} \ \left[  \hat{P}_{i_{N}} \ e^{-i \hat{H} \frac{t}{N}} \  ...  \  e^{-i \hat{H} \frac{t}{N}}   \ \hat{P}_{i_{1}} \ e^{-i \hat{H} \frac{t}{N}}  \right] \ | \chi_{0} \rangle.
\end{align}

\noindent Define the \textit{history operators} $\hat{C}_{i}$, 

\begin{equation} \label{HistoryOperator}
\hat{C}_{i} \equiv \hat{C}_{i_{1},...,i_{N}} \equiv   \hat{P}_{i_{N}} e^{-i \hat{H} \frac{t}{N}}  ...   e^{-i \hat{H} \frac{t}{N}} \hat{P}_{i_{1}} e^{-i \hat{H} \frac{t}{N}},
\end{equation}

\noindent and note that 

\begin{equation} \label{HistorySum}
\sum_{i_{1},...,i_{N}} \hat{C}_{i_{1},...,i_{N}} = e^{-i\hat{H}t}.
\end{equation}

\noindent Eqn. (\ref{Wfcn}) can be rewritten

\begin{equation} \label{StateHistory}
| \chi(t) \rangle =    \sum_{i_{1},..., i_{N}} \hat{C}_{i_{1},...,i_{N}}   | \chi_{0} \rangle.
\end{equation}

\noindent This expansion is completely general and places no restrictions on the evolution of the state beyond those already imposed by the Schrodinger evolution. Moreover, nothing prevents the use of different PVM's at different times, though there will be no need for this generalisation here. A sequence of PVM operators constitutes a \textit{history}, and can be identified by its associated sequence of indices $(i_{1},...,i_{N})$. The set of all such sequences of projectors, associated with set of index sequences $\{  (i_{1},...,i_{N}) \}$, constitutes a \textit{history space}. A history $(i_{1},...,i_{N})$ is \textit{realised} iff $\hat{C}_{i_{1},...,i_{N}}   | \chi_{0} \rangle \neq 0$.
\footnote{or, more precisely, iff $\left| \hat{C}_{i_{1},...,i_{N}}   | \chi_{0} \rangle \right| > \ \epsilon$ where $\epsilon$ is the very small but finite threshold below which the weight of the history is drowned out by the `noise' of the miniscule but finite residual interference terms that remain between effectively decohered branches.} 
For ease of notation, I will denote the history $(i_{1},...,i_{N})$ by $i$, and the history space $\{  (i_{1},...,i_{N}) \}$ by $\{ i \}$. Note further that two histories $i\equiv (i_{1},...,i_{N})$ and $i' \equiv (i'_{1},...,i'_{N})$ differ if they differ if they differ with respect to \textit{any} $i_{k}$, where $1 \leq k \leq N$.

Following Gell-Mann and Hartle, a \textit{coarse graining} $\{ \bar{i} \}$ of the history space $\{ i \}$ is a partitioning of $\{ i \}$  such that every history $i$ belongs to exactly one element $\bar{i}$ of the partition, entailing

\begin{equation}
\sum_{\bar{i}} \hat{C}_{\bar{i}} =  \sum_{i} \hat{C}_{i} =  e^{-i\hat{H}t}
\end{equation}

\noindent where

\begin{equation}
\hat{C}_{\bar{i}} \equiv  \sum_{i \in \bar{i}} \hat{C}_{i} =  \sum_{i_{1},...,i_{N} \in \bar{i}} \hat{C}_{i_{1},...,i_{N}}.  
\end{equation} 

\noindent The coarse-grained history operators $\hat{C}_{\bar{i}}$ are sums of alternating sequences of PVM and time evolution operators, but may not themselves expressible as such sequences - that is, they may not be expressible in the form (\ref{HistoryOperator}). Note that if the projection operators $\{ \hat{P}^{'}_{i} \}$ in a PVM each can be expressed as the sum of some projection operators in some other PVM $\{ \hat{P}_{j} \}$, then the history operators formed using the $\hat{P}^{'}_{i}$ clearly are coarse-grainings of the history operators formed using the $\hat{P}_{j}$.

The weights  $\left| \hat{C}_{\bar{i}}| \chi_{0} \rangle \right|^{2}$ of the different coarse-grained histories $(\bar{i})$ can be construed as probabilities for these histories only if they satisfy the axioms of probability theory. In particular, for two histories $\bar{i}$ and $\bar{i}'$, it is necessary that 

\begin{equation} \label{Exclusive}
 \text{Pr}(\bar{i} \   \text{or} \  \bar{i}') =  \text{Pr}(\bar{i}) +  \text{Pr}(\bar{i}'),
\end{equation}

\noindent Since

 \begin{equation}
  \text{Pr}(\bar{i} \   \text{or} \  \bar{i}') =   \langle \chi_{0} | (\hat{C}^{\dagger}_{\bar{i}} + \hat{C}^{\dagger}_{\bar{i}'} ) ( \hat{C}_{\bar{i}} + \hat{C}_{\bar{i}'} ) | \chi_{0} \rangle, 
 \end{equation}

 \begin{equation}
  \text{Pr}(\bar{i} ) =   \langle \chi_{0} | \hat{C}^{\dagger}_{\bar{i}}   \hat{C}_{\bar{i}} | \chi_{0} \rangle, 
 \end{equation} 
 
 \noindent and 
 
  \begin{equation}
  \text{Pr}(\bar{i}' ) =   \langle \chi_{0} | \hat{C}^{\dagger}_{\bar{i}'}   \hat{C}_{\bar{i}'} | \chi_{0} \rangle, 
 \end{equation}

\noindent eqn. (\ref{Exclusive}) amounts to the requirement that

\begin{equation} \label{Consistency}
  \langle \chi_{0} | (\hat{C}^{\dagger}_{\bar{i}} + \hat{C}^{\dagger}_{\bar{i}'} ) ( \hat{C}_{\bar{i}} + \hat{C}_{\bar{i}'} ) | \chi_{0} \rangle 
= \langle \chi_{0} | \hat{C}^{\dagger}_{\bar{i}} \hat{C}_{\bar{i}} | \chi_{0}\rangle + \langle \chi_{0} | \hat{C}^{\dagger}_{\bar{i}'} \hat{C}_{\bar{i}'} | \chi_{0}\rangle. 
\end{equation}

\noindent But, in general

\begin{align}
&\langle \chi_{0} | (\hat{C}^{\dagger}_{\bar{i}} + \hat{C}^{\dagger}_{\bar{i}'} ) ( \hat{C}_{\bar{i}} + \hat{C}_{\bar{i}'} ) | \chi_{0} \rangle \\
& = \langle \chi_{0} | \hat{C}^{\dagger}_{\bar{i}} \hat{C}_{\bar{i}} | \chi_{0}\rangle + \langle \chi_{0} | \hat{C}^{\dagger}_{\bar{i}'} \hat{C}_{\bar{i}} | \chi_{0}\rangle  + \langle \chi_{0} | \hat{C}^{\dagger}_{\bar{i}} \hat{C}_{\bar{i}'} | \chi_{0}\rangle + \langle \chi_{0} | \hat{C}^{\dagger}_{\bar{i}'} \hat{C}_{\bar{i}'} | \chi_{0}\rangle,
\end{align}

\noindent so for (\ref{Exclusive}) or (\ref{Consistency}) to be satisfied, it is necessary and sufficient that the condition, known as \textit{weak decoherence}, 

\begin{equation}
 \text{Re}\left( \langle \chi_{0} | \hat{C}^{\dagger}_{\bar{i}} \hat{C}_{\bar{i}'} | \chi_{0}\rangle \right) \approx 0 \  \text{for} \ \bar{i} \neq \bar{i}'
\end{equation}

\noindent be satisfied (since the imaginary components of the cross-terms cancel each other automatically). In practice, though, in cases where this condition is satisfied, the stronger condition,

\begin{equation} \label{MedDec}
D(\bar{i},\bar{i}') \equiv \langle \chi_{0} | \hat{C}^{\dagger}_{\bar{i}'} \hat{C}_{\bar{i}} | \chi_{0}\rangle \approx 0 \  \text{for} \ \bar{i} \neq \bar{i}',
\end{equation}

\noindent known as \textit{medium decoherence}, usually is also satisfied. The term $D(\bar{i},\bar{i}')$ is known as the \textit{decoherence functional}. A set $\{\bar{i} \}$ of coarse-grained histories is \text{consistent} if $D(\bar{i},\bar{i}') \approx 0$ for $\bar{i} \neq \bar{i}'$. The probability of a history $\bar{i}$ in this case is given by 

\begin{equation}
\text{Pr}(\bar{i}) = D(\bar{i},\bar{i})=  \langle  \chi_{0}|   \hat{C}^{\dagger}_{\bar{i}} \hat{C}_{\bar{i}}   | \chi_{0} \rangle = \left| \hat{C}_{\bar{i}}| \chi_{0} \rangle \right|^{2},
\end{equation}

\noindent and 

\begin{equation}
\sum_{\bar{i}} \text{Pr}(\bar{i}) = 1.
\end{equation}

\noindent Beyond the above conditions, though, it is usually also required that individual histories belonging to different coarse-grained histories satisfy either the weak, or more often the medium, decoherence condition, so that

\begin{equation}
D(i,i') \equiv \langle \chi_{0} | \hat{C}^{\dagger}_{i'} \hat{C}_{i} | \chi_{0}\rangle \approx 0 \  \text{for} \ i \in \bar{i} \ \text{and} \ i' \in \bar{i}' \ with  \  \bar{i} \neq \bar{i}'.
\end{equation}

\noindent This can be written more explicitly as

\begin{equation} \label{DecoherenceHistories}
D(i,i') \equiv \langle \chi_{0} | \hat{C}_{i'_{1},...,i'_{N}}^{ \dagger} \hat{C}_{i_{1},...,i_{N}}   | \chi_{0}\rangle \approx 0 \ \text{for} \ i \in \bar{i} \ \text{and} \ i' \in \bar{i}' \ \text{with}  \  \bar{i} \neq \bar{i}' .
\end{equation}

\noindent This then implies that 

\begin{equation}
D(\bar{i},\bar{i}') \equiv \langle \chi_{0} | \left( \sum_{i'_{1},...,i'_{N} \in \bar{i}'} \hat{C}_{i'_{1},...,i'_{N}}^{ \dagger} \right) \left( \sum_{i_{1},...,i_{N} \in \bar{i}}\hat{C}_{i_{1},...,i_{N}} \right)   | \chi_{0}\rangle \approx 0 \ \text{for}  \  \bar{i} \neq \bar{i}' ,
\end{equation}

\noindent which is just another way of writing eqn. (\ref{MedDec}).

\subsubsection{Decoherent Histories in Bohm's Theory}

Moving to Bohm's theory, Hartle has claimed that the probabilities associated with decoherent histories and those associated with the histories prescribed by the Bohmian trajectories themselves predict different probabilities \cite{hartle2004bohmian}. Hartle's analysis defines histories in the Bohmian context directly in terms of the trajectories of Bohmian configurations, and compares the probabilities of histories defined in this manner, as specified by Bohm's theory, to the probabilities of corresponding histories within the decoherent histories framework. I argue here that the requirement of configuration space decoherence offers a more perspicuous way of understanding the connection between the decoherent histories framework and Bohm's theory, andmoreover ensures agreement between the probabilities predicted by the decoherent histories framework within the contexts of the Bare/Everett theory and of the Bohm theory. 

However, what Hartle does not observe in his analysis is that if the histories as defined in the context of the decoherent histories framework satisfy the configuration space decoherence requirement, rather than merely the decoherence requirement, with respect to the initial quantum state $| \chi_{0} \rangle$, the probabilities of histories as predicted by the Bare/Everett theory (which is at least formally equivalent to the decoherent histories framework as Hartle defines it) and the Bohm theory will be the same. That is, the proper way to understand the significance of the decoherent histories formalism for Bohm's theory is through the result that the probabilities the probabilities predicted by the Bohm theory for the histories $ \bar{i} $, as defined in the decoherent histories framework,  will be the same as those in the ordinary decoherent histories formalism if the following stronger decoherence condition on histories is satisfied:

\begin{equation} \label{BohmDH}
D_{X}(\bar{i},\bar{i}') \equiv \langle \chi_{0} | \hat{C}^{\dagger}_{\bar{i}'}| X \rangle \langle X | \hat{C}_{\bar{i}} | \chi_{0}\rangle \approx 0 \  \text{for all} \ X \ \text{if} \ \bar{i} \neq \bar{i}',
\end{equation}

\noindent where $| X \rangle$ is a eigenstate of configuration. The condition (\ref{BohmDH}) entails (\ref{MedDec}), but more importantly, guarantees that the coarse-grained branches $ \hat{C}_{\bar{i}} | \chi_{0}\rangle$ are disjoint (in the sense of the $\epsilon$-support defined above) in the configuration space of the system, which in turn ensures that the configuration $Q$ of the system selects only one such branch (again, putting aside for the moment the worry about Bohmian effective collapse in the case of continuous pointer bases). 

Typically, it will also be the case that the condition requiring states associated with individual subhistories within distinct coarse-grained histories to be disjoint, namely,

\begin{equation}
D_{X}(i,i') \equiv \langle \chi_{0} | \hat{C}_{i'_{1},...,i'_{N}}^{ \dagger} | X \rangle \langle X | \hat{C}_{i_{1},...,i_{N}}   | \chi_{0}\rangle \approx 0 \  \text{for all} \ X, \ \text{with} \ i \in \bar{i}, \ i' \in \bar{i}' \ \text{and}  \  \bar{i} \neq \bar{i},' 
\end{equation}

\noindent is satisfied. This then implies

\begin{equation}
D_{X}(\bar{i},\bar{i}') \equiv \langle \chi_{0} | \left( \sum_{i'_{1},...,i'_{N} \in \bar{i}'} \hat{C}_{i'_{1},...,i'_{N}}^{ \dagger} \right) | X \rangle \langle X | \left( \sum_{i_{1},...,i_{N} \in \bar{i}}\hat{C}_{i_{1},...,i_{N}} \right)   | \chi_{0}\rangle \approx 0 \ \text{for}  \  \bar{i} \neq \bar{i}' ,
\end{equation}

\noindent which is just another way of writing eqn. (\ref{BohmDH}).

\subsection{Branching of the Quantum State}

The Everett or Many Worlds interpretation is often said to associate an array of dynamically independent `worlds' with the branches of the total quantum state. But what is exactly is meant by the claim that the quantum state has a branching structure?

 The discussion here closely follows that in Chapter 3 of Wallace's \cite{wallace2012emergent}.   First, define the weight of a projector $P_{i}$ at some time $t$ as
 
 \begin{equation}
 W_{i}(t) \equiv  \left|  \hat{P}_{i} \  e^{-i\hat{H}t} | \chi_{0} \rangle \right|^{2} =  \langle \chi_{0} | e^{i \hat{H}t}   \hat{P}_{i} \  e^{-i\hat{H}t} | \chi_{0} \rangle
 \end{equation}
 
 \noindent The transition weight between projector $\hat{P}_{i}$ at time $t$ and projector $\hat{P}^{'}_{i'}$ at time $t'$ (which may belong to a completely different PVM from $\hat{P}_{i}$), where $t'>t$, is defined as 
 
 \begin{align}
 T(i,t;i',t') &  \equiv \frac{ \left|    \hat{P}^{'}_{i'} e^{-i\hat{H}(t'-t)}   \hat{P}_{i} \  e^{-i\hat{H}t}  | \chi_{0} \rangle \right |^{2} }{\left|  \hat{P}_{i} \  e^{-i\hat{H}t} | \chi_{0} \rangle \right|^{2}} \\
 & =   \frac{ \langle \chi_{0} |   e^{i\hat{H} t}  \hat{P}_{i}   e^{i\hat{H} (t'-t)}    \hat{P}^{'}_{i'}     e^{-i\hat{H}(t'-t)}   \hat{P}_{i}   e^{-i\hat{H}t}  | \chi_{0} \rangle }{  \langle \chi_{0} | e^{i \hat{H}t}   \hat{P}_{i} \  e^{-i\hat{H}t} | \chi_{0} \rangle }.
 \end{align}

\noindent In cases where $\left|  \hat{P}_{i} \  e^{-i\hat{H}t} | \chi_{0} \rangle \right|^{2} = 0$, the above expression for the transition weight is not defined, so define $T(i,t;i',t')=0$. If a history $(i_{1},...,i_{N})$ is realised, it is straightforward to see that $T(i_{n},t_{n};i_{n+1},t_{n+1}) \neq 0$ for all  $ 0 \leq n \leq N $.

The evolution of the quantum state exhibits branching relative to the given PVMs if it is the case that

\begin{quote}
\textbf{Branching Condition:} if $T(i ,t;i',t') \neq 0$ and $T(j,t;i',t') \neq 0$, then $i=j$,
\end{quote}

\noindent so that no two distinct projectors at an earlier time have non-zero transition weights into the same projector at a later time.

\subsubsection{Branching and Decoherence}

%If we allow the PVM's at different times to vary in $\ref{Wfcn}$, the expansion in $\ref{Wfcn}$ can be generalised to 

%\begin{align} \label{WfcnGeneral}
%| \chi(t) \rangle &=   \left(\sum_{\alpha_{N}} \hat{P}^{N}_{\alpha_{N}}\right) e^{-i \hat{H} \frac{t}{N}}  ...    e^{-i \hat{H} \frac{t}{N}}   \left(\sum_{\alpha_{1}} \hat{P}^{1}_{\alpha_{1}}\right) e^{-i \hat{H} \frac{t}{N}} | \chi_{0} \rangle \\
%& =  \sum_{\alpha_{1},...,\alpha_{N}} \ \left[  \hat{P}^{N}_{\alpha_{N}} \ e^{-i \hat{H} \frac{t}{N}} \  ...  \  e^{-i \hat{H} \frac{t}{N}}   \ \hat{P}^{1}_{\alpha_{1}} \ e^{-i \hat{H} \frac{t}{N}}  \right] \ | \chi_{0} \rangle,
%\end{align}

%\noindent where $\hat{P}^{k}_{\alpha_{k}}$ is a projector belonging to the PVM $\{ \hat{P}^{k}_{\alpha_{k}} \}$ associated with time $k$. 

If a space of histories $(i_{1},...,i_{N})$ is decoherent in the sense specified by (\ref{DecoherenceHistories}), then the set of operators

\begin{equation}
\hat{P}^{k}_{i_{1},...,i_{k}}  \equiv \hat{C}_{i_{1},...,i_{k}} | \chi_{0} \rangle \langle \chi_{0} |  \hat{C}^{\dagger}_{i_{1},...,i_{k}}, 
\end{equation}

\noindent one for each sequence $(i_{1},...,i_{k})$, form an approximate PVM. This can be seen from the fact that the first condition to be a PVM,

\begin{equation}
\sum_{i_{1},...,i_{k}} \hat{P}^{k}_{i_{1},...,i_{k}}  = \hat{I},
\end{equation}

\noindent is satisfied straightforwardly as a consequence of (\ref{HistorySum}) (where the resulting evolution operators cancel to give the identity), and the second condition

\begin{equation}
\hat{P}^{k}_{i_{1},...,i_{k}} \hat{P}^{k}_{i'_{1},...,i'_{k}} \approx  \delta_{i_{1},i'_{1}}  \ ... \  \delta_{i_{1},i'_{1}} \ \hat{P}^{k}_{i_{1},...,i_{k}}
\end{equation}

\noindent follows straightforwardly from the definition of the $\hat{P}^{k}_{i_{1},...,i_{k}}$ and from the decoherence condition (\ref{DecoherenceHistories}). 

It then follows that the evolution of the quantum state exhibits branching - in accordance with the Branching Condition above - with $\hat{P}_{i}$ in the expression for the transition weight above equal to $\hat{P}^{k}_{i_{1},...,i_{k}} $ at some earlier time $t=k \Delta t$ and the PVM $P_{i'}$ in the transition weight equal to $ \hat{P}^{k+m}_{i'_{1},...,i'_{k},i'_{k+1},...,i'_{k+m}} $ at some later time $t'=(k+m) \Delta t$ (where $i$ and $i'$ are to be regarded as collective indices for $(i_{1},...,i_{k})$ and $(i'_{1},...,i'_{k},i'_{k+1},...,i'_{k+m})$, respectively). The Branching Condition is satisfied for arbitrary $m$, as can be seen by using the expansion 
$| \chi(t) \rangle =    \sum_{i_{1},..., i_{N}} \hat{C}_{i_{1},...,i_{N}}   | \chi_{0} \rangle$ and the definition  of the $\hat{P}^{k}_{i_{1},...,i_{k}}$; specifically, the transition weight between $\hat{P}^{k}_{i_{1},...,i_{k}} $ and $ \hat{P}^{k+m}_{i'_{1},...,i'_{k},i'_{k+1},...,i'_{k+m}} $  will be zero unless $(i'_{1},...,i'_{k}) = (i_{1},...,i_{k}) $, thereby entailing that there is a unique projector at any earlier time that contributes to the weight of the projector $ \hat{P}^{k+m}_{i'_{1},...,i'_{k},i'_{k+1},...,i'_{k+m}} $  at any later time, namely $\hat{P}^{k}_{i'_{1},...,i'_{k}}$. 

Condensing notation so that $\textbf{i}^{k} \equiv (i_{1},...,i_{k})$, we have that

\begin{equation}
T[\textbf{i}^{k},k \Delta t; \textbf{j}^{k+m},(k+m)\Delta t] \approx 0 \ \text{unless $\textbf{i}^{k}$ is the initial segment of $\textbf{j}^{k+m}$}
\end{equation}

\noindent thereby entailing that only one projector in the time-$k$ PVM contributes to the weight of any projector in the time-$(k+m)$ PVM.

% In particular, expanding the quantum state as

%\begin{equation}
%| \chi(t) \rangle =    \sum_{i_{1},..., i_{N}} \hat{C}_{i_{1},...,i_{N}}  \left( \sum_{i_{0}} \hat{P}_{i_{0}}  \right)  | \chi_{0} \rangle
%\end{equation}

%\noindent the branching criterion entails that

%\begin{quote}
%\small
%if $T(i_{n},t_{n};i_{n+1},t_{n+1}) \neq 0$ and $T(i'_{n},t_{n};i_{n+1},t_{n+1}) \neq 0$, then $i_{n}=i'_{n}$, 
%\end{quote}

%\noindent for all $ 0 \leq n < N $, where $t_{n} \equiv \frac{n}{N} t$. 

%Note that if the histories $\left( i_{1},...,i_{N} \right)$ satisfy the decoherence condition, that is, if

%\begin{equation}
 %\langle \chi_{0}|  \hat{C}^{\dagger}_{i'_{1},...,i'_{N}}   \hat{C}_{i_{1},...,i_{N}} |\chi_{0} \rangle  \approx 0 \ \text{if} \ i_{k} \neq i'_{k} \ \text{for any} \ 1\leq k \leq N,
%\end{equation}

%\noindent then the histories also satisfy the branching condition. 

\section{DS Reduction of the CM Model to the Bare/Everett Model}  \label{CMEverett}

In Chapter 1, I discussed the case of the DS reduction of the CM Model to the Bare/Everett model without environment.  However, as Wallace argues in \cite{wallace2012emergent}, in cases where the degrees of freedom in question are associated with macroscopic degrees of freedom such as centers of mass of macroscopic bodies, there are a number of significant problems with attempts to model the Newtonian behavior of these systems as isolated from external degrees of freedom:

\begin{itemize}
\item It is unrealistic to treat macroscopic degrees of freedom as isolated, given that actual macroscopic systems interact constantly with external degrees of freedom in their environment.
\item Narrow wave packets are needed to underpin the appearance of classical trajectories. However, wave packets do not necessarily remain narrow on acceptably long timescales to account for the appearance of trajectories, even for macroscopic systems. As Wallace observes, systems whose classical description incorporates dynamics that are chaotic may have initially narrow wave packets spreading to macroscopic coherence lengths on relatively short timescales. For instance, Zurek and Paz have argued that in the case of Saturn's moon Hyperion, such chaotic effects may produce wavefunctions with macroscopic coherence lengthsfrom an initially narrow wave packet on a time scale of about 10 years \cite{zurek1995quantum}.
\end{itemize}

\noindent I turn, then, to the more realistic model of classical behavior that incorporates the interaction between the macroscopic degrees of freedom and the environment. 

In the case of the Bare/Everett model with environment, the DS reduction is more subtle than it was in the example given in Chapter 1, given the additional effects of environmental decoherence at play. As in the case considered in Chapter 1, in the case where the environment is incorporated, the relevant bridge map for the reduction of the NM model described above to the corresponding Bare QM model is given by the expectation values of the position and momentum operators of the central system, in this case consisting of the centers of mass of some macroscopic bodies. However, in the more realistic case where the environment is taken into account, the domain of states does not consist simply of narrow wave packets in the Hilbert space of the macroscopic degrees of freedom; rather, it consists of so-called `branch states' in the total Hilbert space of the centers of mass and environment. These branch states are components of the total quantum state of the macroscopic degrees of freedom and their environment, and are determined by the stucture that decoherence gives to this state. As we will see shortly, the decoherent histories formalism will prove an especially useful tool for describing branch states. Much of the material concerning branching of the quantum state relative to an approximate coherent state PVM is adapted from Chapter 3 of Wallace's \cite{wallace2012emergent}.

While the domain of classical behavior in the quantum model here does not consist simply of narrow wave packets in $\mathcal{H}_{S}$, narrow wave packets still have a crucial role to play in the reduction of models of macroscopic classical behavior to the Bare QM model with environment. Their significance is two-fold. First, they comprise the so-called `decoherence-preferred' basis of states in $\mathcal{H}_{S}$ - that is, the states that suffer minimal entanglement with the environment, which also form a basis for $\mathcal{H}_{S}$. Second, as observed in Chapter 1, they are the states that exhibit approximately Newtonian behavior through Ehrenfest's Theorem, allbeit only on timescales for which wave packet spreading can be ignored. Nevertheless, we will see in section \ref{NewtonianTrajectories} that it is important to distinguish between the sort of narrow wave packet that suffers minimal entanglement under interaction with the environment - a `pointer state' wave packet - and the typically more inclusive category of wave packets that approximately follow a Newtonian trajectories - an `Ehrenfest' wave packet.

\subsection{Decoherence and Branching in Phase Space} 

Define `quasiclassical' histories as those in which the state of $S$ relative to a particular branch is always localised about a particular phase space point $z \equiv (q,p)$, but in which that phase space point does not necessarily traverse an approximately Newtonian trajectory. Define `classical' histories as those quasiclassical histories  in which the state of $S$ relative to a particular branch does traverse an approximately Newtonian trajectory (to within some margin of error and relative to some timescale). 
%Of course, a history can only be described as classical relative to some time scale and some margin of error, which will correspond to some coarse-graining of the history space associated with sequences of approximate time-indexed coherent state PVMs; for, relative to a fixed coarse-graining wave packet spreading ultimately produces non-classical trajectories on some time scale and likewise, relative to a fixed time scale, there will typically be some degree of coarse-graining (that is, there will be some partition of phase space with small enough cells) relative to which the realised histories are non-classical. 
In this subsection, we will see how environmental decoherence leads generically (or at least in a very wide set of cases) to quasiclassical histories; in the next, we will discuss how those quasiclassical histories also turn out in certain circumstances to be classical. 

\subsubsection{Decoherence Relative to a Coherent State Basis}

Let us examine the evolution of the quantum state of the system $SE$, given only the assumption that the pointer states of $S$ - that is, states of $S$ which suffer least entanglement upon interaction with the environment $E$ - are the coherent states $| q,p \rangle$. For brevity, I will use the condensed notation $z\equiv(q,p)$, $ | z \rangle \equiv | q,p \rangle$. An arbitrary quantum state of $SE$ at some initial time $t=0$ can be expressed in the form,

\begin{equation}
| \chi_{0}  \rangle = \int dz_{0} \ \alpha(z_{0}) \ | z_{0}  \rangle \otimes  | \xi(z_{0}) \rangle.
\end{equation}

\noindent Let us examine the evolution of an individual element of this superposition, $| z_{0} \rangle \otimes | \xi(z_{0}) \rangle$. Note that, initially, the $| \xi(z_{0}) \rangle$ need not be orthogonal. However, after a very brief time $\tau$ (on the order of some typical decoherence timescale), as a result of environmental scattering,

\begin{equation}
 | z_{0} \rangle \otimes | \xi(z_{0}) \rangle   \overset{\tau}{\Rightarrow}     | z_{0} \rangle \otimes | \phi(z_{0}) \rangle,
\end{equation}

\noindent where $\langle \phi(z'_{0}) | \phi(z_{0}) \rangle \approx 0$ for $z_{0}$ and $z'_{0}$ sufficiently different. On the longer time scale $\Delta t$ characteristic of the Hamiltonian $\hat{H}_{S}$, over which $\hat{H}_{S}$ induces significant changes on $\mathcal{H}_{S}$, we have that

\begin{equation}
  | z_{0} \rangle \otimes | \phi(z_{0}) \rangle  \overset{\Delta t}{\Longrightarrow}  \int dz_{1}  \beta(z_{0}, z_{1}) | z_{1} \rangle \otimes | \phi(z_{0}, z_{1}) \rangle
\end{equation}

\noindent where $\langle \phi(z'_{0}, z'_{1}) | \phi(z_{0}, z_{1}) \rangle \approx 0$ for $z_{0}$ and $z'_{0}$, or $z_{1}$ and $z'_{1}$,  sufficiently different. Moreover, we should expect that if $\langle \phi(z'_{0}) | \phi(z_{0}) \rangle \approx 0$, then it will turn out to be the case that $\langle \phi(z'_{0}, z'_{1}) | \phi(z_{0}, z_{1}) \rangle \approx 0$ irrespective of the values of $z_{1}$ and $z^{'}_{1}$ since the environmental particles whose scattering caused the initial decoherence process will now be spread widely across the environment's configuration space; for this reason, later scatterings of certain particles in the environment are generally unlikely to disrupt the orthogonality of environmental states resulting from earlier scatterings of other particles. Likewise, evolving each element of this last superposition by $\Delta t$, we have

\begin{equation}
  | z_{1} \rangle \otimes | \phi(z_{0}, z_{1}) \rangle  \overset{\Delta t}{\Longrightarrow}  \int dz_{2}  \beta(z_{0},z_{1},z_{2}) | z_{2} \rangle \otimes | \phi(z_{0}, z_{1},z_{2}) \rangle,
\end{equation}

\noindent where $\langle \phi(z'_{0}, z'_{1}, z'_{2}) | \phi(z_{0}, z_{1},z_{2}) \rangle \approx 0$ for $z_{0}$ and $z'_{0}$, or $z_{1}$ and $z'_{1}$, or  $z_{2}$ and $z'_{2}$, sufficiently different. Iterating this process $N$ times, we obtain,

\begin{equation}
 | z_{0} \rangle \otimes | \phi(z_{0}) \rangle  \overset{N \Delta t}{\Longrightarrow}  \int dz_{1}  ... \int dz_{N} \ B(z_{0}, z_{1},...,z_{N}) \ | z_{N} \rangle \otimes | \phi(z_{0},z_{1}, ..., z_{N}) \rangle
 \end{equation}

\noindent where

\begin{equation}
B(z_{0}, z_{1},...,z_{N}) \equiv \beta(z_{0}, z_{1}) \ \beta(z_{0}, z_{1}, z_{2}) \ ... \ \beta(z_{0}, ..., z_{N-1},z_{N}).
\end{equation}

\noindent By linearity of the Schrodinger evolution, we then have

\begin{equation} \label{StateEvolution}
| \chi(t) \rangle =    \int dz_{0} dz_{1} ... dz_{N} \ \alpha(z_{0}) B(z_{0}, z_{1},...,z_{N}) \  | z_{N} \rangle \otimes | \phi(z_{0}, z_{1},...,z_{N}) \rangle,
\end{equation}

\noindent with 

\begin{equation}
\langle \phi(z'_{0}, z'_{1},...,z'_{N}) | \phi(z_{0}, z_{1},...,z_{N}) \rangle \approx 0 \ \text{for} \ z_{k} \ \text{and} \ z'_{k} \ \text{ sufficiently different for any} \ 0 \leq k \leq N.
\end{equation}

%\noindent Taking $\Delta t \rightarrow 0$ yields a path integral over phase space trajectories $z(t)$, so that 

%\begin{equation}
%| \chi(t) \rangle =     \int \mathcal{D}z \ C[z(t)] \  | z_{t} \rangle \otimes | \phi[z(t)] \rangle, 
%\end{equation}

%\noindent where $C[z(t)] \equiv \lim_{\Delta t \rightarrow 0} \ \beta(z_{0}; z_{1}) \ \beta(z_{1}; z_{2}) \ ... \ \beta(z_{N-1}; z_{N}) $.

\noindent These last two equations provide a completely general expression for the quantum state of $SE$ evolved up to some arbitrary time $N \Delta t$, under the assumption that the coherent states $| z \rangle$ are the decoherence-preferred states of $S$ under its interaction with $E$.

\vspace{5mm}

%Let us now consider a general evolution of a product state at $t=0$ consisting of a single narrow wave packet $|q_{0},p_{0} \rangle$ and an arbitary environment state $| \phi \rangle$. Assuming that the wave packet spreads, as it typically will, the evolution of the state up to some time $t_{1}$ is 

%\begin{equation}
%|q_{0},p_{0} \rangle \otimes | \phi \rangle   \longrightarrow  |q_{0},p_{0} \rangle \otimes
%\end{equation}

It will prove helpful for our purposes to examine this evolution also from the perspective of the decoherent histories formalism. The approximate PVM of eq. (\ref{PhaseSpacePVM}) can be extended to an approximate PVM on $\mathcal{H}_{S}\otimes \mathcal{H}_{E}$ simply by tensoring the projectors with the identity $\hat{I}_{E}$ on $\mathcal{H}_{E}$, taking the operators $\{ \hat{\Pi}_{i} \otimes \hat{I}_{E} \}$ to constitute the extended approximate PVM. Using this PVM, the evolution of a state on $\mathcal{H}_{S}\otimes \mathcal{H}_{E}$ can be expressed generally as

\begin{align} \label{PhaseSpaceSum}
 | \chi(t) \rangle & = \sum_{i_{1},...,i_{N}} \left( \hat{\Pi}_{i_{N}} \otimes \hat{I}_{E}  \right) e^{-i\hat{H}\frac{t}{N}} ...  \left( \hat{\Pi}_{i_{1}} \otimes \hat{I}_{E}  \right) e^{-i\hat{H} \frac{t}{N}}  \left( \hat{\Pi}_{i_{0}} \otimes \hat{I}_{E}  \right) | \chi_{0} \rangle \\
& = \sum_{i_{0}, i_{1},...,i_{N}} \hat{C}_{i_{0}, i_{1},...,i_{N}} | \chi_{0} \rangle
\end{align}

\noindent where

\begin{equation}
\hat{C}_{i_{0}, i_{1},...,i_{N}} | \chi_{0} \rangle \equiv  \left( \hat{\Pi}_{i_{N}} \otimes \hat{I}_{E}  \right) e^{-i\hat{H}\frac{t}{N}} ... e^{-i\hat{H} \frac{t}{N}}  \left( \hat{\Pi}_{i_{1}} \otimes \hat{I}_{E}  \right) e^{-i\hat{H} \frac{t}{N}}  \left( \hat{\Pi}_{i_{0}} \otimes \hat{I}_{E}  \right).
\end{equation}

\noindent This history operator corresponds to a coarse-grained trajectory in phase space that successively traverses the sequence $(\Sigma_{i_{0} }, \Sigma_{i_{1}},...,\Sigma_{i_{N}})$ of cells in the phase space partition. Using the defintion of the operators $\hat{\Pi}_{i}$, we can rewrite the history operators as follows:

\begin{align}
& \hat{C}_{i_{0}, i_{1},...,i_{N}} | \chi_{0} \rangle \\
&\equiv  \int_{\Sigma_{i_{0}}} \int_{\Sigma_{i_{1}}} ... \int_{\Sigma_{i_{N}}}dz_{1} ...  dz_{N}    \left(  | z_{N} \rangle  \langle z_{N}  | \otimes \hat{I}_{E}  \right) e^{-i\hat{H}\frac{t}{N}} ... e^{-i\hat{H} \frac{t}{N}}  \left(  | z_{1} \rangle  \langle z_{1}  | \otimes \hat{I}_{E}  \right)  e^{-i\hat{H} \frac{t}{N}}  \left(  | z_{0} \rangle  \langle z_{0}  | \otimes \hat{I}_{E}  \right) | \chi_{0} \rangle \\
& \equiv \int_{\Sigma_{i_{0}}}  \int_{\Sigma_{i_{1}}} ... \int_{\Sigma_{i_{N}}} dz_{1} ...  dz_{N} \  \hat{C}_{z_{0}, z_{1},...,z_{N}} | \chi_{0} \rangle,
\end{align}

\noindent where 

\begin{equation}
\hat{C}_{z_{0}, z_{1},...,z_{N}}  \equiv  \left(  | z_{N} \rangle  \langle z_{N}  | \otimes \hat{I}_{E}  \right) e^{-i\hat{H}\frac{t}{N}} ...   \left(  | z_{1} \rangle  \langle z_{1}  | \otimes \hat{I}_{E}  \right)  e^{-i\hat{H} \frac{t}{N}} \left(  | z_{0} \rangle  \langle z_{0}  | \otimes \hat{I}_{E}  \right). 
\end{equation}

\noindent Moreover, note that 

\begin{align}
\scriptsize
\hat{C}_{z_{0},z_{1},...,z_{N}} | z_{0}\rangle & = | z_{N} \rangle \otimes \bigg[ \sum_{i} | e_{i} \rangle  \langle z_{N}, e_{i}  | e^{-i\hat{H}\frac{t}{N}}  \left(  | z_{N-1} \rangle  \langle z_{N-1}  | \otimes \hat{I}_{E}  \right) \\
&...  \left(  | z_{1} \rangle  \langle z_{1}  | \otimes \hat{I}_{E}  \right)  e^{-i\hat{H} \frac{t}{N}} \left(  | z_{0} \rangle  \langle z_{0}  | \otimes \hat{I}_{E}  \right) | z_{0} \rangle \bigg]  \\
& =  | z_{N} \rangle \otimes \left[ \sum_{i} | e_{i} \rangle  \langle z_{N}, e_{i}  | e^{-i\hat{H}\frac{t}{N}} \hat{C}_{z_{0},z_{1},...,z_{N-1}} \ | z_{0}\rangle \right] \\
&=  | z_{N} \rangle \otimes | \tilde{\phi}(z_{0},z_{1},...,z_{N}) \rangle
\end{align}

\noindent where $\{ e_{i} \}$ is an arbitrary basis for $\mathcal{H}_{E}$ and

\begin{equation}
 | \tilde{\phi}( z_{0}, z_{1},...,z_{N}) \rangle \equiv  \sum_{i} | e_{i} \rangle  \langle z_{N}, e_{i}  | e^{-i\hat{H}\frac{t}{N}} \hat{C}_{z_{0},z_{1},...,z_{N-1}} \ | z_{0}\rangle. 
\end{equation}

\noindent The vectors $ | \tilde{\phi}(z_{0},  z_{1},...,z_{N}) \rangle$ will not in general be normalised. If $w(z_{0}, z_{1},...,z_{N}) \equiv   \langle  \tilde{\phi}( z_{0}, z_{1},...,z_{N}) | \tilde{\phi}( z_{0},z_{1},...,z_{N}) \rangle$, then we can define

\begin{equation}
| \phi(z_{0}, z_{1},...,z_{N}) \rangle \equiv \frac{1}{w^{1/2}(z_{0}, z_{1},...,z_{N})} | \tilde{\phi}( z_{0},z_{1},...,z_{N}) \rangle,
\end{equation}

\noindent which are normalised. In terms of these normalised states, we then have   

\begin{equation}
\hat{C}_{z_{0},z_{1},...,z_{N}} | z_{0}\rangle =  w^{1/2}(z_{0}, z_{1},...,z_{N}) | z_{N} \rangle \otimes | \phi(z_{0},z_{1},...,z_{N}) \rangle,
\end{equation}

%\begin{align}
%& \hat{C}_{i_{0},i_{1},...,i_{N}} | \chi_{0} \rangle \\
%&\approx  \int_{\Sigma_{i_{0}}} \int_{\Sigma_{i_{1}}} ... \int_{\Sigma_{i_{N}}} dz_{0} dz_{1} ...  dz_{N} \  \alpha(z_{0}) w^{1/2}(z_{0},z_{1},...,z_{N}) | z_{N} \rangle \otimes | \phi(z_{0}, z_{1},...,z_{N}) \rangle
%\end{align},

\noindent and 

\begin{align}
 | \chi(N \Delta t) \rangle & = e^{-i \hat{H} N \Delta t}  | \chi_{0}\rangle \\
 & =  \int \int ... \int dz_{0} dz_{1} ...  dz_{N} \ \hat{C}_{z_{0},z_{1},...,z_{N}} | \chi_{0}\rangle     \\
 & = \int \int ... \int dz_{0} dz_{1} ...  dz_{N} \  \alpha(z_{0}) w^{1/2}(z_{0},z_{1},...,z_{N}) | z_{N} \rangle \otimes | \phi(z_{0}, z_{1},...,z_{N}) \rangle
 \end{align}

\noindent Note further that if $\langle \phi( z'_{0}, z'_{1},...,z'_{N}) | \phi( z_{0}, z_{1},...,z_{N}) \rangle \approx 0$ for $z_{k}$ and $z'_{k}$ sufficiently different, for any $0 \leq k \leq N$, then $\langle \chi_{0} | \hat{C}_{i'_{0},i'_{1},...,i'_{N}}^{ \dagger} \hat{C}_{i_{0},i_{1},...,i_{N}}   | \chi_{0}\rangle \approx 0 $ for $i_{k} \neq i'_{k}$ for $0 \leq k \leq N$ - assuming that the partitioning of phase space is such that $z_{k}$ and $z'_{k}$ are `sufficiently different' when they belong to different elements of the phase space partition. 

\subsubsection{Branching}

From the above analysis, we can conclude that because the history space associated with sequences of coherent state projectors $\hat{\Pi}_{i_{k}}$ is decoherent, the quantum state has branching structure relative to the history space associated with sequences of projectors $\hat{P}^{n}_{i_{0}, i_{1},...,i_{n}} \equiv \frac{1}{w(i_{0},i_{1},...,i_{n})}\hat{C}_{i_{0} i_{1},...,i_{n}} | \chi_{0} \rangle \langle \chi_{0} | \hat{C}^{\dagger}_{i_{0},i_{1},...,i_{n}}$, for successive $n$, where $w(i_{0}, i_{1},...,i_{n})\equiv \langle \chi_{0} | \hat{C}^{\dagger}_{i_{0},i_{1},...,i_{n}} \hat{C}_{i_{0},i_{1},...,i_{n}} | \chi_{0} \rangle$. Using the abbreviation $\textbf{i}^{n} \equiv (i_{0},i_{1},...,i_{n})$, we can write 

\begin{align} \label{PSBranchingSum}
 | \chi(t) \rangle & = \sum_{\textbf{i}^{1},...,\textbf{i}^{N}} \left( \hat{P}_{\textbf{i}^{N}} \otimes \hat{I}_{E}  \right) e^{-i\hat{H}\frac{t}{N}} ...  \left( \hat{P}_{\textbf{i}^{1}} \otimes \hat{I}_{E}  \right) e^{-i\hat{H} \frac{t}{N}}  \left( \hat{P}_{\textbf{i}^{0}} \otimes \hat{I}_{E}  \right) | \chi_{0} \rangle \\
& = \sum_{\textbf{i}^{0},\textbf{i}^{1},...,\textbf{i}^{N}} \hat{K}_{\textbf{i}^{0}, \textbf{i}^{1},...,\textbf{i}^{N}} | \chi_{0} \rangle.
\end{align}

\noindent where $\hat{K}_{\textbf{i}^{0},\textbf{i}^{1},...,\textbf{i}^{N}} | \chi_{0} \rangle \equiv  \left( \hat{P}_{\textbf{i}^{N}} \otimes \hat{I}_{E}  \right) e^{-i\hat{H}\frac{t}{N}} ...  \left( \hat{P}_{\textbf{i}^{1}} \otimes \hat{I}_{E}  \right) e^{-i\hat{H} \frac{t}{N}}   \left( \hat{P}_{\textbf{i}^{0}} \otimes \hat{I}_{E}  \right)  | \chi_{0} \rangle$. Thus, the set of histories in this history space corresponds to the set of sequences (of sequences) of the form $(\textbf{i}^{0},\textbf{i}^{1},...,\textbf{i}^{N})$. To say that the quantum state has branching structure relative to this history space entails that any two realised histories that agree on their $n^{th}$ index - that is, that share the index $\textbf{i}^{n}$ - agree with respect to all previous indices $\textbf{i}^{k}$ with $0 \leq k < n$. This is indeed the case since, as one can show quite straightforwardly, the only realised histories are those for which $\textbf{i}^{n}$ is an initial sequence of $\textbf{i}^{n+1}$ for all $1\leq n \leq N-1$. Thus, any two histories that agree on $\textbf{i}^{n}$ necessarily agree on all previous indices since all of these indices are uniquely determined by $\textbf{i}^{n}$.

Thus, we can write the state evolution in terms of a sum restricted only to histories $(\textbf{i}^{0}, \textbf{i}^{1},...,\textbf{i}^{N})$ that obey the restriction that each subsequence $\textbf{i}^{n}$ is the initial sequence of the next, $\textbf{i}^{n+1}$. Abbreviating this condition by writing $\textbf{i}^{n} \subset \textbf{i}^{n+1}$, we have 

\begin{align} \label{PSBranchingSum}
 | \chi(t) \rangle  \approx \sum_{\textbf{i}^{n} \subset \textbf{i}^{n+1}} \hat{K}_{\textbf{i}^{0}, \textbf{i}^{1},...,\textbf{i}^{N}} | \chi_{0} \rangle,
\end{align}

\noindent where (again) the condition below the summation symbol indicates that only sequences for which each element of the sequence is a subsequence of all elements that come after it are included in the sum.  

Note moreover that the history space associated with the set of sequences $(i_{0},i_{1},...,i_{N})$, formed using the approximate coherent state PVM, is not branching since it is entirely possible (because of wave packet spreading) that two realised (but mutually decoherent) histories agree on some index $i_{n}$, with $1 \leq n \leq N$, but not with respect to some or any of the previous indices $i_{k}$, where $0 \leq k < n$.

Finally, note that the history space associated with the set of sequences $(i_{0},i_{1},...,i_{N})$ is a coarse-graining of the history space associated with the set of sequences $(\textbf{i}^{0}, \textbf{i}^{1},...,\textbf{i}^{N})$, since 

\begin{equation}
 \hat{C}_{i_{0},i_{1},...,i_{N}} | \chi_{0} \rangle = \hat{C}_{\textbf{i}^{N}} | \chi_{0} \rangle 
 = \sum_{\textbf{i}^{0},\textbf{i}^{1},...,\textbf{i}^{N-1}} \hat{K}_{\textbf{i}^{0}, \textbf{i}^{1},...,\textbf{i}^{N}} | \chi_{0} \rangle, 
\end{equation}

\noindent where the reader should note that only the indices $(\textbf{i}^{1},...,\textbf{i}^{N-1})$ have been summed over. 

%Thus, one can express the branch states $\hat{C}_{i_{1},...,i_{N}} | \chi_{0} \rangle$ in the form

%\begin{align}
%& \hat{C}_{i_{1},...,i_{N}} | \chi_{0} \rangle =  \int_{\Sigma_{i_{1}}} ... \int_{\Sigma_{i_{N}}}dz_{1} ...  dz_{N} \ \\
%&  \ \ \ \ \alpha(z_{0}) \ \beta(z_{0}, z_{1}) \ \beta(z_{1}, z_{2}) \ ... \ \beta(z_{N-1},z_{N}) \  | z_{N} \rangle \otimes | \phi(z_{0}, z_{1},...,z_{N}) \rangle.
%\end{align}

%\noindent Note further that if $\langle \phi(z'_{0}, z'_{1},...,z'_{N}) | \phi(z_{0}, z_{1},...,z_{N}) \rangle \approx 0$ for $z_{k}$ and $z'_{k}$ sufficiently different for any $0 \leq k \leq N$, then $\langle \chi_{0} | \hat{C}_{i'_{1},...,i'_{N}}^{ \dagger} \hat{C}_{i_{1},...,i_{N}}   | \chi_{0}\rangle \approx 0 $ for $i_{k} \neq i'_{k}$ for $0 \leq k \leq N$. Here, the partition of the phase space should be chosen so that $z_{k}$ and $z'_{k}$ are sufficiently different when they belong to different elements of the phase space partition associated with the indices $i_{k}$.

\vspace{20mm}

Returning to the history space associated with sequences of coherent state PVM's (as opposed to the history space associated with sequences of such sequences), the sum in eq. (\ref{PhaseSpaceSum}) includes all histories $(i_{1},...,i_{N})$ in the history space associated with the approximate PVM of eq. (\ref{PhaseSpacePVM}). Typically, not all histories in the history space associated with the set of sequences $(i_{1},...,i_{N})$  will be realised - that is, for many of these histories, we will have 

\begin{equation}
\hat{C}_{i_{1},...,i_{N}} | \chi_{0} \rangle  \approx 0. 
\end{equation}

\noindent For this reason, one can to a good approximation restrict the sum in eq. (\ref{PhaseSpaceSum}) to only those histories that are realised. If we denote the set of realised histories (where realised histories are defined as those whose weights surpass some arbitrarily specified but small threshold $\epsilon$) as $I_{r}$, then we have,

\begin{equation} \label{Realised}
 | \chi(t) \rangle \approx \sum_{(i_{1},...,i_{N}) \in I_{r}} \hat{C}_{i_{1},...,i_{N}} | \chi_{0} \rangle.
\end{equation}

\noindent What can we say, then, about which histories tend to be realised in a model like the Bare/Everett model described above, in which the central system $S$ consists of some macroscopic centers of mass interacting through a conservative time-independent potential, as well as with some external environment $E$? As we will see in the next subsection, under certain conditions and for a certain measure of coarse graining, the realised histories will be those that traverse approximately Newtonian trajectories.

\subsection{The Occurrence of Classical Newtonian Trajectories} \label{NewtonianTrajectories}

Thus far, decoherence has explained why a branch of the total quantum state should be such that the state of $S$ relative to that branch evolves quasiclassically. However, the analysis above has not offered any suggestion as to why the state of $S$ relative to each branch should traverse a \textit{classical} trajectory - i.e., why its evolution should conform approximately to Newtonian equations of motion. It is here that Ehrenfest's Theorem enters the analysis. 

\subsubsection{Ehrenfest's Theorem for Open Systems}

As we saw in Chapter 1, Ehrenfest's Theorem applies only to systems that evolve unitarily in a pure state. However, as I explain here, the theorem can be straightforwardly generalised to certain models of open quantum systems. In situtations of the sort considered in this thesis, the interaction between system and environment is typically sufficiently weak that the dissipation and renormalisation terms in (\ref{CaldLegg}) can be ignored (see, for instance, \cite{wallace2012emergent}, section 3.6) so that only a `pure decoherence' term, $- i \Lambda \left[ \hat{X},\left[ \hat{X}  , \hat{\rho}_{S} \right] \right]$, remains: 

\begin{equation} \label{DecoherenceMaster}
i \frac{d \hat{\rho}_{S}}{dt} = [ \hat{H}_{S}, \hat{\rho}_{S} ]  - i \Lambda \left[ \hat{X},\left[ \hat{X}  , \hat{\rho}_{S} \right] \right]
\end{equation}

\noindent \footnote{I have not encountered this generalisation of Ehrenfest's Theorem elsewhere in the literature on decoherence, though the calculation required for the generalisation is sufficiently straightforward, and the motivation sufficiently obvious, that I will refrain from claiming any originality for the result.}. In this model, the time derivative of the expectation value of momentum $\langle \hat{P} \rangle = Tr [{\hat{\rho}_{S} \hat{P}}] $ can be calculated by multiplying both sides of this equation by the operator $\hat{P}$ and then taking the trace of both sides. Using the commutation relations $[ \hat{X}, \hat{P} ]$ and the cyclic property of the trace, one can show that

\begin{equation}
Tr \left\{ \left[ \hat{X},\left[ \hat{X}  , \hat{\rho}_{S} \right] \right] \hat{P} \right\} = 0. 
\end{equation}

\noindent Thus, 

\begin{equation}
i \frac{d \langle \hat{P}\rangle}{dt} =  Tr  \left\{ [ \hat{H}_{S}, \hat{\rho}_{S} ]  \hat{P} \right\}.
\end{equation}

\noindent But this equation is the same as the equation for the evolution of $\langle \hat{P}\rangle$ under the unitary dynamics prescribed by $\hat{H}_{S}$, the primary difference from the closed system case being that $\hat{\rho}_{S}$ in this case will not in general be pure. Just as in the derivations of Ehrenfest's Theorem in the case of a closed system governed solely by the Hamiltonian $ \hat{H}_{S} = \frac{\hat{P}^{2}}{2M} + V(X)$ - which does not rely on the purity of $\hat{\rho}_{S}$ - we then have

\begin{equation}
\frac{d \langle \hat{P}\rangle}{dt} = - \left\langle \hat{ \frac{\partial V(X)}{\partial X}} \right\rangle.
\end{equation}

\noindent thus furnishing a generalisation of Ehrenfest's Theorem to systems whose interaction with their environment is characterised approximately by `pure decoherence.' 

Specialising now to the case where $\rho_{S}(X,X') \equiv \langle X' | \hat{\rho}_{S} | X \rangle$ represents a state with narrow ensemble width - i.e., such that the diagonal elements  $\rho_{S}(X,X)$ are narrowly peaked about some particular value of $X$ -  we have that $\langle f(\hat{X}) \rangle \approx f(\langle \hat{X} \rangle )$, so that

\begin{equation} \label{EhrenfestOpenNarrow}
\frac{d \langle \hat{P} \rangle}{dt} \approx - \frac{\partial V(\langle \hat{X} \rangle)}{\partial \langle \hat{X} \rangle}, 
\end{equation}

\noindent thereby furnishing an approximate version of Newton's Second Law for expectation values. The relation $\frac{d \langle \hat{X} \rangle}{dt} = \frac{\langle \hat{P} \rangle}{M}$ also continues to hold as long as the interaction $\hat{H}_{int}$ depends only on position $\hat{X}$ and not on the momentum $\hat{P}$, as is the case in all models considered here - and notably, in particular, in the Caldeira-Legett model. Note that the expectation value in this relation is taken relative to the total quantum state of system $S$, not its state relative to a particular branch of the quantum state. While `branch-relative' expectation values will prove crucial to my analysis of Newtonian behavior below, it is crucial to recognise that the expectation values here average \textit{across} multiple mutually decohered branches. 

From this generalisation of Ehrenfest's Theorem, we can conclude that position and momentum expectation values even of decohering systems will follow approximately classical trajectories, so long as the internal dynamics $\hat{H}_{S}$ does not cause $\rho_{S}(X,X)$  to become too spread out; if it does become too spread out relative to the characteristic length scales on which the potential $V(X)$ varies, the relation $\left\langle \hat{ \frac{\partial V(X)}{\partial X}} \right\rangle \approx \frac{\partial V(\langle \hat{X} \rangle)}{\partial \langle \hat{X} \rangle} $ will cease to hold, and approximate Newtonian behavior will no longer be guaranteed.

\subsubsection{The Dual Role of Narrow Wave Packets}

At this point, it is worth underscoring the fact that narrow wave packets play a dual role in accounting for the classical behavior of macroscopic open quantum systems: first, among them are the states of $S$ - the coherent states - that are minimally entangled with the environment on timescales short relative to the characteristic timescales associated with $\hat{H}_{S}$; second, they evolve in approximately Newtonian fashion, even when $S$ is open. 

However, it is also crucial to distinguish two corresponding senses of narrowness here: one that relates to considerations of environmental decoherence, which ensures \textit{quasi}-classicality of the state evolution (at least on the Bare/Everett model), and the other that relates to approximate Newtonian behavior, which ensures approximate classicality of the evolution. The former requires narrowness in the sense of not being wider than the typical coherence length of the system under its interaction with the environment, which is characteristed by the position-space width of a coherent pointer state $| q,p \rangle$. The latter requires narrowness of $\rho_{S}(X,X)$ relative to the dimensions of the potential $V(X)$; assuming that $V(X)$ varies significantly only on macroscopic length scales, this notion of narrowness will include wave packets that are narrow in the first sense as well as those whose widths exceed typical coherence lengths of the system $S$. 

As a result of these considerations, there is a regime of behavior in which wave packets may spread to width beyond the coherence length - thereby inducing decoherence and the branching that goes along with it - but remain narrow with respect to the relevant dimensions of the potential $V(X)$. This will be a regime in which it is both the case that: 1) on any given branch, expectation values of position and momentum evolve along approximately deterministic, Newtonian trajectories, to within a certain margin of error associated with the distribution $\rho_{S}(X,X)$ ; 2) branching still occurs on length scales below this margin of error, which one might characterise as being associated with `quantum fluctuations' around an average classical, deterministic trajectory. 

Thus, macroscopic classical behavior is underwritten quantum mechanically by a wave packet for $S$ that spreads out, as it does so rapidly becoming replaced by an incoherent superposition of localised states. Each of the localised states in the incoherent mixture will in turn spread out and be replaced by its own incoherent superposition of localised states. Eventually, an initially narrowly peaked $\rho_{S}(X,X)$ will spread incoherently to a width comparable to the scales on which $V(X)$ varies significantly, and at this point Newtonian behavior cannot be expected even as an approximation. 

\subsubsection{Factors Affecting the Rate of Wave Packet Spreading}

The primary factors that determine the rate of wave packet spreading in $S$ are 

\begin{enumerate}
\item \underbar{the size of the masses $M$ appearing in $\hat{H}_{S}$}: other factors being equal, larger masses typically correlate to slower spreading; a simple back-of-the-envelope calculation for the case of a free particle indicates that it would take an initially narrow wave packet of an object with macroscopic mass, localised on atomic length scales, longer than the age of the universe to spread to macroscopic size (say 1cm.);
\item \underbar{the presence of chaotic effects}: these can significantly accelerate the rate of wave packet spreading;  in systems with macroscopic mass where $\hat{H}_{S}$ is strongly chaotic (i.e., in which closely spaced initial conditions diverge on short timescales), a wave packet initially localised on atomic length scales may spread to macroscopic length scales over much shorter time periods than in cases where the effects of chaos are weak or absent; for example, Zurek and Paz have argued convincingly that Saturn's moon Hyperion, which tumbles chaotically in its orbit, should exhibit macroscopic divergences from Newtonian predictions on a timescale of about 10 years \cite{zurek1995quantum};
\item \underbar{pure decoherence}: the pure decoherence term characterising the influence of $E$ on $S$ in the master equation $(\ref{CaldLegg})$ typically will also increase the rate of wave packet spreading; while this term constantly suppresses the \textit{coherent} spreading of $S$'s state, it actually results in a slight increase in the rate of \textit{ensemble} spreading in $S$ (see, for example, \cite{schlosshauer2008decoherence}, p. 145 ).
\end{enumerate}

 \subsubsection{Parallel Coarse-Grained Classical Histories}
 
 Consider an arbitrary initial superposition at time $t=0$:

 \begin{equation}
| \chi_{0}  \rangle = \int dz_{0} \ \alpha(z_{0}) \ | z_{0}  \rangle \otimes  | \phi(z_{0}) \rangle.
\end{equation}

\noindent Let us examine the evolution of a single element of this superposition, 

\begin{equation}
| z_{0}  \rangle \otimes  | \phi(z_{0}) \rangle.
\end{equation}

\noindent The position space density matrix associated with this state, $\rho^{z_{0}}_{S}(X,X',0)$, is pure and has both a narrow coherence length and ensemble width, which are both equal to the position space width of the packet  $| z_{0}  \rangle$. Over some time $\Delta t$, the wave packet will spread both under the influence of the internal dynamics $\hat{H}_{S}$ and as a result of the influence of the environment; however, as long as the ensemble width remains narrow by comparison with the potential $V(X)$, the expectation value of $X$ in this incoherent superposition will continue to evolve along an approximately Newtonian trajectory to within a margin of error determined by the the ensemble width. The total state of $SE$ will evolve to a state of the form

 \begin{equation}
\int_{\Sigma_{1}} d z_{1} \beta_{1}( z_{0}, z_{1})  | z_{1}  \rangle \otimes  | \phi(z_{0}, z_{1}) \rangle,
\end{equation}
 
 \noindent such that $\langle \phi(z'_{0}, z'_{1}) | \phi(z_{0}, z_{1}) \rangle \approx 0$ for $z_{0}$ and $z'_{0}$, or $z_{1}$ and $z'_{1}$,  sufficiently different, and such that the integral over $z_{1}$ can, to a good approximation, be restricted to a small cell $\Sigma_{1}$ containing the classical future evolute of $z_{0}$ up to time $\Delta t$. The density matrix $\rho^{z_{0}}_{S}(X,X',\Delta t)$ associated with this state will have narrow coherence length equal to the width of a pointer coherent state, while the ensemble width will have spread beyond this; thus, it will be more spread out along the diagonal direction than along the off-diagonal direction. A single element $| z_{1}  \rangle \otimes  | \phi(z_{0}, z_{1}) \rangle$ of this last superposition will evolve over the next interval $\Delta t$ into a superposition of the form
$\int_{\Sigma_{2}} d z_{2} \beta_{2}( z_{0}, z_{1}, z_{2})  | z_{2}  \rangle \otimes  | \phi(z_{0}, z_{1}, z_{2}) \rangle$, where $\Sigma_{2}$ is the classical future evolute of of $z_{0}$ up to time $2 \Delta t$. Thus, the total state to which $| z_{0}  \rangle \otimes  | \phi(z_{0}) \rangle$ has evolved by time $2\Delta t$ is 
 
\begin{equation}
\int_{\Sigma_{1}}  \int_{\Sigma_{2}} d z_{1}  d z_{2} \beta_{2}( z_{0}, z_{1}, z_{2}) \beta_{1}( z_{0}, z_{1})  | z_{2}  \rangle \otimes  | \phi(z_{0}, z_{1}, z_{2}) \rangle,
\end{equation}

\noindent Iterating this procedure, then, the single initial state $| z_{0}  \rangle \otimes  | \phi(z_{0}) \rangle$ will evolve over some time $N \Delta t$ into a state of the form 
 
\begin{equation} \
\int_{\Sigma_{1}} ... \int_{\Sigma_{N}} d z_{1} ... d z_{N}    B(z_{0}, z_{1},...,z_{N})   | z_{N}  \rangle \otimes  | \phi(z_{0}, z_{1},...,z_{N}) \rangle,
\end{equation}
 
\noindent where

\begin{equation}
B(z_{0}, z_{1},...,z_{N}) \equiv  \beta_{N}( z_{0}, z_{1},...,z_{N})  \beta_{N-1}( z_{0}, z_{1},...,z_{N-1}) ... \beta_{1}( z_{0}, z_{1}),
\end{equation}
 
 \noindent and
 
 \begin{equation}
\langle \phi(z'_{0}, z'_{1},...,z'_{N}) | \phi(z_{0}, z_{1},...,z_{N}) \rangle \approx 0 \ \text{for} \ z_{k} \ \text{and} \ z'_{k} \ \text{ sufficiently different for any} \ 0 \leq k \leq N.
\end{equation}

\noindent  The function $B(z_{0}, z_{1},...,z_{N})$ differs substantially from zero only when $(z_{0}, z_{1},...,z_{N})$ all lie close to the classical trajectory associated with $z_{0}$. The density matrix $\rho^{z_{0}}_{S}(X,X', N\Delta t)$ will be an incoherent mixture of localised wave packets, with coherence length equal to the position width of one coherent pointer state and the ensemble width growing progressively larger for increasing values of $N$. As we have seen, decoherence of the form we have been discussing lends the evolution of the quantum state a branching structure. However, even as this branching occurs, the histories associated with the different branches of the total state will remain concentrated around a single classical trajectory for as long as the ensemble width $\rho^{z_{0}}_{S}(X,X', N\Delta t)$ remains sufficiently narrow relative to the dimensions of the potential $V(X)$. For systems in which chaotic effects on wave packet spreading are significant, the timescales on which this narrowness is maintained will be significantly shorter than for those in which chaotic effects can be neglected. Nevertheless, even for chaotic systems there will be some time scale on which Newtonian predictability holds to within some reasonable margin of error. 

The preceding analysis has shown that if we restrict our consideration to timescales less than those on which the ensemble width of $S$ becomes comparable to the relevant dimensions of the potential $V(X)$, then the evolution of a single initial wave packet $| z_{0}  \rangle \otimes  | \phi(z_{0}) \rangle$ gives rise to multiple decohered branches, with the histories associated with different branches of the quantum state all concentrated around the single classical trajectory associated with $z_{0} $. Because the ensemble width will typically grow with time, the average deviation of these branches from the classical trajectory also will tend to increase with time. Eventually, when the ensemble width becomes comparable to the scale on which the potential $V(X)$ changes significantly, it will be the case note only that the individual branches will be be prone to deviate more widely from the mean, but also that the mean itself will cease to evolve approximately classically. Thus, the time scales on which one should generally expect Newtonian behavior are limited by the scale on which the ensemble width of an initial narrow wave packet product state $| z_{0}  \rangle \otimes  | \phi(z_{0}) \rangle$ expands to a scale comparable to the spatial dimensions of $V(X)$. Before this point, one still may observe deviations from classicality that can be characterised as fluctuations $\textit{around}$ some mean classical behavior; but after this time, classicality ceases to hold even in the mean. 

Given that for a single wave packet, the evolution is 

\begin{align}
e^{-i\hat{H}N \Delta t}  \left( | z_{0}  \rangle \otimes  | \phi(z_{0}) \rangle  \right) & \approx \int_{\Sigma_{1}(z_{0})} ... \int_{\Sigma_{N}(z_{0})} d z_{1} ... d z_{N}    B(z_{0}, z_{1},...,z_{N})   | z_{N}  \rangle \otimes  | \phi(z_{0}, z_{1},...,z_{N}) \rangle \\
& \approx  \int_{\Sigma_{1}(z_{0})} ... \int_{\Sigma_{N}(z_{0})} d z_{1} ... d z_{N} \hat{C}_{z_{0},z_{1},...,z_{N}} | z_{0} \rangle, 
\end{align}

\noindent the linearity of the Schrodinger evolution entails that

\begin{align}
 | \chi(N \Delta t) \rangle &= e^{-i\hat{H}N \Delta t} | \chi_{0} \rangle \\
& \approx   \int dz_{0} \ \alpha(z_{0})  \int_{\Sigma_{1}(z_{0})} ... \int_{\Sigma_{N}(z_{0})} d z_{1} ... d z_{N}    B(z_{0}, z_{1},...,z_{N})   | z_{N}  \rangle \otimes  | \phi(z_{0}, z_{1},...,z_{N}) \rangle \\
& =  \int dz_{0} \ \alpha(z_{0})  \int_{\Sigma_{1}(z_{0})} ... \int_{\Sigma_{N}(z_{0})} d z_{1} ... d z_{N}  \ \hat{C}_{z_{0}, z_{1},...,z_{N}}  | z_{0} \rangle 
\end{align}

\noindent where $(\Sigma_{1}(z_{0}),..., \Sigma_{N}(z_{0}))$ are all concentrated around the classical trajectory whose initial condition is $z_{0}$, and where

\begin{equation} \label{WavePacketDec}
\langle z_{0} | \hat{C}^{\dagger}_{ z'_{0},z'_{1},...,z'_{N}} \hat{C}_{z_{0}, z_{1},...,z_{N}}  | z_{0} \rangle \approx 0 \ \text{for} \ z_{k} \ \text{and} \ z'_{k} \ \text{sufficiently different for any} \ 0 \leq k \leq N,
\end{equation}

\noindent since 

\begin{equation}
\langle \phi(z'_{0}, z'_{1},...,z'_{N}) | \phi(z_{0}, z_{1},...,z_{N}) \rangle \approx 0 \ \text{for} \ z_{k} \ \text{and} \ z'_{k} \ \text{sufficiently different for any} \ 0 \leq k \leq N, 
\end{equation}

\noindent which in turn, as we have seen, will be satisfied as a result of environmental scattering. 

%\begin{equation}
%| \chi(N \Delta t) \rangle  = \sum_{i_{1},...,i_{N}}  \hat{C}_{i_{1},...,i_{N}} | \chi_{0} \rangle  
%\end{equation}

%\noindent 

%\begin{equation}
%\langle \chi_{0} |   \hat{C}^{\dagger}_{i'_{1},...,i'_{N}}    \hat{C}_{i_{1},...,i_{N}} | \chi_{0} \rangle \approx 0 \ \text{if} \ i_{k} \neq i'_{k} \ \text{for} \ 1 \leq k \leq N,
%\end{equation}

%\noindent

%\begin{equation}
%| \chi(N \Delta t) \rangle  \approx \sum_{j_{0}} \hat{C}_{j_{0}, j_{1}^{c},...,j_{N}^{c}} | \chi_{0} \rangle
%\end{equation}

%Alternatively, we may write the state evolution as a sum over history operators defined in terms of approximate coherent state PVMs $\{ \hat{P}^{k}_{i_{k}} = \int_{\Sigma_{i_{k}}} dz_{k} \ | z_{k} \rangle \langle z_{k} | \}$, where we allow the partitioning of phase space associated with each PVM to vary with time.

%\begin{align}
%& | \chi(N \Delta t) \rangle = e^{-i\hat{H}N \Delta t} \sum_{i_{0}} \hat{P}_{i_{0}}| \chi_{0} \rangle \\
%& \approx   \sum_{i_{0}} \int_{\Sigma_{i_{0}}} dz_{0} \ \alpha(z_{0})  \int_{\Sigma^{c}_{1}(z_{0})} ... \int_{\Sigma^{c}_{N}(z_{0})} d z_{1} ... d z_{N}    B(z_{0}, z_{1},...,z_{N})   | z_{N}  \rangle \otimes  | \phi(z_{0}, z_{1},...,z_{N}) \rangle \\
%& =  \sum_{i_{0}} \int_{\Sigma_{i_{0}}} dz_{0} \ \alpha(z_{0})  \int_{\Sigma^{c}_{1}(z_{0})} ... \int_{\Sigma^{c}_{N}(z_{0})} d z_{1} ... d z_{N}  \  \\
%&=   \sum_{i_{0},i_{1},...,i_{N}} \ \int_{\Sigma_{i_{0}}}  ... \int_{\Sigma_{i_{N}}} \ dz_{0} ... dz_{N} \ \hat{C}_{ z_{1},...,z_{N}}  | z_{0} \rangle
%\end{align}

Rather than examine the evolution of the total state at the level of individual wave packets, we can examine it on a more coarse-grained scale, in terms of histories defined using approximate coherent state PVM's. If the cells of the partitions $\{ \Sigma_{i_{k}}  \}$ associated with approximate coherent state PVMs $\{ \hat{\Pi}_{i_{k}} \}$ are of a certain intermediate size - somewhere between the phase space volume of a single coherent state $| z \rangle$ and the phase space volumes associated with macroscopic differences of position and momentum - then the associated history space will be decoherent and the realised histories, while all close to a Newtonian trajectory (assuming the ensemble width of $\rho^{z_{0}}_{S}$ remains sufficiently narrow by comparison with $V(X)$) will also exhibit branching behavior. The state evolution can be expressed in the form,

\begin{equation}
| \chi(N \Delta t) \rangle = \sum_{i_{0},i_{1},...,i_{N}} \hat{C}_{i_{0},i_{1},...,i_{N}} | \chi_{0} \rangle 
\end{equation}

\noindent where

\begin{align}
\hat{C}_{i_{0},i_{1},...,i_{N}} | \chi_{0} \rangle & = \hat{\Pi}_{i_{N}}e^{-i\hat{H} \Delta t} ...  \hat{\Pi}_{i_{1}}  e^{-i\hat{H} \Delta t} \hat{\Pi}_{i_{0}}  | \chi_{0} \rangle \\  
& = \int_{\Sigma_{i_{0}}} \int_{\Sigma_{i_{1}}} ... \int_{\Sigma_{i_{N}}} d z_{0}  d z_{1} ... d z_{N} \ \hat{C}_{z_{0}, z_{1},...,z_{N}}  | \chi_{0} \rangle 
\end{align}

\noindent and, as a consequence of (\ref{WavePacketDec}),

\begin{equation}
\langle \chi_{0} |   \hat{C}^{\dagger}_{i'_{0},i'_{1},...,i'_{N}}    \hat{C}_{i_{0},i_{1},...,i_{N}} | \chi_{0} \rangle \approx 0 \ \text{if} \ i_{k} \neq i'_{k} \ \text{for} \ 0 \leq k \leq N,
\end{equation}

\noindent and, in addition, all realised histories are such that the sequence of regions $(\Sigma_{i_{0}}, \Sigma_{i_{1}},...,\Sigma_{i_{N}} )$ associated with each history fall close, to within some reasonable margin of error,  to some Newtonian trajectory (though they will exhibit fluctuations around this trajectory associated with their branching behavior). As we have seen, by virtue of satisfying this decoherence condition, the quantum state will possess a branching structure relative to the history space associated with a time-indexed sequence of PVM's that take the form $\{\hat{P}^{k}_{i_{0},...,i_{k}} \equiv  \hat{C}_{i_{0},...,i_{k}} | \chi_{0} \rangle   \langle \chi_{0} | \hat{C}_{i_{0},...,i_{k}}  \}$, where $\{\hat{P}^{k}_{i_{0},...,i_{k}} \}$ is the PVM assigned to time $k \Delta t$. 

However, by choosing a history space such that the phase space partitions $\{ \Sigma'^{k}_{j_{k}} \}$ used to define the associated coherent state PVM's are allowed to depend on the time index $k$ and such that the partition cells  $\Sigma'^{k}_{j_{k}}$ are large in comparison with the regions $\Sigma_{N}(z_{0})$ of phase space over which a coherent wave packet spreads on the timescale in question, and also large in comparison with the elements of the more fine-grained partition $\{ \Sigma_{i_{k}} \}$ just discussed, we can find a set of time-indexed PVM's $ \{ \hat{\Pi}^{'k}_{j_{k}} \}$ such that the quantum state evolution approximately takes the form of parallel, non-interfering classical histories on phase space - that is, such that at any given time $N \Delta t$ within the appropriate timescale,

\begin{align}
| \chi(N \Delta t) \rangle =  \sum_{j_{0},j_{1},...,j_{N}} \hat{C}_{ j_{1},...,j_{N}} | \chi_{0} \rangle \approx \sum_{j_{0}} \hat{C}_{j_{0}, j_{1}^{c},...,j_{N}^{c}} | \chi_{0} \rangle. 
\end{align}

\noindent where 

\begin{align}
\hat{C}_{j_{0}, j_{1},...,j_{N}} | \chi_{0} \rangle & \equiv  \hat{C}_{j_{1},...,j_{N}} \hat{\Pi}'^{0}_{j_{0}} \\
&= \hat{\Pi}'^{N}_{j_{N}}e^{-i\hat{H} \Delta t} ...  \hat{\Pi}'^{1}_{j_{1}}  e^{-i\hat{H} \Delta t}   \hat{\Pi}'^{0}_{j_{0}}  | \chi_{0} \rangle \\
&=  \int_{\Sigma'^{0}_{j_{0}}}   \int_{\Sigma'^{1}_{j_{1}}} ... \int_{\Sigma'^{N}_{j_{N}}}  dz_{0} d z_{1} ... d z_{N} \ \hat{C}_{z_{0},z_{1},...,z_{N}}  | \chi_{0} \rangle \\
& =   \int_{\Sigma'^{0}_{j_{0}}} \int_{\Sigma'^{1}_{j_{1}}} ... \int_{\Sigma'^{N}_{j_{N}}} d z_{0} d z_{1} ... d z_{N} \ B(z_{0}, z_{1},...,z_{N})   | z_{N}  \rangle \otimes  | \phi(z_{0}, z_{1},...,z_{N}) \rangle,
\end{align}

\noindent  %and where the  regions $( \Sigma'^{1}_{j_{1}^{c}},...,\Sigma'^{N}_{j_{N}^{c}})$ and the partitions to which they belong are chosen so that these regions contain most (according to the usual Liouville measure on phase space) of the classical trajectories that evolve from the initial conditions in some $\Sigma^{0}_{j_{0}}$. 
and where the only realised histories will be those indexed by sequences of the form $(j_{0}, j_{1}^{c},...,j_{N}^{c})$, where the superscript $c$ indicates that a given index $j^{c}_{k}$ is associated with a partition cell $\{ \Sigma'^{k}_{j^{c}_{k}}$ that contains the classical phase space points future-evolved up to $k \Delta t$ from the initial region $\Sigma^{0}_{j_{0}}$ (so, the value $j^{c}_{k}$ implicitly depends on the value of $j_{0}$ at the beginning of the sequence).

These histories, of course, satisfy the decoherence condition:

\begin{equation}
\langle \chi_{0} |   \hat{C}^{\dagger}_{j_{0},j'_{1},...,j'_{N}}    \hat{C}_{j_{0},j_{1},...,j_{N}} | \chi_{0} \rangle \approx 0 \ \text{if} \ j_{k} \neq j'_{k} \ \text{for} \ 0 \leq k \leq N.
\end{equation}

\noindent In most cases, this is a consequence of the fact that $\hat{C}_{j_{0},j_{1},...,j_{N}} | \chi_{0} \rangle \approx 0$ or $\hat{C}_{j'_{0},j'_{1},...,j'_{N}} | \chi_{0} \rangle \approx 0$ - i.e., one of the histories simply isn't realised. The condition is satisfied non-trivially between two realised classical histories by virtue of $(\ref{WavePacketDec})$.

Except at $t=0$, when the initial superposition becomes decohered, the quantum state evolution exhibits branching relative to the history space $(\textbf{j}^{0}, \textbf{j}^{1},...,\textbf{j}^{N})$ only in a trivial sense: for a given $\textbf{j}_{k}$, with $0<k<N $, the transition amplitude \\
$T(\textbf{j}_{k},k\Delta t; \textbf{j}_{k+1}, (k+1) \Delta t)$ for any projector $\hat{P}^{k}_{\textbf{j}_{k}}$ into  $\hat{P}^{k+1}_{\textbf{j}_{k+1}}$ is zero for all but one value of $j_{k+1}$, whereas archetypal branching behavior involves non-zero transition amplitudes from one projector into \textit{more than one} future projector. As we will see presently, the uniqueness of the future projectors into which the present projector has non-zero transition amplitude helps to account for the appearance of determinism on a coarse-grained level.

%associated history space do not exhibit any branching

 %total quantum state takes the form of parallel classical histories that do not branch. that is, transition amplitude for a given projector $\{ \hat{P}'^{k}_{j_{k}} \}$ at some time is non-nozero for one and only one projector $\{ \hat{P}'^{k+m}_{j_{k+m}} \}$ at any future time $(k+m)\Delta t$. The 

 %such that the size of each partitioning is large in comparison with the measure of spreading of the wave packets in phase space, then there will be no branching with respect to the more coarse-grained history space associated with this partitioning, except at the initial moment when the initial (potentially coherent) superposition is decohered. Instead, the state evolution consists of multiple parallel histories, one associated with each initial index $j_{0}$, corresponding to some initial set $\Sigma_{i_{0}}$ of initial conditions. Thus, on this coarse-graining, the state evolution can be written as 

\vspace{10mm}

The appropriate bridge map $B_{MW}^{CM}$ (where the MW is for `Many Worlds') for the reduction of the given model of $CM$ to the given model of the Bare/Everett theory is given by the expectation value of the extended position and momentum operators,  $\hat{X} \otimes \hat{I}_{E}$ and  $\hat{P} \otimes \hat{I}_{E}$ for $S$:

\

\noindent \underbar{\textit{Bridge Map:}}

\begin{equation}
\small
\begin{split} 
& B_{MW}^{CM}: \mathcal{H}_{S} \otimes \mathcal{H}_{E}  \longrightarrow \Gamma_{N} \\
& \\
& B_{MW}^{CM}:   | \chi \rangle \longmapsto \left(  \langle \chi |    \hat{X} \otimes \hat{I}_{E}  | \chi  \rangle, \langle \chi |   \hat{P} \otimes \hat{I}_{E}  | \chi \rangle \right).
\end{split}
\end{equation}

\noindent At any given time $N \Delta t$, the domain of states from which we should expect approximately classical behavior is the set of branch states defined by decoherence and the appropriate level of coarse-graining that serves to mask the branching behavior that occurs on smaller scales:

\

\noindent \underbar{\textit{Domain:}}

\

\begin{equation}
d_{CM} = \left\{  | \chi(N \Delta t) \rangle \in \mathcal{H}_{S} \otimes \mathcal{H}_{E} \ \bigg| \  | \chi(N \Delta t) \rangle  =  \frac{1}{W^{1/2}(j_{0},j^{c}_{1},...,j^{c}_{N})} \hat{C}_{j_{0},j_{1}^{c}...,j^{c}_{N}} | \chi_{0} \rangle   \right\}.
\end{equation}

\noindent Through this bridge map, the dynamics associated on a given branch will induce a dynamics on the phase space $\Gamma$:

\noindent \underbar{\textit{Bridge Rule:}}

\begin{equation}
\small
\left( X'(N \Delta t), P'(N  \Delta t)  \right) \equiv   \left(  \langle \chi(N \Delta t) |    \hat{X} \otimes \hat{I}_{E}  | \chi(N \Delta t)  \rangle, \langle \chi(N \Delta t) |   \hat{P} \otimes \hat{I}_{E}  | \chi( N \Delta t) \rangle \right) .
\end{equation}

\noindent  Note that, as a consequence of the coarse-graining, the branch states here do not actually `branch' in the sense of one projector's weight contributing to the weights of many projectors at a later time, but rather evolve in such a way that one projector contributes only to a single, unique future projector, in this respect serving to mimic dynamics that are effectively deterministic.

%where the sequence of integration regions $(\Sigma_{1},...,\Sigma_{N})$ fall approximately along a Newtonian trajectory beginning in $\Sigma_{0}$, the phase space $\epsilon$-support of $| Z_{0}\rangle$. In this case, the only realised histories will be histories such that $( Z_{N},..., Z_{1})$ fall approximately along some Newtonian trajectory, and the branch-relative expectation values, $\left( X'(N \Delta t), P'(N \Delta t)  \right)$, will follow an approximately Newtonian trajectory as the value of $N$ is increased. 
\

%\noindent \underbar{\textbf{DS Reduction:}}

%\

%\noindent In this case, the DSR condition (\ref{DSRApprox}) requires that 

%\begin{equation}
%B_{MW}^{CM}(D_{MW}( |\eta_{0} \rangle)) \approx D_{CM}( B_{MW}^{CM}( |\eta_{0} \rangle) )
%\end{equation}

%\noindent for some domain of states in $\mathcal{H}$ and over some timescale $\tau$. The domain of states $|\eta_{0} \rangle$ satisfying this approximate equality consists of product states of the form $| Z_{0} \rangle \otimes |\phi(Z_{0}) \rangle$ - that is, the states that give rise to the distinct branches. The reduction timescale $\tau$ for approximate macroscopic deterministic Newtonian behavior is limited by the chaotic effects on the spreading of wave packets. Depending on the chosen margin of approximation in (\ref{CMMWStrongAnalogy}) below, this timescale will be approximately equal to the timescale on which wave packets become sufficiently spread out that they no longer can be expected under Ehrenfest's Theorem to follow classical trajectories. 

\

\noindent The foregoing considerations entail that expecation values of $\hat{X} \otimes \hat{I}_{E}$ and $\hat{P} \otimes \hat{I}_{E}$, which I abbreviate $\langle \hat{X} \rangle$ and $\langle \hat{P} \rangle$ respectively, will, for states in the specified domain, follow an approximately Newtonian trajectory. The dynamical equations of the image model are 

\
 
\noindent \underbar{\textit{Image Model:}}

\
\begin{equation}
\frac{d \langle \hat{P}_{i}  \rangle}{dt} \approx - \frac{\partial V(\langle \hat{X} \rangle) }{\partial \langle \hat{X}_{i} \rangle}
\end{equation}
\

\begin{equation}
\frac{d \langle \hat{X}_{i} \rangle}{dt} \approx \frac{1}{M_{i}} \langle \hat{P}_{i} \rangle.
\end{equation}

\noindent Recall that if the image model holds then the DSR condition is satisfied. The analogue model is obtained straighforwardly from the image model by the bridge rule substitutions $(X',P') \equiv \left( \langle \hat{X} \rangle, \langle \hat{P} \rangle \right) $:

\

\noindent \underbar{\textit{Analogue Model:}}

\
\begin{equation}
\frac{d P'_{i}}{dt} \approx - \frac{\partial V(X') }{\partial X'_{i}}
\end{equation}
\

\begin{equation}
\frac{d X'_{i} }{dt} \approx \frac{1}{M_{i}} P'_{i}.
\end{equation}

\noindent Note finally that the condition of strong analogy is simply a rephrasing of the DSR condition and requires that 

\

\noindent \underbar{\textit{`Strong Analogy':}} 

\

\begin{equation} \label{CMMWStrongAnalogy}
\begin{split}
&| X(N \Delta t) - X'(N \Delta t) | < \delta_{X} \\
&| P(N \Delta t) - P'(N \Delta t) | < \delta_{P}.
\end{split}
\end{equation}

\noindent for $0 \leq N \Delta t \leq \tau$. This condition will be satisfied as long as the laws of the image model hold to good approximation. The margins of error $\delta_{X} $ and $\delta_{P} $ are determined by the size of the partition cells - i.e. the measure of coarse-graining - associated with elements of the history space $\left\{ \left(j_{0}, j_{1},..., j_{N}  \right) \right\}$. Given these margins of error, the timescale $\tau$ will depend on the sizes of the mass parameters in $\hat{H}_{S}$, on the strength of chaotic effects associated with this Hamiltonian, and on the value of the coefficient $\Lambda$ characterising the strength of environmental decoherence (recall that the decoherence term in (\ref{DecoherenceMaster}) can increase the rate of ensemble spreading while constantly maintaining the coherence length below a certain threshold).

\section{DS Reduction of the CM Model to the Bohm Model} \label{CMBohm}

%Many accounts of classical behavior focus on the case where a system such as $S$ is isolated and describable in terms of a pure state. However, because of the possibility of quantum entanglement with external degrees of freedom, isolated quantum systems do not provide a good approximation to the sort of macroscopic system that we wish to describe.

As in the case of the Bare/Everett theory, it is instructive when considering the reduction of classical models of macroscopic systems to the Bohm theory to begin by considering the idealised case in which the relevant macroscopic degrees of freedom $S$ are isolated from any environment. I consider two existing approaches to reducing classical mechanics to Bohm's theory that for the most rely on the assumption that the relevant degrees of freedom in question are isolated. I refer to these approaches as the `narrow wave packet approach' and the `quantum potential approach.' I then explain the need to consider the environment in explaining macroscopic classical behavior in Bohm's theory. Finally, I provide a template for the DS reduction of the $N$-center-of-mass classical model described above to the corresponding Bohm model, including the environment. 

%I will continue to confine my analysis to systems and timescales for which the spreading of wave packets is negligible. If the classical Hamiltonian $H_{S}$ is regular (i.e., not chaotic), the masses $M_{i}$ being macroscopically large (~ 1kg.) will suffice to enforce this assumption over relatively long timescales. However, as suggested above, if $H_{S}$ is chaotic, then spreading effects due to chaos may trump the large masses on relatively short timescales. In either case, I restrict my attention to timescales over which
%deterministic Newtonian equations provide a good approximation to the system's behavior.

\subsection{The Narrow Wave Packet Approach}

Let us first consider the evolution of a narrow wave packet $|q,p\rangle$ in $S$ with average position $q$ and average momentum $p$ at some time $t_{0}$. Given our assumptions about the non-spreading of wave packets, Ehrenfest's Theorem, which states that for a general wave function,

\begin{equation} \label{Ehrenfest}
m \frac{d^{2}  \langle \hat{x} \rangle}{dt^{2}} = - \langle \frac{\partial V(\hat{x})}{\partial \hat{x}} \rangle
\end{equation}

\noindent dictates that such a wave packet will approximately satisfy the stronger condition

\begin{equation} \label{StrongEhrenfest}
m \frac{d^{2} \langle \hat{x}\rangle}{dt^{2}} \approx -  \frac{\partial V(\langle \hat{x} \rangle)}{\partial \langle \hat{x} \rangle}
\end{equation}

\noindent and that the wave packet therefore will follow an approximately classical trajectory as long as it remains sufficiently narrow. More specifically, it will follow the classical trajectory whose position and momentum at time  $t_{0}$ are $q$ and $p$. 

If the wave packet follows a classical trajectory, then by equivariance, essentially all of the Bohmian trajectories associated with that wave packet will also be approximately classical; by the phrase `essentially all of the Bohmian trajectories,' I am referring to an ensemble of possible Bohmian trajectories corresponding to different possible initial configurations. Thus, it seems initially that the system $S$ being in a narrow wave packet suffices to ensure that Bohmian trajectories are classical. Among others, Bowman has been a strong advocate of this approach, although, as I discuss below, he begins to incorporate effects of the environment after considering this result for the base of isolated systems.

However, narrow wave packets constitute only a very restricted subset of possible solutions to the Schrodinger equation, and the most general solution will not necessarily be narrowly peaked in both position and momentum space. The most general solution will rather be a superposition of narrow wave packets, of the form

\begin{equation} \label{CoherentStateExpansion}
|\Psi \rangle = \int \ dq \ dp \ \alpha(q,p) \ |q,p \rangle,
\end{equation}

\noindent where each $|q,p\rangle$ traverses its own classical trajectory. 

But such a solution will not, in general, yield a classical trajectory for the Bohmian configuration. For example, consider the simple case where $S$ consists of a single center of mass with a free Hamiltonian $\hat{H}_{S}=\frac{\hat{P}^{2}}{2m}$; the mass may be macroscopic, though this does not affect my conclusion in this instance. Let the wave function of this system initially take the form of two spatially separated wave packets moving toward each other with opposite average momenta, so that they overlap at some point in time and then pass through each other:

\begin{equation}
|\Psi \rangle= \frac{1}{\sqrt{2}}[|q_{1},p \rangle + |q_{2},-p \rangle].
\end{equation}

\noindent Initially, the set of Bohmian trajectories associated in the ensemble with each of these packets will, by equivariance, follow the same classical path that its wave packet follows. However, this will cease to be true when the packets overlap. Because Bohmian trajectories associated with a single pure state can never cross, when the packets overlap and pass through each other, the trajectories will not be able to follow suit. Instead, they will reverse direction and leave the region of overlap in the packet in which they did \textit{not} begin. This reversal of direction represents a highly non-classical effect on the trajectories in $S$, since if the trajectories were classical, they would simply follow a straight line path with the same wave packet all the way through.

We can see more generally that this sort of non-classical behavior on the part of the Bohmian trajectory will occur whenever the expansion (\ref{CoherentStateExpansion}) of the wave function in terms of spatially localized wave packets (e.g., coherent states) contains wave packets that are initially separated and later come to overlap in configuration space. If $\alpha(q,p)$ and $\alpha(q',p')$ are non-zero for any two initially non-overlapping wave packets $|q,p \rangle$ and $|q',p' \rangle$ whose future evolutions cause them to overlap in configuration space at some point in time, the Bohmian trajectories associated with $|\Psi \rangle$ will become non-classical. Thus, non-classical Bohmian trajectories can result from a very wide variety of wave functions, even when the mass is large.

Since generic states of $S$ yield nonclassical trajectories for the Bohmian configuration, we cannot explain the emergence of classical trajectories at the macroscopic scale in terms of an isolated set of macroscopic degrees of freedom without excluding a very broad class of solutions to the Schrodinger equation as viable physical descriptions of the system in question. Since one would have to exclude any  wave functions that contain a pair of wave packets that intersect in configuration space, the set of wave functions that one must discard will depend heavily on the dynamics of the particular system, which makes such an exclusion seem especially \textit{ad hoc}. 

Bowman has suggested invoking environmental decoherence - and therefore abandoning the assumption of isolation - as the explanation for a restriction to narrow wave packets; while the analysis that I provide below agrees with this approach, Bowman's analysis overlooks certain subtleties involving the need specifically for configuration space decoherence, and is less comprehensive in that it does not consider the stucture of the total pure state of the closed system $SE$, but only that of the mixed state of the open system $S$.
 
\vspace{5mm}

As in the case of Bare/Everett theory, another problem with attempting to model classical behavior using an isolated macroscopic system is that the assumption of isolation is highly unrealistic. While microscopic environmental degrees of freedom are often ignored in classical descriptions of macroscopic systems, the extreme succeptibility of macroscopic superpositions in these systems to entanglement with their environment requires us to consider the effect of the environment when we are examining such systems at the quantum level. In section \ref{Environment}, we shall see in more detail what sort of effect interaction with the environment can have on such a system.

\subsection{The Quantum Potential Approach}
%CITE MATZKIN AS WELL

The most popular approach to explaining Newtonian behavior on the basis of Bohm's theory is the so-called `quantum potential' approach (see, e.g., \cite{holland1995quantum} Ch.6, \cite{allori2002seven}, \cite{bohm1995undivided} ). Consider a closed system, such as the system $S$ we have been discussing, and assume initially, as before, that there is no interaction or entanglement with any external environment $E$. If one plugs in the polar decomposition of the wave function, $\psi(x,t) = R(x,t) \exp[\frac{i S(x,t)}{\hbar}]$, where $R$ and $S$ are real, into the time-dependent Schrodinger equation, one obtains the following pair of coupled differential equations for $R$ and $S$ (one corresponding to the real part of Schrodinger's equation and one to the imaginary part):

\begin{equation}\label{quantumHJ}
\frac{\partial S}{\partial t} + \frac{1}{2m}(\nabla S)^2 + V - \frac{\hbar^2}{2m} \frac{\nabla^{2}R}{R} = 0,   
\end{equation}

\begin{equation}\label{continuityQM}
\frac{\partial R^{2}}{\partial t} +  \nabla \cdot(R^{2} \frac{\nabla S}{m}) = 0.
\end{equation}

\noindent The first of these equations is the Hamilton-Jacobi equation, but with an additional `quantum potential' term, $Q \equiv - \frac{\hbar^2}{2m} \frac{\nabla ^{2}R}{R} $, added to the usual classical potential $V$. The second is a continuity equation for the probability distibution $R^{2}$. In the limit $Q \rightarrow 0$, the solution $S$ becomes a solution to the classical Hamilton-Jacobi equation, and the trajectories that it determines through the guidance equation (which also appears in classical Hamilton-Jacobi theory),

\begin{equation} \label{guidance}
\dot{q_{i}} = \frac{1}{m} \nabla S(x,t),
\end{equation}

\noindent are Newtonian in form.

Moreover, from (\ref{quantumHJ}) and (\ref{guidance}) one can deduce the following Bohmian version of Newton's Second Law:

\begin{equation}\label{SecondLawBohm}
m \frac{d^{2} q}{dt^{2}} = -\nabla{V} - \nabla{Q}.
\end{equation}

\noindent The term $-\nabla{Q}$, which I denote $Q'$ for short, is sometimes referred to as the `quantum force.' Here, the classical equation of motion is retrieved if $Q'$ vanishes.

Thus, the Bohmian trajectories $q(t)$ of an isolated quantum system like $S$ are approximately classical in form whenever the conditions $Q' \rightarrow 0$ and $Q \rightarrow 0$, sometimes called the `canonical conditions' for classicality in pilot wave theory, are satisfied; in fact, they will be classical in form as long as $Q$ is a constant. Here, just as in classical mechanics, one has macroscopic centers of mass traversing classical trajectories. For this reason, it is tempting in Bohm's theory to characterise macroscopic classical motion in terms of an isolated quantum system like $S$, in which the conditions $Q' \rightarrow 0$ and $Q \rightarrow 0$ are satisfied.

\vspace{5mm}

There are two primary flaws with this approach.

\textbf{1)} First, the quantum potential cannot be relied upon to remain small even in isolated macroscopic systems with large mass.

It may appear on the surface that in order for the center of mass to follow a classical trajectory, it suffices for the system to have large mass because the quantum potential $Q=-\frac{\hbar^{2}}{2m}\frac{\nabla^{2}R}{R}$, which characterizes deviations of Bohmian trajectories from classicality, becomes negligible as the mass $m$ becomes macroscopically large. However, it may happen as a rather generic phenomenon that the term $\frac{\nabla^{2}R}{R}$ in the quantum potential becomes large enough to cancel out the effects of $m$ being large. As we saw in the last section, when two initially separated wave packets converge in configuration, the no-crossing rule for Bohmian trajectories (associated with a single pure state) prevents these trajectories from following the wave packet in which they initially lay. The Bohmian trajectories become highly non-classical, and, returning to equation (\ref{quantumHJ}) or (\ref{SecondLawBohm}), it is easy to see that  this non-classical effect must be attributed to the quantum potential or force.

\textbf{2)} The second reason that the quantum potential approach does not work is that the quantum potential and force, as they occur in the above equations, and as they ordinarily are presented, are only well-defined for systems in a pure state. When the system $S$ is open to the environment $E$, the above equations involving the quantum potential and quantum force on system $S$ no longer apply. Yet it is these equations that form the basis for the quantum potential approach to explaining classical behavior of $S$ in Bohm's theory. Thus, the quantum potential approach, as it usually presented, does not apply to the systems for which we ought to expect actual classical behavior to occur.  Some accounts of classical motion in pilot wave theory do attempt incorporate the effects of environmental decoherence into a quantum potential approach \cite{allori2002seven}, \cite{bohm1995undivided}. However, these accounts do not offer any specific suggestions as to how to quantum potential should be defined in the context of open systems, or why it should be negligible in the context of such systems. One can define the quantum potential for the closed system $SE$, but since behavior of the microscopic degrees of freedom in the environment \textit{should} be non-classical in nature (consider an electron in one of the atoms of the macroscopic body), we would not expect this quantum potential to be zero. 

\vspace{5mm}

While it may be possible to modify the quantum potential approach so as to address these criticisms, no such modification has yet been given in the literature.

\subsection{Other Approaches}

Unlike the most other accounts of classicality in Bohm's theory, which rely on the quantum potential, Bowman has suggested in  \cite{bowman2001wave} and \cite{bowmanFoundations} that narrow wave packets and Ehrenfest's theorem lie at the root of classical behavior; moreover, he asserts, as I do below, that decoherence lies at the root of explaining why macroscopic systems can reliably be expected to be in states which are effectively narrow wave packets. However, Bowman's analysis is approached from the perspective of S's being a open system, without analysing the full dynamics of the closed system $SE$ in which $S$ is contained. Moreover, he does not recognise the insufficiency or decoherence \textit{simpliciter} to account for classical behavior in Bohm's theory, and the need specifically for configuration decoherence. The analysis that I provide in a number of respects extends Bowman's approach to fill these gaps. 

Appleby, also, has examined the behavior of Bohmian trajectories in the context specifically of a quantum Brownian motion model, investigating through extensive calculation the effects of environmental decoherence on the velocities of Bohmian configurations, when the system under investigation begins in an approximate energy eigenstate. While his results are consistent with the classical evolutions of the Bohmian configuration of the central degrees of freedom, they do not necessarily imply it. From a template-based perspective, Appleby's analysis focuses on a set of narrowly prescribed parameters for a rather specific model, obscuring the general mechanisms at work in the occurrence of macroscopic classical behavior \cite{appleby1999bohmian}, \cite{appleby1999generic}. 

\vspace{5mm}

Below, I offer what I believe to be a more transparent account of macroscopic classical motion that makes no reference to the quantum potential or force, and that fills certain gaps or corrects certain flaws in other decoherence-based accounts. Before doing so, however, I explain why it is essential to consider the effects of environmental decoherence in explaining the classical trajectories for macroscopic systems, and how doing so can address the difficulties encountered in the case of isolated systems. 
%Perhaps most importantly, taking the environment into account enables us to avoid the need to place \textit{ad hoc} restrictions on the allowable states of the system $S$, but shows how for a general initial state, the dynamics of the total system will carry $S$ into a state that restricts $S$'s Bohmian trajectory to be classical.

\subsection{The Need to Consider the Environment} \label{Environment}

Above, I considered an example where $S$ was an isolated free system in which two initially separated wave packets overlap, passing through each other, and in which the overlap of the packets caused the associated Bohmian trajectories to become non-classical. Returning to that example, let us now abandon our assumption that $S$ is isolated and allow the center of mass to become entangled with the external degrees of freedom in the environment $E$. Suppose that at every time the wave function of the closed system $SE$ consisting of the center of mass and its environment (which may consist of photons, neutrinos, or other particles of matter) takes the form

\begin{equation}
|\Psi \rangle = \frac{1}{\sqrt{2}}[|q_{1},p \rangle \otimes |\phi_{1} \rangle  + |q_{2},-p \rangle \otimes|\phi_{2} \rangle],
\end{equation}

\noindent for some values of $q_{1}$, $q_{2}$, and $p$, where the states $|\phi_{1} \rangle$ and $|\phi_{2} \rangle$ belong to $E$'s Hilbert space $\mathcal{H}_{E}$ and are assumed always to have disjoint supports in $E$'s configuration space $\mathbb{Q}_{E}$. The disjointness of the supports can be expressed as the condition

\begin{equation} \label{ConfigDec}
\langle \phi_{1}|y \rangle \langle y|\phi_{2} \rangle \approx 0 \ \text{for all} \ y \in \mathbb{Q}_{E}
\end{equation}

\noindent where $|y \rangle$ is a position (or more accurately, configuration) eigenstate of the environment. Note that this is the condition for configuration space decoherence, which implies, but is not equivalent to, the condition that $|\phi_{1} \rangle$ and $|\phi_{2} \rangle$ are orthogonal, the condition for decoherence. 

Because of the disjointness of the supports of $|\phi_{1} \rangle$ and $|\phi_{2} \rangle$, the two wave packets comprising $|\Psi \rangle$ will never overlap in the total configuration space of the combined system $SE$, although the profiles of these wave packets with respect to the configuration space of the center of mass will overlap. The motion of Bohmian trajectories in the total system $SE$'s configuration space follows the motion of the wave packets in this space. Since these packets never overlap, there is never any reversal of direction of the Bohmian trajectories. In the center of mass' configuration space, the Bohmian trajectories of the center of mass associated with the two wave packets pass right through each other, and continue in a straight line. Thus, they remain classical, even though the profiles of the wave packets with respect to the center of mass configuration space overlap.

This example illustrates how configuration space decoherence suppresses the non-classical effects of wave packet overlap in $S$, by virtue of the fact that it prevents wave packets from overlapping in the total configuration space of the entire system $SE$. Note that the above analysis applies irrespective of whether the environment contains $1$ degree of freedom or $10^{23}$ degrees of freedom. An example of the former case is a single high frequency photon or electron destroying macroscopic coherence of a wave function in position space. Macroscopic superpositions are extremely sensitive to interactions with environmental degrees of freedom that in a classical description of the same system would exert negligible influence - providing one among several reasons why the environment cannot be ignored when attempting to provide a pilot wave, or any quantum, description of macroscopic systems. In cases where the environment consists of a very large number of degrees of freedom, this fact typically serves to make the destruction of coherence effectively irreversible.

\vspace{5mm}

The fact that we require a special kind of decoherence, characterised by equation (\ref{ConfigDec}), to enforce classicality of the central system's trajectory also implies a constraint on the evolution of the environment's configuration. Since the environmental states correlated to different wave packets for the central system have disjoint supports, by equivariance, the Bohmian configuration of the environment must lie in one of these supports. Equivariance further entails that this configuration must lie in the support of the environmental state which is correlated to the wave packet in which the Bohmian configuration of the central system lies. In this manner, decoherence, which we originally invoked in order to enforce classicality of the central system's Bohmian trajectory, also causes the Bohmian configuration of the environment $E$ to become correlated strongly with that of the central system $S$ (though this is only the case because the pointer states in $S$ are narrowly peaked in $S$'s configuration space).

\subsection{A DS Template for the Reduction of the CM Model to the Bohm Model} 

As we saw in section \ref{CMEverett}, under the dynamical assumptions made there, a fairly generic \footnote{The time-reversibility of the quantum dynamics requires us to impose some restrictions, which I will not discuss further here.} quantum state at $t=0$, which can be written

\begin{equation}
| \chi_{0} \rangle = \int dz_{0} \ \alpha(z_{0}) \  | z_{0} \rangle \otimes |\phi(z_{0}) \rangle,
\end{equation}

\noindent can be expected to evolve into a state of  the form,

\begin{equation}
\begin{split}
 |\chi(N \Delta t) \rangle & \approx \int dz_{0} \ \alpha(z_{0})  \int_{\Sigma_{1}(z_{0})} ... \int_{\Sigma_{N}(z_{0})} d z_{1} ... d z_{N}  \ \hat{C}_{ z_{1},...,z_{N}}  | z_{0} \rangle \\
& \approx \sum_{(i_{1},...,i_{N}) \in I_{r}} \hat{C}_{i_{1},...,i_{N}} | \chi_{0} \rangle \\
&  \approx \sum_{j_{0}} \hat{C}_{j_{0}, j_{1}^{c},...,j_{N}^{c}} | \chi_{0} \rangle
\end{split}
\end{equation}

\noindent where, again, the sequence of integration regions $(\Sigma_{1}(z_{0}),...,\Sigma_{N}(z_{0}))$ falls approximately along a Newtonian trajectory starting at $z_{0}$; likewise, the more coarse-grained sequence of regions $(\Sigma_{i_{1}},...,\Sigma_{i_{N}})$ falls approximately along Newtonian trajectories beginning in some region $\Sigma_{i_{0}}$, and the still more coarse-grained sequences $(\Sigma'^{1}_{j^{c}_{1}},...,\Sigma'^{N}_{j^{c}_{N}})$ fall along the Newtonian trajectories beginning in some region $\Sigma'^{0}_{j_{0}}$. However, in the context of the Bohm theory, the fact that these sets of histories are decoherent does not suffice to induce irreversible effective collapse of the state $| \chi_{0} \rangle$ , and so as a criterion for classical Newtonian behavior on the Bohm model we must impose the stronger condition of configuration space decoherence \footnote{Note that while this condition may be logically stronger than simple decoherence, the question as to whether it is stronger in practice depends on the dynamics of the system and more specifically on whether the state comes to satisfy the ordinary decoherence condition, without configuration space decoherence, \textit{on the way} to satisfying the configuration space decoherence condition.}:

\begin{equation} \label{DecHist}
\begin{split}
& \langle \chi_{0} |   \hat{C}^{\dagger}_{ z'_{N},..., z_{1}}| X,y \rangle \langle X,y| \hat{C}_{ z_{N},..., z_{1}} | \chi_{0} \rangle \approx 0 \\
& \ \text{for all $X,y$,} \ \text{if $z_{i}$ and $z'_{i}$ are sufficiently different, for any $1\leq i \leq N$.}
\end{split}
\end{equation}

\noindent In fact, we should impose the even stronger condition,

\begin{equation} \label{DecEnv}
\begin{split}
& \langle \phi(z'_{N},...,z'_{0}) | y \rangle \langle y|  \phi(z_{N},..., z_{0}) \rangle \approx 0  \\
&\ \text{for all $y$,} \ \text{if $z_{i}$ and $z'_{i}$ are sufficiently different, for any $0 \leq i \leq N$,}
\end{split}
\end{equation}

\noindent which will ensure irreversibility of the collapse, given that the number of degrees of freedom in the environment is very large. As a consequence of these relations, configuration space decoherence also holds among histories in the more coarse-grained history spaces:

\begin{equation}
\langle \chi_{0} | \hat{C}^{\dagger}_{i'_{1},...,i'_{N}} | X,y \rangle \langle X,y |  \hat{C}_{i_{1},...,i_{N}} | \chi_{0} \rangle \approx 0 \  \text{for all $X$, $y$,} \ \text{if $i_{k} \neq i'_{k}$, for any $0 \leq k \leq N$.}
\end{equation}

\noindent 

\begin{equation}
\langle \chi_{0} | \hat{C}^{\dagger}_{j'_{0},j'^{c}_{1},...,j'^{c}_{N}} | X,y \rangle \langle X,y |  \hat{C}_{j_{0},j^{c}_{1},...,j^{c}_{N}} | \chi_{0} \rangle \approx 0 \  \text{for all $X$, $y$,} \ \text{if $j_{0} \neq j'_{0}$.}
\end{equation}

\noindent Assuming that the quantum state has the structure specified by these relations, the disjointness in configuration space entails that the total configuration of $SE$ will not drift between branches (taking for granted the No Drift Conjecture), and that all but one branch effectively can be  ignored in assessing the Bohmian configuration's dynamics. At the most coarse-grained level, we can ignore all but one branch, so the state $|\chi(N \Delta t) \rangle$ can be replaced by the effectively collapsed state corresponding to a single branch, so that

\begin{align} \label{EffectiveState}
|\chi_{eff}(N \Delta t) \rangle & = \frac{1}{W(j_{0},j^{c}_{1},...,j^{c}_{N})} \hat{C}_{j_{0},j^{c}_{1},...,j^{c}_{N}} | \chi_{0} \rangle \\
& = \frac{1}{W(j_{0},j^{c}_{1},...,j^{c}_{N})}  \int_{\Sigma^{0}_{j_{0}}} \int_{\Sigma^{1}_{j^{c}_{1}}} ... \int_{\Sigma^{N}_{j^{c}_{N}}} d z_{0} d z_{1} ... d z_{N} \ B(z_{0}, z_{1},...,z_{N})   | z_{N}  \rangle \otimes  | \phi(z_{0}, z_{1},...,z_{N}) \rangle
\end{align}

\noindent for some $j_{0}$.      
%As the time - that is, as $N$ - increases, $ |\chi_{eff}(N \Delta t) \rangle$ remains approximately in a product state, with $| Z_{N} \rangle$ traversing an approximately Newtonian trajectory, and $| \phi(Z_{N},..., Z_{0}) \rangle$ becoming correlated to $| Z_{N} \rangle$'s past trajectory. 

Turning to configuration space, $\mathbb{Q} = \mathbb{Q}_{S} \oplus  \mathbb{Q}_{E}$, we have for the $\epsilon$-support in $\mathbb{Q}$ of the branch $\frac{1}{W(j_{0},j^{c}_{1},...,j^{c}_{N})} \hat{C}_{j_{0},j^{c}_{1},...,j^{c}_{N}} | \chi_{0} \rangle$

\begin{equation}
SE_{j_{0},j^{c}_{1},...,j^{c}_{N}} \equiv \text{supp}_{\epsilon} \big[ \hat{C}_{j_{0},j^{c}_{1},...,j^{c}_{N}} | \chi_{0} \rangle \big] = \bigcup_{z_{0}\in \Sigma^{0}_{j_{0}},...,z_{N}\in \Sigma^{N}_{j^{c}_{N}}} \left(  S_{z_{N}} \times E_{z_{N},..., z_{0}} \right), 
\end{equation}

\noindent where $S_{z_{N}} \equiv \text{supp}_{\epsilon} \big( | z_{N} \rangle \big)$ and $E_{z_{N},..., z_{0}} \equiv \text{supp}_{\epsilon} \big(|\phi(z_{N},..., z_{0}) \rangle \big)$. After effective collapse, we have that 

\begin{equation}
 Q \in SE_{  j_{0},j^{c}_{1},...,j^{c}_{N}   } \subset \mathbb{Q} 
 \end{equation}
 
\noindent for some branch $( j_{0},j^{c}_{1},...,j^{c}_{N} )$. Defining $E_{ j_{0},j^{c}_{1},...,j^{c}_{N} } \equiv \bigcup_{z_{0}\in \Sigma^{0}_{j_{0}},...,z_{N}\in \Sigma^{N}_{j^{c}_{N}}}  E_{z_{N},..., z_{0}}$, and $S_{\Sigma^{N}_{j^{c}_{N}}} \equiv \bigcup_{z_{N}\in \Sigma^{N}_{j^{c}_{N}}}  S_{z_{N}}$, it follows that

\begin{align}  \label{Configs}
& Q_{S}(N \Delta t) \in  S_{\Sigma^{N}_{j^{c}_{N}}}  \subset \mathbb{Q}_{S}  \\
& q_{E}(N \Delta t) \in E_{ j_{0},j^{c}_{1},...,j^{c}_{N} }  \subset  \mathbb{Q}_{E}.
\end{align}

\noindent Since the region $S_{\Sigma^{N}_{j^{c}_{N}}}$ proceeds roughly along a Newtonian trajectory for increasing $N$, the configuration $ Q_{S}(N \Delta t)$ will do the same. Furthermore, since 

\begin{equation}
E_{ j_{0},j^{c}_{1},...,j^{c}_{N} } \cap E_{ j'_{0},j'^{c}_{1},...,j'^{c}_{N} }=  \emptyset \ \text{if $j_{0} \neq j'_{0}$},
\end{equation}

\noindent the configuration $q_{E}$ becomes correlated through the dynamics to the entire past trajectory of the wave packet  $| z_{N} \rangle$, and since $| z_{N} \rangle$ has narrow support in $S$'s configuration space, also to the entire past trajectory of $Q_{S}$.

In order to frame this analysis within the DS approach to reduction, we can adopt the bridge map

\

\noindent \underbar{\textit{Bridge Map:}}

\

\begin{equation}
\begin{split} 
& B_{BM}^{CM}: \mathcal{H} \times \mathbb{Q}  \longrightarrow \Gamma_{N} \\
& \\
& B_{BM}^{CM}:  \left( |\chi \rangle, Q \right) \longmapsto \left(Q_{S}, M  \dot{Q}_{S}   \right), 
\end{split}
\end{equation}

\noindent which is, notably, distinct from the bridge map employed in the reduction to the Bare/Everett model (the subscript $BM$ stands for `Bohmian Mechanics'). Also note that the bridge map only depends on the Bohmian configuration $Q_{S}$ of the macroscopic system $S$, and not on the quantum state $| \chi \rangle$, and not on $q_{E}$. 
\

%\noindent \underbar{\textbf{DSR Condition:}}

%\

%\begin{equation}
%B_{BM}^{CM}(D_{BM}( |\eta_{0} \rangle, Q)) \approx D_{CM}( B_{BM}^{CM}( |\eta_{0} \rangle, Q) )
%\end{equation}

%\noindent for some domain of states in the low level state space $\mathcal{H} \times \mathbb{Q}$ and some timescale $\tau$. As in the Bare/Everett model, the domain of states satisfying this approximate equality consists of product states of the form $| Z_{0} \rangle \otimes |\phi(Z_{0}) \rangle$ and configurations $Q$ lying in the $\epsilon$-support of this state. Because the Newtonian behavior of the trajectory $Q_{S}$ relies so heavily on the Newtonian behavior of $| Z_{0} \rangle$, the timescale on which $Q_{S}$ ceases to behave approximately according to Newtonian equations of motion is roughly the timescale on which wave packets can no longer be approximated as narrow. So, as in the reduction to the Bare/Everett model, the reduction timescale $\tau$ for approximate macroscopic deterministic Newtonian behavior is limited by the chaotic effects on the spreading of wave packets.  

The dynamical equations of the image model are

\
 
\noindent \underbar{\textit{Image Model:}}

\
\begin{equation}
\frac{d}{dt} (M_{i} \dot{Q}_{S,i}) \approx - \frac{\partial V(X) }{\partial X} \big|_{X=Q_{S}}
\end{equation}
\

\begin{equation}
\frac{d Q_{S,i}}{dt} \approx \frac{1}{M_{i}} (M_{i} \dot{Q}_{S,i}),
\end{equation}

\noindent where, recall, the validity of the image model is equivalent to satisfying the DSR condition. The domain of the image model is 

\

\noindent \underbar{\textit{Domain:}}

\

\begin{equation}
\begin{split}
d_{CM} = & \bigg\{  \left( | \chi \rangle, Q  \right)  \in  \big( \mathcal{H}_{S} \otimes \mathcal{H}_{E} \big) \times \big( \mathbb{Q}_{S} \oplus  \mathbb{Q}_{E}  \big) \ \bigg|  \\
&  |\chi(N \Delta t) \rangle  = \sum_{j_{0}} \hat{C}_{j_{0},j^{c}_{1},...,j^{c}_{N}} | \chi_{0} \rangle  \bigg\},
\end{split}
\end{equation}

 %Q \in \text{supp}_{\epsilon}[ \hat{C}_{j_{0},j^{c}_{1},...,j^{c}_{N}} | \chi_{0} \rangle  ] 

\noindent where the decoherence of branches and other restrictions on the state $\sum_{j_{0}} \hat{C}_{j_{0},j^{c}_{1},...,j^{c}_{N}} | \chi_{0} \rangle $ that were discussed should be taken as implicit. Note that I have made no mention of the Bohmian configuration in the specification of the domain because there is no need to; if the quantum state lies within the specified domain, the dynamics of the Bohm model ensure that the Bohmian configuration, whatever it happens to be, will follow an approximately Newtonian trajectory. Note also that no restriction to a particular branch has been made, either in the domain or in the bridge map, again because none is needed; the dynamics of the Bohmian configuration  automatically entail this restriction through the process of effective collapse.

With the bridge rule substitutions:

\

\noindent \underbar{\textit{Bridge Rule:}}

\

\begin{equation}
\left( X'(N \Delta t), P'(N \Delta t)  \right) \equiv  \left(Q_{S}(N \Delta t), m  \dot{Q}_{S}(N \Delta t)   \right). 
\end{equation}

\noindent the analogue model is obtained straighforwardly from the image model:

\

\noindent \underbar{\textit{Analogue Model:}}

\
\begin{equation}
\frac{d P'_{i}}{dt}  \approx - \frac{\partial V(X) }{\partial X_{i}} \big|_{X = X'}
\end{equation}
\

\begin{equation}
\frac{d X'_{i}}{dt} \approx \frac{1}{M_{i}} (P'_{i}).
\end{equation}

\

\noindent The `strong analogy' condition 
\footnote{I continue to use the term `strong analogy' only to emphasise the parallels of DS reduction with the GNS account;  `approximate agreement' would more adequately reflect the nature the relationship in question.}  
states,

\

\noindent \underbar{\textit{`Strong Analogy':}} 

\

\begin{equation} \label{CMMWStrongAnalogy}
\begin{split}
&| X(N \Delta t) - X'(N \Delta t) | < \delta_{X} \\
&| P(N \Delta t) - P'(\Delta t) | < \delta_{P},
\end{split}
\end{equation}

\noindent for $0 \leq t\leq \tau$, and is satisfied as long as the image model holds, which we should expect to be the case on the same timescale as in the Everettian/Bare-QM case.
\vspace{5mm}

%address hartle's account

\section{Summary}

From the analyses of reduction performed in this chapter, we can see that the occurrence of approximate classical behavior at the macroscopic scale, whether in the Bare/Everett or in the Bohm model, is grounded in the following sequence of insights:

\begin{enumerate}
\item In the Bare-QM model, environmental decoherence is responsible for quasi-classicality - that is, approximate localisation - of the the branch-relative state of $S$; this localisation permits us to speak of a `trajectory' for $S$ relative to a particular branch of the total quantum state. In the Bohm model, quasi-classicality comes automatically as a result of the localisation of the particle configurations. 
\item Ehrenfest's Theorem, as generalised to the case of open systems, is responsible for approximate classicality - that is, conformity to Newtonian equations - of   the trajectory of the localised relative state of $S$ over certain timescales. In the case of the Bohm model, this combined with the further requirement of configuration space decoherence ensures approximate classicality of $S$'s trajectory on the same timescales. 
\item Three factors determine the rate of wave packet spreading, and therefore the timescales on which classicality reliably holds in $S$: the mass of $S$, and the strength of chaotic effects on $S$'s evolution, and pure decoherence. For systems in which chaotic effects are strong, the timescales on which classicality holds are typically much shorter than for systems in which these effects can be ignored, in both the Bare/Everett and Bohm models (in both models ,\textit{quasi}-classicality, unlike classicality, is ensured for all times irrespective of chaotic effects). Nevertheless, even for systems in which chaotic effects are small or nonexistent, the relatively small measure of wave packet spreading that does occur still causes branching when the state evolution is examined on sufficiently small scales of length and momentum. 

\end{enumerate} 

\noindent We will see in Chapter 4 that the same basic pattern of reasoning - decoherence or definiteness of configurations ensuring quasi-classicality, Ehrenfest's Theorem further imposing approximate classicality for narrow wave packets,  wave packet spreading limiting the timescales on which approximate classicality holds -  will apply, in broad outline, to the reduction of classical field theory models to quantum field theory ones.

%In the preceding chapter, I have provided templates for two separate DS reductions: first, of a Newtonian model of macroscopic deterministic behavior to the corresponding Bare/Everett model of the same system; second, of the same Newtonian model to the corresponding Bohmian model. In my analysis, I have taken for granted assumptions of two kinds: first, those that have the status of unproven conjectures, and second, those that require system-specific details, and a corresponding customising of the template, for their justification.  In the first category, I have taken for granted (as most analyses of Bohm's theory do implicitly) that the effective collapse mechanism in Bohm's theory extends to cases of continuously indexed branches, and that the system is such that the center of mass Bohmian configurations obey a guidance equation of the usual form. In the second category, I have taken for granted that the pointer states for environmental decoherence are coherent states, and that the dynamics of the system in question are such that wave packets do not spread too rapidly; in both cases, more specific information about the system is needed before these assumptions can be fully justified. 

\chapter{Quantum Field Theory - Preliminaries}
\label{ch3-corrections}

Chapters 4 and 5 provide templates for the DS reduction of certain models of classical field theory and of certain models of nonrelativistic quantum mechanics to relativisitic quantum field theory. While quantum field theory is usually presented in the manifestly covariant Heisenberg picture, or in terms of path integrals, it will prove particularly helpful in the analysis of DS reduction within these contexts to approach the reduction from the perspective of the less conventional - and also less fully developed - Schrodinger picture of quantum field theory, most obviously because it formulates the models quantum field theory as dynamical systems. In addition, the Schrodinger picture of QFT facilitates an analysis of decoherence and effective collapse in QFT that closely parallels the analysis of these subjects in the context of NRQM. Finally, the Schrodinger picture of QFT forms the basis for Bohmian versions of QFT that have been proposed in the literature.  In section \ref{QFTSchrod}, I offer a basic introduction to various models of quantum field theory in the Schrodinger picture. In section \ref{QFTBohm}, I present the Bohmian versions of several models of QFT. For an extensive overview of QFT in the Schrodinger picture, the reader may consult Hatfield's \cite{hatfield1992quantum}, Jackiw's \cite{jackiw1995diverse}, and Jackiw and Floreannini's \cite{jackiwt1988schrodinger}. For reviews of the major approaches to Bohmian QFT, see Struyve and Westman's \cite{struyve2006new}, Colin and Struyve's \cite{colin2007dirac}, and Struyve's \cite{struyve2007field}.

I emphasise here that, as regards the question of mathematical rigor, there are a number of approaches one can take to QFT. Arguably the most mathematically rigorous of the approaches to QFT is the algebraic approach, though currently there do not exist any formulations of realistic, interacting, quantum field theories within the framework of algebraic QFT; thus, for the moment at least, in return for the added mathematical sophistication of the algebraic approach we must incur the rather severe cost of not being able to link the mathematical models to empirical data, or generally generally to reproduce the empirical successes of QFT as it is practiced most physicists. It is for this reason that I adopt an alternative, but admittedly more heuristic, approach to foundational analysis of QFT that has been developed more closely in line with QFT as it is practiced by most physicists and taught in most textbooks - namely, what David Wallace has dubbed the `cutoff' approach to QFT, which involves performing a foundational analysis on QFTs not in the case where the models are taken to incorporate an infinite number of degrees of freedom, but rather in the case where the model in question has been regularised by some cutoff (typically either a lattice or strict bounds on momentum) so that it incorporates only a large-but-finite number of degrees of freedom, thereby making the model more mathematically and conceptually tractable. For further details of the cutoff approach, the reader may consult  \cite{wallaceNaivete}. For an introduction to and overview of the algebraic approach, the reader may consult \cite{halvorson2006algebraic}.

\section{QFT in the Schrodinger Picture} \label{QFTSchrod}

In this section, I present the Schrodinger picture models for free scalar and fermionic quantum field theories, and for interacting scalar quantum field theory (with $\lambda \phi^{4}$ interaction term) and relativistic QED. As in the case of NRQM, quantum field theories can be modelled in terms of a state space 

\begin{equation}
S=\mathcal{H}
\end{equation}

\noindent for some Hilbert space $\mathcal{H}$, and some Schrodinger dynamics on that Hilbert space, 

\begin{equation}
i \frac{\partial }{\partial t} | \Psi \rangle = \hat{H} | \Psi \rangle,
\end{equation}

\noindent for some hermitian Hamiltonian $\hat{H}$ on $\mathcal{H}$, and some $| \Psi \rangle \in \mathcal{H}$. Note that the although all QFT's that I consider are relativistically covariant at the level of the amplitudes that they predict, the Schrodinger picture destroys \textit{manifest} Lorentz covariance of the theory by specialising to a particular reference frame with a particular time parameter $t$. (Bohmian QFT's, on the other hand, destroy more than merely the manifest Lorentz invariance; the dynamics at the level of the beables breaks fundamental Lorentz invariance as well, while maintaining Lorentz invariance at the level of the theory's empirical predictions.)

The choice of the Hilbert space corresponds to a particular choice of representation for the quantum field theory's commutation or anti-commutation relations. If the number of degrees of freedom in the quantum field theory is infinite - that is, if no cutoffs are imposed in the infrared and ultraviolet -  then there will be infinitely many unitarily inequivalent representations of these relations. Partly for this reason, algebraic approaches to quantum field theory attempt to model these theories without specialising to a particular representation on some Hilbert space, as is done in the Schrodinger picture, but rather solely on the basis of the algebraic properties of the theory's observables. I do not adopt this approach here; for more on the algebraic approach to QFT, the reader may consult, for instance, \cite{haag1964algebraic} or \cite{halvorson2006algebraic}. The approach to quantum field theory taken here is to address the various difficulties generated by QFT's infinite degrees of freedom by adopting a large but finite UV cutoff throughout my analysis (and where necessary, an infrared cutoff as well). The reader should assume these cutoffs to be implicit in my notation. For a discussion of the conceptual foundations of the cutoff approach to QFT, see Wallace's \cite{wallaceNaivete}. Most material here is adapted from the QFT texts of Peskin and Schroeder, Hatfield, and Srednicki. 

\subsection{Free Scalar Field Theory} \label{Scalar}

The Hamiltonian for a free scalar quantum field theory, also known as Klein-Gordon theory, is 

\begin{equation}
\hat{H}_{KG} = \frac{1}{2} \int d^{3} x \ \left[  \hat{\pi}^{2}(x) \ + \ \left( \nabla \hat{\phi}(x) \right)^{2} + m^{2} \hat{\phi}^{2}(x) \right],
\end{equation}

\noindent where $\hat{\phi}(x)$ and $\hat{\pi}(x)$ are, respectively, field operators and field momentum operators associated with each point in 3-space $x$, satisfying the canonical commutation relations

\begin{equation} \label{CCR}
\left[ \hat{\phi}(x), \hat{\pi}(y) \right] = i \delta^{3}(x-y),
\end{equation}

\noindent with all other commutators zero. The Hilbert space of the theory, however it is defined, must carry a representation of these commutation relations. 

\subsubsection{Particle Representation}
 
Define the creation and annihilation operators, 

\begin{equation}
\hat{a}(k) = \int d^{3} x \ e^{ikx} \left[  E_{k} \hat{\phi}(x) + i \hat{\pi}(x) \right]
\end{equation}

\begin{equation}
\hat{a}^{\dagger}(k) = \int d^{3} x \ e^{-ikx} \left[  E_{k} \hat{\phi}(x) - i \hat{\pi}(x) \right],
\end{equation}

\noindent with $E_{k} \equiv \sqrt{k^{2} + m^{2}}$. For later use, the inverse of these relations is 

\begin{equation}
\hat{\phi}(x) = \int \widetilde{d^{3}k}  \ \left[ e^{-ikx} \hat{a}(k) \ + \ e^{ikx} \hat{a}^{\dagger}(k) \right]
\end{equation}

\begin{equation}
\hat{\pi}(x) = \int \widetilde{d^{3}k } \ iE_{k}  \ \left[ e^{-ikx} \hat{a}(k) \ - \ e^{ikx} \hat{a}^{\dagger}(k) \right].
\end{equation}

\noindent where, following Srednicki's notation, 

\begin{equation}
\widetilde{d^{3}k} \equiv \frac{d^{3}k}{(2 \pi)^{3} 2 E_{k}}.
\end{equation}

\noindent One can prove on the basis of $(\ref{CCR})$  that  the $\hat{a}^{\dagger}(k)$ and $\hat{a}(k)$ satisfy the commutation relations

\begin{equation}
\left[ \hat{a}(k), \hat{a}^{\dagger}(k') \right] = (2 \pi)^{3} 2 E_{k} \delta^{3}(k-k').
\end{equation}

\noindent $\hat{H}_{KG}$ then can be rewritten 

\begin{equation}
\begin{split}
\hat{H}_{KG} &= \frac{1}{2} \int \widetilde{d^{3}k} \ E_{k} \ \left[ \hat{a}^{\dagger}(k) \hat{a}(k) \ + \ \hat{a}(k) \hat{a}^{\dagger}(k)  \right] \\
&= \int \widetilde{d^{3}k} \ E_{k} \  \left[  \hat{a}^{\dagger}(k) \hat{a}(k) \ + \  (2 \pi)^{3} E_{k} \delta^{3}(0)  \right]
\end{split}
\end{equation}

\noindent The term $ (2 \pi)^{3} E_{k} \delta^{3}(0)$ (although divergent with the theory's UV cutoff) is a constant and so has no detectable effect, so for convenience we can redefine $\hat{H}_{KG} $ as

\begin{equation}
\hat{H}_{KG} = \int \widetilde{d^{3}k} \ E_{k} \  \hat{a}^{\dagger}(k) \hat{a}(k).
\end{equation}

\noindent If $| 0 \rangle$ is the ground state, or vacuum, of $\hat{H}_{KG}$, so that $\hat{H}_{KG}| 0 \rangle = 0$, then it is easily proven that  the eigenstates of $\hat{H}_{KG}$ are states of the form 

\begin{equation}
| k_{1},...,k_{n} \rangle \equiv \hat{a}^{\dagger}(k_{n}) ... \hat{a}^{\dagger}(k_{1}) | 0 \rangle
\end{equation}

\noindent for $n = 1,2,3,....$. That is, 

\begin{equation}
\hat{H}_{KG} | k_{1},...,k_{n} \rangle = \left( E_{k_{1}}+ ... + E_{k_{n}} \right) | k_{1},...,k_{n} \rangle.
\end{equation}

\noindent The states $\{| 0 \rangle, | k_{1},...,k_{n} \rangle \}$ (being eigenstates of a Hermitian operator) form an orthonormal basis for the state space $\mathcal{H}$ of the theory, known as the Fock basis. In this basis, the identity operator takes the form,

\begin{equation}
\hat{I} = | 0 \rangle \langle 0 | + \sum_{n=1}^{\infty} \int d^{3}k_{1} ... d^{3}k_{n}  \ | k_{1},...,k_{n} \rangle \langle k_{1},...,k_{n} |.
\end{equation}

\noindent Thus, a general state $| \Psi \rangle \in \mathcal{H}$ can be expressed in the form

\begin{equation} \label{KGState}
| \Psi \rangle = \psi_{0} | 0 \rangle  + \sum_{n=1}^{\infty} \int d^{3}k_{1} ... d^{3}k_{n}  \ \tilde{\psi}_{n}(k_{1},...,k_{n}) \ | k_{1},...,k_{n} \rangle 
\end{equation}

\noindent  where $\psi_{0} \equiv \langle 0 | \Psi \rangle $  and $\tilde{\psi}_{n}(k_{1},...,k_{n}) \equiv  \langle k_{1},...,k_{n} | \Psi \rangle$. The inner product of two states in this representation is 

\begin{equation}
\langle \Phi | \Psi \rangle = \phi_{0} \psi_{0} + \sum_{n=1}^{\infty} \int d^{3}k_{1} ... d^{3}k_{n}   \ \tilde{\phi}^{*}_{n}(k_{1},...,k_{n}) \ \tilde{\psi}_{n}(k_{1},...,k_{n}).
\end{equation}

\noindent The Schrodinger equation for free Klein-Gordon QFT,

\begin{equation}
i \frac{\partial }{\partial t} | \Psi \rangle = \hat{H}_{KG}  | \Psi \rangle,
\end{equation}

\noindent entails 

\begin{equation}
\begin{split}
& i \frac{\partial }{\partial t} \psi_{0} = 0 \\
& i \frac{\partial }{\partial t} \tilde{\psi}_{n}(k_{1},...,k_{n},t)  = \left( E_{k_{1}}+ ... + E_{k_{n}} \right) \tilde{\psi}_{n}(k_{1},...,k_{n},t). 
\end{split}
\end{equation}

\noindent This, in turn, entails that the general time-dependent solution takes the form

\begin{equation}
| \Psi(t) \rangle = \psi_{0}(t ~\shorteq~ 0) | 0 \rangle  + \sum_{i=1}^{n} \int d^{3}k_{1} ... d^{3}k_{n}  \ \tilde{\psi}_{n}(k_{1},...,k_{n},t ~\shorteq~ 0) \ e^{-i \left( E_{k_{1}}+ ... + E_{k_{n}} \right) t}   \ | k_{1},...,k_{n} \rangle. 
\end{equation}

\noindent Turning attention from $k$-space to $x$-space, define the n-particle position space wave function as 

\begin{equation}
\begin{split}
 \psi_{n}(x_{1},...x_{n}) & \equiv \int \widetilde{d^{3}k_{1}} ... \widetilde{d^{3}k_{n}} \ \tilde{\psi}_{n}(k_{1},...,k_{n}) \ e^{-i(k_{1}x_{1} + ... + k_{n}x_{n})} \\
& = \langle 0 | \hat{\phi}(x_{1})... \hat{\phi}(x_{n}) | \Psi \rangle.
\end{split}
\end{equation}

\noindent The inner product between two states in this representation takes the form

\begin{equation}
\langle \Phi | \Psi \rangle = \phi_{0} \psi_{0} + \sum_{n=1}^{\infty} \int d^{3}x_{1} ... d^{3}x_{n}   \ \phi^{*}_{n}(x_{1},...,x_{n}) \ \psi_{n}(x_{1},...,x_{n}).
\end{equation}

\noindent The Klein-Gordon Schrodinger equation entails that 

\begin{equation}
\begin{split}
&  i \frac{\partial }{\partial t} \psi_{0} = 0 \\
&  i \frac{\partial }{\partial t} \psi_{n}(x_{1},...x_{n},t) =    \left( \sqrt{ \nabla_{x_{1}}^{2} + m^{2} }  + ... +      \sqrt{ \nabla_{x_{n}}^{2} + m^{2} }  \right)   \psi_{n}(x_{1},...x_{n},t)
\end{split}
\end{equation}

\noindent where $\sqrt{\nabla^{2} + m^{2}} f(x) \equiv C \ \int dk \ \sqrt{k^{2} + m^{2}} \ \tilde{f}(k) \ e^{-ikx}$ (C is a convention-dependent normalisation constant for the Fourier integral).
 
The $k$-space and $x$-space expansion coefficients $ \tilde{\psi}_{n}(k_{1},...,k_{n},t)$ and $\psi_{n}(x_{1},...x_{n},t)$ are the $n$-particle momentum and position space wave functions, respectively. Ultimately, the justification for this association must come from a demonstration that these functions characterise the behavior of $n$-particle systems - either quantum or classical - in the appropriate circumstances. At the present stage of the analysis, though, one major motivation for identifying the operator $\hat{a}^{\dagger}(k)$ as creating a single particle of momentum $k$ is that, associating $k$ with momentum and $E_{k}$ with energy, the state $\hat{a}^{\dagger}(k )|0 \rangle$ exhibits the appropriate relativistic relationship between energy and momentum: namely $E^{2}-k^{2}=m^{2}$.

\subsubsection{Field Representation}

An alternative to the Fock basis for the Hilbert space $\mathcal{H}$ is the basis of simulataneous eigenstates of the field operators $\hat{\phi}(x)$. The field operators $\hat{\phi(x)}$ form a complete set of commuting operators. The eigenstates of this complete set are the states $| \phi \rangle$ which are simultaneous eigenstates of all field operators $\hat{\phi(x)}$:

\begin{equation}
\hat{\phi}(x) | \phi \rangle = \phi(x)  | \phi \rangle \ \text{for all} \ x,
\end{equation} 

\noindent where, note, $\phi(x)$ is a number representing an eigenvalue of the operator $\hat{\phi}(x)$. The field eigenstates constitute an eigenbasis $\{ | \phi \rangle \}$, known sometimes as the field basis. (It is worth noting here that in the case of an inifinite number of degrees of freedom, the Hilbert space is non-separable since it does not admit a countable orthonormal basis; however, given the cutoff approach adopted here, the number of degrees of freedom will typically be taken to be large-but-finite; the reader should keep this in mind when interpreting my admittedly heuristic use of notation here.) Being orthonormal, the states $\{ | \phi \rangle \}$ satisfy the relation

\begin{equation}
\langle \phi' | \phi \rangle = \delta[\phi - \phi'],
\end{equation} 

\noindent where $ \delta[\phi - \phi']$ is the functional delta function (for a review of functional calculus, see Hatfield \cite{hatfield1992quantum}). In this basis, the identity operator on $\mathcal{H}$ takes the form

\begin{equation}
\hat{I} = \int \mathcal{D} \phi \ | \phi \rangle \langle \phi |,
\end{equation}

\noindent where $\int \mathcal{D} \phi$ designates a functional integral over field configurations. The quantum state $| \Psi \rangle$ can thus be represented as

\begin{equation}
| \Psi \rangle =  \int \mathcal{D} \phi \  \Psi[\phi]  \  | \phi \rangle, 
\end{equation}

\noindent where $\Psi[\phi] \equiv \langle \phi | \Psi \rangle$. Note that this is simply an alternative, equivalent representation of $|\Psi \rangle$ to the representation provided by $(\ref{KGState})$. We can use this expression for the state $|\Psi \rangle$ to determine the action of the operator $\hat{\phi}$ on an arbitrary state in $\mathcal{H}$:

\begin{equation}
\hat{\phi}(x) | \Psi \rangle =  \int \mathcal{D} \phi \  \phi(x) \  \Psi[\phi]  \  | \phi \rangle.
\end{equation}

\noindent The momentum operator $\hat{\pi}(x)$ must be defined to act on $\Psi[\phi]$ in such a way as to produce a representation of the commutation relations (\ref{CCR}). Define the action of $\hat{\pi}(x)$ by

\begin{equation}
\hat{\pi}(x) | \Psi \rangle =  \int \mathcal{D} \phi \  -i \frac{\delta}{\delta  \phi(x) }  \  \Psi[\phi]  \  | \phi \rangle
\end{equation}

\noindent where $\frac{\delta}{\delta  \phi(x) }$ is the functional derivative with respect to the variable $\phi(x)$. It is then straightforward, using the rules of functional differentiation, to check that 

\begin{equation}
\left[  \hat{\phi}(x), \hat{\pi}(y)  \right] \ | \Psi \rangle =  \int \mathcal{D} \phi   \ -i \left( \phi(x)  \frac{\delta}{\delta  \phi(y) }  -  \frac{\delta}{\delta  \phi(y) }  \phi(x)  \right) \  \Psi[\phi]  \  | \phi \rangle = \int \mathcal{D} \phi \ i\delta^{3}(x - y)  \ \Psi[\phi]  \  | \phi \rangle
= i\delta^{3}(x - y)  | \Psi \rangle.
\end{equation}

\noindent Employing the representation of the identity, we obtain an expression for the inner product of two states:

\begin{equation}
\langle \Phi | \Psi \rangle = \int \mathcal{D} \phi \ \Phi^{*}[\phi] \Psi[\phi]
\end{equation}

\noindent In the field representation, the Schrodinger equation for free Klein-Gordon field theory,

\begin{equation}
i \frac{\partial }{\partial t} | \Psi \rangle = \hat{H}_{KG}  | \Psi \rangle,
\end{equation}

\noindent entails 

\begin{equation}
i \frac{\partial }{\partial t} \Psi[\phi,t] =  \left[ - \frac{\delta^{2}}{\delta  \phi(x)^{2} } + \left( \nabla \phi(x)\right)^{2}  + m^{2} \phi^{2}(x)    \right]  \Psi[\phi,t] .
\end{equation}

\noindent Because the Hamiltonian is time-independent, the solution takes the form

\begin{equation}
 \Psi[\phi,t] = \Psi[\phi] e^{-iEt},
\end{equation}

\noindent and the corresponding time-independent Schrodinger equation takes the form

\begin{equation}
 \left[ - \frac{\delta^{2}}{\delta  \phi(x)^{2} } + \left( \nabla \phi(x)\right)^{2}  + m^{2} \phi^{2}(x)    \right]  \Psi[\phi] = E  \Psi[\phi].
\end{equation}

\noindent where $E$ is an energy eigenvalue. Note that the Schrodinger equation is more straightforwardly solved in the momentum space particle representation, where the Hamiltonian is diagonal.

\subsubsection{Transforming Between Field and Particle Representations}

It is possible to transform between the field and particle representations by means of straightforward insertions of the identity operator.

Transforming first from the field to the particle representation, we have 

\begin{equation}
\psi_{0} = \langle 0 |\Psi \rangle = \int \mathcal{D} \phi \ \langle 0 | \phi \rangle \langle \phi |\Psi \rangle = \int \mathcal{D} \phi \ \Psi^{*}_{0}[\phi] \ \Psi[\phi]
\end{equation}

\begin{equation}
\tilde{\psi}_{n}(k_{1},...,k_{n}) = \langle k_{1},...,k_{n}|\Psi \rangle = \int \mathcal{D} \phi \ \langle k_{1},...,k_{n}| \phi \rangle \langle \phi |\Psi \rangle = \int \mathcal{D} \phi \ \Psi^{*}_{k_{1},...,k_{n}}[\phi] \ \Psi[\phi]
\end{equation}  

\begin{equation}
\psi_{n}(x_{1},...,x_{n}) = \langle  x_{1},...,x_{n}|\Psi \rangle = \int \mathcal{D} \phi \ \langle 0| \hat{\phi}(x_{1}),..., \hat{\phi}(x_{n})| \phi \rangle \langle \phi |\Psi \rangle = \int \mathcal{D} \phi \ \phi(x_{1})... \phi(x_{n}) \ \Psi^{*}_{0}[\phi] \ \Psi[\phi],
\end{equation}  

\noindent where $\Psi^{*}_{0}[\phi]$ is the conjugate vacuum state wave functional and $\Psi^{*}_{k_{1},...,k_{n}}[\phi]$ the conjugate wave functional of an $n$-particle state with momenta $(k_{1},...,k_{n})$. Explicit expressions for the former and certain of the latter with low particle numbers can be found in Hatfield \cite{hatfield1992quantum}. Note that the wave functional for an $n$-particle state localised about $(x_{1},...,x_{n})$ is  simply $\phi(x_{1})... \phi(x_{n}) \ \Psi_{0}[\phi]$.

Transforming in the reverse direction, from the particle to the field representation, we have
\begin{equation}
\begin{split}
\Psi[\phi] &= \langle \phi | \Psi \rangle = \langle \phi | \left(  | 0 \rangle \langle 0 | + \sum_{i=1}^{n} \int d^{3}k_{1} ... d^{3}k_{n}  \ | k_{1},...,k_{n} \rangle \langle k_{1},...,k_{n} | \right) | \Psi \rangle    \\
& =   \Psi_{0}[\phi] \ \psi_{0} + \sum_{i=1}^{n} \int d^{3}k_{1} ... d^{3}k_{n}  \ \Psi_{k_{1},...,k_{n}}[\phi]  \ \tilde{\psi}_{n}(k_{1},...,k_{n}), 
\end{split}
\end{equation}

\noindent where I have inserted the identity operator in the momentum space particle representation rather than in the field representation.

\subsection{Theory of the Free EM Field} \label{EMFree}

The Hamiltonian for a free electromagnetic field quantised in Coulomb gauge, $\nabla \cdot \hat{\vec{A}} = 0$, is 

\begin{equation}
\hat{H}_{EM} = \int d^{3} x \  \left[ \hat{\vec{E}}^{2} \ + \  \hat{\vec{B}}^{2}  \right] = \int d^{3} x \ \left[ \hat{\vec{E}}^{2}(x) \ + \left( \nabla \times \hat{\vec{A}}(x) \right)^{2} \right],
\end{equation}

\noindent where the vector potential operator $\hat{\vec{A}}(x) $ and electric field operator $ \hat{\vec{E}}$ satisfy the canonical commutation relations

\begin{equation} \label{EMCCR}
\left[ \hat{A}_{i}(x), \hat{E}_{j}(y) \right] = -i \delta^{T}_{ij}(x-y),
\end{equation}

\noindent where $\delta^{T}_{ij}(x-y)$ is transverse delta function, defined by 

\begin{equation}
\delta^{T}_{ij}(x-y) \equiv \left( \delta_{ij} - \frac{\partial_{i} \partial_{j}}{\nabla^{2}} \right) \delta^{3}(x-y)  = \int \frac{d^{3} k}{(2 \pi)^{3}} \ e^{ik \cdot(x-y)} \left( \delta_{ij} - \frac{k_{i}k_{j}}{|k|^{2}} \right).
\end{equation}

\noindent It is straightforward to see that, when integrated against an arbitrary vector vield $v_{i}(x)$, the transverse delta function returns the transverse component of this vector field  (where, recall, this is defined as the vector field whose Fourier transform is obtained by projecting out the component of $v_{i}(x)$'s Fourier transform parallel to $\vec{k}$, for each $\vec{k}$.) So, 

\begin{equation}
v_{i}^{T}(x) =    \int d^{3}y \  \delta^{T}_{ij}(x-y) \ v_{j}(y).
\end{equation}

\noindent As in the case of the free scalar field, the EM field state has both particle and field representations. I consider the particle representation first. 

\
\subsubsection{Particle Representation}

To extract a particle representation from the model, expand the operators $\hat{\vec{A}}(x)$ and $\hat{\vec{E}}(x)$ as 

\begin{equation} \label{VecPot}
\hat{\vec{A}}(x) = \int \widetilde {d^{3} k} \ \sum_{\lambda=1,2} \  \vec{\epsilon}(k,\lambda) \   \left(  \hat{a}(k,\lambda) \ e^{ik \cdot x} \ +  \ \hat{a}^{\dagger}(k,\lambda) \ e^{-ik \cdot x} \ \right)
\end{equation}

\begin{equation} \label{VecPot}
\hat{\vec{E}}(x) = \int \widetilde{ d^{3} k} \ \sum_{\lambda=1,2} \  \vec{\epsilon}(k,\lambda) \ i | k |  \left( - \hat{a}(k,\lambda) \ e^{ik\cdot x} \ +  \ \hat{a}^{\dagger}(k,\lambda) \ e^{-ik \cdot x} \ \right)
\end{equation}

\noindent where again, 

\begin{equation}
\widetilde{d^{3}k} \equiv \frac{d^{3}k}{(2 \pi)^{3} 2 E_{k}},
\end{equation}

\noindent with $E_{k} = | k |$ and the $\vec{\epsilon}(k,\lambda)$ two linear polarisation vectors satifying transversality, $\vec{k} \cdot \vec{\epsilon}(k,\lambda)=0 $, and orthogonality $\vec{\epsilon}(k,\lambda) \cdot \vec{\epsilon}(k,\lambda') = \delta_{\lambda \lambda'}$. The expansion (\ref{VecPot}) reproduces the canonical commutation relations for  $\hat{\vec{A}}(x) $ and  $\hat{\vec{E}}(x) $ if 

\begin{equation}
\left[\hat{a}(k,\lambda), \hat{a}^{\dagger}(k',\lambda') \right] = (2 \pi)^{3} 2 E_{k} \delta^{3}(k-k') \delta_{\lambda \lambda'},
\end{equation}

\noindent with all other commutators among the $a$'s and $a^{\dagger}$'s zero. The Hamiltonian of the theory then takes the form, after dropping dropping an infinite (or rather divergent with the UV cutoff) constant,

\begin{equation}
\hat{H}_{EM} =   \sum_{\lambda = 1,2}  \ \int \widetilde{d^{3}k} \ E_{k} \ \hat{a}^{\dagger}(k',\lambda') \hat{a}(k,\lambda).
\end{equation}

\noindent The identity in this representation takes the form

\begin{equation}
\hat{I} = | 0_{EM} \rangle \langle 0_{EM} | +  \sum_{\lambda_{1},...,\lambda_{n} = 1}^{2} \ \sum_{n=1}^{\infty} \int d^{3}k_{1} ... d^{3}k_{n}  \  \hat{a}^{\dagger}(k_{n},\lambda_{n}) ... \hat{a}^{\dagger}(k_{1},\lambda_{1}) |0_{EM} \rangle \langle 0_{EM} | \ \hat{a}(k_{1},\lambda_{1}) ... \hat{a}^{\dagger}(k_{n},\lambda_{n})
\end{equation}

\noindent and a general state $| \Phi \rangle \in \mathcal{H}_{EM}$ can be expressed in the form

\begin{equation} \label{EMParticleState}
| \Phi \rangle = \phi_{0} | 0 \rangle  +  \sum_{\lambda_{1},...,\lambda_{n} = 1}^{2} \sum_{n=1}^{\infty} \int d^{3}k_{1} ... d^{3}k_{n}  \ \tilde{\phi}^{\lambda_{1},...,\lambda_{n}}_{n}(k_{1},...,k_{n}) \ \  \hat{a}^{\dagger}(k_{n},\lambda_{n}) ... \hat{a}^{\dagger}(k_{1},\lambda_{1}) |0_{EM} \rangle
\end{equation}

\noindent with $\tilde{\phi}^{\lambda_{1},...,\lambda_{n}}_{n}(k_{1},...,k_{n})$ the $n$-photon momentum space wave function. Inserting this into the Schrodinger equation as an initial condition, the general solution for the time evolution of the state is

\begin{equation} \label{EMParticleEvolution}
\small
| \Phi \rangle = \phi_{0} | 0 \rangle  +  \sum_{\lambda_{1},...,\lambda_{n} = 1}^{2} \sum_{n=1}^{\infty} \int d^{3}k_{1} ... d^{3}k_{n}  \ \tilde{\phi}^{\lambda_{1},...,\lambda_{n}}_{n}(k_{1},...,k_{n}, t ~\shorteq~ 0) \ e^{-i\left( (|k_{1}| + ... + |k_{n}| \right) t} \  \hat{a}^{\dagger}(k_{n},\lambda_{n}) ... \hat{a}^{\dagger}(k_{1},\lambda_{1}) |0_{EM} \rangle
\end{equation}

\noindent We could go on to consider the position space representation for the free EM field state in a manner analogous to the analysis for the free scalar field, but I will have no use for it in my later analysis.

\subsubsection{Field Representation}

An alternative representation of the quantum state of the free EM field is provided by the field eigenstates $| \vec{A} \rangle$, which satisfy

\begin{equation}
\hat{\vec{A}}(x) | \vec{A} \rangle = \vec{A}(x)  | \vec{A} \rangle \ \text{for all} \ x,
\end{equation} 

\noindent where, note, $\vec{A}(x)$ is a number representing an eigenvalue of the operator $\hat{\vec{A}}(x)$. Since $\nabla \cdot \hat{\vec{A}} = 0$ and $\nabla \cdot \hat{\vec{A}} | \vec{A} \rangle  = \nabla \cdot \vec{A} | \vec{A} \rangle $, all eigenvalue field configurations are also transverse: $ \nabla \cdot \vec{A} = 0$. As a reminder that the field configurations are transverse, I will designate them as $\vec{A}^{T}$. The field eigenstates constitute an eigenbasis $\{ | \vec{A}^{T} \rangle \}$; being orthonormal, they satisfy the relation

\begin{equation}
\langle \vec{A}^{T'} | \vec{A}^{T}  \rangle = \delta[\vec{A}^{T} -\vec{A}^{T'} ],
\end{equation} 

\noindent where $ \delta[\vec{A}^{T} -\vec{A}^{T'}]$ is the functional delta function, satifying

\begin{equation}
f[\vec{A}^{T}_{0}] = \int \mathcal{D} A^{T} \  f[\vec{A}^{T}]  \ \delta[\vec{A}^{T} -\vec{A}_{0}^{T}]
\end{equation}

\noindent where $\int \mathcal{D} A^{T} \equiv \int \mathcal{D} A \ \delta[\nabla \cdot \vec{A}]$ designates a functional integral only over transverse field configurations; note that the delta functional $\delta[\nabla \cdot \vec{A}]$ enforces the restriction to transverse field configurations in the integral. In the field basis, the identity operator on $\mathcal{H}$ takes the form

\begin{equation}
\hat{I} = \int \mathcal{D} A^{T} \ | \vec{A}^{T} \rangle \langle \vec{A}^{T} |,
\end{equation}

\noindent so the quantum state $| \Phi \rangle$ thus can be represented as

\begin{equation}
| \Phi \rangle =  \int \mathcal{D}  A^{T} \  \Phi[\vec{A}^{T}]  \  | \vec{A}^{T} \rangle, 
\end{equation}

\noindent where $\Psi[\vec{A}^{T}] \equiv \langle \vec{A}^{T} | \Phi \rangle$.  We can use this expression for the state $|\Phi \rangle$ to determine the action of the operator $\hat{\vec{A}}^{T}(x)$ on an arbitrary state in $\mathcal{H}$:

\begin{equation}
\hat{A}^{T}_{i}(x) | \Phi \rangle =  \int \mathcal{D} A^{T} \  A^{T}_{i}(x) \  \Phi[\vec{A}^{T}]  \  | \vec{A}^{T} \rangle.
\end{equation}

\noindent The canonically conjugate electric field operator $\hat{E}_{i}(x)$ must be defined to act on $\Phi[\vec{A}^{T}]$ in such a way as to produce a representation of the commutation relations (\ref{EMCCR}). Define the action of $\hat{\vec{E}}(x)$ by

\begin{equation}
\hat{E}_{i}(x) | \Phi \rangle =  \int \mathcal{D} A^{T} \  i \frac{\delta}{\delta  A^{T}_{i}(x) }  \  \Phi[\vec{A}^{T}]  \  | \vec{A}^{T} \rangle
\end{equation}

\noindent where $\frac{\delta}{\delta  A^{T}_{i}(x) } \equiv \left( \delta_{ij} - \frac{\partial_{i} \partial_{j}}{\nabla^{2}}  \right) \frac{\delta}{\delta  A_{i}(x) }  $ is the functional derivative with respect to $A^{T}_{j}(x)$. It is then straightforward, using the rules of functional differentiation, to check that 

\begin{equation}
\begin{split}
\left[  \hat{A}_{i}(x), \hat{E}_{j}(y)  \right] \ | \Phi \rangle &=  \int \mathcal{D} A^{T}   \ i \left( A^{T}_{i}(x)  \frac{\delta}{\delta  A^{T}_{j}(y) }  -  \frac{\delta}{\delta A_{j}^{T}(y) }  A^{T}_{i}(x)  \right) \  \Phi[\vec{A}^{T}]  \  | \vec{A}^{T} \rangle  \\
& = - i \delta^{T}_{ij}(x - y) \ \int \mathcal{D} A^{T}   \ \Phi[\vec{A}^{T}]  \  | \vec{A}^{T} \rangle,
= - i\delta^{T}_{ij}(x - y)  | \Phi \rangle,
\end{split}
\end{equation}

\noindent where I have used the result $\frac{\delta}{\delta A_{j}^{T}(y) }  A^{T}_{i}(x) = \delta^{T}_{ij}(x-y)$, which can be proven straightforwardly using the rules of functional differentiation and the definition of the transverse projection operator. Employing the representation of the identity, we obtain an expression for the inner product of two states:

\begin{equation}
\langle \Phi | \Psi \rangle = \int \mathcal{D} A^{T} \ \Phi^{*}[\vec{A}^{T}] \Psi[\vec{A}^{T}]
\end{equation}

\noindent In the field representation, the Schrodinger equation for free EM field theory,

\begin{equation}
i \frac{\partial }{\partial t} | \Phi \rangle = \hat{H}_{EM}  | \Phi \rangle,
\end{equation}

\noindent entails 

\begin{equation}
i \frac{\partial }{\partial t} \Phi[\vec{A}^{T},t] = \int d^{3}x \  \left[ - \frac{\delta^{2}}{\delta  \vec{A}^{T}(x)^{2} } + \left( \nabla \times  \vec{A}^{T} \right)^{2}      \right]  \Phi[\vec{A}^{T},t],
\end{equation}

\noindent where $\frac{\delta^{2}}{\delta  \vec{A}^{T}(x)^{2} } \equiv  \sum_{i=1}^{3}  \frac{\delta^{2}}{\delta  A_{i}^{T}(x)^{2} } $ is the fuctional Laplacian. Because the Hamiltonian is time-independent, the solution takes the form

\begin{equation}
 \Phi[\vec{A}^{T},t] = \Phi[\vec{A}^{T}] e^{-iEt},
\end{equation}

\noindent and the Schrodinger equation takes the form

\begin{equation}
  \int d^{3}x \ \left[ - \frac{\delta^{2}}{\delta  \vec{A}^{T}(x)^{2} } + \left( \nabla \times  \vec{A}^{T} \right)^{2} \right] \Phi[\vec{A}]   = E \Phi[\vec{A}^{T},t].
\end{equation}

\noindent where $E$ is the energy eigenvalue. Note that the Schrodinger equation is much more straightforwardly solved in the momentum space particle representation, where the Hamiltonian is diagonal.

\subsubsection{Coherent States of the EM Field}

A particular set of states of the electromagnetic field, known as coherent states, will prove central to the discussion in the next two chapters. Before defining a coherent state of the field, though, it is helpful to define the notion of a coherent state for a single mode $k$ of the field, which in the free theory has the dynamics of a simple quantum harmonic oscillator. A coherent state $| \alpha_{k} \rangle$ of the mode $(k, \lambda)$ is defined as an eigenstate of the annihilation operator $\hat{a}_{k}$:

\begin{equation}
\hat{a}_{k} | \alpha \rangle_{k} = \alpha | \alpha \rangle_{k},
\end{equation} 

\noindent where, because $\hat{a}_{k} $ is not Hermitian, the eigenvalue $\alpha$ may be complex. Inserting a complete set of energy eigenstates for the mode (which are just the eigenstates of the quantum harmonic oscillator associated with that mode), $\sum_{n} | n_{k} \rangle \langle n_{k}|$, one has after some calculation that

\begin{equation}
 | \alpha \rangle_{k}  = e^{-\frac{1}{2} |\alpha|^{2}} \sum_{n_{k}}  \left(\frac{ \alpha^{n}}{\sqrt{n!}} \right) | n_{k} \rangle. 
\end{equation}

\noindent  Noting that $\hat{a}_{k} = \frac{1}{\sqrt{2 \omega_{k}}} \left( \omega_{k}\hat{\tilde{A}}_{k} + i \hat{\tilde{E}}_{k} \right)  $ and designating $\alpha =  \frac{1}{\sqrt{2 \omega_{k}}} \left( \omega_{k} \tilde{A}_{k}^{0} + i \tilde{E}_{k}^{0} \right)$    For the wave function of the mode $k$, we have

\begin{equation}
\phi(\tilde{A}_{k}) \equiv \langle \tilde{A}_{k} | \alpha \rangle_{k}  = \left(\frac{\omega_{k}}{\pi} \right)^{\frac{1}{4}} e^{-\frac{\omega_{k}}{2} \left( \tilde{A}_{k} -  \tilde{A}_{k}^{0} \right)^{2} + i \tilde{E}_{k}^{0} \tilde{A}_{k}  }
\end{equation}

\noindent Having defined a coherent state of a single mode of the field, we can now define a coherent state of the whole field, which consists simply of the tensor product of coherent state wave functions for all modes: 

\begin{equation}
| \alpha \rangle = \bigotimes_{k,\lambda} |\alpha \rangle^{\lambda}_{k}  
\end{equation}

\noindent As a functional of the whole field configuration, specified by the field Fourier transform $\tilde{\vec{A}}(k)$, the coherent state takes the form

\begin{equation}
\begin{split}
 \alpha [\tilde{\vec{A}}]  & =  \prod_{k, \lambda} \left(\frac{\omega_{k}}{\pi} \right)^{\frac{1}{4}} e^{-\frac{\omega_{k}}{2} \left( \tilde{A}_{k} -  \tilde{A}_{k}^{0} \right)^{2} + i \tilde{E}_{k}^{0} \tilde{A}_{k}  } \\
& = \left( \prod_{k, \lambda} \left(\frac{\omega_{k}}{\pi} \right)^{\frac{1}{4}} \right)     e^{- \int d^{3} k \ \frac{\omega_{k}}{2} \left( \tilde{A}_{k} -  \tilde{A}_{k}^{0} \right)^{2} + i \tilde{E}_{k}^{0} \tilde{A}_{k}  }.
\end{split}
\end{equation}

\noindent This functional over the field Fourier transform $\vec{\tilde{A}}(k)$, describes a product of states each narrowly peaked about a particular configuration $\vec{\tilde{A}}^{0}(k)$ of the field Fourier transform and simultaneously (to within constraints established by the canonical field commutators), a particular configuration of $\tilde{E}^{0}(k)$ of the field momentum Fourier transform. This state, in turn, corresponds to a wave functional narrowly peaked about a particlar spatial field configuration $\vec{A}^{0}(x)$ and a particular spatial field momentum configuration $\vec{E}^{0}(x)$. In later chapters, I designate the coherent state centered on field configuration $\vec{A}(x)$ and field momentum configuration $\vec{E}(x)$, $| \vec{A}, \vec{E} \rangle$. 

%\noindent Alternatively, one may express the coherent state as a functional of the spatial field configuration $\vec{A}(x)$:

%\begin{equation}
%\begin{split}
% \alpha [\vec{A}]  = det(\sqrt{\nabla^{2}}) 
%\end{split}
%\end{equation}

%FINISH, PUT INTO SPATIAL FIELD FORM

\subsection{Free Fermionic Field Theory} \label{Fermion}

The Hamiltonian for a free fermionic field theory, also known as free Dirac field theory, is 

\begin{equation}
\hat{H}_{D} = \int d^{3}x \ \hat{\psi}^{\dagger}(x) \left(  -i \vec{\alpha} \cdot \nabla + \beta m \right)\hat{\psi}(x) 
\end{equation}

\noindent  where $\hat{\psi}(x)$ and $\hat{\psi}^{\dagger}(x)$ are, respectively, 4-spinor field operators and their canonically conjugate field momentum operators associated with each point in 3-space x, $\alpha_{i} = \gamma_{0}\gamma_{i}$, $i=1,2,3$ and $\beta = \gamma_{0}$, where the $\gamma$'s are Dirac matrices. The field operator and conjugate momentum operator in this theory are stipulated to satisfy the canonical anti-commutation relations,

\begin{equation} \label{CAR}
\{ \hat{\psi}^{a}(x), \hat{\psi}^{\dagger b}(y)   \} = i \delta^{ab} \delta^{3}(x-y),
\end{equation}

\noindent where the 4-spinor indices on the field operators have been made explicit. The Hilbert space of the theory, however it is defined, must carry a representation of these anticommutation relations.

\subsubsection{Particle Representation}

Designating the energy eigenstates of the free Dirac equation $u^{r}(k)$, $v^{s}(k)$ with $r,s=1,2$, we have

\begin{equation}
\left(  -i \vec{\alpha} \cdot \nabla + \beta m \right) u^{r}(k)  = E_{k} u^{r}(k),
\end{equation}

\begin{equation}
\left(  -i \vec{\alpha} \cdot \nabla + \beta m \right) v^{s}(k)  = -E_{k} v^{s}(k),
\end{equation}

\noindent where $E_{k} = \sqrt{k^{2} + m^{2}}$. Expand the field operators $\hat{\psi}(x)$ and $\hat{\psi}^{\dagger}(x)$ in terms of these eigenspinors, 

\begin{equation}
\hat{\psi}(x) = \int \widetilde{d^{3} k} \ \sum_{r} \left(  \hat{b}^{r}(k) \ u^{r}(k) \  e^{ikx} \ + \  \hat{c}^{\dagger r}(k) \ v^{r}(k) \ e^{-ikx}  \right) 
\end{equation}

\begin{equation}
\hat{\psi}^{\dagger}(x) =  \int \widetilde{d^{3} k} \ \sum_{r} \left(  \hat{b}^{\dagger r}(k) \ u^{\dagger r}(k) \  e^{- ikx} \ + \  \hat{c}^{ r}(k) \ v^{\dagger r}(k) \ e^{-ikx}  \right) 
\end{equation}

\noindent where, in the context of the fermionic theory  $\widetilde{d^{3} k} $ has been redefined as

\begin{equation}
\widetilde{d^{3} k} = \frac{d^{3} k}{\sqrt{2 E_{k}}}.
\end{equation}

\noindent Stipulating the operators $\hat{b}^{r}(k) $,  $\hat{b}^{\dagger r}(k)$, $\hat{c}^{ r}(k)$, and $\hat{c}^{\dagger r}(k)$ to satisfy the anticommutation relations

\begin{equation}
\begin{split}
& \{ \hat{b}^{r}(k) ,\hat{b}^{s \dagger}(k') \} = (2 \pi)^{3} \delta_{rs} \delta^{3}(k-k') \\
&  \{ \hat{c}^{r}(k) ,\hat{c}^{s \dagger}(k') \} = (2 \pi)^{3} \delta_{rs} \delta^{3}(k-k'), 
\end{split}
\end{equation}

\noindent with all other anticommutators zero, the above expansions of the the field operators reproduces the anticommutators $(\ref{CAR})$. In terms of creation and annihilation operators, and subtracting the usual infinite constant, the Hamiltonian is 

\begin{equation}
\hat{H}_{D} = \sum_{r} \int \frac{d^{3} k}{(2 \pi)^{3}} \ E_{k} [\hat{b}^{r \dagger}(k) \hat{b}^{r}(k)+  \hat{c}^{r \dagger}(k) \hat{c}^{r}(k)].
\end{equation}

\noindent Designating the ground state of this Hamiltonian $| 0_{D }\rangle$, we can define the $n$-particle, $l$-antiparticle states $\hat{c}^{s_{l} \dagger}(p_{l}) ... \hat{c}^{s_{1} \dagger}(p_{1}) \  \hat{b}^{r_{n} \dagger}(k_{n}) ... \hat{b}^{r_{1} \dagger}(k_{1}) | 0_{D }\rangle$, we can write the identity operator on the Hilbert space $\mathcal{H}_{D}$ as in earlier field theories (the expression is cumbersome and straightforward generalisation of those that I have written down earlier, so I forego writing it down here). Since the eigenstates of $\hat{H}_{D}$ constitute a basis, we can expand any state in $\mathcal{H}_{D}$ in terms of these states: 

\begin{equation}
\begin{split}
| \Psi \rangle = & \ \ \psi_{0} | 0_{D} \rangle \ \\
& + \ \sum_{n=1}^{\infty} \sum_{r_{1},...,r_{n}} \ \int d^{3} k_{1} ... d^{3} k_{n} \ \tilde{\psi}_{n,0}^{r_{1},...,r_{n}} (k_{1}, ... , k_{n})     \ \hat{b}^{\dagger,r_{n}}_{k_{n}} ... \hat{b}^{\dagger,r_{1}}_{k_{1}} |0_{D} \rangle \\
 & + \ \sum_{l=1}^{\infty} \sum_{s_{1},...,s_{l}}  \ \int d^{3} p_{1} ... d^{3} p_{l}  \ \tilde{\psi}_{0,l}^{s_{1},...,s_{l}} (p_{1}, ... , p_{l})    \  \hat{c}^{\dagger,s_{l}}_{p_{l}} ... \hat{c}^{\dagger,s_{1}}_{p_{1}} |0_{D} \rangle \  \\ 
&+ \sum_{n,l=1}^{\infty} \sum_{\substack{r_{1},...,r_{n},\\ s_{1},...,s_{l}}} \ \int d^{3} k_{1} ... d^{3} k_{n} \ d^{3} p_{1} ... d^{3} p_{l}  \ \tilde{\psi}_{n,l}^{\substack{r_{1},...,r_{n},\\s_{1},...,s_{l}}} (k_{1}, ... , k_{n}; p_{1}, ... , p_{l}) \hat{c}^{\dagger,s_{l}}_{p_{l}} ... \hat{c}^{\dagger,s_{1}}_{p_{1}}  \hat{b}^{\dagger,r_{n}}_{k_{n}} ... \hat{b}^{\dagger,r_{1}}_{k_{1}} |0_{D} \rangle, 
\end{split}
\end{equation}

\noindent where $\tilde{\psi}_{n,l}^{\substack{r_{1},...,r_{n},\\ s_{1},...,s_{l}}} (k_{1}, ... , k_{n}; p_{1}, ... , p_{l}) \equiv\langle 0 | \hat{b}^{r_{1}}_{k_{1}} ... \hat{b}^{r_{n}}_{k_{n}}  \hat{c}^{s_{1}}_{p_{1}} ... \hat{c}^{s_{l}}_{p_{l}}  | \Psi >$ and likewise for the $(n,0)$, $(0,l)$ and $(0,0)$ coefficients. Thus, the quantum state of a Dirac field with $n$ particles and $l$ antiparticles is encoded in the $2^{n+l}$ functions $\tilde{\psi}_{n,l}^{\substack{r_{1},...,r_{n},\\ s_{1},...,s_{l}}} (k_{1}, ... , k_{n}; p_{1}, ... , p_{l})$. A general state of the field is given by an arbitrary (normalized) superposition of these states for all positive integral values of $n$ and $l$ and of the vaccuum.

The Schrodinger equation for the free fermionic field is 

\begin{equation}
i \frac{\partial }{\partial t} | \Psi \rangle = \hat{H}_{D}  | \Psi \rangle,
\end{equation}

\noindent In the momentum space particle representation, this yields the following uncoupled (as in the previous free theories we have considered) dynamical equations for  $\tilde{\psi}_{n,l}^{\substack{r_{1},...,r_{n},\\ s_{1},...,s_{l}}} (k_{1}, ... , k_{n}; p_{1}, ... , p_{l})$ and the other momentum space coefficients:

\begin{equation}
\footnotesize
\begin{split}
&  i \frac{\partial}{\partial t} \psi_{0} = 0  \\
&   i \frac{\partial}{\partial t}\tilde{\psi}_{n,0}^{r_{1},...,r_{n}} (k_{1}, ... , k_{n},t) =     \left( \sqrt{|k_{1}|^{2} + m^{2}} + ... +  \sqrt{|k_{n}|^{2} + m^{2}} \right) \tilde{\psi}_{n,0}^{r_{1},...,r_{n}} (k_{1}, ... , k_{n},t)  \\
&   i \frac{\partial}{\partial t}\tilde{\psi}_{0,l}^{s_{1},...,s_{l}} (s_{1}, ... , s_{l},t) =     \left( \sqrt{|p_{1}|^{2} + m^{2}} + ... +  \sqrt{|p_{l}|^{2} + m^{2}} \right) \tilde{\psi}_{0,l}^{s_{1},...,s_{l}} (p_{1}, ... , p_{l},t)  \\
& \begin{split}
i \frac{\partial}{\partial t}\tilde{\psi}_{n,l}^{\substack{r_{1},...,r_{n},\\ s_{1},...,s_{l}}} (k_{1}, ... , k_{n}; p_{1}, ... , p_{l},t) = &    \bigg( \sqrt{|k_{1}|^{2} + m^{2}} + ... +  \sqrt{|k_{n}|^{2} + m^{2}} \\ 
& + \sqrt{|p_{1}|^{2} + m^{2}} + ... +  \sqrt{|p_{l}|^{2} + m^2}  \bigg) \  \tilde{\psi}_{n,l}^{\substack{r_{1},...,r_{n}, \\ s_{1},...,s_{l}}} (k_{1}, ... , k_{n}; p_{1}, ... , p_{l},t)  
\end{split}
\end{split}
\end{equation}

\noindent with solutions

\begin{equation}
\footnotesize
\begin{split}
& \psi_{0} =  \psi_{0}(t ~\shorteq~ 0)   \\
&  \tilde{\psi}_{n,0}^{r_{1},...,r_{n}} (k_{1}, ... , k_{n},t) =     e^{-i \left( \sqrt{|k_{1}|^{2} + m^{2}} + ... +  \sqrt{|k_{n}|^{2} + m^{2}} \right) t}\ \tilde{\psi}_{n,0}^{r_{1},...,r_{n}} (k_{1}, ... , k_{n},t ~\shorteq~ 0)   \\
&   \tilde{\psi}_{0,l}^{s_{1},...,s_{l}} (p_{1}, ... , p_{l},t) =     e^{i \left( \sqrt{|p_{1}|^{2} + m^{2}} + ... +  \sqrt{|p_{l}|^{2} + m^{2}} \right) t}\ \tilde{\psi}_{0,l}^{s_{1},...,s_{l}} (p_{1}, ... , p_{l},t ~\shorteq~ 0)  \\
& \begin{split}
\tilde{\psi}_{n,l}^{\substack{r_{1},...,r_{n},\\ s_{1},...,s_{l}}} (k_{1}, ... , k_{n}; p_{1}, ... , p_{l},t) = &  e^{ -i ( \sqrt{|k_{1}|^{2} + m^{2}} + ... +  \sqrt{|k_{n}|^{2} + m^{2}}  
 - \sqrt{|p_{1}|^{2} + m^{2}} - ... -  \sqrt{|p_{l}|^{2} + m^2}  ) t }   \tilde{\psi}_{n,l}^{\substack{r_{1},...,r_{n}, \\ s_{1},...,s_{l}}} (k_{1}, ... , k_{n}; p_{1}, ... , p_{l}, t ~\shorteq~ 0)  
\end{split}
\end{split}
\end{equation}

\noindent The Schrodinger equation for free Dirac theory can also be given a position representation. For example, defining $\psi_{n,l}^{\substack{a_{1},...,a_{n}, \\ b_{1},...,b_{l}}}(\vec{x}_{1}, ... , \vec{x}_{n}; \vec{y}_{1}, ... , \vec{y}_{l} ) \equiv \langle 0_{D} | \hat{\psi}_{a_{1}}(\vec{x_{1}}) ... \hat{\psi}_{a_{n}}(\vec{x_{n}}) \hat{\psi}^{\dagger}_{b_{1}}(\vec{y_{1}}) ... \hat{\psi}^{\dagger}_{b_{l}} (\vec{y_{l}}) | \Psi>$, we have

\begin{equation}
\begin{split}
 i\frac{\partial}{\partial t} \psi^{\substack{a_{1},...,a_{n}, \\ b_{1},...,b_{l}}}(\vec{x}_{1}, ... , \vec{x}_{n}&; \vec{y}_{1}, ... , \vec{y}_{l};t)  = \bigg\{   \big[ -i \vec{\alpha} \cdot \vec{\nabla}_{1}  + \beta m   \big]^{a_{1} c_{1}}  \delta^{a_{2}c_{2}} \ ...  \ \delta ^{a_{n}c_{n}}  +  \delta^{a_{1}c_{1}}  \big[  -i \vec{\alpha} \cdot \vec{\nabla}_{2}  + \beta m   \big]^{a_{2} c_{2}}\delta^{a_{3}c_{3}} \ ... \ \delta^{a_{n}c_{n}} \\
&+ ...  + \delta^{a_{1}c_{1}} \ ... \  \delta^{a_{n-1}c_{n-1}}  \big[ -i \vec{\alpha} \cdot  \vec{\nabla}_{n}  + \beta m   \big]^{a_{n} c_{n}} \\
&+ \ \big[ -i \vec{\alpha} \cdot \vec{\nabla}_{1}  + \beta m   \big]^{b_{1} d_{1}}  \delta^{b_{2}d_{2}} \ ...  \ \delta ^{b_{l}d_{l}}  +  \delta^{b_{1}d_{1}}  \big[  -i \vec{\alpha} \cdot \vec{\nabla}_{2}  + \beta m   \big]^{b_{2} d_{2}}\delta^{b_{3}d_{3}} \ ... \ \delta^{b_{l}d_{l}} \\
&+ ...  + \delta^{b_{1}b_{1}} \ ... \  \delta^{b_{l-1}d_{l-1}}  \big[ -i \vec{\alpha} \cdot  \vec{\nabla}_{l}  + \beta m   \big]^{b_{l} d_{l}} \bigg\} \   \psi^{\substack{c_{1},...,c_{n}, \\ d_{1},...,d_{l}}}(\vec{x}_{1}, ... , \vec{x}_{n}; \vec{y}_{1}, ... , \vec{y}_{l};t)
\end{split}
\end{equation}

\noindent and likewise for the other coefficients. Note that in both the momentum and position representations, the free Schrodinger equation keeps states with different $(n,l)$ uncoupled.

\subsubsection{Field Representation}

Due to the anticommuting, rather than commuting, nature of fermionic field operators, there do not exist states in $\mathcal{H}_{D}$ that are simultaneously eigenstates of all the field operators $\hat{\psi}(x)$ at all positions $x$. Thus, in the fermionic case, there is no field eigenbasis in the same way that there is in the case of bosonic quantum field theory. However, there do exist representations of the canonical anticommutation relations on a Hilbert space of fermionic functionals, otherwise understood as elements of an infinite-dimensional Grassman algebra (or rather very high-dimensional with the cutoffs). There exist strong formal analogies between the functional formulation of fermionic field theory and the functional formulation of bosonic field theory; however, because the particle representation of fermionic field theories will suffice for my purposes, I do not review the functional formulation of fermionic field theory here. For two distinct Grassman algebra representations of the fermionic field, see \cite{hatfield1992quantum} and \cite{floreanini1988functional}.

\subsection{Interacting Scalar Field Theory} \label{IntScalar}

The Hamiltonian for scalar field theory with a $\hat{\phi}^{4}(x) $ interaction term is 

\begin{equation}
\hat{H}_{int} = \frac{1}{2} \int d^{3} x \ \left[  \hat{\pi}^{2}(x) \ + \ \left( \nabla \hat{\phi}(x) \right)^{2} + m^{2} \hat{\phi}^{2}(x) + \frac{1}{4!} \lambda \hat{\phi}^{4}(x)  \right].
\end{equation}

\noindent The Hilbert space in this theory is the same as in the case of the free Klein-Gordon theory, as the presence of an interaction term in the dynamics does not alter the state space, only the evolution of states within it. The perturbative calculation of amplitudes such as S-matrix elements produces the famous divergences of quantum field theory at order $\lambda^{2}$ and higher in the perturbation expansion. 

However, through the process of renormalisation, the amplitudes predicted by the theory can be made finite by absorbing the divergences of the theory into the definitions of $m$, $\lambda$ and the normalisation of the field and field momenum operators $\hat{\phi}(x)$ and  $\hat{\pi}(x)$. Specifically, we can rewrite $\hat{H}_{int}$ in the form

\begin{equation}
\hat{H}_{int} = \hat{H}_{int}^{r} +  \hat{H}_{CT} = \hat{H}_{KG}^{r} + \hat{H}_{I}^{r} +  \hat{H}_{CT} ,
\end{equation}

\noindent where $\hat{H}_{int}^{r} =  \hat{H}_{KG}^{r} + \hat{H}_{I}^{r} $ and,

\begin{equation}
\begin{split}
& \hat{H}_{KG}^{r} \equiv  \frac{1}{2} \int d^{3} x \ \left[  \hat{\pi}_{r}^{2}(x) \ + \ \left( \nabla \hat{\phi}_{r}(x) \right)^{2} + m_{r}^{2} \hat{\phi}_{r}^{2}(x) \right] \\
& \hat{H}_{I}^{r}  \equiv \frac{1}{4!}  \int d^{3} x \ \lambda_{r} \hat{\phi}_{r}^{4}(x) \\
& \hat{H}_{CT} \equiv \frac{1}{2} \int d^{3} x \ \left[ -  \frac{\delta Z}{Z}\ \hat{\pi}_{r}^{2}(x) \ + \  \delta Z  \left( \nabla \hat{\phi}_{r}(x) \right)^{2} + \ \delta m^{2} \  \hat{\phi}_{r}^{2}(x) +  \frac{1}{4!} \ \delta \lambda \ \hat{\phi}_{r}^{4}(x)  \right],
\end{split}
\end{equation}

\noindent where $\hat{\phi}(x) = Z^{\frac{1}{2}} \hat{\phi}_{r}(x)$, $\hat{\pi}(x) = Z^{-\frac{1}{2}} \hat{\pi}_{r}(x)$, so that $\left[ \hat{\phi}_{r}(x), \hat{\pi}_{r}(y) \right] = i \delta^{3}(x-y)$ ), and where  $\delta Z \equiv Z-1$, $\delta m \equiv Zm -m_{r}$, and $\delta \lambda \equiv Z^{2} \lambda -\lambda_{r}$; the field renormalisation $Z$ is defined by $Z= |\langle \Omega |\hat{\phi}(0) | \lambda_{0} \rangle|^{2} $, where $|  \Omega \rangle$ is the vacuum of the fully interacting Hamiltonian $\hat{H}_{int}$ and $|  \lambda_{0} \rangle$ is an eigenstate of $\hat{H}_{int}$ with field momentum zero, so $\hat{\vec{P}} |  \lambda_{0} \rangle = 0$ (for further discussion of the field renormalisation constant and its significance, see \cite{peskin1996introduction}).

 The renormalised mass $m_{r}$ and renormalised coupling $\lambda_{r}$ are finite values fixed by measurements of particular amplitudes of the theory, which establish the so-called renormalisation conditions of the theory. If $\lambda_{r}$ is sufficiently small, the renormalisation conditions can be used to perturbatively compute the values of the counterterms order by order in $\lambda_{r}$. 

The Schrodinger equation for $\lambda \hat{\phi}^{4}$ theory is 

\begin{equation} \label{ISFTSchrod}
i \frac{\partial}{\partial t} | \Psi \rangle = \hat{H}_{int} | \Psi \rangle. 
\end{equation}

\noindent Splitting $\hat{H}_{int}$ into the renormalised free Hamiltonian $\hat{H}_{KG}^{r}$, whose solutions we know, and the perturbative interaction term $\hat{V} \equiv  \hat{H}_{I}^{r} +  \hat{H}_{CT}$, we can write

\begin{equation}
i \frac{\partial}{\partial t} | \Psi \rangle = \left[  \hat{H}_{KG}^{r} + \hat{V}  \right] | \Psi \rangle.
\end{equation}

\noindent Expressed in this form, it is possible to solve for the perturbed energy eigenvalues and energy eigenstates using the traditional Rayleigh-Schrodinger perturbation theory familiar from nonrelativistic quantum mechanics. That is, designating the perturbed energy  $E_{k_{1},...,k_{n}}$, the unperturbed energy $E^{(0)}_{k_{1},...,k_{n}}$, the order $\lambda_{r}$ correction $E^{(1)}_{k_{1},...,k_{n}}$, the order $\lambda_{r}^{2}$ correction $E^{(2)}_{k_{1},...,k_{n}}$, and so on, we have

\begin{equation}
E_{k_{1},...,k_{n}} = E^{(0)}_{k_{1},...,k_{n}} \ + \ E^{(1)}_{k_{1},...,k_{n}} \ + \ E^{(2)}_{k_{1},...,k_{n}} \ +  ... 
\end{equation}

\noindent and likewise for eigenstates of the Hamiltonian,

\begin{equation}
| k_{1},...,k_{n} \rangle = | k_{1},...,k_{n} \rangle^{(0)} + | k_{1},...,k_{n} \rangle^{(1)} + | k_{1},...,k_{n} \rangle^{(2)} + ...
\end{equation}

\noindent where $| k_{1},...,k_{n} \rangle^{(0)}$ is the eigenstate of the renormalised free Hamiltonian $\hat{H}^{r}_{KG}$ with momenta $(k_{1},...,k_{n})$, and $| k_{1},...,k_{n} \rangle$ the corresponding eigenstate of the perturbed Hamiltonian $\hat{H}_{int}$. Rayleigh-Schrodinger theory enables us to calculate the corrections to the energy eigenvalues and eigenstates order by order in $\lambda_{r}$. For example,

\begin{equation}
E^{(1)}_{k_{1},...,k_{n}} = \ \  ^{(0)}\langle k_{1},...,k_{n} |  \hat{V} | k_{1},...,k_{n} \rangle^{(0)}
\end{equation}

\begin{equation}
| k_{1},...,k_{n} \rangle^{(1)} =  \sum_{E^{(0)}_{p_{1},...,p_{m}} \neq E^{(0)}_{k_{1},...,E_{k_{n}}} }  \frac{  ^{(0)}\langle p_{1},...,p_{m} |   \hat{V} | k_{1},...,k_{n} \rangle^{(0)}}{E^{(0)}_{k_{1},...,k_{n}} - E^{(0)}_{p_{1},...,p_{m}} } \ | p_{1},...,p_{m} \rangle^{(0)}.
\end{equation}

\noindent Hatfield has performed this calculation up to first order for the vacuum state and for one- and -two particle states. (As far as I am aware, no calculation to 1-loop order in this framework has been published.)
% \frac{1}{2} \int d^{3} x \ \left[  \hat{\pi}^{2}(x) \ + \ \left( \nabla \hat{\phi}(x) \right)^{2} + m^{2} \hat{\phi}^{2}(x) + \frac{1}{4!} \lambda \hat{\phi}^{4}(x)  \right].

\subsection{Quantum Electrodynamics} \label{QED}

The Hamiltonian for quantum electrodynamics, quantised canonically in Coulomb gauge, is

\begin{equation}
\hat{H}_{QED}= \int d^{3}x \   \big[   \hat{\psi}^{\dagger} ( -i \vec{\alpha} \cdot \nabla + \beta m ) \hat{\psi} + \frac{1}{2} (\hat{\vec{E}}^{2}  +  \hat{\vec{B}}^{2} ) + e \hat{\psi}^{\dagger} \vec{\alpha} \cdot \hat{\vec{A}} \hat{\psi} \big] + \frac{e^{2}}{8 \pi} \int  d^{3}x \ d^{3}y  \ \frac{\hat{\rho}(x) \hat{\rho}(y)}{|x - y|},
\end{equation}

\noindent where $\hat{\rho}(x) \equiv e\hat{\psi}^{\dagger}(x) \hat{\psi}(x)$.  The commutation relations for the fermionic and bosonic field operators are the same as in the respective free field versions of these theories, and the fermionic and bosonic operators commute with each other. The state space of the theory is the tensor product of the fermionic Hilbert space and the electromagnetic Hilbert space:

\begin{equation}
\mathcal{H}_{QED} = \mathcal{H}_{D} \otimes  \mathcal{H}_{EM}.
\end{equation}

\noindent Note that the operators $\hat{\psi}$, $\hat{\psi}^{\dagger}$, and therefore $\hat{\rho}$, are operators on $ \mathcal{H}_{D}$, while $\hat{\vec{A}}$,  $\hat{\vec{B}}$, and $\hat{\vec{E}}$ are operators on $ \mathcal{H}_{EM}$. In the Hamiltonian, they should be understood as operators extended to the full Hilbert space by tensoring with the identity on the Hilbert on which they are not originally defined: e.g., $\hat{\psi}^{\dagger} \otimes \hat{I}_{EM}$, or $\hat{I}_{D} \otimes \hat{\vec{A}}$. As in the case of the interacting scalar field, we can rewrite the Hamiltonian by splitting it into renormalised and counterterm parts:

\begin{equation}
\hat{H}_{QED} = \hat{H}_{QED}^{r} +  \hat{H}_{CT} = \hat{H}_{D}^{r} + \hat{H}_{EM}^{r} + \hat{H}_{I}^{r} + \hat{H}_{C}^{r}  +  \hat{H}_{CT} ,
\end{equation}

\noindent where 

\begin{equation}
\begin{split}
& \hat{H}_{D}^{r} \equiv  \ \int d^{3}x \   \big[   \hat{\psi}_{r}^{\dagger} ( -i \vec{\alpha} \cdot \nabla + \beta m_{r} ) \hat{\psi}_{r}    \\
& \hat{H}_{EM}^{r} \equiv \   \int d^{3}x \    \frac{1}{2} \left( \hat{\vec{E}}_{r}^{2}  +  \hat{\vec{B}}_{r}^{2} \right) \\
& \hat{H}_{I}^{r}  \equiv   \   e_{r} \int d^{3}x \    \hat{\psi}^{\dagger}_{r} \vec{\alpha} \cdot \hat{\vec{A}}_{r} \hat{\psi}_{r}    \\
&   \hat{H}_{C}^{r}  \equiv  \    \frac{e_{r}^{2}}{8 \pi} \int d^{3}x \ d^{3}y \frac{\hat{\rho}_{r}(x) \hat{\rho}_{r}(y)}{|x - y|}  \\
& \hat{H}_{CT} \equiv \frac{1}{2} \int d^{3} x \ \left[  \hat{\psi}_{r}^{\dagger} ( -i \delta_{2} \vec{\alpha} \cdot \nabla + \beta \delta m ) \hat{\psi}_{r} +    \frac{1}{2} \  \delta_{3}  \ \left( \hat{\vec{E}}_{r}^{2}  +  \hat{\vec{B}}_{r}^{2} \right)  \ + \  \delta e \ \hat{\psi}^{\dagger}_{r} \vec{\alpha} \cdot \hat{\vec{A}}_{r} \hat{\psi}_{r}  \right],
\end{split}
\end{equation}

\noindent where $\hat{\psi}(x) = Z_{2}^{1/2} \hat{\psi}_{r}(x) $, $\hat{A}_{i} = Z_{3}^{1/2} \hat{A}^{r}_{i}$, $\delta m = Z_{2} m - m_{r}$, and $\delta e \equiv Z_{2} Z_{3}^{1/2} e - e_{r}$. The field renormalisation constant $Z_{2}$ is defined by the relation $\langle \Omega| \hat{\psi}(0) | p,s \rangle = \sqrt{Z_{2}}u^{s}(p)$ (again, see \cite{peskin1996introduction} for further discussion of the field renormalisation's signficance). As in the case of interacting scalar field theory, the renormalised mass and coupling are defined by suitable renormalisation conditions, which I do not include since they go beyond what is needed for the analysis provided in the next two chapters; I do, however, assume that the renormalisation conditions fix $m_{r}$ at the usual value for the electron mass and $e_{r}$ at the usual value for the electron charge.   

The Schrodinger equation for QED is 

\begin{equation} \label{QEDSchrod}
i \frac{\partial }{\partial t} | \Xi \rangle = \hat{H}_{QED} | \Xi \rangle,
\end{equation}

\noindent where $ | \Xi \rangle \in \mathcal{H}_{QED}$. To apply perturbation theory in the coupling $e_{r}$, we can rewrite this in the form

\begin{equation}
i \frac{\partial }{\partial t} | \Xi \rangle = \left[ \hat{H}^{r}_{D} + \hat{H}^{r}_{EM} +  \hat{H}^{r}_{C} + \hat{V} \right] | \Xi \rangle,
\end{equation}

\noindent where $\hat{V} \equiv \hat{H}^{r}_{I} + \hat{H}_{CT} $ describes the interaction between the fermionic and bosonic degrees of freedom of the theory. Note that the renormalised Coulomb portion of the Hamiltonian, $\hat{H}^{r}_{C}$, introduces a nonlinear interaction among the momentum modes within $\mathcal{H}_{D}$, but does not cause any interaction between the fermionic degrees of freedom described by states in $\mathcal{H}_{D}$ and the bosonic degrees of freedom described by states in $\mathcal{H}_{EM}$ since it is an operator only on $\mathcal{H}_{D}$. 

As we have seen, multiple representations of states in both factor spaces are possible; to each pair of bases$\{ | f_{i}\rangle \in \mathcal{H}_{D} \}$, $ \{ | b_{j}\rangle \in \mathcal{H}_{EM} \}$, there corresponds a basis $\{ | f_{i}\rangle \otimes | b_{j}\rangle \in \mathcal{H}_{QED} \}$ of  $\mathcal{H}_{QED}$, and therefore a distinct representation of the state $| \Xi \rangle$:

\begin{equation}
| \Xi \rangle = \sum_{i,j} \ c_{i,j} \ | f_{i}\rangle \otimes | b_{j}\rangle,
\end{equation} 

\noindent where the sum also may be understood as an integral and the indices as continuous variables for the case of continuously indexed bases. 

Applying the Rayleigh Schrodinger formulas in a manner directly analogous to their application in the case of self-interacting scalar field theory, we may obtain a perturbative expansion of the energies and energy eignenstates of the unperturbed Hamiltonian $\hat{H}^{r}_{0} \equiv  \hat{H}^{r}_{D}  +  \hat{H}^{r}_{C} + \hat{H}^{r}_{EM}$. Note that the unperturbed eigenstates will be product states of fermionic and bosonic eigenstates. The bosonic eigenstates are simply the eigenstates of the free renormalised EM field Hamiltonian $ \hat{H}^{r}_{EM}$; the fermionic eigenstates, because of the presence of the Coulomb term $\hat{H}^{r}_{C}$, will not be the simple momentum states, but rather will consist of both a discrete set of bound $n$-particle, $l$-antiparticle states and a continuum of unbound $n$-particle, $l$-antiparticle states, for each pair of non-negative integers $(n,l)$.

\section{Bohmian QFT} \label{QFTBohm} 

In this section, I review two Bohmian models of quantum field theory, one for scalar field theory and the other for QED. Both models are deterministic models with field beables. For a deterministic model of QED with particle beables, see Colin's \cite{colin2007dirac}. For a stochastic model of QFT, see for instance Bell's \cite{bell2004speakable}. 

\subsection{Scalar Field Theory}

A Bohmian model of scalar field theory can be formulated by close analogy with the case of NRQM; the guidance equations and posited probability distributions are not affected by the presence of the $\lambda \phi^{4}$ interaction term, so the following points apply to both the interacting and free theories. Assume a field configuration beable $\eta(x)$. With the choice of field beable, the natural representation of the quantum state to use is the field or wave functional representation $\Psi[\phi,t]$.  Expanding the wave functional in polar form $\Psi[\phi,t] = R[\phi,t] e^{iS[\phi,t]}$, the Schrodinger equation (\ref{ISFTSchrod}) yields a continuity equation and Hamilton-Jacobi equation as in the NRQM case. From this, one can extract that the probability 
\begin{equation}
P[\phi,t] = |\Psi[\phi,t] |^{2}
\end{equation}

\noindent is preserved equivariantly by the guidance equation,

\begin{equation}
\frac{\partial \eta(x,t)}{\partial t} = \frac{\delta S[\phi,t]}{\delta \phi(x)} \bigg|_{\phi(x) = \eta(x)}.
\end{equation}

\noindent That is, if the probability over the possible beable configurations is $|\Psi[\phi,t ~\shorteq~ 0] |^{2}$ at time $t=0$, then it will be  $|\Psi[\phi,t] |^{2}$ at all later times. 

\subsubsection{Effective Collapse}

For effective collapse to take place in this theory, the branches of the wave function, as defined by the ordinary decoherence condition, must be disjoint specifically with respect to the field configuration space. In the decoherent histories framework, this disjointness can be expressed as the condition that 

\begin{equation}
 \langle  \Psi_{0}  |\hat{C}_{i'_{n},...,i'_{1}} |  \phi \rangle  \langle \phi | \hat{C}_{i_{n},...,i_{1}} | \Psi_{0} \rangle     \ \text{for all} \  \phi \  \in \mathbb{Q} \ \text{and for} \ i_{k} \neq i'_{k}\ \text{for any} \ 1 \leq k \leq n
\end{equation}

\noindent thus guaranteeing that the branches defined by decoherence are disjoint specifically with respect to the beable configuration space. 

%mention wallace result for one and two particle states, which are not disjoint in the configuration space. 

\subsection{Struyve and Westman's Minimalist Model of QED}

Struyve and Westman proposed a Bohmian model of QED in which the sole beables are those associated with the transverse components $\vec{A}^{T}$ of the electromagnetic field. The state space of the theory consists of the Hilbert space of QED, $\mathcal{H}_{QED}$, discussed above, in addition to the field beable configuration space $\mathcal{Q}_{EM}$:

\begin{equation}
S = \mathcal{H}_{QED} \times \mathbb{Q}_{EM}.
\end{equation}

\noindent  Thus, a full specification of the state in this model is $( | \Xi \rangle, \vec{a}^{T}) \in  \mathcal{H}_{QED} \times \mathbb{Q}_{EM}$ (where I have used a lowercase $a$ to distinguish the beable $ \vec{a}^{T}$ from the eigenvalue $ \vec{A}^{T}$).  The dynamics of the quantum state of the model, $| \Xi \rangle$, has been specified in eqn. (\ref{QEDSchrod}). The model is thus fully specified once we have provided a guidance equation for the beable $ \vec{a}^{T}$ and a probability distribution over the beables that is equivariant with respect to the dynamics. Before reproducing Struyve and Westman's guidance equation here, a few preliminary remarks will be necessary. 

Struyve and Westman begin by Fourier transforming the transverse field and field momentum operators as follows:

\begin{equation}
\hat{\vec{A}}^{T}(x) =  \frac{1}{(2 \pi)^{3}} \sum_{\lambda=1}^{2} \int d^{3}k  \ e^{ikx} \ \vec{\epsilon}_{\lambda}(k) \hat{q}_{\lambda}(k) 
\end{equation}

\begin{equation}
\hat{\vec{E}}^{T}(x) =  \frac{1}{(2 \pi)^{3}} \sum_{\lambda=1}^{2} \int d^{3}k  \ e^{-ikx} \ \vec{\epsilon}_{\lambda}(k) \hat{\pi}_{\lambda}(k) 
\end{equation}

\noindent with

\begin{equation}
\left[   \hat{q}_{\lambda}(k), \hat{\pi}_{\lambda'}(k')  \right] = \delta_{\lambda,\lambda'} \ \delta^{3}(q-q'),
\end{equation}

\noindent and all other commutators for the EM Hilbert space zero. Note that these are simply the canonical commutation relations (\ref{EMCCR}) as applied to the Fourier transforms of the transverse field and field momentum operators. The operator  $\hat{q}_{\lambda}(k)$ is the Fourier transform of the field operator corresponding to polarisation $\lambda$; $\hat{\pi}_{\lambda'}(k') $ is the Fourier transform of the transverse momentum operator corresponding to polarisation $\lambda'$. Also, $\vec{\epsilon}_{\lambda}(k)$ is a transverse polarisation vector, with $\vec{\epsilon}_{\lambda}(k) \cdot \vec{k} = 0$, $\sum_{\lambda=1}^{2} \epsilon_{\lambda}^{i}(k)\epsilon_{\lambda}^{j}(k) = \delta_{ij} - \frac{k_{i}k_{j}}{|k|^{2}}$, $\vec{\epsilon}_{\lambda}(k) \cdot \vec{\epsilon}_{\lambda'}(k) = \delta_{\lambda, \lambda'}$,  $\vec{\epsilon}_{\lambda}(k)=\vec{\epsilon}_{\lambda}(-k)$. The operators $\hat{\vec{A}}^{T}$ and $ \hat{\vec{E}}^{T}$ are Hermitian, so $\hat{q}_{\lambda}(k) = \hat{q}_{\lambda}^{\dagger}(-k) $ and $\hat{\pi}_{\lambda}(k) = \hat{\pi}_{\lambda}^{\dagger}(-k)$.

Struyve and Westman take as a basis for the Hilbert space the eigenstates $| q_{1}, q_{2}  \rangle$, where $\hat{q}_{\lambda}(k) | q_{1}, q_{2}  \rangle = q_{\lambda}(k)| q_{1}(k), q_{2}(k)  \rangle$ for all $k$. That is, each state $| q_{1}(k), q_{2}(k)  \rangle$ is a simultaneous eigenstate of $\hat{q}_{\lambda}(k)$ for every $k$ and $\lambda$, and is also an eigenstate of the Fourier transformed field operator $\hat{\tilde{\vec{A}}}(k)$, whose eigenvalues yield a particular Fourier transform function $\tilde{\vec{A}}(k)= \sum_{\lambda} \vec{\epsilon}_{\lambda}(k) q_{\lambda}(k) $. Moreover, the eigenvalues $q_{\lambda}(k)$  satisfy the relation $q_{\lambda}(k)= q^{*}_{\lambda}(-k)$, as do the eigenvalues of the Fourier transformed field operator, $\tilde{\vec{A}}(k) =  \tilde{\vec{A}}^{*}(-k)$.

Struyve and Westman then posit beables for the values of $\left(q_{1}(k), q_{2}(k) \right)$, for every $k$. Since these values are equivalent to specifying the value of the field Fourier transform, they are equivalent (at least mathematically, if arguably not ontologically) to specifying the spatial configuration of the field itself. As preparation for writing down the guidance equation, let the QED quantum state $| \Xi \rangle$ be written as 

\begin{equation}
| \Xi \rangle =  \sum_{q_{1},q_{2}}  \sum_{f}    \Xi(f;q_{1},q_{2}) | f  \rangle \otimes | q_{1}, q_{2}  \rangle
\end{equation}

\noindent where $f$ indexes any basis for the fermionic Hilbert space $\mathcal{H}_{D}$ and  $\Xi[f;q_{1},q_{2}] \equiv \langle f; q_{1}, q_{2} | \Xi \rangle $, with $| f; q_{1}, q_{2} \rangle \equiv | f  \rangle \otimes | q_{1}, q_{2}  \rangle$. The probability distribution over the beables $\left(q_{1}(k), q_{2}(k) \right)$ that Struve and Westman posit  is 

\begin{equation}
P[q_{1},q_{2};t] = \sum_{f} | \Xi[f;q_{1},q_{2}]|^{2}.
\end{equation} 

\noindent Note that this is simply the ordinary Born Rule probability integrated over the fermionic degrees of freedom. Writing the state expansion coefficient in polar form, we have $ \Xi[f;q_{1},q_{2};t] = R[f;q_{1},q_{2};t] e^{i S[f;q_{1},q_{2};t]}$. The guidance equation, which is explicitly consitructed to be equivariant with respect to the probability distribution $P$, then can be written in the form

\begin{equation}
\frac{\partial q_{l}(k,t)}{\partial t} = \frac{1}{P(q_{1},q_{2},t)} \sum_{f} | \Xi[f;q_{1},q_{2}]|^{2} \frac{\delta S[f;q_{1},q_{2};t]}{\delta q^{*}_{l}(k)}  \bigg|_{q_{1}, q_{2}}.
\end{equation}

\noindent The evolution of $\left(q_{1}(k), q_{2}(k) \right)$ determined by this guidance equation determines the evolution of the beable field configuration $\vec{a}^{T}(x)$ through the relation

\begin{equation}
\vec{a}^{T}(x,t) = \frac{1}{(2 \pi)^{3}} \sum_{\lambda=1}^{2} \int d^{3}k  \ e^{ikx} \ \vec{\epsilon}_{\lambda}(k) q_{\lambda}(k,t).
\end{equation}

\noindent The full dynamics of the theory are thereby specified. 

% express probability, guidance equation, in terms of $\vec{a}^{T}$

Struyve and Westman further make a point of noting that while the only beables in this theory are associated with the transverse electromagnetic field, one may, without any change to the theory's predictions, attribute beables to the fermionic degrees freedom simply by choosing some  
Hermitian operator on $\mathcal{H}_{D}$ and defining the beable as the expectation value of this operator conditional on the EM field beable configuration. For example, take the charge density operator $\hat{\rho}(x) = e \hat{\psi}^{\dagger}(x) \hat{\psi}(x)$. The beable associated with this operator is 

\begin{equation}
\rho(x,t) \equiv \frac{\sum_{f,f'} \Xi[f;q_{1},q_{2}]  \ \ \rho_{f,f'}(x) \ \ \Xi[f';q_{1},q_{2}]  }{P[q_{1},q_{2};t]} \bigg|_{q_{1}(t), q_{2}(t)}
\end{equation}

\noindent where $\rho_{f,f'}(x) \equiv    \langle f' | \hat{\rho}(x) | f \rangle$. This equation entails that $\rho(x,t)$ has no dynamics of its own and is merely `along for the ride,' with its value and evolution determined entirely by the quantum state $| \Xi \rangle$ and the beables $\left(q_{1}(k,t), q_{2}(k,t) \right)$.

%\section{Decoherence and Effective Collapse in QFT} \label{DecQFT}

% \begin{equation}
%| \Xi(t) \rangle = \sum_{\bar{\alpha}} \hat{C}_{\bar{\alpha}} | \Xi_{0} \rangle = \sum_{\alpha_{1}, ... , \alpha_{n}} \hat{C}_{\alpha_{n},...,\alpha_{1}} | \Xi_{0} \rangle,
%\end{equation}

%DESCRIBE PARTICLE CREATION ANNIHILATION AS BRANCHING, EFFECTIVE COLLAPSE PROCESS

\subsubsection{Effective Collapse in the QED Minimalist Model}

Assume that the total state $| \Xi \rangle$ of QED consists of a number of approximately orthogonal branches, so that

\begin{equation} \label{MinEffCollState}
| \Xi \rangle = \sum_{i} c_{i} | \Xi_{i} \rangle
\end{equation}

\noindent where $  \langle \Xi_{i} | \Xi_{j} \rangle \approx 0$ for $i \neq j$. For effective collapse onto one of these branches to take place, the probability distribution at any given point $(q_{1},q_{2}$ in the configuration space $\mathbb{Q}_{EM}$ should be the probability distribution associated with just one of the branches. That is, effective collapse requires that 

\begin{equation} \label{MinEffCollCons}
 P[q_{1},q_{2}] \approx     P_{i}[q_{1},q_{2}]  \ \text{for some $i$, for all $(q_{1},q_{2}) \in \mathbb{Q}_{EM}$}.
\end{equation}

\noindent Expanding $P[q_{1},q_{2}] $ for the state (\ref{MinEffCollState}), we have 

\begin{equation}
\begin{split}
& P[q_{1},q_{2}]  \equiv \sum_{f} \left| \Xi[f;q_{1},q_{2}] \right|^{2} = \sum_{f} \big| \sum_{i}  \Xi_{i} [f;q_{1},q_{2}] \big|^{2} \\
& = \sum_{i} P_{i}[q_{1},q_{2}] +\sum_{i \neq j}  \sum_{f}   \Xi_{i}^{*} [f;q_{1},q_{2}]  \Xi_{j} [f;q_{1},q_{2}]
\end{split}
\end{equation}

\noindent If we impose the two conditions

\begin{equation}
 \sum_{f}   \Xi_{i}^{*} [f;q_{1},q_{2}]  \Xi_{j} [f;q_{1},q_{2}] \approx 0 \ \text{for all} \ \left(q_{1},q_{2} \right) \in \mathbb{Q} \ \ \text{for} \ i \neq j,
\end{equation} 

\begin{equation}
P_{i}[q_{1},q_{2};t] P_{j}[q_{1},q_{2};t] \approx 0  \ \text{for all} \ \left(q_{1},q_{2} \right) \in \mathbb{Q} \ \text{for} \ i \neq j
\end{equation}

\noindent the condition (\ref{MinEffCollCons}) will be satisfied. Thus, in the case of the minimalist model, there are two conditions for effective collapse. Alternatively, but equivalently, one can formulate these two conditions in terms of spatial configurations $\vec{A}^{T}$, in which case we have

\begin{equation}
 \sum_{f}   \Xi_{i}^{*} [f;\vec{A}^{T}]  \Xi_{j} [f;\vec{A}^{T}] \approx 0 \  \ \text{for all} \ \vec{A}^{T}  \in \mathbb{Q}  \ \text{for} \ i \neq j,
\end{equation} 

\begin{equation}
P_{i}[\vec{A}^{T}] P_{j}[\vec{A}^{T}] \approx 0  \ \text{for all} \ \vec{A}^{T}  \in \mathbb{Q} \ \text{for} \ i \neq j.
\end{equation}

\noindent This case illustrates how the effective collapse conditions in Bohm's theory depend on the choice of beable. 

%\noindent In the decoherent histories framework, the corresponding condition for decoherence of branches in the minimalist model is then

%\begin{equation}
%\begin{split} 
% & \sum_{f}   \big| \langle q_{1}, q_{2}; f |\hat{C}_{i'_{n},...,i'_{1}} | \Xi_{0} \rangle \big|^{2}    \sum_{f}   \big| \langle q_{1}, q_{2}; f| \hat{C}_{i_{n},...,i_{1}} | \Xi_{0} \rangle   \big|^{2} \ \approx \ 0 \\
% & \ \text{for all} \  \left(q_{1},q_{2} \right) \in \mathbb{Q} \ \text{and for} \ i_{k} \neq i'_{k}\ \text{for any} \ 1 \leq k \leq n.
%\end{split}
%\end{equation}

%\noindent Alternatively, but equivalently, one can formulate this as the requirement that 

%\begin{equation}
%\begin{split} 
% & \sum_{f}   \big| \langle \vec{A}^{T}; f |\hat{C}_{i'_{n},...,i'_{1}} | \Xi_{0} \rangle \big|^{2}    \sum_{f}   \big| \langle  \vec{A}^{T}; f| \hat{C}_{i_{n},...,i_{1}} | \Xi_{0} \rangle   \big|^{2} \ \approx \ 0 \\
% & \ \text{for all} \  \vec{A}^{T} \in \mathbb{Q} \ \text{and for} \ i_{k} \neq i'_{k}\ \text{for any} \ 1 \leq k \leq n.
%\end{split}
%\end{equation}

%\noindent The minimalist model thus illustrates how the particular form of decoherence required for effective collapse in a Bohmian theory depends essentially on the choice of beable. It is disjointness of the marginal Born distributions associated with the different branches, specifically with respect to the configuration space of beables, that generates effective collapse. 

\chapter{The Classical Domain of Relativistic Quantum Electrodynamics}
\label{ch4}

In this chapter, I provide templates for the DS reduction of certain models of classical electrodynamics to the Bare/Everett and Bohm models of quantum electrodynamics. Classical electrodynamics (CED) describes a vast and disparate array of systems, ranging from the propagation of electromagnetic waves through various media, to the behavior of electrical circuits, to the motions of and radiation produced by elementary particles in an electromagnetic field. In this chapter, I will be considering the reduction of classical to quantum electrodynamics for a very particular kind of system: a small number of elementary charges, such as electrons, interacting with an electromagnetic field in empty space. One can find instances of such applications in particle accelerators, where the laws of classical electrodynamics are used to guide beams of charged elementary particles, and also to describe the radiation emitted by these particles \cite{rosenzweig2003fundamentals}.  

Before beginning to formulate the DS reduction for these cases, I discuss a challenge to the internal consistency of CED that was raised by Frisch, and that was addressed by Muller, Belot, Vickers and Zuchowski. While I agree that these authors are collectively successful in rebutting Frisch's particular concern about the internal consistency of CED, it is important to keep in mind that the claim of the internal consistency of CED still has yet to be proven. For the purposes of my analysis here, the question of internal consistency of CED does not enter in any crucial way since the reductions that I perform are reductions not of the full theory of CED, whose consistency is at issue, but of approximations to CED in which either the charge/current distribution or the electromagnetic field is independently prescribed, and whose internal consistency is not in doubt. I designate the approximation to CED in which the charge/current distribution is independently prescribed and the electromagnetic field solved for using Maxwell's equations as the `Classical Maxwell' model; likewise, I designate the approximation to CED in which the electromagnetic field is independently prescribed and the motion of charge distribution solved for using the Lorentz Force Law the `Classical Lorentz' model.

Corresponding to the two classical models of CED that I consider in the context of DS reduction are two models of QED that likewise are only approximations to the full theory of QED; I consider both models in the context of both the Everettian/Bare and Bohmian interpretations. The first model, which I call the` Quantum Maxwell' model, takes the evolution of the fermionic degrees of freedom as classically prescribed and solves for the evolution of the bosonic quantum state using an effective Hamiltonian where the fermionic charge current is given. The second quantum model, which I call the `Quantum Lorentz' model, takes the evolution of the bosonic degrees of freedom as classically prescribed and solves for the evolution of the fermionic degrees of freedom. As we will see, the Quantum Maxwell and Lorentz models are variations on so-called `semiclassical' models of quantum theory. 

After introducing the Quantum Maxwell and Lorentz models, I go on to derive analogues to Ehrenfest's Theorem for these models; as I explain shortly, the extraction of approximately classical behavior from these models procedes along lines that are largely analogous to the reduction of nonrelativistic particle theories considered in Chapter 2, and just as Ehrenfest's Theorem plays a crucial role there, so the appropriate analogues to this theorem will here. Subsequently, I  consider the domains of both the Quantum Maxwell and Lorentz models in the full theory of QED and conclude that they must be product states obeying a number of other restrictions that I go on to specify. Finally, on the basis of these considerations, I frame the reduction of the classical to quantum models in the DS approach. 

%The grounding of these models in product states of the full QED model is obvious though, as far as I am aware, original, as are the added constraints that I place on these product states. The derivation of the Ehrenfest Theorem for the Quantum Lorentz model is also original; the Ehrenfest Theorem for the Quantum Maxwell model is trivial since the Heisenberg equations of motion for the electromagnetic field operators take the exact form of Maxwell's equations for the corresponding classical fields. Generally speaking, by using the term `Ehrenfest Theorem' in the generalised sense, I mean to refer to any relation among quantum mechanical expectation values that enables the deduction of approximate classical trajectories for these expectation values for certain domains of states - typically states that are  peaked around particular values of the associated classical variables. Finally, the framing of these CED/QED reductions within the DS approach, both in the Bare/Everett and Bohmian cases, is original. 

By analogy with the reduction of classical to quantum models in the nonrelativistic case, we will see that the reduction of classical to quantum models in the relativistic, field-theoretic context can be understood as proceeding according to the same basic outline, with modifications:

\begin{enumerate}
\item \textbf{Quasiclassicality:} In the Quantum Maxwell and Quantum Lorentz models, examined the Everttian/Bare context, quasi-classicality  can be understood to follow from localisation (either in physical space or in field configuration space) that results from decoherence induced by degrees of freedom external to those under consideration. In the Bohmian context, quasiclassicality in the associated Quantum Maxwell model follows automatically from the localisation of the electromagnetic field beables; quasiclassicality in the associated Quantum Lorentz Model is a more complicated matter, given that the Bohmian minimalist model of QED does not necessarily incorporate any beables for the fermionic degrees of freedom, so any appearance of quasiclassicality on the part of these degrees of freedom is necessarily, in a sense, an illusion created by the behavior of the bosonic field beables. 
\item \textbf{Ehrenfest Theorems:} The appropriate generalisation or analogue of Ehrenfest's Theorem in the Quantum Maxwell or Lorentz model ensures, for special states, approximate classicality of quasiclassical trajectories. By analogy with the nonrelativistic case, these special states are also narrow, coherent wave packets: to be more specific, in the fermionic sector, they are states of definite particle number in which all particles are narrowly peaked in both particle position and momentum; in the bosonic sector, they are coherent states $|\vec{A}, \vec{E} \rangle$ of the electromagnetic field narrowly peaked about some classical field configuration and momentum $(\vec{A}, \vec{E})$. Moreover, whereas effective collapse of the quantum state in the Everett/Bare version of these models is ensured by simple decoherence, in the Bohmian case, disjointness of branches of the state with respect to the beable configuration space, which amounts to a particular form of decoherence that is logically stronger than the one required for effective collapse in the Everett/Bare case, is again required.
\item \textbf{Wave Packet Spreading and Branching:} By \textit{disanalogy} with the nonrelativistic case, the internal dynamics of the electromagnetic field, as it turns out, will not generallty cause wave packet coherent states in this space to spread; this is a consequence of the fact that the free electromagnetic field in fact consists of many distinct harmonic oscillators, and a coherent state of the EM field is built from coherent states for all the individual oscillators, and the harmonic oscillator is the one system for which coherent states retain their width. Moreover, because the Maxwell field couples linearly to its fermionic source term, and because the operator Maxwell equations are linear in the electromagnetic field operators, in the Quantum Maxwell model one can always expect expectation values of the electric and magnetic field operators to exactly obey the classical Maxwell's equations. On the other hand, the analogy with the nonrelativistic case does continue to hold in the case of the Quantum Lorentz model insofar as the internal dynamics of the fermionic sector does typically cause wave packets in states of definite particle number to spread out. One should expect that on appropriate timescales, this spreading will result in branching as the coherent superposition that results from the spreading is decohered by interaction either with the electromagnetic field or with other fermions.

 \ \ \ \     In the Quantum Maxwell model, the factors that affect the rate of bosonic wave packet spreading are simply those that affect fermionic wave packet spreading, since bosonic packets do not tend to spread of their own accord but only indirectly by virtue of their interaction with fermionic degrees of freedom (for example, a fermionic state consisting of two widely separated wave packets for a single particle will tend to generate a superposition of very different classical electromagnetic field configurations, where each field configuration can be regarded as being generated by a different one of the quasi-classical fermionic wave packets). In the Quantum Lorentz model, on the other hand, the factors affecting wave packet spreading are likely to include those that affect wave packet spreading in the nonrelativistic case: namely the mass of the particles in question (which, unlike in the nonrelativistic instances I considered, will be small in the cases I consider here, on the order of the mass of, say, an electron); it is possible if not likely that chaotic effects also will promote wave packet spreading in these relativistic cases, though I do not explore this question in any depth. 
\end{enumerate}

\noindent The derivation of the Ehrenfest Theorem in the quantum Lorentz model below is original; in the case of the quantum Maxwell model, as we will see, it is trivial. The discussion of the reduction of the quantum Maxwell and Lorentz models to QED, and of the role of decoherence therein, is also original. Finally, the framing of the reduction of the classical Maxwell and Lorentz models to their quantum counterparts in the context of the DS approach is also original.

\section{Models of Classical Electrodynamics}

The full theory of classical electrodynamics can be formulated in a Hamiltonian framework, with state space equal to the Cartesian product of $N$-particle phase space $\Gamma_{Np}$  and the phase space $ \Gamma_{EM}$ of the transverse electromagnetic field (continuing to assume the Coulomb gauge condition $\nabla \cdot \vec{A} = 0$):

\begin{equation}
S=\Gamma_{Np} \times \Gamma_{EM}.
\end{equation}

\noindent The equations of motion of the theory take the form

\begin{equation} \label{HamiltonParticle}
\begin{split}
&\frac{d \vec{q}_{i}}{dt} = \frac{\partial H_{CED}}{\partial \vec{p}_{i}} \\
&\frac{d \vec{p}_{i}}{dt} = - \frac{\partial H_{CED}}{\partial \vec{q}_{i}}
\end{split}
\end{equation}

\begin{equation} \label{HamiltonField}
\begin{split}
&\frac{\partial \vec{A}^{T}}{\partial t} = \frac{\delta H_{CED}}{\delta \vec{E}^{T}} \\
&\frac{\partial \vec{E}^{T}}{\partial t} = - \frac{\delta H_{CED}}{ \delta \vec{A}^{T}}
\end{split}
\end{equation}

\noindent where $\vec{p}_{i} = \gamma m \dot{\vec{q}}_{i} + e\vec{A}(\vec{q}_{i})$, and the Hamiltonian $H_{CED}$, formulated  in the Coulomb gauge $\nabla \cdot \vec{A} = 0$, is a funcion of the particle positions and canonical momenta $q_{i}$ and $p_{i}$, and a functional of the field configuration $ \vec{A}^{T}(x)$ and field canonical field momentum  $\vec{E}^{T}(x)$, which is also the transverse electric field:   

\begin{equation}
H_{CED}(\vec{q}, \vec{p}; \vec{A}, \vec{E}_{T}] =  \sum_{i} \ \sqrt{(\vec{p}_{i}-e\vec{A}(\vec{q}_{i}))^{2} + m^{2} } + \frac{e^{2}}{4\pi} \sum_{i\neq j} \frac{1}{|\vec{q}_{i} - \vec{q}_{j}|} \ + \ \frac{1}{2} \   \int d^{3}x \ \left[ \vec{E}_{T}^{2} + \vec{B}^{2} \right]. 
\end{equation}

\noindent Jackson, in his canonical text on classical electrodynamics, notes that most applications of classical electrodynamics are not applications of this model, which represents the complete theory of classical electrodynamics, but rather applications of one of two kinds of model that approximate the full laws of classical electrodynamics in cases where back reaction effects between charges and field can be neglected  \cite{jackson1999classical}: 

\begin{itemize}
\item \textbf{Lorentz Model:} First, models in which some background electric and magnetic fields $\vec{E}(\vec{x},t)$ and $\vec{B}(\vec{x},t)$ are independently prescribed, and the motion $\vec{x}(t)$ of charged particles is determined on the basis of the Lorentz Force Law. In this model, the state space is N-particle phase space:

\begin{equation}
S_{L} = \Gamma_{Np}.
\end{equation}

\noindent The dynamics are determined by the effective classical Hamiltonian 

\begin{equation}
H_{L}^{eff}(\vec{q},\vec{p}) = \sum_{i} \ \sqrt{(\vec{p}_{i}-e \boldsymbol{\vec{A}}(\vec{q}_{i}),t)^{2} + m^{2} } + \frac{e^{2}}{4\pi} \sum_{i\neq j} \frac{1}{|\vec{q}_{i} - \vec{q}_{j}|} ,
\end{equation}

\noindent in which the kinetic term for the electromagnetic field has been ignored and a prescribed time evolution for the electromagnetic $(\boldsymbol{\vec{A}}(x,t), \boldsymbol{\vec{E}}(x,t))$ - written in boldface to underscore the fact  that it is an independently specified function - inserted instead into the Hamiltonian. With this Hamiltonian, the particle Hamilton equations (\ref{HamiltonParticle}) yield the Lorentz Force Law:

\begin{equation}
\frac{d}{dt} [ \gamma m \frac{d \vec{q}(t)}{dt} ] = e \boldsymbol{\vec{E}}(\vec{q}(t),t) + e\frac{d \vec{q}(t)}{dt}  \times \boldsymbol{\vec{B}}(\vec{q}(t),t),
\end{equation}

\noindent where $\vec{E}(x,t) = \vec{E}_{T}(x,t) + \vec{E}_{L}(x,t)$, and $ \vec{E}_{L}(x)$ is a longitudinal solution to the Gauss equation $\nabla \cdot  \vec{E}_{L}(x) = \rho(x,t) = \sum_{i} e \delta^{3}(x-q_{i}(t))$. Note that $\vec{E}_{L}$ is completely determined by the locations of the charges and therefore does not constitute an independent degree of freedom of the theory, either electromagnetic or fermionic.

\item \textbf{Maxwell Model:} Second, models in which some possibly time-dependent charge and current distributions $\rho(\vec{x},t) \equiv \sum_{i} e\delta^{3}(\vec{x}-\vec{q}_{i}(t))$ and $\vec{j}(x,t) \equiv \sum_{i} e \dot{\vec{q}}_{i} \delta^{3}(\vec{x}-\vec{q}_{i}(t)) $ are prescribed, and the resulting electromagnetic field must be determined on the basis of Maxwell's equations. In this model, the state space is the electromagnetic phase space of field configurations and momenta:

\begin{equation}
S_{M} = \Gamma_{EM}.
\end{equation}

\noindent The dynamics are determined by the effective classical Hamiltonian 

\begin{equation}
H_{M}^{eff}[\vec{A},\vec{E}] =  \  \int d^{3}x \ \left[ \vec{E}_{T}^{2} + \vec{B}^{2} \right]   + \sum_{i} \ \sqrt{ \left(\boldsymbol{\vec{p}_{i}(t)}   -e\vec{A}(\boldsymbol{\vec{q}_{i}(t)},t), \right)^{2} + m^{2} }
\end{equation}

\noindent where $\boldsymbol{\vec{q}_{i}}(t)$ and $\boldsymbol{\vec{p}_{i}}(t)$ have been written in boldface to indicate that they are independently prescribed functions. The field Hamilton equations (\ref{HamiltonField}) yield Maxwell's equations:

\begin{equation}
\frac{\partial \vec{E}}{\partial t} = -\nabla \times \vec{B} + 4 \pi \boldsymbol{\vec{j}}, 
\end{equation}

\begin{equation}
\frac{\partial \vec{B}}{\partial t} = \nabla \times \vec{E},
\end{equation}

\noindent where the Maxwell equation $\nabla \cdot \vec{B} = 0$ is automatically entailed by the definition $ \vec{B} \equiv \nabla \times \vec{A} $, and the Gauss Law $\nabla \cdot \vec{E} = \rho$ does not concern the electromagnetic degrees of freedom $\vec{E}^{T}$ and $\vec{B}$, but instead, as a consequence of the Coulomb gauge condition, reflects only the dynamics of the particle degrees of freedom.

\end{itemize}

\noindent It is on these two models of classical electrodynamics, which should be seen as approximations to the full dynamics of CED, that I shall focus while discussing the DS reduction to quantum electrodynamics; the reduction of classical systems exhibiting back reaction effects is beyond the scope of my analysis. In the next two sections, I provide distinct templates for the reduction of these models to QED.

While these two models apply to a wide variety of systems, the specific applications that I will be considering in my analysis concern the interaction of charged elementary particles with the electromagnetic field. As regards applications of the Lorentz model, the physics used to guide beams of charged subatomic particles at relativistic energies is nothing other than the classical, relativistic, Lorentz force law for a single particle in an electromagnetic field. As regards the description of radiation emitted by a prescribed, generally time-dependent, charge distribution consisting of a single charged particle, one successful application of the Maxwell model is the prediction of the angular and frequency distribution of synchroton radiation by means of the Lienard-Wiechert potentials (which in turn are derived from Maxwell's equations), and also to the description of low-frequency bremsstrahlung \cite{jackson1999classical}. It is these sorts of relatively simple, but very important, applications of CED that I seek to understand on the basis of QED.

While there can be little doubt that classical models have been applied successfully to the description of subatomic charged particles and their electromagnetic interactions, the reader may nevertheless wonder \textit{why} this should be the case, given that subatomic particles such as electrons and protons are typically the types of systems which we expect to behave \textit{non}classically since, because of their small mass, their wave packets tend to spread on relatively short time scales. In short, the microscopicness of these systems would seem to preclude any robust classical behavior on the their part. 

However, a little further thought shows that this is not the case. When particles such as electrons are moving sufficiently quickly, the centers of their wave packets may traverse a substantial distance in the time it takes their wave packets to spread. In this time, the position and momentum expectation values of these wave packets will approximately satisfy Newtonian equations of motion. To make this assumption more plausible, let us do a back-of-the-envelope calculation to show that the timescales on which wave packets of a particle like an electron (mass $\sim 10^{31} kg.$) which is moving sufficiently fast (say $10^{6}$ m/s.)  spread are sufficiently long to allow for trajectory-like, Newtonian behavior to within a reasonable margin of error (say, $10^{-3}$m), over the typical length of particle tracks observed in a detector (say $\sim 1m.$). The timescale on which a free electron Gaussian wave packet spreads can be determined using the expression for the width $a$ of the wave packet over time: 

\begin{equation}
a = \sqrt{a_{0}^{2} + \frac{4 \hbar^{2} t^{2}}{m^{2} a_{0}^{2}}}.
\end{equation} 

\noindent If the initial packet width is on the order of $10^{-5}m.$ (we do not want to make it too narrow initially, or else it will spread too quickly), so that $a_{0} \sim 10^{-5} m.$, then inserting $a \sim 10^{-3} m.$, $m \sim 10^{-31}$kg., and $\hbar \sim 10^{-34} kg.m.^{2}/s.$, we have

\begin{equation}
t=   \sqrt{\frac{m^{2}a_{0}^{2}}{4 \hbar^{2}} (a^{2} - a_{0}^{2})} = \frac{m a_{0}}{2 \hbar} \sqrt{ (a^{2} - a_{0}^{2})} \approx \frac{m a_{0} a}{2 \hbar} \sim \frac{10^{-31} 10^{-5} 10^{-3}}{10^{-34}} \sim 10^{-5}s.
\end{equation} 

\noindent If the velocity is semi-relativistic, say $10^{6} m./s$, then the length of the electron path over which Newtonian trajecories can be expected to within a margin of error (trajectory width) of $10^{-6}m$ is then given approximately by

\begin{equation}
D=vt \sim 10^{6}m./s 10^{-5} s. \sim 10^{1} m.
\end{equation}

\noindent Thus, for particles travelling at velocities that are sufficiently fast, but not necessarily strongly relativistic, Newtonian trajectories can be expected to persist over length scales of tens of meters, to within a path width of a millimeter. Thus, behavior in this domain can be approximated as classical, though an order-of-magnitude adjustment of initial wave packet width or velocity could disrupt this classicality.

\section{Worries about the Consistency of Classical Electrodynamics}

Before proceeding to discuss the reduction of CED to QED, it is worth taking a moment to consider concerns about the internal consistency of CED that have recently received a significant amount of attention in the philosophy of physics literature. While I do not engage in any depth with this debate here, these concerns are worth taking note of if only to flag the worry that the theory whose reduction I am considering contains internal contradictions, which of course will have bearing on the possibility of reducing it to some more fundamental theory. While my discussion here is limited to reduction in the context of basic applications of the Lorentz and Maxwell models (quantum and classical), any completely comprehensive account of the manner in which the exact CED model - not just the approximations I consider here - reduces to the full QED model first must place the consistency of CED on firm footing. Here, I discuss one worry about the internal consistency of CED recently raised by Frisch and the responses it has elicited; I focus here on Zuchowski's reply to Frisch's argument, which draws on those of Muller, Belot and Vickers, while also diverging from them in important respects. 

Frisch has argued that CED is inconsistent because, he claims, it prescribes two mutually contradictory ways of setting up energy conservation in its description of an accelerating charged particle \cite{Frisch2005}. Frisch starts from what Zuchowski has dubbed Jackson's `two-step procedure,' in which one either begins with some prescribed electromagnetic fields and calculates the resulting particle trajectory from the Lorentz Force Law or one begins with some prescribed charge and current distribution and calculates the resulting electromagnetic field. However, as Zuchowski has observed, Frisch's inconsistency claim seems to result at least partially from a failure to recognise that this two-step procedure does not represent the theory of CED itself, but rather two separate approximations to CED that work well only in certain contexts and whose solutions do not represent exact solutions to the full theory of CED. As with the three-body problem in classical mechanics, solutions to the fully coupled, exact equations of CED have thus far resisted any exact, closed-form statement and so most applications of the theory have, as a result, relied on approximations such as the Maxwell and Lorentz models.

First, for a charged particle moving in some prescribed electric and magnetic fields $\vec{E}_{0}$ and $\vec{B}_{0}$ (importantly, that are not generated by the particle itself, but that may be generated by other charges and currents), the energy transfer from the field to the particle is determined by the relation

\begin{equation} \label{CEDCons}
\frac{1}{2} m \dot{\vec{x}}^{2} = \int \vec{F}_{ext} (\vec{E}_{0},\vec{B}_{0}) \cdot d\vec{x}
\end{equation}

\noindent where $\vec{F}_{ext} (\vec{E}_{0},\vec{B}_{0})$ is the Lorentz force associated with the given fields, $\vec{x}$ is the position of the particle, and the integration extends over the volume over which the particle's charge is distributed. However, this prescription does not take account of the fact that charged particles radiate when they accelerate (a fact that is verified empirically and that is also expected on a theoretical basis from Maxwell's equations), and so lose energy in the process. This energy loss is sometimes accounted for by the addition of another term to the energy balance, associated with an internal or `self' force, due to the fields generated by the particle itself (which are, in turn, determined by Maxwell's equations):

\begin{equation} \label{CEDConsSelf}
\frac{1}{2} m \dot{\vec{x}}^{2} = \int \vec{F}_{ext} (\vec{E}_{0},\vec{B}_{0}) \cdot d\vec{x} + \int \vec{F}_{int} (\vec{E}_{p},\vec{B}_{p}) \cdot d \vec{x}.
\end{equation}

\noindent Here, $\vec{F}_{int}$ is the Lorentz force associated with the fields $\vec{E}_{p}$ and $\vec{B}_{p}$, which are generated by the charged particle itself and which are determined using Maxwell's equations with the charged particle as source. 

The conservation relations (\ref{CEDCons}) and (\ref{CEDConsSelf}) are clearly in contradiction with each other. On this basis, Frisch concludes that CED, which he understands to be a theory that neglects the energy loss associated with the self force in describing the motion of the particle, does not conserve energy. Frisch further concludes on this basis that the Lorentz equation (which on its own neglects the self force) is inconsistent with Maxwell's equations (which entail the presence of the fields associated that generate the self-force ), and since these are the central equations of CED, he concludes that CED itself is inconsistent. However, as Zuckowski notes, Frisch's notion of inconsistency seems to deviate from the more conventional definition, which regards equations as inconsistent only when they have no solutions; Frisch does not offer any proof of the claim that there are no solutions to the combined Maxwell and Lorentz equations, but only the observation that there are two mutually contradictory, and arguably permissible, ways of setting up energy conservation.

Muller, Belot, Vickers and Zuchowski have all offered extended critiques of Frisch's inconsistency argument \cite{MullerCED}, \cite{BelotCED}, \cite{VickersCED}. I will not review them all here as this has already been done in Zuchowski's \cite{ZuchowskiCED}. However, Zuchowski has shown that Frisch's argument against the constitency of CED fails because one can show, without invoking any approximations, that full CED does indeed respect energy conservation; when one considers the fully coupled equations of CED, rather than the approximate two-step procedure, there is no contradiction or ambiguity about energy conservation; this apparent contradiction is merely an artefact of the approximation scheme that Frisch considers, which he incorrectly identifies with the full theory of CED. The original proof of energy conservation for an accelerating charged particle was given by Kiessling in \cite{Kiessling}; Zuchowski presents a simplified version of it specialised to the nonrelativistic case. 

As Zuchowski emphasizes, her refutation of Frisch's argument to the effect that CED is inconsistent should not be taken to entail that CED is in fact consistent - only that if it happens not to be consistent, then it is not for the reasons that Frisch cites. I have restricted my attention in this thesis to the reduction of the classical Maxwell and Lorentz model in part because there is not space here to fully and properly engage the more general and much more difficult question of how to reduce the full interacting CED model. If CED is indeed inconsistent, then it would be unreasonable to attempt a reduction of the exact theory of CED to QED; rather, we would have to satisfy ourselves with analysing the reduction of those particular approximations to the full theory, such as the Lorentz and Maxwell models , that are internally consistent and that do generate clear empirically well-confirmed predictions. Pending further developments on this matter, I have limited myself here to considering the DS reduction only of two successful approximations to CED that are mathematically well-understood and empirically well-confirmed. Because these models can indeed be treated in the framework of dynamical systems theory, and because, as I demonstrate shortly, one can construct quantum analogues to the classical Maxwell and Lorentz models that are likewise approximations to the full QED model, we are in a position to see whether the classical and quantum versions of these simplified approximations relate to each other in the way that DS reduction requires. 

Of course, once one has considered the question of the internal consistency of CED, it becomes natural to inquire as to the internal consistency of QED. On this question, the matter of the so-called `Landau pole' in the renormalisation group flow of the `physical' electromagnetic coupling is sometimes cited as a potential source of internal inconsistency in QED. In order to evade such issues here, one can, without significant alteration to the theory's low-momentum predictions, simply set the high-momentum cutoff for the theory at a scale very much larger than the momenta involved in the pheneomena under consideration, but less than the momentum at which the renormalised coupling diverges  (for further discussion of the Landau pole, see for instance \cite{BanksQFT}, section 9.9, and \cite{LandauPolePRL})   \footnote{Thanks to David Wallace for a helpful discussion on this point.}.  

As Zuchowski also makes a point of emphasising, the sheer technical difficulty of devising a closed-form solution to the full model of CED - i.e., of solving the Maxwell and Lorentz equations simultaneously, rather than iteratively and perturbatively - in no way entails that solutions to this model do not exist. As an illustration of this point, she refers back repeatedly to the case of the three-body problem in classical mechanics. Although this problem has for more than a century resisted the efforts of mathematicians and physicists to devise a closed-form solution, it is widely acknowledged that these difficulties do not entail that the model is inconsistent - that is, that it has no solutions. However, the consistency of the full CED model clearly needs to be proven, and the nature of its solutions better understood, before a fully comprehensive DS reduction of CED to QED can be carried out.

\section{The Quantum Maxwell and Lorentz Models} \label{QuantumMaxLor}

As discussed above, the classical Lorentz and Maxwell models each have quantum counterparts. In the quantum Maxwell model, it is the evolution of the fermionic degrees of freedom that is prescribed; in the quantum Lorentz model, it is the evolution of the electromagnetic degrees of freedom that is prescribed. 

\

\noindent \underbar{\textbf{Quantum Maxwell Model}}

\

The quantum counterpart to the Maxwell model has as its state space the electromagnetic Hilbert space,

\begin{equation}
S_{M} = \mathcal{H}_{EM}
\end{equation}

\noindent  and its dynamics are determined by the effective Schrodinger equation,

\begin{equation}
i \frac{\partial }{\partial t} | \Phi \rangle = \hat{H}_{M} | \Phi \rangle
\end{equation}

\noindent where

\begin{equation} \label{MaxwellSemi}
\hat{H}_{M}  | \Phi \rangle  \equiv  \int d^{3}x \   \big[ \frac{1}{2} (\hat{\vec{E}}^{2}_{T}  +  \hat{\vec{B}}^{2} ) + \boldsymbol{\vec{j}} \cdot \hat{\vec{A}}  \big]   | \Phi \rangle,
\end{equation}

\noindent and $\boldsymbol{\vec{j}(x)}$ is some prescribed classical (that is, c-number- rather than operator-valued) source current. This model, which I refer to as the quantum Maxwell model, and is an example of a `semiclassical' model of quantum field theory, in which one set of degrees of freedom is treated quantum mechanically - in this case, the bosonic degrees of freedom -  while the other is treated classically - in this case, the fermionic degrees of freedom. (The reader may consult, for instance, \cite{peskin1996introduction}, p. 32-33, for a brief introduction to semi-classical models of quantum field theory.) 

Much of the discussion in section $\ref{LMEhrenfest}$ will concern the matter of how to reduce the classical Maxwell model to the quantum Maxwell model. For the moment, though, I will briefly discuss the reduction of the quantum Maxwell model to the full QED model, before returning to this matter in section \ref{QFTDecoherence}.  

In order that the evolution of the bosonic degrees of freedom in QED may be approximated by some unitary evolution - i.e. in terms of some Schrodinger equation, as it is in (\ref{MaxwellSemi}) - it is necessary that the underlying QED state in $\mathcal{H}_{D} \otimes \mathcal{H}_{EM}$ remain a product state over the timescale for which the quantum Maxwell model is expected to apply. In addition, it is also necessary that the fermionic degrees of freedom not vary too rapidly in time - that is, that the fermionic state not involve energies that are too high. To see the reason for this requirement, consider the time derivative of the QED product state $| \Psi(t) \rangle \otimes | \Phi(t) \rangle$, as determined by the full QED Schrodinger equation:

\begin{equation}
\frac{\partial}{\partial t} \left( | \Psi(t) \rangle \otimes | \Phi(t) \rangle \right) = \hat{H}_{QED}  \left( | \Psi(t) \rangle \otimes | \Phi(t) \rangle \right).
\end{equation}

\noindent Now since $\frac{\partial}{\partial t} \left( | \Psi(t) \rangle \otimes | \Phi(t) \rangle \right) = \frac{\partial}{\partial t} \left( | \Psi(t) \rangle \right) \otimes | \Phi(t) \rangle +  | \Psi(t) \rangle  \otimes \frac{\partial}{\partial t} \left( | \Phi(t) \rangle \right)$, if we stipulate that the fermionic degrees of vary slowly - i.e., $\frac{\partial}{\partial t}  | \Psi(t) \rangle  \approx 0$ - then  $\frac{\partial}{\partial t} \left( | \Psi(t) \rangle \otimes | \Phi(t) \rangle \right) \approx  | \Psi(t) \rangle  \otimes \frac{\partial}{\partial t} \left( | \Phi(t) \rangle \right)$. If moreover, $| \Psi(t) \rangle$ and $| \Phi(t) \rangle$ are such that

\begin{equation} \label{HQEDApprox}
\hat{H}_{QED}  \left( | \Psi(t) \rangle \otimes | \Phi(t) \rangle \right) \approx  | \Psi(t) \rangle \otimes \hat{H}_{M}^{eff}  \left( | \Phi(t) \rangle \right),
\end{equation}

\noindent for some Hermitian operator $\hat{H}_{M}^{eff}$ on $\mathcal{H}_{EM}$, then it follows that 

\begin{equation}
| \Psi(t) \rangle \otimes \frac{\partial}{\partial t}  | \Phi(t) \rangle \approx | \Psi(t) \rangle \otimes \hat{H}_{M}^{eff}  | \Phi(t) \rangle
\end{equation}

\noindent and from this, that

\begin{equation}
\frac{\partial}{\partial t}  | \Phi(t) \rangle \approx \hat{H}_{M}^{eff}  | \Phi(t) \rangle 
\end{equation}

\noindent thereby yielding an approximately unitary, pure state evolution for the bosonic degrees of freedom. To derive the semiclassical quantum Maxwell model discussed above, we need  $\hat{H}_{M}^{eff} \approx \hat{H}_{M}$. 

A full reduction of the semiclassical Maxwell model to the full QED model goes beyond the scope of my analysis here. Nevertheless, I will endeavour to provide some level of heuristic insight into how it comes about. Consider for the moment just the renormalised portion $\hat{H}^{r}_{QED} = \hat{H}^{r}_{D} + \hat{H}^{r}_{C} + \hat{H}^{r}_{EM} + \hat{H}^{r}_{I} $ of $\hat{H}_{QED}$, ignoring the counterterm portion $\hat{H}_{CT}$.

 If $| \Psi(t) \rangle$ is slowly varying, then the energies associated with the purely fermionic portion $\hat{H}^{r}_{D} + \hat{H}^{r}_{C}$ of the renormalised QED Hamiltonian will be low, so that by comparison with other terms in the Hamiltonian, $(\hat{H}^{r}_{D} + \hat{H}^{r}_{C}) | \Psi(t) \rangle \approx 0$. If in addition, $| \Psi(t) \rangle$ is an approximate eigenstate of the interaction Hamiltonian $\hat{H}^{r}_{I} $, so that  $\hat{\psi}^{r \dagger}(x) \vec{\alpha} \cdot \hat{\psi}^{r}(x) \cdot \hat{\vec{A}}(x) | \Psi(t) \rangle \otimes | \Phi(t) \rangle  \approx \vec{j}(x) \cdot \hat{\vec{A}}(x) | \Psi(t) \rangle \otimes | \Phi(t) \rangle $, then 
 
 \begin{align}
 \hat{H}^{r}_{I} | \Psi(t) \rangle \otimes | \Phi(t) \rangle & =  \left[\int d^{3} x \ e_{r} \hat{\psi}^{r \dagger} \vec{\alpha} \cdot \hat{\vec{A}}^{r} \hat{\psi}^{r} \right] | \Psi(t) \rangle \otimes  | \Phi(t) \rangle \\
 & \approx \left[\int d^{3} x \ \vec{j} \cdot \hat{\vec{A}}^{r} \right] | \Psi(t) \rangle \otimes  | \Phi(t) \rangle. 
 \end{align}

\noindent From this it follows that

\begin{align}
\hat{H}^{r}_{QED}  | \Psi(t) \rangle \otimes  | \Phi(t) \rangle & \approx  \left(  \hat{H}^{r}_{EM} + \hat{H}^{r}_{I} \right)  | \Psi(t) \rangle \otimes  | \Phi(t) \rangle \\
& \approx | \Psi(t) \rangle \otimes \left[\int d^{3} x \     \frac{1}{2} (\hat{\vec{E}}^{r 2}_{T}  +  \hat{\vec{B}}^{r 2} ) +  \vec{j} \cdot \hat{\vec{A}}^{r} \right]   | \Phi(t) \rangle.
\end{align}

\noindent in accordance with $(\ref{HQEDApprox})$, with $\hat{H}_{M}^{eff} \approx \hat{H}_{M}$, except that here I have employed only the renormalised portion $\hat{H}^{r}_{QED}$ of the QED Hamiltonian rather than the full Hamiltonian $\hat{H}^{r}_{QED} + \hat{H}^{r}_{CT} $, and in $ \hat{H}_{M}$ appearing in (\ref{MaxwellSemi}) the superscripts $r$ are omitted. Noting that we have employed in this discussion only the renormalised portion $\hat{H}^{r}_{QED}$ of the QED Hamiltonian rather than the full Hamiltonian $\hat{H}^{r}_{QED}=\hat{H}^{r}_{QED} + \hat{H}^{r}_{CT} $, we are still left with the issue of how to address the presence of the counterterms and to account for their effect. I defer this issue to future research, acknowledging that modifications to the present discussion may be required in light of the effects of the counterterms.

%\begin{equation}
%i \frac{\partial }{\partial t} \left( \boldsymbol{| \Psi(t) \rangle} \otimes | \Phi \rangle \right)  = \hat{H}_{M}^{eff}  \left( \boldsymbol{| \Psi(t) \rangle} \otimes | \Phi \rangle \right),
%\end{equation}

%\noindent with $\boldsymbol{| \Psi(t) \rangle}$ a prescribed evolution for the fermionic state that remains unentangled with the electromagnetic state throughout its evolution, and where the effective Hamiltonian $ \hat{H}_{M}^{eff}% is defined by the relation

%\begin{equation}
%\hat{H}_{M}^{eff}  \left( \boldsymbol{| \Psi(t) \rangle} \otimes | \Phi \rangle \right)  \equiv  \int d^{3}x \   \left[ \frac{1}{2} (\hat{\vec{E}}^{r 2}_{T}  +  \hat{\vec{B}}^{r 2} ) + e \hat{\psi}^{r \dagger} \vec{\alpha} \hat{\psi}^{r}  \cdot \hat{\vec{A}}^{r} \right]  \left( \boldsymbol{| \Psi(t) \rangle} \otimes | \Phi \rangle \right) 
%\end{equation}

%\noindent where all field operators appearing in the effective Hamiltonian are the renormalised operators appearing in the renormalised portion $\hat{H}^{r}_{QED}$ full QED Hamiltonian. 

A full reduction of the classical Maxwell model to QED may take as an intermediary step the reduction of the quantum Maxwell model to QED. In doing so, it must explain why and in what domains the quantum Maxwell model successfully approximates QED. I have offered some heuristic and preliminary remarks on this matter, though they are no doubt incomplete. In particular, I have suggested that the domain of QED in which the quantum Maxwell model applies should be restricted to a domain of states that are approximately product states, and moreover that the fermionic factor of these product states should be slowly varying (so, contain only low energies by comparison with those contained in the bosonic factor). I offer some further suggestions on this matter in section \ref{QFTDecoherence}.

\

\noindent \underbar{\textbf{Quantum Lorentz Model}}

\

The quantum counterpart to the classical Lorentz model has as its state space the fermionic Hilbert space,

\begin{equation}
S_{L} = \mathcal{H}_{D}
\end{equation}

\noindent  and its dynamics are determined by the effective Schrodinger equation

\begin{equation} \label{LSchrod}
i \frac{\partial | \Psi \rangle}{\partial t} = \hat{H}_{L}  | \Psi \rangle,
\end{equation}

\noindent where 

\begin{equation}
\begin{split}
 \hat{H}_{L} | \Psi(t) \rangle  \equiv   \int d^{3}x \  & \bigg[   \hat{\psi}^{\dagger} ( -i \vec{\alpha} \cdot \nabla + \beta m ) \hat{\psi} + e \ \hat{\psi}^{ \dagger} \vec{\alpha} \cdot  \boldsymbol{\vec{A}}  \hat{\psi}  \\
& \ + \  \frac{e^{2}}{8 \pi} \ \int d^{3}y \frac{\hat{\rho}(x) \hat{\rho}(y)}{|x - y|}    \bigg] | \Psi(t) \rangle
\end{split}
\end{equation}

\noindent with $ \boldsymbol{ \vec{A}(x)}$ a prescribed classical electromagnetic field. Like the quantum Maxwell model, this model, which I call the quantum Lorentz model, is semiclassical; however, in this model it is the electromagnetic field that is treated as classical and the fermionic field that is treated as quantum in nature. 

All of the considerations that applied to the reduction of the quantum Maxwell model to QED also apply, \textit{mutatis mutandis}, to the reduction of the quantum Lorentz model to QED. In particular, while we must require as in the Maxwell model that the underlying QED state be approximately a product state, it should now be the bosonic degrees of freedom, described by the state $| \Phi \rangle$, that are slow and so contain only low energies. Moreover, it is now the bosonic state $| \Phi \rangle$ that should be an approximate eigenstate of the interaction Hamiltonian, so that $\hat{\psi}^{r \dagger}(x) \vec{\alpha} \cdot  \hat{\vec{A}}(x) \hat{\psi}^{r}(x) | \Psi(t) \rangle \otimes | \Phi(t) \rangle  \approx \hat{\psi}^{r \dagger}(x) \vec{\alpha}\cdot \vec{A}(x) \hat{\psi}^{r}(x)  | \Psi(t) \rangle \otimes | \Phi(t) \rangle $, where the reader should note that the vector potential on the right-hand side is now a c-number rather than an operator.

As in the case of the Maxwell model, a full reduction of the classical Lorentz model may take the reduction of the quantum Lorentz model to QED as an intermediary step. However, in doing so it must explain why and in what domains the quantum Lorentz model successfully approximates QED. We have seen that the QED states that approximately instantiate the quantum Lorentz model should be approximate product states in which the bosonic degrees of freedom are low-energy. As I discuss further in section \ref{QFTDecoherence}, the set of QED states that evolve approximately as product states (again, over some limited timescale) should include the set of product coherent states (and states close to these); so, product coherent states for which the bosonic state is low-energy and an approximate eigenstate of the interaction Hamiltonian (as it will be if it is sufficiently narrowly peaked in field configuration space) seem likely to be among the elements of the Lorentz model's domain.  

\section{Ehrenfest Theorems for Quantum Maxwell and Lorentz Models} \label{LMEhrenfest}

\subsection{Ehrenfest Theorem for the Quantum Maxwell Model}

As in Chapter 2's discussion of the classical domain of NRQM, an Ehrenfest Theorem for the electromagnetic field in the quantum Maxwell model is crucial to explaining classical behavior of this field. To derive the Ehrenfest Theorem in this case, it will prove convenient to work in the Heisenberg picture. In the Heisenberg picture, one finds, if one employs $\hat{H}_{M}$ as the system's Hamiltonian and uses the equal-time commutation relations $[\hat{A}^{T}_{i}(x,t), \hat{E}_{j}(x',t)] = -i\delta_{ij}^{T}(x - x')$, that the Heisenberg equations of motion for the operators $\vec{A}$ and $\vec{E}$ take the form:

\begin{equation}
\dot{\vec{A}}^{T}(x,t) = \frac{\partial \hat{\vec{A}}^{T}(x,t)}{\partial t} = i [ \hat{H}_{M},  \hat{\vec{A}}^{T}(x,t) ] = \hat{\vec{E}}^{T }(x,t)
\end{equation}

\noindent from which it immediately follows that

\begin{equation}
\dot{\vec{B}} = \frac{\partial \hat{\vec{B}}}{\partial t} = \nabla \times \frac{\partial \hat{\vec{A}}^{T}}{\partial t} = \nabla \times \hat{\vec{E}}^{T}
\end{equation}

\noindent giving the operator form of Maxwell's magnetic induction equation. Likewise, 

\begin{equation}
\dot{\vec{E}}^{T}= \frac{\partial \hat{\vec{E}}^{T}}{\partial t} = i [ \hat{H}_{M},  \hat{\vec{E}}^{T} ] = - \nabla \times \hat{\vec{B}} + 4 \pi \boldsymbol{\vec{j}},
\end{equation}

\noindent where $\boldsymbol{\vec{j}}$ is the prescribed source current appearing in $\hat{H}_{M}$. Taking expectation values with respect to an arbitrary state in $\mathcal{H}_{EM}$, we find

\begin{equation}
\frac{\partial \langle \hat{\vec{B}} \rangle}{\partial t} =  \nabla \times \langle \hat{\vec{E}} \rangle
\end{equation}

\noindent and

\begin{equation}
\frac{\partial \langle \hat{\vec{E}} \rangle}{\partial t} =  -\nabla \times \langle \hat{\vec{B}} \rangle   +    4 \pi \boldsymbol{\vec{j}} .
\end{equation}

\noindent Thus, no matter what the state of the EM field, the expectation values of the electric and magnetic fields obey classical equations of motion, with $\boldsymbol{\vec{j}}$ standing in as a prescribed classical source current. That is, no restriction need be placed on the quantum state in $\mathcal{H}_{EM}$ in order for expectation values of the electric and magnetic fields to behave classically. This fact can be traced back to the fact that the electromagnetic field couples linearly to the Fermi field, and to the fact that the operator Maxwell equations are linear in the field and field momentum operators. However, while general states in the Maxwell model produce classical evolutions for expectation values of the electromagnetic field, generally it is still only states $| \vec{A}, \vec{E}  \rangle$ that are narrowly peaked in field configuration position and momentum - of which the EM coherent states are a subset - that can reasonably be called `classical.' For while other states may yield classical evolutions for expectation values of the EM field operators, these states are widely distributed about these expectation values and, as we will discuss, prone to entanglement with fermionic degrees of freedom. This entanglement then can be expected to cause the branch-relative expectation values of the field operators to evolve nonclassically, even though the total expectation values always evolve classically. As we will see in the next section, the universal classical evolution of expectation values cannot be extended to the Dirac field. 

As discussed in Introduction to this chapter, DS reductions of classical models to quantum ones typically exhibit a common sequence of steps: 1) in the Bare/Everett model, decoherence ensures quasiclassicality, while in the Bohm model it is ensured by the definiteness of the field beable configuration; 2) some analogue of Ehrenfest's Theorem then ensures classicality, but potentially only for a restricted domain of states; 3) while decoherence will usually serve to enforce quasi-classicality at all times, wave packet spreading can disrupt classicality; the factors that affect wave packet spreading depend most strongly on the internal dynamics of the degrees of freedom under consideration, though may also depend on the nature of the interaction with external degrees of freedom. 

In the case of the reduction of the classical Maxwell model to the  Bare/Everett version of the quantum Maxwell model, we should expect degrees of freedom external to the EM field - in particular, the fermions - to help to enforce quasiclassicality of field states in systems where we observe the field to behave classically. I discuss this further in section \ref{QFTDecoherence}. Ehrenfest's Theorem, however, does not in this particular case require any restriction to a particular domain of states in order to ensure that expectation values evolve classically. Rather, the relevance of coherent states, or of narrow wave packets more generally, to the classical behavior of electromagnetic fields is derived not from the need for classicality, but solely from the need for \textit{quasi}classicality. As regards wave packet spreading, in the case of the quantum Maxwell model, the free dynamics of the EM field do not cause coherent packets to spread. However, it is possible the external classical source current may furnish an external cause of spreading.

\subsection{Ehrenfest Theorem for the Lorentz Model}

In this section, I derive an Ehrenfest Theorem for the quantum Lorentz model, which will be essential to reducing the classical Lorentz model. I will specialise here to the classical model of two electrons, though the analysis can be generalised straightforwardly to higher numbers. Focusing then on the domain of 2-particle, 0-antiparticle states in the quantum Lorentz model, 

\begin{equation}
| \Psi(t) \rangle = \int d^{3} x_{1} \ d^{3} x_{2} \ \psi_{2,0}^{a_{1} a_{2}}(x_{1},x_{2}   ) \ \hat{\psi}_{r}^{ \dagger a_{1}}(x_{1})\hat{\psi}_{r}^{ \dagger a_{2}}(x_{2}) \ |  0^{r}_{D} \rangle
\end{equation}

\noindent one can show through straighforward manipulation of the canonical anticommutation relations that the effective Schrodinger for the Lorentz model, (\ref{LSchrod}), entails that

\begin{equation} \label{DiracSchrod2e}
\small
\begin{split}
 i\frac{\partial}{\partial t} \psi_{2,0}^{a_{1} a_{2}}(x_{1},x_{2}) & =  \bigg\{   \big[ \vec{\alpha} \cdot(  -i \vec{\nabla}_{1} + e_{r} \boldsymbol{\vec{A}}(x_{1}) )  + \beta m_{r}   \big]^{a_{1} c_{1}}  \delta^{a_{2}c_{2}} \\
 & +  \delta^{a_{1}c_{1}}  \big[ \vec{\alpha} \cdot(  -i \vec{\nabla}_{2} + e_{r} \boldsymbol{ \vec{A}}(x_{2}) )  + \beta m_{r}   \big]^{a_{2} c_{2}} + \frac{e_{r}^{2}}{4 \pi}  \frac{1}{ \left| x_{1} - x_{2} \right|}   \bigg\}  \psi_{2,0}^{c_{1} c_{2}}(x_{1},x_{2})
\end{split}
\end{equation}

\noindent for the two-particle wave function $\psi_{2,0}^{a_{1} a_{2}}(x_{1},x_{2},t) \equiv \langle 0_{D} |  \hat{\psi}_{r}^{a_{1}}(x_{1})  \hat{\psi}_{r}^{a_{2}}(x_{2}) | \Psi(t) \rangle   $, where an infinite constant corresponding to the quantum Coulomb self-energy has been left out.

\vspace{5mm}

As a first step to reducing the classical Lorentz model to the quantum Lorentz model, I will prove the following generalisation of Ehrenfest's Theorem for two relativistic Dirac fermions interacting via a Coulomb potential in a background electromagnetic field (the generalisation to higher numbers of fermions is straighforward): 

\begin{equation}
\begin{split}
& \textbf{Ehrenfest Theorem (2 fermions in background EM field):} \\
&  \frac{d}{dt} \langle \hat{\vec{p}}_{1} \rangle  =  \int d^{3} x_{1} \ d^{3} x_{2} \ \rho(x_{1},x_{2}) \boldsymbol{\vec{E}^{1}}(x_{1}, x_{2}) + \int d^{3} x_{1} \ d^{3} x_{2} \ \vec{j}(x_{1},x_{2}) \times \boldsymbol{ \vec{B}}^{1}(\vec{x}_{1}).
\end{split}
\end{equation}

\noindent where 

\begin{equation}
\vec{j}^{1}(x_{1},x_{2}) \equiv - e_{r} \sum_{a_{1} a_{2}} \psi_{2,0}^{\dagger a_{1} a_{2}}(x_{1},x_{2})   \vec{\alpha}^{a_{1} b_{1}} \delta^{a_{2} b_{2}} \  \psi_{2,0}^{b_{1} b_{2}},(x_{1},x_{2}) 
\end{equation}

\begin{equation}
\rho(x_{1},x_{2}) \equiv - e_{r} \sum_{a_{1} a_{2}}   \psi_{2,0}^{\dagger a_{1} a_{2}}(x_{1},x_{2},t)  \psi_{2,0}^{a_{1} a_{2}}(x_{1},x_{2}),
\end{equation}

\begin{equation}
\vec{E}^{1}(x_{1}, x_{2}) \equiv  \vec{E}^{1}_{T}(x_{1}, x_{2}) + \vec{E}^{1}_{L}(x_{1}, x_{2})
\end{equation}

\begin{equation}
 \vec{E}^{1}_{L}(x_{1}, x_{2}) \equiv \nabla_{1} \left( \frac{e_{r}^{2}}{4 \pi}  \frac{1}{ \left| x_{1} - x_{2} \right|}  \right),
\end{equation}

\begin{equation}
 \vec{E}^{1}_{T}(x_{1}) \equiv - \frac{\partial}{\partial t} \vec{A}(x_{1}) =  - \frac{d}{d t} \vec{A}(x_{1})
\end{equation}

\begin{equation}
\vec{B}^{1}(x_{1}) \equiv  \nabla_{1} \times \vec{A}(x_{1}),
\end{equation}

\begin{equation}
 \langle \hat{\vec{x}}_{1} \rangle  =  \sum_{a_{1} a_{2}}  \int d^{3} x_{1}d^{3} x_{2} \   \psi_{2,0}^{\dagger a_{1} a_{2}}(x_{1},x_{2}) \left( x_{1} \right)  \psi_{2,0}^{a_{1} a_{2}}(x_{1},x_{2}).
\end{equation}

\begin{equation}
 \langle \hat{\vec{p}}_{1} \rangle  =  \sum_{a_{1} a_{2}}  \int d^{3} x_{1}d^{3} x_{2} \   \psi_{2,0}^{\dagger a_{1} a_{2}}(x_{1},x_{2}) \left( -i \nabla_{1} \right)   \psi_{2,0}^{a_{1} a_{2}}(x_{1},x_{2}).
\end{equation}

\noindent Note that the Ehrenfest relation, though notationally reminiscent of the classical Lorentz Force Law, does not on its own imply classical evolutions for electron wave packet trajectories. The proof of this particular Ehrenfest Theorem that I provide employs the 2-electron Dirac Hamiltonian in prescribed background field that appears on the right-hand side of (\ref{DiracSchrod2e}),

\begin{equation}
 \hat{H}_{2e} = \bigg\{   \big[ \vec{\alpha} \cdot(  -i \vec{\nabla}_{1} \ + \ e_{r} \boldsymbol{\vec{A}}(x_{1}) )  \ + \  \beta m_{r}   \big]^{a_{1} c_{1}}  \delta^{a_{2}c_{2}} 
\  + \  \delta^{a_{1}c_{1}}  \big[ \vec{\alpha} \cdot(  -i \vec{\nabla}_{2} + e_{r} \boldsymbol{ \vec{A}}(x_{2}) ) \  + \  \beta m_{r}   \big]^{a_{2} c_{2}} \ + \ \frac{e_{r}^{2}}{4 \pi}  \frac{1}{ \left| x_{1} - x_{2} \right|}  \bigg\}
\end{equation}

\noindent and designates the Coulomb term $V_{c} \equiv \ \frac{e_{r}^{2}}{4 \pi}  \frac{1}{ \left| x_{1} - x_{2} \right|}  $.  The Heisenberg equation of motion for the 2-electron model with this Hamiltonian entails

\noindent \textbf{Proof:} 

\begin{equation}
\begin{split}
& \frac{d}{dt} \langle \hat{\vec{p}}_{1} + e_{r} \vec{A}(x_{1}) \rangle = i \langle \big[ \hat{H}_{2e}, \hat{\vec{p}}_{1} + e_{r} \vec{A}(x_{1})  \big]  \rangle \\
& = i \langle \big[ \hat{H}_{2e}, \hat{\vec{p}}_{1}  \big]  \rangle +i e_{r} \langle \big[ \hat{H}_{2e},  \vec{A}(x_{1})  \big]  \rangle \\
& =  i \langle \big[ e_{r} \vec{\alpha} \cdot \vec{A}(x_{1}), \hat{\vec{p}}_{1}  \big]  \rangle + i \langle \big[ V_{C}, \hat{\vec{p}}_{1}  \big]  \rangle + i  \langle \big[ e_{r} \vec{\alpha} \cdot \hat{\vec{p}}_{1} ,  \vec{A}(x_{1})  \big]  \rangle \\
& = i \langle ie_{r} \alpha_{j} \nabla_{1} A_{j}(x_{1}) \rangle + i \langle i \nabla_{1} V_{C} \rangle + i\langle -ie_{r} \left( \vec{\alpha} \cdot \nabla_{1} \right) \vec{A}(x_{1}) \rangle \\
& = -e_{r} \int d^{3} x_{1}d^{3} x_{2} \ \psi_{2,0}^{\dagger a_{1} a_{2}}(x_{1},x_{2}) \  \alpha^{a_{1} b_{1}}_{j}  \ \delta^{a_{2} b_{2}} \  \psi_{2,0}^{b_{1} b_{2}}(x_{1},x_{2}) \ \nabla_{1} A_{j}(x_{1}) \\
& \ + \ e_{r} \int d^{3} x_{1}d^{3} x_{2} \   \psi_{2,0}^{\dagger a_{1} a_{2}}(x_{1},x_{2})  \psi_{2,0}^{a_{1} a_{2}}(x_{1},x_{2})  \ \nabla_{1} V_{C}    \\
& \ + \ e_{r}  \ \int d^{3} x_{1}d^{3} x_{2} \   \psi_{2,0}^{\dagger a_{1} a_{2}}(x_{1},x_{2}) \  \alpha^{a_{1} b_{1}}_{j}   \ \delta^{a_{2} b_{2}} \    \psi_{2,0}^{b_{1} b_{2}}(x_{1},x_{2}) \ (\nabla_{1})_{j} \vec{A}(x_{1})
\end{split}
\end{equation}

\noindent where repeated indices have been summed over implicitly, and in going from the third to the fourth line I have used the identity $[ \hat{p}_{i}, F(\hat{\vec{x}}) ] = - i \frac{\partial F}{\partial x_{i}}$. Now, employing the identity $[\vec{j} \times (\nabla \times \vec{A})]_{i}  = j_{j} \partial_{i} A_{j} - j_{j} \partial_{j} A_{i}$, we have

\begin{equation}
\begin{split}
& \frac{d}{dt} \langle \hat{\vec{p}}_{1} \rangle - e_{r} \int d^{3} x_{1} \ d^{3} x_{2} \      \psi_{2,0}^{\dagger a_{1} a_{2}}(x_{1},x_{2})   \psi_{2,0}^{a_{1} a_{2}}(x_{1},x_{2})    \  \vec{E}^{1}_{T}(x_{1}) \\  & =  \int d^{3} x_{1} \ d^{3} x_{2} \ \rho(x_{1},x_{2}) \vec{E}^{1}_{L}(x_{1}, x_{2}) + \int d^{3} x_{1} d^{3} x_{2} \ \vec{j}(x_{1},x_{2}) \times \vec{B}^{1}(\vec{x}_{1}).
\end{split}
\end{equation}

\noindent or,

\begin{equation}
\small
\begin{split}
&  \frac{d}{dt} \langle \hat{\vec{p}}_{1} \rangle -  \int d^{3} x_{1} \ d^{3} x_{2} \ \rho(x_{1},x_{2}) \vec{E}^{1}_{T}(x_{1}, x_{2})  =  \int d^{3} x_{1} \ d^{3} x_{2} \ \rho(x_{1},x_{2}) \vec{E}^{1}_{L}(x_{1}, x_{2}) + \int d^{3} x_{1} \ d^{3} x_{2}\ \vec{j}(x_{1},x_{2}) \times \vec{B}^{1}(\vec{x}_{1}).
\end{split}
\end{equation}

\noindent Finally, this yields

\begin{equation}
 \frac{d}{dt} \langle \hat{\vec{p}}_{1} \rangle  =  \int d^{3} x_{1} \ d^{3} x_{2} \ \rho(x_{1},x_{2}) \vec{E}^{1}(x_{1}, x_{2}) + \int d^{3} x_{1} \ \vec{j}(x_{1},x_{2}) \times \vec{B}^{1}(\vec{x}_{1}),
\end{equation}

\noindent which completes the proof.

\vspace{10mm}

Now, given Ehrenfest's theorem, if the wave packet $ \psi_{2,0}^{a_{1} a_{2}}(x_{1},x_{2})$ is an approximate product state, so $ \psi_{2,0}^{a_{1} a_{2}}(x_{1},x_{2}) \approx \psi_{1,0}^{a_{1}}(x_{1})\psi_{1,0}^{ a_{2}}(x_{2})$, narrowly peaked in both position and momentum for both particles $1$ and $2$, so that both $\rho(x_{1},x_{2})$ and $\vec{j}^{1}(x_{1},x_{2}) $ are substantially different from zero only in a small volume around $\left( \langle \hat{x}_{1}\rangle, \langle \hat{x}_{2}\rangle \right)$,
\footnote{Note that by  $\langle \hat{x}_{1}\rangle$ here, I mean the quantity  $\int d^{3}x_{1} d^{3}x_{2}  \  \psi_{2,0}^{\dagger c_{1} c_{2}}(x_{1},x_{2}) \ \big( x_{1} \big) \ \psi_{2,0}^{c_{1} c_{2}}(x_{1},x_{2})$, and likewise for  $\langle \hat{x}_{2}\rangle$. The matter of how to define the position operator $\hat{x}_{1}$ outside of the expectation value in relativistic quantum theories is one of the notoriously difficult problems in the foundations of relativistic quantum theory. One well-known approach to defining such an operator is the so-called Newton-Wigner method; see, for instance, \cite{halvorson2001reeh}. }
we can write 

\begin{equation}
\begin{split}
&  \frac{d}{dt} \langle \hat{\vec{p}}_{1} \rangle \approx \left( \int d^{3} x_{1} d^{3} x_{2} \ \rho(x_{1},x_{2}) \right) \vec{E}^{1}(\langle \hat{x}_{1}\rangle, \langle \hat{x}_{2}\rangle) + \left(  \int d^{3} x_{1} d^{3} x_{2}  \  \vec{j}^{1}(x_{1},x_{2}) \right) \times \vec{B}^{1}(\langle \hat{x_{1}}\rangle) \\
&=  (-e_{r}) \vec{E}^{1}(\langle \hat{x}_{1} \rangle, \langle \hat{x}_{2} \rangle) + (-e_{r})  \frac{d\langle \hat{x}_{1} \rangle} {dt} \times  \vec{B}^{1}(\langle \hat{x}_{1}\rangle),
\end{split}
\end{equation}

\noindent since $\int d^{3} x_{1} d^{3} x_{2} \ \rho( x_{1}, x_{2}  ) = -e_{r}$ 
and, as we will see, $ \int d^{3} x_{1} d^{3} x_{2}  \ \vec{j}^{1}( x_{1}, x_{2} ) = -e_{r}  \frac{d\langle \hat{x}_{1} \rangle} {dt}$.
\footnote{Note that $\rho( x_{1}, x_{2}  )$, being a function on configuration space, does not correspond to ordinary spatial charge density except in the 1-particle case.} 
As mentioned above, the relation 

\begin{equation}
  \frac{d}{dt} \langle \hat{\vec{p}}_{1} \rangle  \approx  (-e_{r})  \vec{E}^{1}\left( \langle \hat{x}_{1} \rangle, \langle \hat{x}_{2} \rangle \right) + (-e_{r})  \frac{d\langle \hat{x}_{1} \rangle} {dt} \times  \vec{B}^{1} \left(\langle \hat{x}_{1}\rangle \right),
\end{equation}

\noindent does not suffice to guarantee that position expectation values follow classical trajectories. What we need in addition is that 

\begin{equation}
 \langle \hat{\vec{p}}_{1} \rangle \approx \gamma m_{r}  \frac{d \langle \hat{\vec{x}}_{1} \rangle}{dt},
\end{equation}

\noindent with $\gamma$ the relativistic Lorentz factor, for narrow wave packets. To show that this is the case, I make use of the identity

\begin{equation}
\langle \hat{\vec{p}}_{1} \rangle = \frac{1}{2} \langle  \hat{H}^{1}_{0} \vec{\alpha} +  \vec{\alpha} \hat{H}^{1}_{0}  \rangle, 
\end{equation}

\noindent where $\hat{H}^{1}_{0} \equiv \alpha \cdot \hat{\vec{p}}_{1} + \beta m_{r} $ is the Dirac Hamiltonian for particle 1 in the absence of electromagnetic potentials. If  $ \psi_{2,0}^{a_{1} a_{2}}(x_{1},x_{2}) \approx \psi_{1,0}^{a_{1}}(x_{1})\psi_{1,0}^{ a_{2}}(x_{2})$ is not only narrowly peaked in $x_{1}$ and $x_{2}$ but $\psi_{1,0}^{a_{1}}(x_{1})$ is approximately an eigenstate of $\hat{\vec{p}}_{1}$ in such a way that its spinor components make it also approximately an eigenstate of $\hat{H}^{1}_{0}$, then 

\begin{equation}
\langle \hat{\vec{p}}_{1} \rangle = \frac{1}{2} \langle  \hat{H}^{1}_{0} \vec{\alpha} +  \vec{\alpha} \hat{H}^{1}_{0}  \rangle \approx \langle \hat{H}^{1}_{0} \rangle \langle \vec{\alpha} \rangle.
\end{equation}

\noindent If $\psi_{1,0}^{a_{1}}(x_{1})$ is centered about momentum $\vec{p}_{0}$, then $\langle \hat{H}_{0} \rangle \approx \sqrt{\vec{p}_{0}^{2} + m^{2}} = \gamma m$,  where $\gamma$ is the Lorentz factor corresponding to the classical 3-momentum $\vec{p}_{0}$ . Also, one can check that $\frac{d\langle \hat{\vec{x}}_{1} \rangle} {dt} = \langle \vec{\alpha}^{1} \rangle = \int d^{3}x_{1} d^{3}x_{2}\ \vec{j}^{1}(x_{1},x_{2}) $. So, for packets peaked narrowly in both position and momentum, we have

\begin{equation}
 \langle \hat{\vec{p}}_{1} \rangle \approx \gamma m_{r}  \frac{d\langle \hat{\vec{x}}_{1} \rangle}{dt},
\end{equation}

\noindent and thus

\begin{equation}
\frac{d}{dt} \big[ m_{r} \gamma \frac{d}{dt} \langle \hat{\vec{x}}_{1} \rangle \big] \approx(-e_{r}) \vec{E}^{1}(\langle \hat{x}_{1} \rangle, \langle \hat{x}_{2} \rangle) + (-e_{r})  \frac{d\langle \hat{x}_{1} \rangle} {dt} \times  \vec{B}^{1}(\langle \hat{x}_{1}\rangle),
\end{equation}

\noindent as required for the electron packet to follow an approximately classical trajectory. The generalisation to the $N$-fermion case proceeds by direct analogy to the 2-particle case.

\section{The Domains of the Classical and Quantum Maxwell and Lorentz Models} \label{QFTDecoherence}

The preceding analysis of the Lorentz model strongly suggests that the states that instantiate approximate classical behavior of the combined fermionic and electromagnetic degrees of freedom are tensor products of an electromagnetic coherent state $| \vec{A}, \vec{E}  \rangle$ and a fermionic state of definite particle number that is itself a product of 1-particle states localised in both position and  momentum that are also approximate eigenstates of the free renormalised 1-particle Dirac Hamiltonian. For notational simplicity, I abbreviate such fermionic $N$-particle states $| q,p\rangle$. Thus, the states that comprise the classical domain of the Lorentz model - as well as of the Maxwell model, although as already discussed, classicality of expectation value evolutions is ensured for any state evolution in this model - are the product wave packet states,

\begin{equation}
 | q,p \rangle \otimes | \vec{A}, \vec{E}  \rangle,
\end{equation}

\noindent and states in the total Hilbert space that are `close' to such states. 

Beyond the fact that they yield approximately classical dynamical evolutions for expectation values, the introduction of product wave packet states is also motivated by considerations of decoherence. Both the quantum Maxwell and Lorentz models presuppose that the total state of the combined fermionic and electromagnetic system remains as a product state, without suffering entanglement, throughout its dynamical evolution. In the dynamics prescribed by the full QED model, however, a generic product state \textit{will} evolve into an entangled state. It is only for special product states in $\mathcal{H}_{D} \otimes \mathcal{H}_{EM}$, if any at all, that the absence of entanglement between fermionic and bosonic Hilbert spaces will persist to a reliable approximation under the influence of the QED dynamics, so that

\begin{equation}
e^{-i \hat{H}_{QED}t} \left( | \Psi_{0} \rangle \otimes | \Phi_{0} \rangle \right) \approx | \Psi(t) \rangle \otimes | \Phi(t) \rangle. 
\end{equation}

\noindent The states satisfying this condition constitute the entanglement-free subspace  of $\mathcal{H}_{QED}$.

It is natural to guess, by analogy with analyses of decoherence in a wide variety of systems, that the decoherence-free subspace of QED includes the product coherent states $| q, p \rangle \otimes |A, E \rangle$ - that is, that

\begin{equation} \label{QEDCoherentState}
e^{-i \hat{H}_{QED}t} \left( | q_{0},p_{0} \rangle \otimes | \vec{A}_{0}, \vec{E}_{0} \rangle \right) \approx | q(t),p(t) \rangle \otimes | \vec{A}(t), \vec{E}(t) \rangle.
\end{equation}

\noindent  A more rigorous quantitative analysis - possibly employing Zurek's well-known `predictability sieve' method - should be given in order to confirm this assumption \cite{paz1999quantum}, \cite{zurek1993coherent}. However, the assumption is already supported to some extent by Anglin and Zurek's  analysis demonstrating that pointer states of the electromagnetic field under a certain model of its interaction with an environment of massive particles should be coherent states $ | \vec{A}(t), \vec{E}(t) \rangle$, and by Zurek's earlier analysis that coherent states of matter degrees of freedom under interaction with an environment - either electromagnetic or fermionic - should, under fairly generic assumptions, be coherent states \cite{anglin1996decoherence}, \cite{zurek1993coherent}. 

However, given the assumption that the pointer states of QED are product coherent states, it should be recongnised that the spreading of wave packets in the N-electron subspace of the fermionic Hilbert space is likely to limit the time-scale on which the approximation of a persisting product state holds, since such spreading is likely to generate entanglement when the N-electron wave packets become sufficiently broad in position space.  (Again, that while the free fermionic dynamics generically causes spreading of the states $|q,p \rangle$, the free electromagnetic dynamics does not cause the coherent states $| \vec{A}_{0}, \vec{E}_{0} \rangle $ to spread; the states $| \vec{A}_{0}, \vec{E}_{0} \rangle $ spread only as a consequence of the interaction between the electromagnetic and fermionic degrees of freedom.)

For the purposes of moving forward with the analysis, I assume that the product coherent states and nearby states in $\mathcal{H}_{QED}$ resist entanglement, and restrict my analysis to timescales for which the full QED dynamics approximately carries coherent product states into other coherent product states. 

%Under the assumption that product coherent states evolve approximately classicallyAn arbitrary initial state of the total quantum field can be expanded as

%\begin{equation}
%| \Xi_{0} \rangle =  \int dq_{0} \ dp_{0} \ \mathcal{D} A_{0} \mathcal{D} E_{0} \ \alpha(q_{0},p_{0};A_{0},E_{0} ] \ | q_{0},p_{0} \rangle \otimes |A_{0},E_{0} \rangle
%\end{equation}

%\noindent Assuming that product coherent states evolve to other coherent states in the manner prescribed by (\ref{QEDCoherentState}), the state evolution takes the form, 

%\begin{equation}
%| \Xi(t) \rangle \approx  \int dq_{0} \ dp_{0} \ \mathcal{D} A_{0} \mathcal{D} E_{0} \ \alpha(q_{0},p_{0};A_{0},E_{0} ] \ | q_{c}(t),p_{c}(t) \rangle \otimes |A_{c}(t),E_{c}(t) \rangle
%\end{equation}

%\noindent each state evolution $| q_{c}(t),p_{c}(t) \rangle \otimes |A_{c}(t),E_{c}(t) \rangle$ corresponds approximately to a branch of the total quantum state, and to the classical evolution of a point $(q,p;A,E) \in \Gamma_{Np} \times \Gamma_{EM}$. 

%However, as we will see when we discuss the nonrelativistic domain of QED in the next chapter, the decoherence free subspace also includes states of the form 
%$ | \Psi_{Ne} \rangle \otimes  | A, E \rangle$, where $ | \Psi_{Ne} \rangle $ is an $N$-electron state of the Fermi field that is not necessarily a coherent state, or narrowly peaked in position and momentum.

\subsection{Dealing with Counterterms}

The dynamics of the fermionic quantum state $| \Psi(t) \rangle$ in the quantum Lorentz model is determined by the effective Hamiltonian $\hat{H}_{L}$; the dynamics of the quantum state in the Maxwell model likewise is determined by $\hat{H}_{M}$. The analysis of section \ref{QuantumMaxLor} showed that these effective Hamiltonians could be extracted in certain state domains as approximations to the renormalised QED Hamiltonian $\hat{H}^{r}_{QED}$.   Yet the dynamics of QED are determined by the full Hamiltonian  $\hat{H}_{QED}^{r} + \hat{H}_{CT} $, which includes the divergent counterterms of  $\hat{H}_{CT}$ . For a complete reduction of the classical Lorentz and Maxwell models to QED, it is necessary to explain why, even though the quantum Lorentz and Maxwell models only seem to be underwritten solely by the renormalised Hamiltonian, these models nevertheless serve as reliable approximations to a dynamics that incorporates both the renormalised and counterterm components of the full QED Hamiltonian. I leave this as a subject for future research. 

%In other words, it is necessary to show that 

%\begin{equation} \label{QEDCoherentState}
% e^{-i \left(\hat{H}^{r}_{QED}+ \hat{H}_{CT}\right)t} \left( | q_{0},p_{0} \rangle \otimes | \vec{A}_{0}, \vec{E}_{0} \rangle \right) \approx  e^{-i \hat{H}^{eff}_{L}t} | q_{0},p_{0} \rangle \otimes e^{-i \hat{H}^{eff}_{M}t} | \vec{A}_{0}, \vec{E}_{0} \rangle.
%\end{equation}

%\noindent 

%But the dynamics of QED are not determined solely by the renormalised portion of the Hamiltonian, but by the whole Hamiltonian, including the counterterm Hamiltonian $\hat{H}_{CT}$. For the classical models to be fully reduced, it is necessary to explain why the counterm portion of the Hamiltonian can effectively be neglected in the classical domains that I have considered. In the domain of the Lorentz model in particular, it is necessary to prove that if $ | \Xi \rangle \in d_{DF}$ with $| \Xi \rangle = | \Psi \rangle \otimes | \Phi \rangle$, then 

%\begin{equation}
%i \frac{\partial }{\partial t} | \Xi \rangle  = \left[ \hat{H}^{r}_{QED} +  \hat{H}_{CT} \right] | \Xi \rangle \approx \left( \hat{H}^{eff}_{L} | \Psi \rangle \right)   \otimes | \Phi \rangle + | \Psi  \rangle \otimes \left( \hat{H}^{eff}_{M} | \Phi \rangle \right).
%\end{equation}

%\noindent I leave it to future research to demonstrate the cancellation of divergences in the Schrodinger evolution of the QFT state. 

\section{DS Reductions of the Classical Maxwell and Lorentz Models}

In this section, I frame the preceding discussion within the context of the DS approach to reduction. 

\subsection{DS Reduction of Classical Maxwell Model}

The bridge map connecting the state spaces of the quantum Maxwell and classical Maxwell models, $B_{M}^{M}$, is simply the expectation values of the renormalised field and field momentum operators:

\

\noindent \underbar{\textit{Bridge Map:}}

\begin{equation}
\begin{split}
& B_{M}^{M}: \mathcal{H}_{EM} \longrightarrow \Gamma_{EM} \\
& B_{M}^{M}: | \Phi \rangle \longmapsto \left( \langle \Phi | \hat{\vec{A}}^{r,T}(x) | \Phi \rangle  ,   \langle \Phi | \hat{\vec{E}}^{r,T}(x)| \Phi \rangle   \right).
\end{split}
\end{equation}

\noindent Because expectation values of the electric and magnetic field operators satisfy Maxwell's equations for any state $| \Phi \rangle \in \mathcal{H}_{EM}$, the domain of the classical Maxwell model in the quantum one consists of all states in $\mathcal{H}_{EM}$: 

\

\noindent \underbar{\textit{Domain:}}

\

\begin{equation}
d_{MC} = \{ | \Phi \rangle \in \mathcal{H}_{EM} \} 
\end{equation}

\noindent However, as we have seen, for the classical Lorentz model to apply as well, a restriction specifically to coherent states $| \vec{A}, \rangle{E} \rangle $ is necessary. Also, it should be noted that since the expectation values used to define the bridge map are not branch-relative, the classical evolution of the expectation value may not reflect the behavior of the field in individual branches of the total state.

The laws of the image theory, which ensue satisfaction of the DSR condition, are 

\

\noindent \underbar{\textit{Image Model:}}

\

\begin{equation}
\frac{\partial \langle \hat{\vec{B}}^{r} \rangle}{\partial t} =  \nabla \times \langle \hat{\vec{E}}^{r,T} \rangle
\end{equation}

\begin{equation}
\frac{\partial \langle \hat{\vec{E}}^{r,T} \rangle}{\partial t} =  -\nabla \times \langle \hat{\vec{B}}^{r} \rangle   +    4 \pi \langle \hat{\vec{j}} \rangle,
\end{equation}

\noindent as proven above. The laws of the analogue theory are then straightforwardly obtained through the bridge rule substitution

\

\noindent \underbar{\textit{Bridge Rules:}}

\

\begin{equation}
\left(\vec{A}'(x), \vec{E}'(x) \right) \equiv \left( \langle \Phi | \hat{\vec{A}}^{r,T}(x) | \Phi \rangle  ,   \langle \Phi | \hat{\vec{E}}^{r,T}(x)| \Phi \rangle   \right)
\end{equation}

\begin{equation}
\boldsymbol{\vec{j}'}(x) \equiv     e_{r} \boldsymbol{\langle \Psi(t) |}  \hat{\psi}^{r,\dagger}(x) \ \vec{\alpha}  \ \hat{\psi}^{r}(x) \boldsymbol{| \Psi(t) \rangle}.
\end{equation}

\noindent Explicitly, the laws of the analogue theory are

\

\noindent \underbar{\textit{Analogue Model:}}

\

\begin{equation}
\frac{\partial \vec{B}'}{\partial t} =  \nabla \times  \vec{E}'
\end{equation}

\begin{equation}
\frac{\partial \vec{E}'}{\partial t} =  -\nabla \times \vec{B}'   +    4 \pi \boldsymbol{\vec{j} }.
\end{equation}

\noindent The `strong analogy' condition, 

\

\noindent \underbar{\textit{`Strong Analogy':}}

\

\begin{equation} \label{MMStrongAnalogy}
\begin{split}
&| \vec{A}(x,t) - \vec{A}'(x,t) | < \delta_{A} \ \forall \ x \\
&| \vec{E}(x,t) - \vec{E}'(x,t) | < \delta_{E} \ \forall \ x 
\end{split}
\end{equation}

\noindent is then satisfied for all times since $\vec{A}'(x,t)=\vec{A}(x,t)$ and $\vec{E}'(x,t)=\vec{E}(x,t)$. In this particular case, the reduction timescale $\tau$ is infinite.

\subsection{DS Reduction of the Classical Lorentz Model} \label{DSLorentzEB}

Let us examine now how the results discussed above fit into the framework of DS reduction.

The bridge map connecting the state spaces of the quantum Lorentz and classical Lorentz models, $B_{L}^{L}$, is simply to take the expectation values of particle position and momentum. Note that the domain of this bridge map is restricted to the 2-particle states in $\mathcal{H}_{D}$. There may be a way to extend the map to all of $\mathcal{H}_{D}$ so that it coincides with the definition I give here on the 2-particle subspace; however, the map I provide serves my purposes here:

\

\noindent \underbar{\textit{Bridge Map:}}

\begin{flalign*}
\small
B_{L}^{L}: \mathcal{H}_{D} \longrightarrow \Gamma_{2p} &&
\end{flalign*}

\begin{equation}
\small
\begin{split} 
 B_{L}^{L}: | \Psi_{2p} \rangle \longmapsto &  \bigg( \langle \hat{x}_{1} \rangle, \langle  \hat{x}_{2} \rangle; \langle \hat{p}_{1} \rangle, \langle \hat{p}_{2} \rangle     \bigg) \\
& = \bigg(   \int d^{3}x_{1} d^{3}x_{2}  \  \psi_{2,0}^{\dagger c_{1} c_{2}}(x_{1},x_{2}) \ \big( x_{1} \big) \ \psi_{2,0}^{c_{1} c_{2}}(x_{1},x_{2}) , \int d^{3}x_{1} d^{3}x_{2} \ x_{2} \psi_{2,0}^{\dagger c_{1} c_{2}}(x_{1},x_{2}) \ \big( x_{2} \big) \  \psi_{2,0}^{c_{1} c_{2}}(x_{1},x_{2}) ; \\
& \int d^{3} x_{1} d^{3} x_{2}  \psi_{2,0}^{\dagger c_{1} c_{2}}(x_{1},x_{2})   \left(  -i\nabla_{1} \right)    \psi_{2,0}^{c_{1} c_{2}}(x_{1},x_{2})  , \int d^{3} x_{1} d^{3} x_{2}  \psi_{2,0}^{\dagger c_{1} c_{2}}(x_{1},x_{2})   \left(  -i\nabla_{2} \right)    \psi_{2,0}^{c_{1} c_{2}}(x_{1},x_{2})  \bigg). 
\end{split} 
\end{equation}

\noindent The domain of the classical 2-electron Lorentz model is the set of 2-electron states that are products of 1-electron states that are narrow wave packets in both position and momentum and approximate eigenstates of the free first-quantised Dirac Hamiltonian:

\

\noindent \underbar{\textit{Domain:}}

\

\begin{equation}
 d_{L} = \{ | \Psi_{2e} \rangle \in \mathcal{H}_{2e}  |   \ | \Psi_{2e} \rangle  = | q_{1},p_{1}  \rangle \otimes |q_{2},p_{2} \rangle  \}.
\end{equation}

\noindent The laws of the image theory are 

\

\noindent \underbar{\textit{Image Model:}}

\begin{equation}
\frac{d}{dt} \big[ m \gamma \frac{d}{dt} \langle \hat{\vec{x}}_{1} \rangle \big] \approx (-e) \langle \hat{\vec{E}}_{r} \rangle (\langle \hat{\vec{x}} \rangle_{1}, \langle \hat{\vec{x}}_{2} \rangle) +  (-e) \frac{d\langle \hat{\vec{x}}_{1} \rangle} {dt} \times \langle \hat{\vec{B}}_{r} \rangle(\langle \hat{\vec{x}}_{1} \rangle),
\end{equation}

\noindent as proven above, and likewise for $\langle \hat{\vec{x}}_{2} \rangle$. The laws of the analogue theory are then straightforwardly obtained through the bridge rule substitutions

\

\noindent \underbar{\textit{Bridge Rules:}}

\

\begin{equation}
\left(x'_{1}, x'_{2}; p'_{1},p'_{2} \right) \equiv  \left( \langle \hat{x}_{1} \rangle, \langle  \hat{x}_{2} \rangle; \langle \hat{p}_{1} \rangle, \langle \hat{p}_{2} \rangle    \right)
\end{equation}

\begin{equation}
\boldsymbol{ \vec{E}_{T}^{'}}(x,t) =  \boldsymbol{ \langle \vec{A}(t), \vec{E}(t) |}\hat{\vec{E}}^{r,T}(x)  \boldsymbol{ | \vec{A}(t), \vec{E}(t) \rangle}
\end{equation}

\begin{equation}
\vec{E}_{L}^{'}(x_{1},t) =  \nabla_{x_{1}}  \left(  \frac{e_{r}}{4 \pi}  \frac{1}{ \left| x_{1} - x_{2} \right|}    \right) 
\end{equation}

\begin{equation}
\vec{E}_{L}^{'}(x_{2},t) =  \nabla_{x_{2}}  \left(  \frac{e_{r}}{4 \pi}  \frac{1}{ \left| x_{1} - x_{2} \right|}    \right) 
\end{equation}

\begin{equation}
\boldsymbol{ \vec{B}'}(x,t) = \nabla \times  \boldsymbol{ \langle \vec{A}, \vec{E} |}\hat{\vec{A}}^{r,T}(x)  \boldsymbol{ | \vec{A}, \vec{E} \rangle}
\end{equation}

\noindent Applying these bridge rules to the dynamical equations of image model above, we obtain the dynamics of the analogue model,

\

\noindent \underbar{\textit{Analogue Model:}}

\begin{equation}
\frac{d}{dt} \big[ m_{r} \gamma \frac{d x'_{1}}{dt}  \rangle \big] \approx (-e_{r})  \boldsymbol{\vec{E}' } (x'_{1},x'_{2}) +  (-e_{r}) \frac{d x'_{1} } {dt} \times   \boldsymbol{\vec{B}'}( x'_{1}),
\end{equation}

\noindent and similarly for $x'_{2}$. Applying the abbreviation, $\left( x'(t),p'(t) \right) \equiv \left(x'_{1}, x'_{2}; p'_{1},p'_{2} \right)$, the `strong analogy' condition, which is ensured by validity of these analogue dynamics, reads 

\

\noindent \underbar{\textit{`Strong Analogy':}}

\

\begin{equation} \label{CMMWStrongAnalogy}
\begin{split}
&| x(t) - x'(t) | < \delta_{x} \\
&| p(t) - p'(t) | < \delta_{p}.
\end{split}
\end{equation}

\noindent for $0 \leq t\leq \tau$, where the reduction timescale $\tau$ is bounded by the timescale on which wave packets in $\mathcal{H}_{2e}$ spread substantially.

\section{DS Reduction of CED to Bohmian QED }

%The effective collapse condition for the minimalist model requires in this case that  

%\begin{equation}
%\small
%\left[ \int dq \ dp \left|  \langle  q,p; \vec{A}^{T}  | q'_{c}(t),p'_{c}(t); A'_{c}(t),E'_{c}(t)  \rangle \right|^{2}  \right] \left[   \int dq \ dp \left|  \langle q,p; \vec{A}^{T}  | q_{c}(t),p_{c}(t); A_{c}(t),E_{c}(t) \rangle \right|^{2} \right]  \approx 0  \ \ \forall \ \vec{A}^{T} \in  \ \mathbb{Q}_{EM}  
%\end{equation}

%\noindent where the phase space integration $\int dq \ dp$ represents one possible basis in which to take the sum $\sum_{f}$ over fermionic variables appearing in the minimalist model effective collapse condition; this condition will hold as long the phase space points $\left( q_{c}(t),p_{c}(t); A_{c}(t),E_{c}(t) \right)$ and  $\left( q'_{c}(t),p'_{c}(t); A'_{c}(t),E'_{c}(t) \right)$ about which the coherent states are peaked are sufficiently different. 

%If the effective collapse condition is saitsfied, then the electromagnetic field configuration beable of the model should, by equivariance, remain in the configuration space support of the packet $|A_{c}(t),E_{c}(t) \rangle$. Since this packet follows an approximately classical trajectory, the beable configuration $\vec{a}^{T}(t)$ as well should be expected to follow a classical trajectory governed by Maxwell's equations.

%\vspace{5mm}

The preceeding discussion has argued that the domains of the classical Maxwell and Lorentz models within $\mathcal{H}_{QED}$ include the coherent product states $|q,p \rangle \otimes | \vec{A},\vec{E} \rangle$. I suggested that these states likely reside in the decoherence free subspace of $\mathcal{H}_{QED}$, and thus likely provide the product states required by both the quantum Maxwell and quantum Lorentz models. Moreover, the fact that these states are narrow wave packets provides the added assurance that they will evolve approximately along classical trajectories, at least on certain limited timescales. In discussing the reduction of the classical Lorentz and Maxwell models to the Bohmian minimalist model of QED, I assume, as in the preceding discussion, that the state, or effective state, of the total system in $\mathcal{H}_{QED}$ is a product coherent state evolving approximately according to classical equations of motion. 

\subsection{DS Reduction of Classical Maxwell Model to the Bohmian Minimalist Model}

The bridge map connecting the state spaces of the Bohmian minimalist and classical Maxwell models, $B_{M}^{M}$, 

\

\noindent \underbar{\textit{Bridge Map:}}

\begin{equation}
\begin{split}
& B_{M}^{M}: \mathcal{H}_{EM} \times \mathbb{Q}_{EM} \longrightarrow \Gamma_{EM} \\
& B_{M}^{M}: \left( | \Phi \rangle, \vec{a}^{T}(x,t) \right) \longmapsto \left( \vec{a}^{T}(x,t)   ,  -  \frac{\partial  \vec{a}^{T}(x,t) }{\partial t}  \right).
\end{split}
\end{equation}

\noindent The domain of $\mathcal{H}_{EM} \times \mathbb{Q}_{EM}$ that instantiates the classical Maxwell model is

\

\noindent \underbar{\textit{Domain:}}

\

\begin{equation}
d_{MC} = \left\{ \left( | \vec{A}, \vec{E} \rangle,   \vec{a}^{T}(x,t) \right)   \in \mathcal{H}_{EM}  \times \mathbb{Q}_{EM}  \bigg| \vec{a}^{T}(x,t) \in \text{supp}_{\epsilon}( \langle \vec{A}^{T} | \vec{A}, \vec{E} \rangle )  \right\} 
\end{equation} 

\noindent since, by equivariance, if the support of the wave packet $| \vec{A}, \vec{E} \rangle$ traverses a Maxwellian evolution, then so must the Bohmian configuration $\vec{a}^{T}(x,t)$. Defining $\vec{b}(x,t) \equiv \nabla \times  \vec{a}^{T}(x,t) $ $\vec{e}^{T}(x,t) \equiv  -  \frac{\partial  \vec{a}^{T}(x,t) }{\partial t}$, the laws of the image theory can be stated

\

\noindent \underbar{\textit{Image Model:}}

\

\begin{equation}
\frac{\partial  \vec{b} }{\partial t} \approx  \nabla \times \vec{e}^{T}
\end{equation}

\begin{equation}
\frac{\partial \vec{e}^{T}}{\partial t} \approx  -\nabla \times \vec{b}   +    4 \pi \langle \hat{\vec{j}} \rangle,
\end{equation}

\noindent where, note, the source current $\langle  \hat{\vec{j}} \rangle$ continues in the Bohmian case to be a function of the quantum state and not of a Bohmian configuration, in part because there are no beables corresponding to the fermionic sources, but more importantly because the beable $ \vec{a}^{T}(x,t)$ evolves approximately classically only by virtue of the fact that the quantum state $| \vec{A}, \vec{E} \rangle$ does, and the classical evolution of $| \vec{A}, \vec{E} \rangle$ in turn is one associated with the source current $\langle \hat{\vec{j}} \rangle$. The bridge rules for this model are 

\

\noindent \underbar{\textit{Bridge Rules:}}

\

\begin{equation}
\left(\vec{A}'(x), \vec{E}'(x) \right) \equiv  \left( \vec{a}^{T}(x,t),  -  \frac{\partial  \vec{a}^{T}(x,t) }{\partial t}  \right)
\end{equation}

\begin{equation}
\boldsymbol{\vec{j}'}(x) \equiv     \boldsymbol{\langle \Psi(t) |}  e\hat{\psi}_{r }^{\dagger}(x) \vec{\alpha} \hat{\psi}_{r}(x) \boldsymbol{| \Psi(t) \rangle}.
\end{equation}

\noindent Applying these bridge rule substitutions to the image model, we find that the laws of the analogue model are

\

\noindent \underbar{\textit{Analogue Model:}}

\

\begin{equation}
\frac{\partial \vec{B}'}{\partial t} \approx  \nabla \times  \vec{E}'
\end{equation}

\begin{equation}
\frac{\partial \vec{E}'}{\partial t} \approx  -\nabla \times \vec{B}'   +    4 \pi \boldsymbol{\vec{j}' }.
\end{equation}

\noindent The `strong analogy' condition, 

\

\noindent \underbar{\textit{`Strong Analogy':}}

\

\begin{equation} \label{MMStrongAnalogy}
\begin{split}
&| \vec{A}(x,t) - \vec{A}'(x,t) | < \delta_{A} \ \forall \ x \\
&| \vec{E}(x,t) - \vec{E}'(x,t) | < \delta_{E} \ \forall \ x 
\end{split}
\end{equation}

\noindent is then satisfied only for times $\tau$ such that the wave packet $| \vec{A}, \vec{E} \rangle$ evolves classically, which in turn occurs only when the electromagnetic and fermionic degrees of freedom remain unentangled - that is, when total effective state of $\mathcal{H}_{D} \times \mathcal{H}_{EM}$ persists approximately as a product state.

\subsection{DS Reduction of Classical Lorentz Model to the Bohmian Minimalist Model}

Since there are no beables corresponding to the fermionic degrees of freedom in the QED minimalist model, the motion of electrons as described in the classical Lorentz model is, in a sense, merely `implied' by the behavior of the beable radiation field. The domain of Bohmian QED in which the classical Lorentz model applies approximately is the domain in which the beable radiation field is approximately one which would be produced by an electron evolving along a classical trajectory. 

The domain of $\mathcal{H}_{QED}$ in which the classical Lorentz model applies is the domain of product coherent states $|q,p \rangle \otimes | \vec{A},\vec{E} \rangle$. Assuming that the state, or effective state, $|\Xi \rangle$ of the full system lies in this domain, then the trajectory $q(t),p(t)$ followed by the fermionic wave packet $|q,p \rangle$, will be approximately classical. The state $| \vec{A},\vec{E} \rangle$ will be centered on a classical EM field configuration corresponding to the radiation field generated by this trajectory. So, 

\begin{equation}
|q_{0}, p_{0} \rangle \otimes | \vec{A}_{0},\vec{E}_{0} \rangle \Longrightarrow |q_{c}(t), p_{c}(t) \rangle \otimes | \vec{A}[q_{c}(t)],\vec{E}_{0}[q_{c}(t)] \rangle
\end{equation}

\noindent where $\vec{A}[q_{c}(t)]$ is a solution to Maxwell's equations (in vector potential form) corresponding to the source distribution $\vec{j}(x,t) = e_{r} \sum^{N}_{i} \dot{\vec{q}}_{c,i}\delta^{3}(x-q_{c,i}(t))$, and where $i$ indexes the different charged particles in an $N$-particle system; $\vec{E}_{0}[q_{c}(t)] \equiv -\frac{\partial \vec{A}[q_{c}(t)] }{\partial t}$ is the transverse electric field associated with  $\vec{A}[q_{c}(t)]$.

Given the assumptions made so far, equivariance requires the beable conifugration $\vec{a}^{T}$ to lie in the support of the wave packet $ | \vec{A}[q_{c}(t)],\vec{E}_{0}[q_{c}(t)] \rangle$, which entails 

\begin{equation}
\vec{a}^{T} \approx  \vec{A}[q_{c}(t)].
\end{equation}

\noindent So the electromagnetic field beable configuration in a sense functions as a record of the fermionic particle trajectory associated with the wave packet $ |q_{c}(t), p_{c}(t) \rangle$, even though there are no fermionic beables associated with the evolution of this packet. On the minimalist model of Bohmian QED, successful applications of the classical Lorentz model are instantiated by a Bohmian system in which the fermionic degrees of freedom are characterised by a wave packet quantum state of definite particle number, and in which, through the dynamics of the model, the electromagnetic field beable $\vec{a}^{T}[q_{c}(t)]$ comes to be configured as a radiation field of the trajectory $q_{c}(t)$. 

Because it is ultimately the wave packet evolution in the fermionic Hilbert space that determines the trajectory $q_{c}(t)$ recorded in the electromagnetic beable configuration $\vec{a}^{T}[q_{c}(t)]$, the DS reduction of the classical Lorentz model to the minimalist Bohmian model will proceed much as the reduction of the classical Lorentz model to the quantum Lorentz model described in section  \ref{DSLorentzEB}. Recall that the reduction there also depended essentially on the fermionic wave packets evolving classically, though for different reasons.

\section{Summary}

In the preceding analysis, I have offered templates for the reduction of two models of classical electrodynamics both to Bare/Everettian versions of these models, and to minimalist Bohmian QED. A number of assumptions of have been taken for granted which require more detailed justification: in particular, a more detailed analysis should demonstrate that the dynamics determined by the full QED Hamiltonian $\hat{H}_{QED} = \hat{H}^{r}_{QED} +  \hat{H}_{CT}$ coincides with the dynamics determined by the quantum Maxwell and Lorentz models in appropriate domains.

\chapter{The Nonrelativistic Domain of Quantum Electrodynamics}
\label{ch5}

In this chapter, I provide two templates for the reduction of the nonrelativistic quantum mechanics of a spin-1/2 particle to relativistic quantum electrodynamics. As a preliminary, I first consider the nonrelativistic domain of both free scalar and free Dirac  quantum field theory; not surprisingly, both theories return the nonrelativistic quantum mechanics of free particles - that is, nonrelativistic quantum mechanics without any potential terms in the Hamiltonian.  I then provide a template for the DS reduction of the nonrelativistic quantum mechanics model of two spin-1/2 charges in a Coulomb potential to the quantum Lorentz model of QED discussed in the previous chapter; I also provide a template for the nonrelativistic quantum mechanics model of a single charge in a background electromagnetic field. Following this, I consider the relation between Bell's Bohmian model of a spin-1/2 particle and the minimalist Bohmian model of QED, and the extent to which a reduction between the two can be effected.

Not much work on the nonrelativistic domain of QFT has been done. What work there is takes an approach completely different from the one taken here, focusing on the nonrelativistic approximation to S-matrix elements (for example, see Peskin and Schroeder \cite{peskin1996introduction}, Colin \cite{ColinNR}, and Beg and Furlong \cite{beg1985lambdavarphi4}). Thus, all of my work here, which takes a completely different approach based on the Schrodinger picture of QFT, is original as far as I am aware, unless explicitly stated otherwise.

\section{The Nonrelativistic Domain of Free Quantum Field Theory}

In this section, I will demonstrate the reduction of two models of nonrelativistic quantum mechanics, the free Schrodinger equation for spinless particles, and the free Pauli equation for particles of spin-1/2, from free Klein-Gordon quantum field theory and free Dirac quantum field theory, respectively.

\subsubsection{Free Scalar Field Theory}

Recall from Chapter 3 that a general state $| \Psi \rangle$ in scalar quantum field theory can be expressed in the form

\begin{equation} \label{KGState}
| \Psi \rangle = \psi_{0} | 0 \rangle  + \sum_{n=1}^{\infty} \int d^{3}k_{1} ... d^{3}k_{n}  \ \tilde{\psi}_{n}(k_{1},...,k_{n}) \ | k_{1},...,k_{n} \rangle 
\end{equation}

\noindent  where $\psi_{0} \equiv \langle 0 | \Psi \rangle $  and $\tilde{\psi}_{n}(k_{1},...,k_{n}) \equiv  \langle k_{1},...,k_{n} | \Psi \rangle$. The Schrodinger equation for free Klein-Gordon QFT,

\begin{equation}
i \frac{\partial }{\partial t} | \Psi \rangle = \hat{H}_{KG}  | \Psi \rangle,
\end{equation}

\noindent entails 

\begin{equation}
\begin{split}
& i \frac{\partial }{\partial t} \psi_{0} = 0 \\
& i \frac{\partial }{\partial t} \tilde{\psi}_{n}(k_{1},...,k_{n},t)  = \left( \sqrt{|k_{1}|^2+m^{2}}+ ... + \sqrt{|k_{n}|^2+m^{2}} \right) \tilde{\psi}_{n}(k_{1},...,k_{n},t). 
\end{split}
\end{equation}

\noindent If we now restrict the state $|\Psi \rangle$ to be a superposition only of states of momentum of magnitude much less than the mass $m$ - that is, if we restrict  $\tilde{\psi}_{m} (k_{1}, ... , k_{m}) = 0 \ \text{for} \ k_{i} << m  \ \forall i$ - then we may make the approximation $\sqrt{|k_{i}|^2+m^{2}} \approx m + \frac{1}{2m} k_{i}^{2}$. It is this approximation that ultimately accounts for the emergence of nonrelativistic behavior. Under this approximation, the Schrodinger equation for the momentum space wave functions take the form

\begin{equation}
\begin{split}
& i \frac{\partial }{\partial t} \psi_{0} = 0 \\
& i \frac{\partial }{\partial t} \tilde{\psi}_{n}(k_{1},...,k_{n},t)  \approx \left( nm + \frac{1}{2m} k_{1}^{2}+ ... +   \frac{1}{2m} k_{n}^{2}   \right) \tilde{\psi}_{n}(k_{1},...,k_{n},t). 
\end{split}
\end{equation}

\noindent Returning to the position representation $\psi_{n}(x_{1},...,x_{n},t) \equiv \langle 0| \hat{\phi}(x_{1}) ... \hat{\phi}(x_{n})| \Psi \rangle$ the Schrodinger equation takes the form

\begin{equation}
i \frac{\partial}{\partial t} \psi_{n}(x_{1},...,x_{n},t) \approx  (  n m  - \frac{1}{2m} \nabla_{1}^{2} - ... - \frac{1}{2m} \nabla_{n}^{2} )\psi_{n}(x_{1},...,x_{n},t).
\end{equation} 

\noindent This equation predicts the same amplitudes (up to an overall phase) as the ordinary Schrodinger equation,

\begin{equation}
i \frac{\partial}{\partial t} \psi_{n}(x_{1},...,x_{n},t) \approx  \left( - \frac{1}{2m} \nabla_{1}^{2} - ... - \frac{1}{2m} \nabla_{n}^{2} \right) \psi_{n}(x_{1},...,x_{n},t),
\end{equation}

\noindent thus accounting for the fact that the latter successfuly approximates nonrelativistic free systems.

\subsubsection{Free Dirac Field Theory} \label{FreeDiracNR}

Recall from Chapter 3 that a general state $| \Psi \rangle$ in free Dirac quantum field theory can be expressed in the form

\begin{equation}
\begin{split}
| \Psi \rangle = & \ \ \psi_{0} | 0_{D} \rangle \ \\
& + \ \sum_{n=1}^{\infty} \sum_{r_{1},...,r_{n}} \ \int d^{3} k_{1} ... d^{3} k_{n} \ \tilde{\psi}_{n,0}^{r_{1},...,r_{n}} (k_{1}, ... , k_{n})     \ \hat{b}^{\dagger,r_{n}}_{k_{n}} ... \hat{b}^{\dagger,r_{1}}_{k_{1}} |0_{D} \rangle \\
 & + \ \sum_{l=1}^{\infty} \sum_{s_{1},...,s_{l}}  \ \int d^{3} p_{1} ... d^{3} p_{l}  \ \tilde{\psi}_{0,l}^{s_{1},...,s_{l}} (p_{1}, ... , p_{l})    \  \hat{c}^{\dagger,s_{l}}_{p_{l}} ... \hat{c}^{\dagger,s_{1}}_{p_{1}} |0_{D} \rangle \  \\ 
&+ \sum_{n,l=1}^{\infty} \sum_{\substack{r_{1},...,r_{n},\\ s_{1},...,s_{l}}} \ \int d^{3} k_{1} ... d^{3} k_{n} \ d^{3} p_{1} ... d^{3} p_{l}  \ \tilde{\psi}_{n,l}^{\substack{r_{1},...,r_{n},\\s_{1},...,s_{l}}} (k_{1}, ... , k_{n}; p_{1}, ... , p_{l}) \hat{c}^{\dagger,s_{l}}_{p_{l}} ... \hat{c}^{\dagger,s_{1}}_{p_{1}}  \hat{b}^{\dagger,r_{n}}_{k_{n}} ... \hat{b}^{\dagger,r_{1}}_{k_{1}} |0_{D} \rangle, 
\end{split}
\end{equation}

\noindent where $\tilde{\psi}_{n,l}^{\substack{r_{1},...,r_{n},\\ s_{1},...,s_{l}}} (k_{1}, ... , k_{n}; p_{1}, ... , p_{l}) \equiv\langle 0 | \hat{b}^{r_{1}}_{k_{1}} ... \hat{b}^{r_{n}}_{k_{n}}  \hat{c}^{s_{1}}_{p_{1}} ... \hat{c}^{s_{l}}_{p_{l}}  | \Psi >$ and likewise for the $(n,0)$, $(0,l)$ and $(0,0)$ coefficients. The Schrodinger equation for the free fermionic field,

\begin{equation}
i \frac{\partial }{\partial t} | \Psi \rangle = \hat{H}_{D}  | \Psi \rangle,
\end{equation}

\noindent entails,

\begin{equation}
\footnotesize
\begin{split}
&  i \frac{\partial}{\partial t} \psi_{0} = 0  \\
&   i \frac{\partial}{\partial t}\tilde{\psi}_{n,0}^{r_{1},...,r_{n}} (k_{1}, ... , k_{n},t) =     \left( \sqrt{|k_{1}|^{2} + m^{2}} + ... +  \sqrt{|k_{n}|^{2} + m^{2}} \right) \tilde{\psi}_{n,0}^{r_{1},...,r_{n}} (k_{1}, ... , k_{n},t)  \\
&   i \frac{\partial}{\partial t}\tilde{\psi}_{0,l}^{s_{1},...,s_{l}} (s_{1}, ... , s_{l},t) =     \left( \sqrt{|p_{1}|^{2} + m^{2}} + ... +  \sqrt{|p_{l}|^{2} + m^{2}} \right) \tilde{\psi}_{0,l}^{s_{1},...,s_{l}} (p_{1}, ... , p_{l},t)  \\
& \begin{split}
i \frac{\partial}{\partial t}\tilde{\psi}_{n,l}^{\substack{r_{1},...,r_{n},\\ s_{1},...,s_{l}}} (k_{1}, ... , k_{n}; p_{1}, ... , p_{l},t) = &    \bigg( \sqrt{|k_{1}|^{2} + m^{2}} + ... +  \sqrt{|k_{n}|^{2} + m^{2}} \\ 
& + \sqrt{|p_{1}|^{2} + m^{2}} + ... +  \sqrt{|p_{l}|^{2} + m^2}  \bigg) \  \tilde{\psi}_{n,l}^{\substack{r_{1},...,r_{n}, \\ s_{1},...,s_{l}}} (k_{1}, ... , k_{n}; p_{1}, ... , p_{l},t)  
\end{split}
\end{split}
\end{equation}

\noindent Specialising to the domain of $n$-electron states, and assuming $|\Psi \rangle$ is a superposition only of states with momentum much less than $m$, the Schrodinger equaiton for the $n$-electron momentum space wave function takes the form

\begin{equation} \label{SchrodDiracMomentum}
i \frac{\partial}{\partial t}\tilde{\psi}_{n,0}^{r_{1},...,r_{n}} (k_{1}, ... , k_{n},t) \approx    \left( nm + \frac{1}{2m} k_{1}^{2}+ ... +   \frac{1}{2m} k_{n}^{2}   \right) \tilde{\psi}_{n,0}^{r_{1},...,r_{n}} (k_{1}, ... , k_{n},t).  
\end{equation}

\noindent Fourier transforming back to the position representation, we have

\begin{equation} 
\begin{split}
&\psi_{n,0}^{a_{1},...,a_{n}}(x_{1}, ... , x_{n}) =  \sum_{r_{1},...,r_{n}} \int \frac{d^{3} k_{1}}{(2 \pi)^{3}} ... \frac{d^{3} k_{n}}{(2 \pi)^{3}} \ \frac{1}{\sqrt{2 E_{k_{1}}}} ... \frac{1}{\sqrt{2 E_{k_{n}}}} \\
& u_{r_{1}}^{a_{1}}(k_{1}) ... u_{r_{n}}^{a_{n}}(k_{n})  \tilde{\psi}_{n,0}^{r_{1},...,r_{n}} (k_{1}, ... , k_{n}) e^{-ik_{1} \cdot x_{1}} ... e^{-i k_{n} \cdot x_{ n}}.
\end{split}
\end{equation}

\noindent Noting that for momenta $k$ such that $\frac{k}{m} << 1$, the basis 4-spinors $u^{r_{i}}(k_{i})$ approximately take the form

\begin{tabular}{p{8cm}p{8cm}}

{\begin{align}
u_{1}(k) & \approx   
\left( \begin{array}{c}
 1\\
0 \\
0 \\
0
\end{array} \right) 
\end{align}}

& 

{\begin{align}
u_{2}(k) & \approx  
\left( \begin{array}{c}
0\\
1 \\
0 \\
0
\end{array} \right) 
\end{align}}

\end{tabular}

\noindent Because the lower two components of the basis spinors $u_{1}(k)$ and  $u_{2}(k)$ can be neglected in the nonrelativistic approximation, we may in this approximation deal exclusively with 2-spinors consisting of the upper two components $\phi_{r}^{\alpha} $:

\begin{tabular}{p{8cm}p{8cm}}

{\begin{align}
\phi_{1} & =   
\left( \begin{array}{c}
 1\\
0 
\end{array} \right) 
\end{align}}

& 

{\begin{align}
\phi_{2} & =  
\left( \begin{array}{c}
0\\
1
\end{array} \right) .
\end{align}}

\end{tabular}

\noindent Note that in the nonrelativistic approximation, the momentum dependence of the basis spinors disappears. In this approximation, the spinor wave function can be approximated

\begin{equation} \label{SpinorWfcn}
\begin{split}
\psi_{n,0}^{\alpha_{1},...,\alpha_{n}}(x_{1}, ... , x_{n}) =  & \sum_{r_{1},...,r_{n} = 1}^{2} \int \frac{d^{3} k_{1}}{(2 \pi)^{3}} ... \frac{d^{3} k_{n}}{(2 \pi)^{3}} \ \frac{1}{\sqrt{2 E_{k_{1}}}} ... \frac{1}{\sqrt{2 E_{k_{n}}}} \\
& \phi_{r_{1}}^{\alpha_{1}} ... \phi_{r_{n}}^{\alpha_{n}}  \tilde{\psi}_{n,0}^{r_{1},...,r_{n}} (k_{1}, ... , k_{n}) e^{-ik_{1} \cdot x_{1}} ... e^{-i k_{n} \cdot x_{ n}}.
\end{split}
\end{equation}

\noindent Translating the momentum space Schrodinger equation (\ref{SchrodDiracMomentum}) to the position representation, we have in the nonrelativistic approximation

\begin{equation}
i \frac{\partial }{\partial t} \psi_{n,0}^{\alpha_{1},...,\alpha_{n}}(x_{1}, ... , x_{n})   \approx \left( - \frac{1}{2m} \nabla_{1}^{2} - ... - \frac{1}{2m} \nabla_{n}^{2} \right) \psi_{n,0}^{\alpha_{1},...,\alpha_{n}}(x_{1}, ... , x_{n}) 
\end{equation}

\subsubsection{Summary}

Thus far, my analysis has shown how the kinetic part of a nonrelativistic Schrodinger equation approximates the dynamics of particular domain of free quantum field theory - namely, low-momentum n-particle states. In the next section, I consider the extension to models of nonrelativistic quantum mechanics whose Hamiltonians include potential terms.

\section{The Nonrelativistic Domain of QED}

In this section I will provide templates for the reduction of two models of nonrelativistic quantum mechanics to the Lorentz model of QED: First, the interaction of two nonrelativistic electrons via a Coulomb potential, and second, the Pauli equation for a single electron in a background electromagnetic field. 

\subsection{DS Reduction of the Nonrelativistic Coulomb Model to QED}

The model of two nonrelativistic electrons interacting via a Coulomb potential takes as its state space the Hilbert space of two nonrelativistic spin-1/2 particles:

\begin{equation}
S = \mathcal{H}^{NR}_{2e}.
\end{equation}

\noindent The dynamics of this model are determined by the Schrodinger equation

\begin{equation}
i \frac{\partial }{\partial t } | \psi \rangle  = \hat{H}^{NR}_{C} | \psi \rangle,
\end{equation}

\noindent  where $| \psi \rangle \in  \mathcal{H}^{NR}_{2e}$ and

\begin{equation}
 \hat{H}^{NR}_{C} = \frac{\hat{p}_{1}^{2}}{2m} + \frac{\hat{p}_{2}^{2}}{2m} + \frac{e^{2}}{4 \pi} \frac{1}{|\hat{x}_{1} - \hat{x}_{2} |}.
\end{equation}

\noindent The low level model to which I reduce this model is the Lorentz model of QED, in which the state of the electromagnetic field is the vaccuum state $| 0_{EM}^{r} \rangle$ of the free renormalised electromagnetic field Hamiltonian, which is simply the coherent state of the electromagnetic field centered around the field configuration $\vec{A}(x) = 0$ and the field momentum configuration $\vec{E}(x) = 0$. In this case, the Lorentz model for two electrons simplifies to 

\begin{equation}
\small
\begin{split}
 i\frac{\partial}{\partial t} \psi_{2,0}^{a_{1} a_{2}}(x_{1},x_{2}) & =   \bigg\{  \big[ \vec{\alpha} \cdot(  -i \vec{\nabla}_{1}  )  + \beta m_{r}   \big]^{a_{1} c_{1}}  \delta^{a_{2}c_{2}} \\
 & +  \delta^{a_{1}c_{1}}  \big[ \vec{\alpha} \cdot(  -i \vec{\nabla}_{2} )  + \beta m_{r}   \big]^{a_{2} c_{2}} + \frac{e_{r}^{2}}{4 \pi}  \frac{1}{ \left| x_{1} - x_{2} \right|}   \bigg\}  \psi_{2,0}^{c_{1} c_{2}}(x_{1},x_{2})
\end{split}
\end{equation}

\noindent In section \ref{FreeDiracNR}, I showed that in the nonrelativistic domain, in which we restrict the wave function $ \psi_{2,0}^{c_{1} c_{2}}(x_{1},x_{2})$ to contain only momenta $k_{1}$ and $k_{2}$ such that $\frac{k_{1}}{m}<<1$ and $\frac{k_{2}}{m}<<1$, the following approximation holds 

\begin{equation}
\begin{split}
& \bigg\{ \big[ \vec{\alpha} \cdot(  -i \vec{\nabla}_{1}  )  + \beta m_{r}   \big]^{a_{1} c_{1}}  \delta^{a_{2}c_{2}}  \bigg\}  \psi_{2,0}^{c_{1} c_{2}, \lambda}(x_{1},x_{2}) \\
& \approx   - \frac{1}{2m_{r}} \nabla_{1}^{2} \  \psi_{2,0}^{a_{1} a_{2}, \lambda}(x_{1},x_{2}).
\end{split}
\end{equation}

\noindent Because the $a=3,4$ components of the basis spinors $u^{a}_{1}(k)$ and $u^{a}_{2}(k) $ are effectively zero for low momenta, we can neglect them and focus on the upper two components $\alpha = 1,2$. 

\begin{equation}
\begin{split}
& \bigg\{ \big[ \vec{\alpha} \cdot(  -i \vec{\nabla}_{1}  )  + \beta m_{r}   \big]^{\alpha_{1} c_{1}}  \delta^{\alpha_{2}c_{2}}  \bigg\}  \psi_{2,0}^{c_{1} c_{2}, \lambda}(x_{1},x_{2}) \\
& \approx   - \frac{1}{2m_{r}} \nabla_{1}^{2} \  \psi_{2,0}^{\alpha_{1} \alpha_{2}, \lambda}(x_{1},x_{2}).
\end{split}
\end{equation}

\noindent Likewise approximating

\begin{equation}
\begin{split}
& \bigg\{   \delta^{\alpha_{1}c_{1}}  \big[ \vec{\alpha} \cdot(  -i \vec{\nabla}_{2}  )  + \beta m_{r}   \big]^{\alpha_{2} c_{2}}  \bigg\}  \psi_{2,0}^{c_{1} c_{2}, \lambda}(x_{1},x_{2}) \\
& \approx   - \frac{1}{2m_{r}} \nabla_{2}^{2} \  \psi_{2,0}^{\alpha_{1} \alpha_{2}, \lambda}(x_{1},x_{2}).
\end{split}
\end{equation}

\noindent we obtain the relation 

\begin{equation}
\small
\begin{split}
 i\frac{\partial}{\partial t} \psi_{2,0}^{\alpha_{1} \alpha_{2}}(x_{1},x_{2})  \approx   \bigg\{     - \frac{1}{2m_{r}} \nabla_{1}^{2}       - \frac{1}{2m_{r}} \nabla_{2}^{2}  + \frac{e_{r}^{2}}{4 \pi}  \frac{1}{ \left| x_{1} - x_{2} \right|}   \bigg\}  \psi_{2,0}^{\alpha_{1} \alpha_{2}}(x_{1},x_{2}),
\end{split}
\end{equation}

\noindent which we recognise as the Schrodinger equation for two nonrelativistic spin-1/2 particles. If we further restrict to the domain of states in which the spin and position degrees of freedom of the two electrons are unentangled, so that $\psi_{2,0}^{\alpha_{1} \alpha_{2}}(x_{1},x_{2}) = \psi(x_{1},x_{2}) s^{\alpha_{1},\alpha_{2}}$, where $s^{\alpha_{1},\alpha_{2}}$ is a $2 \times 2$ matrix independent of position, then we retrieve the more conventional-looking equation,

\begin{equation}
\small
\begin{split}
 i\frac{\partial}{\partial t} \psi(x_{1},x_{2},t)  \approx   \bigg\{     - \frac{1}{2m_{r}} \nabla_{1}^{2}       - \frac{1}{2m_{r}} \nabla_{2}^{2}  + \frac{e_{r}^{2}}{4 \pi}  \frac{1}{ \left| x_{1} - x_{2} \right|}   \bigg\}  \psi(x_{1},x_{2},t),
\end{split}
\end{equation}

\noindent which is simply the Schrodinger equation for two electrons, without regard to spin. 

\vspace{5mm}

I now frame these results within the context of DS reduction. 

The bridge map connecting the state spaces of the Lorentz model of QED and the nonrelativistic Coulomb model, $B_{L}^{C}$, is simply to project the fermionic wave function onto the first two spinor components $\alpha = 1,2$, using the projection operator $P_{a}^{ \alpha}$:

\

\noindent \underbar{\textit{Bridge Map:}}

\begin{equation}
\begin{split}
& B_{L}^{C}: \mathcal{H}_{2e} \longrightarrow \mathcal{H}_{2e}^{NR} \\
& B_{L}^{C}:  \psi_{2,0}^{a_{1} a_{2}}(x_{1},x_{2})  \longmapsto P_{a_{1}}^{\alpha_{1}} P_{a_{2}}^{\alpha_{2}}  \psi_{2,0}^{a_{1} a_{2}}(x_{1},x_{2}) .
\end{split}
\end{equation}

\noindent The domain of the 2-particle NR Coulomb model in the Lorentz model of QED consists of 2-particle states with momenta below the nonrelativistic cutoff: 

\

\noindent \underbar{\textit{Domain:}}

\

\begin{equation}
d_{NRC} = \left\{ | \Psi \rangle \in \mathcal{H}_{D} \ | \  | \Psi \rangle =  \int d^{3} x_{1} d^{3} x_{2} \  \psi_{2,0}^{a_{1} a_{2}, \lambda}(x_{1},x_{2}) \hat{\psi}^{a_{1}, \lambda}(x_{1})\hat{\psi}^{a_{2}, \lambda}(x_{2}) | 0_{D}^{r} \rangle, \ \text{with} \ \lambda << m_{r}  \right\}, 
\end{equation}

\noindent with $ \hat{\psi}^{a_{i}, \lambda}(x_{i})$ is a fermionic field operator containing Fourier components only up to momentum $\lambda$. The laws of the image theory, which also constitute the requirement for DS reduction, are 

\

\noindent \underbar{\textit{Image Model:}}

\

\begin{equation}
 i\frac{\partial}{\partial t} \left[ P_{a_{1}}^{\alpha_{1}} P_{a_{2}}^{\alpha_{2}}  \psi_{2,0}^{a_{1} a_{2}}(x_{1},x_{2},t) \right] \approx   \bigg\{     - \frac{1}{2m_{r}} \nabla_{1}^{2}       - \frac{1}{2m_{r}} \nabla_{2}^{2}  + \frac{e_{r}^{2}}{4 \pi}  \frac{1}{ \left| x_{1} - x_{2} \right|}   \bigg\}     \left[ P_{a_{1}}^{\alpha_{1}} P_{a_{2}}^{\alpha_{2}}  \psi_{2,0}^{a_{1} a_{2}}(x_{1},x_{2},t) \right]
\end{equation}

\noindent as proven above. The laws of the analogue theory are then straightforwardly obtained through the bridge rule substitution

\

\noindent \underbar{\textit{Bridge Rules:}}

\

\begin{equation}
 \psi^{'\alpha_{1} \alpha_{2}}(x_{1},x_{2},t) \equiv P_{a_{1}}^{\alpha_{1}} P_{a_{2}}^{\alpha_{2}}  \psi_{2,0}^{a_{1} a_{2}}(x_{1},x_{2},t)
\end{equation}

\noindent Explicitly, the laws of the analogue theory are

\

\noindent \underbar{\textit{Analogue Model:}}

\

\begin{equation}
 i\frac{\partial}{\partial t}  \psi^{' \alpha_{1} \alpha_{2}}(x_{1},x_{2},t)  \approx   \bigg\{     - \frac{1}{2m_{r}} \nabla_{1}^{2}       - \frac{1}{2m_{r}} \nabla_{2}^{2}  + \frac{e_{r}^{2}}{4 \pi}  \frac{1}{ \left| x_{1} - x_{2} \right|}   \bigg\}      \psi^{' \alpha_{1} \alpha_{2}}(x_{1},x_{2},t). 
\end{equation}

\noindent The `strong analogy' condition, 

\

\noindent \underbar{\textit{`Strong Analogy':}}

\

\begin{equation} \label{MMStrongAnalogy}
\left|    \psi^{\alpha_{1} \alpha_{2}}(x_{1},x_{2},t)  - \psi^{'\alpha_{1} \alpha_{2}}(x_{1},x_{2},t) \right|_{\mathcal{H}_{2e}^{NR}} < \delta 
\end{equation}

\noindent is then satisfied for all times for which the state continues to contain only nonrelativistic momenta.

\subsection{DS Reduction of Nonrelativistic Pauli Model to QED}

The state space of the Paul Theory is the Hilbert space of a single nonrelativistic electron:

\begin{equation}
S_{h} = \mathcal{H}^{NR}_{1e}.
\end{equation}

\noindent The dynamics of this model are determined by the Schrodinger equation

\begin{equation}
i \frac{\partial }{\partial t } | \psi \rangle  = \hat{H}^{NR}_{P} | \psi \rangle,
\end{equation}

\noindent  where $| \psi \rangle \in  \mathcal{H}^{NR}_{1e}$ and

\begin{equation}
 \hat{H}^{NR}_{P} =  \frac{1}{2m} \left[ \vec{\sigma} \cdot  \left( \hat{\vec{p}} + e \vec{A}(x)  \right) \right]^{2}.
\end{equation}

\noindent Note that because there is only one particle in this model, the Coulomb term of the QED Lorentz model does not appear. The model to which I reduce has as its state space the 1-electron subspace of the Lorentz model,

\begin{equation}
S_{l} = \mathcal{H}_{1e};
\end{equation}

\noindent and for its dynamics, the QED Lorentz model dynamics restricted to the 1-electron subspace

\begin{equation} \label{DiracEqn}
\small
\begin{split}
 i\frac{\partial}{\partial t} \psi_{1,0}^{a}(x) =   \big[ \vec{\alpha} \cdot(  -i \vec{\nabla} + e_{r} \boldsymbol{\vec{A}}(x) )  + \beta m_{r}   \big]^{a c}   \psi_{1,0}^{c}(x). 
\end{split}
\end{equation}

\noindent Write $\psi_{1,0}^{c}(x)$ in terms of two 2-spinors $ \phi(x)$ and
$\chi(x)$:

\begin{align} \label{DiracSplit}
\psi_{1,0}^{c}(x)  & =   
\left( \begin{array}{c}
 \phi(x)\\
\chi(x)
\end{array} \right).
\end{align}

\noindent Employing the Dirac representation of the matrices $\alpha_{i}$ and $\beta$,

\[ \alpha_{i} = \left( \begin{array}{ccc}
0 & \sigma_{i} \\
\sigma_{i} & 0 
 \end{array} \right).\]

\[ \beta = \left( \begin{array}{ccc}
1 & 0 \\
0 & -1
 \end{array} \right),\]
 
 \noindent we can write the equation (\ref{DiracEqn}), in terms of the 2-spinors as the two coupled equations,
 
 \begin{equation}
 i \frac{\partial }{\partial t} \phi(x) = \vec{\sigma} \cdot \left( -i \nabla + e_{r} \vec{A}(x)    \right) \chi(x)   + m_{r} \phi(x)
 \end{equation}

 \begin{equation}
 i \frac{\partial }{\partial t} \chi(x) = \vec{\sigma} \cdot \left( -i \nabla + e_{r} \vec{A}(x)    \right) \phi(x)   - m_{r} \chi(x)
 \end{equation}
 
 \noindent In momentum space, these equations read,
 
  \begin{equation}
 E \tilde{\phi}(k) = \vec{\sigma} \cdot \left( - \vec{k} + e_{r} \vec{A}(-k)    \right) \tilde{\chi}(k)   + m_{r} \tilde{\phi}(k)
 \end{equation}

 \begin{equation}
 E \tilde{\chi}(k) = \vec{\sigma} \cdot \left( -\vec{k} + e_{r} \vec{A}(-k)    \right) \tilde{\phi}(k)   - m_{r} \tilde{\chi}(k).
 \end{equation}
 
 \noindent In the nonrelativistic approximation, where $\phi(x)$ and $\chi(x)$ contain only momenta $k<< m_{r}$, $E \approx m_{r}$, the second of these equations gives 
 
 \begin{equation}
\tilde{\chi}(k) \approx \frac{1}{2 m_{r}}  \vec{\sigma} \cdot \left( -\vec{k} + e_{r} \vec{A}(-k)    \right) \tilde{\phi}(k). 
 \end{equation}

\noindent Equations (\ref{DiracSplit}) to (\ref{NRPauli}) follow, with variations, a well-known derivation of the Pauli equation from the Dirac equation that can be found in many texts and lectures on relativistic quantum mechanics \cite{DineNRDirac}. Substituting this relation into the first of the 2-spinor equations, we find 

 \begin{equation}
 E \tilde{\phi}(k) \approx  \frac{1}{2 m_{r}} \vec{\sigma} \cdot \left( - \vec{k} + e_{r} \vec{A}(-k)    \right)^{2}  \tilde{\phi}(k)  + m_{r} \tilde{\phi}(k).
 \end{equation}

\noindent Transforming back to position space, we have

 \begin{equation}
i \frac{\partial }{\partial t} \phi(x,t)  \approx    \left[ m_{r} +  \frac{1}{2 m_{r}} \vec{\sigma} \cdot \left( -i \nabla  + e_{r} \vec{A}(x)   \right)^{2} \right] \phi(x,t).
 \end{equation}

\noindent We can ignore the term $m_{r} $ corresponding to the rest energy since this just contributes the overall phase of the spinor $\phi(x,t)$. This yields the Pauli equation: 

 \begin{equation} \label{NRPauli}
i \frac{\partial }{\partial t} \phi^{\alpha}(x,t)  \approx    \frac{1}{2 m_{r}} \left\{  \left[ \vec{\sigma} \cdot \left( -i \nabla  + e_{r} \vec{A}(x)   \right)\right]^{2}  \right\}^{\alpha \beta}  \phi^{\beta}(x,t).
 \end{equation}

\noindent where the 2-spinor indices have been made explicit. 

\vspace{5mm}

I now frame these results within the context of DS reduction. 

The bridge map connecting the state spaces of the Lorentz model of QED and of the 1-particle nonrelativistic Pauli model, $B_{L}^{P}$, is simply to project the fermionic 1-particle wave function onto the first two spinor components $\alpha = 1,2$, using the projection operator $P_{a}^{ \alpha}$:

\

\noindent \underbar{\textit{Bridge Map:}}

\begin{equation}
\begin{split}
& B_{L}^{C}: \mathcal{H}_{1e} \longrightarrow \mathcal{H}_{1e}^{NR} \\
& B_{L}^{C}:  \psi_{1,0}^{a_{1}}(x_{1})  \longmapsto P_{a_{1}}^{\alpha_{1}}  \psi_{1,0}^{a_{1}}(x_{1}) .
\end{split}
\end{equation}

\noindent The domain of the 1-particle NR Pauli model in the QED Lorentz model state space consists of 1-particle states with momenta below the nonrelativistic cutoff: 

\

\noindent \underbar{\textit{Domain:}}

\

\begin{equation}
d_{NRP} = \left\{ | \Psi \rangle \in \mathcal{H}_{D} \ \big| \  | \Psi \rangle =  \int d^{3} x_{1}  \  \psi_{1,0}^{a_{1}, \lambda}(x_{1}) \hat{\psi}^{a_{1}, \lambda}(x_{1}) | 0_{D}^{r} \rangle, \ with \ \lambda << m_{r}  \right\}, 
\end{equation}

\noindent where $ \hat{\psi}^{a_{1}, \lambda}(x_{1})$ is a fermionic field operator containing Fourier components only up to momentum $\lambda$. The laws of the image theory are 

\

\noindent \underbar{\textit{Image Model:}}

\

\begin{equation}
 i\frac{\partial}{\partial t} P_{a_{1}}^{\alpha_{1}}  \psi_{1,0}^{a_{1}}(x_{1},t)  \approx     \frac{1}{2 m_{r}} \left\{  \left[ \vec{\sigma} \cdot \left( -i \nabla  + e_{r} \vec{A}(x)   \right)\right]^{2}  \right\}^{\alpha_{1} \beta_{1}}     P_{a_{1}}^{\beta_{1}}   \psi_{1,0}^{a_{1} }(x_{1},t)
\end{equation}

\noindent as proven above. The laws of the analogue theory are then straightforwardly obtained through the bridge rule substitution

\

\noindent \underbar{\textit{Bridge Rules:}}

\

\begin{equation}
 \psi^{'\alpha_{1}}(x_{1},t) \equiv P_{a_{1}}^{\alpha_{1}}  \psi_{1,0}^{a_{1}}(x_{1},t)
\end{equation}

\noindent Explicitly, the laws of the analogue theory are

\

\noindent \underbar{\textit{Analogue Model:}}

\

\begin{equation}
 i\frac{\partial}{\partial t}  \psi^{'\alpha_{1}}(x_{1},t)  \approx     \frac{1}{2 m_{r}} \left\{  \left[ \vec{\sigma} \cdot \left( -i \nabla  + e_{r} \vec{A}(x)   \right)\right]^{2}  \right\}^{\alpha_{1} \beta_{1}}     \psi^{'\beta_{1}}(x_{1},t)
\end{equation}

\noindent The `strong analogy' condition, 

\

\noindent \underbar{\textit{`Strong Analogy':}}

\

\begin{equation} \label{MMStrongAnalogy}
\left|    \psi^{\alpha_{1}}(x_{1},t)  - \psi^{'\alpha_{1}}(x_{1},t) \right|_{\mathcal{H}_{1e}^{NR}} < \delta 
\end{equation}

\noindent is then satisfied for all times for which the state continues to contain only nonrelativistic momenta.

\section{The Connection Between Bohmian NRQM and Bohmian QED}

In this section, I discuss the reduction of Bohmian nonrelativistic quantum mechanics of a spin-1/2 particle to the minimalist Bohmian model of QED. Note that in contrast to previous reductions concerning a Bohmian theory, in which the high level theory was classical, in the present case both theories are Bohmian. I briefly present Bell's model of a non-relativistic spin-1/2 system, and consider in what sense if any it is possible to effect a reduction of this model to the minimalist model of QED, especially given that in Bell's model, the beables correspond to the positions of fermionic particles and in the minimalist QED model, there are no beables corresponding to the fermionic field, only to the electromagnetic field. The answer that I arrive at is that while in the strictest sense, the electromagnetic beables of the minimalist model do not instantiate the evolution of the fermionic particle beables, in the nonrelativistic domain where both models are applicable,  the electromagnetic field beables do perform the same function as Bell's particle beables - namely, to select a branch of the total quantum state with probability given by the relevant Born Rule coefficient $|c_{i}|^{2}$.

\subsection{Bell's Model of Spin-1/2 Particles}

The state space of Bell's model of a nonrelativistic spin-1/2 particle is

\begin{equation}
S = \mathcal{H}_{1p}^{NR} \times \mathbb{Q}_{1p}
\end{equation}

\noindent where $\mathbb{Q}_{1p}$ is the configuration space of a single particle in 3-space. The dynamics of $| \psi \rangle \in \mathcal{H}_{1p}^{NR} $ are specified by the Pauli Schrodinger equation

\begin{equation}
i \frac{\partial }{\partial t } \psi^{\alpha}(x,t)  = \frac{1}{2m} \left\{ \left[ \vec{\sigma} \cdot  \left( -i\nabla + e \vec{A}(x)  \right) \right]^{2} \right\}^{\alpha \beta} \psi^{\beta}(x,t) 
\end{equation}

\noindent Expanding the wave function in polar form, $\psi^{\alpha}(x,t) = R^{\alpha}(x,t) e^{iS^{\alpha}(x,t)}$, the guidance equation for the configuration $q \in \mathbb{Q}_{1p}$ is

\begin{equation}
\dot{q} = \sum_{\alpha} \frac{1}{m} \left(  \nabla S_{\alpha}(x) - e \vec{A}(x) \right).
\end{equation}

\noindent These dynamics are equivariant with respect to the probability distribution,

\begin{equation}
\rho(x,t) \equiv \sum_{\alpha} \psi^{ * \alpha }(x,t) \psi^{ \alpha }(x,t).
\end{equation}

\noindent In this model there are no beables corresponding to the spin degrees of freedom \cite{bell2004speakable}. For other Bohmian models of nonrelativistic spin-1/2 particles, see Holland's \cite{holland1995quantum}. The generalisation to N-particles, and the inclusion of a Coulomb potential, is straightforward.

\subsection{Connecting Bell's Model and the Minimalist QED model}

In the Pauli equation, the electromagnetic field has no independent dynamics of its own, but is rather independently prescribed so there is no obvious role for electromagnetic field beables within this model. Yet there is a sense in which the transverse electromagnetic field configuration $\vec{a}^{T}(x,t)$ can serve the same functional role as Bell's particle beables $q(t)$, even if it is a completely different sort of mathematical object, and associated with bosonic rather than fermionic degrees of freedom. Consider the process of entanglement between a spin-1/2 particle and some external degrees of freedom, such as occurs in a measurement. As in Chapter 2, let the initial state $|\chi_{0} \rangle$ of $AB$ be a product state, so that $|\chi_{0} \rangle = |\psi_{0} \rangle \otimes |\phi_{0} \rangle$. Also, let $\{ |a_{i} \rangle \} $ be a (possibly overcomplete) basis $A$'s Hilbert space. Then, if $|\psi_{0} \rangle = \sum_{i} c_{i} |a_{i} \rangle$, the unitary evolution of the quantum state takes the following form

\begin{equation}
( \sum_{i} c_{i} |a_{i} \rangle)|\phi_{0} \rangle  \Longrightarrow  \sum_{i}  c_{i} |\theta_{i}\rangle |\phi_{i}\rangle
\end{equation}

\noindent where $ |\theta_{i} \rangle$ are abitrary states. Assuming that the environment consists both of electromagnetic and fermionic degrees of freedom, then as long as the states $|\phi_{i}\rangle$ are disjoint both with respect to the electromagnetic configuration space and with respect to the configuration space of the fermionic beables associated with environmental degrees of freedom, either type of beable will succeed in selecting a branch of the wave function with probability $|c_{i}|^{2}$. In this sense, the electromagnetic field beables may fill the functional role of Bell's particle beables in nonrelativistic contexts even though the electromagnetic field in models of nonrelativistic quantum mechanics is not associated with any set of independent degrees of freedom.

\vspace{5mm}

An alternative Bohmian model of QED that does attach beables to the fermionic degrees of freedom, such as Colin's Dirac Sea model, may succeed in filling not only the functional role of the beables in the Bell model, but in replicating (approximately) the detailed motions of these beables in nonrelativistic contexts. However, assuming that both Bell's model and the minimalist QED model can be regarded as empirically adequate in their domains, and that the domain of QED encompasses the domain of nonrelativistic quantum mechanics, these facts together suggest that the particle beables of Bell's model are expendable in the process of extending to the more encompassing domain of QED, and that they do not play an indispensable role in the empirical success of the nonrelativistic Bohmian model. What it suggests instead is that the part of the model that does the essential predictive work, and which is most strongly corroborated by the model's empirical success, is the structure associated with the quantum state.

\subsection{Metaphysical Considerations Regarding the Adequacy of Minimalist Models}

I have outlined, in a schematic way, how the electromagnetic field beables may serve to fill the functional role of the fermionic particle beables in nonrelativistic pilot wave models - namely, by selecting one of the decoherence-defined branches of the total quantum state - and thereby to ensure the empirical adequacy of the minimalist model in the particular domain that is well-described by the nonrelativistic  models. However, the legitimacy of the minimalist QED model can be challenged on the basis of at least two distinct metaphysical arguments. First, as Tim Maudlin would argue, it is not sufficient that the electromagnetic field beables solely to fulfill the functional role of the particle beables of the non-relativistic model; even if these beables do succeed in selecting a branch of the quantum state, electromagnetic field beables, unlike fermionic particle beables, are simply not the `sort of stuff' out of which it is possible to reconstruct the world of our experience. Second, and unlike Maudlin (who has written extensively in defense of pilot wave approaches to quantum theory - though only ones employing the `right sort' of beable), Everettians like Brown and Wallace would reject the minimalist model on the grounds that electromagnetic field beables, like all other beables in all other non-collapse pilot wave models, are epiphenomenal and therefore superfluous; it simply does not matter how they behave because the quantum state, they claim, is sufficient to recover the world of our experience (as well as the experience of inahbitants of other branches of the quantum state). 

Before delving into these metaphysical worries about minimalist models, let us first consider in further detail Struyve and Westman's account of how this model is supposed to reproduce the ordinary quantum predictions. I quote them directly:

\begin{quote}
\begin{singlespace}
While there are no variables representing matter, the wave functional which guides the field $\vec{A}^{T}(x)$, still contains the fermionic degrees of freedom. As such, the field configuration $\vec{A}^{T}(x)$ will in certain cases behave as if there was matter present. For example, it might look like radiation that has been scattered off some matter distribution, or like thermal radiation emitted by such a distribution. In this way, it was argued that the model is empirically adequate, because there will be an image of macroscopic matter distributions in the radiation field. Nevertheless such a model seem rather far removed from our everyday experience of the world and probably takes minimalism too far. \cite{StruyvePWQFT} 
\end{singlespace}
\end{quote}

\noindent As we can see from this last line, Struyve himself has doubts about the plausibility of the minimalist  model as a serious candidate for describing physical reality - though to read between the lines, it seems that his reasons are rooted less in doubts about the model's empirical adequacy than they are in deeper metaphysical worries that are more intrinsic to the model itself, akin perhaps to those that Maudlin raises. Nevertheless, the essential idea of Struyve and Westman's model is that the electromagnetic field beables behave in a sense \textit{as if} there were fermionic matter, thus giving the appearance of such matter even though no beables need be associated with the fermionic degrees of freedom in these models. After all, our experience of a projectile moving through the air is not a direct interaction with the fermionic matter making up the projectile, but is invariably mediated by the radiation that is scattered off of it and that enters our retina, or by the electromagnetic forces of repulsion between ourselves and the projectile, say when we catch it. As far as our experience of the projectile is concerned, the fermions constituting the projectile play no direct role and are, at least from a certain point of view, dispensable. There is no need for them because the electromagnetic degrees of freedom that more directly determine our experience of the projectile behave as if the fermions were there. 

In his paper, `Why Bohm's Theory Solve's the Measurement Problem,' Maudlin writes of the nonrelativistic pilot wave theory that `the particle positions are the heart of the theory, they specify the world as we know it. Further, without the \textit{the effective wave function cannot be defined}... .' That is, Maudlin requires that the beables be such that we may find an image of the world in them. One benefit of particle beables such as the ones encountered in nonrelativistic pilot wave theories is that it is relatively straightforward to see how one would find an image of the world in them. The cats that we see are cat-shaped bunches of particles, tables table-shaped bunches of particles, and the reason that we are able to see them is that the configurations of the particles in our brains become correlated through the dynamics to the configurations of these objects (much as would be the case in a classical mechanical account). As Maudlin writes, `If we want to know what happened to the measuring device (e.g. which way the pointer went), we look at it, thereby correlating positions of particles in our brains with the pointer position. If getting the state of our brain correlated with previously unknown external conditions is not getting information about the world, then nothing is' \cite{MaudlinBohmMeasProb}. 

However, as is evident from Struyve's comments on the minimalist model above, the manner in which one might hope to find an image of the world in a minimalist model of Bohmian QED - if indeed the sort of analysis that Struyve sketches does go through - is far less direct and less transparent. In a separate article, `Descying the World in the Wave Function,' Maudlin applies a distinction between what Sellars calls the `Manifest Image' and the `Scientific Image' of the world to the interpretation of quantum mechanics. Maudlin describes the distinction as follows:

\begin{quote}
\begin{singlespace}
The Manifest Image is the world in which we first find outselves, a world of people and actions, characters and habits, tables and chairs. The Scientific Image, in contrast, is a world of postulated theoretical entities, atoms and quarks and electromagnetic fields  \cite{MaudlinDescrying}. 
\end{singlespace}
\end{quote} 

\noindent The distinction, says Maudlin, is illustrated by Eddington's famous passage about `The Two Tables':

\begin{quote}
\begin{singlespace}
One of the has been familiar to me from my earliest years. It is a commonplace object of that environment which I call the world. How shall I describe it? It has extension; it comparatively permanent; it is coloured; above all it is \textit{substantial}...

Table No. 2 is my scientific table. It is a more recent acquaintance and I do not feel so familiar with it. It does not belong to the world previously mentioned - that world which spontaneously appears around me when I open my eyes ... My scientific table is mostly emptiness. Sparsely scattered in that emptiness are numerous electric charges rushing about with great speed; but their combined bulk amounts to less than a billionth of the bulk of the table itself ... 

I need not tell you that modern physics has by delicate test and remorseless logic assured me that my second scientific table is the only one which is really there - wherever `there' may be. \cite{EddingtonTwoTables}, ix-xii
\end{singlespace}
\end{quote}
 
\noindent Of course, Eddington's first table belongs to the Manifest Image and the second to the Scientific Image. 

Maudlin is interested in how the Scientific Image offered by each of the various interpretations of quantum mechanics, and particularly pilot wave interpretations, serves to undergird the Manifest Image that presents itself most immediately to us. For this to occur, he notes, there must be an isomorophism between some portion of the Scientific Image and the Manifest Image. As a supporter of the Bohm theory, he argues that unlike the Everett and GRW theories, which posit the existence only of the wave function evolving on some very high-dimensional space, the Bohm theory makes the connection between the Manifest Image - a world of localised objects existing in three dimensional space - and the Scientific Image - on Bohm's theory, a set of localised point particles guided by a wave function - especially transparent. As he writes, 

\begin{quote}
\begin{singlespace}
The justification for Bohm's choice of beables is simple and powerful. It is relatively easy to discover an isomorphism between the Manifest Image and a Scientific Image which contains particles with determinate positions. It is not a hard task to construct a passable \textit{doppelganger} for the world revealed by experience using particles in motion. Cats in the Manifest Image correspond to cat-shaped collections of particles in the Scientific \cite{MaudlinDescrying}. 
\end{singlespace}
\end{quote}

\noindent In a separate article, his contribution to the Many Worlds conference volume of Saunders, Barrett, Kent and Wallace, he argues that so-called monist interpretations of quantum mechanics like the Everett and GRW theories, which assume that the wave function is a complete description of the state of a system - i.e. that no additional variables are needed - make the link between the Manifest and Scientific Images far more opaque than does the Bohm theory. The Scientific Image that they posit is one of a complex valued function evolving on a very high-dimensional space, not in the three-dimensional space that characterises the Manifest image, so that the three dimensions of our experience must emerge from the theory in a relatively complicated and indirect way. In he Bohm theory, which simply posits from the outset the existence of particles evolving in 3-D space  \cite{MaudlinManyWorlds}, the connection is much clearer.

Turning to the case of pilot wave versions of quantum field theory, he writes 

\begin{quote}
\begin{singlespace}
If the acceptance of field theory demands a new choice of beables, at least we now understand what those beables are for. They must be used to fashion, within the Scientific Image, a structure which corresponds to the Manifest Image \cite{MaudlinDescrying}. 
\end{singlespace}
\end{quote}

\noindent Yet while Maudlin regards the simplicity of the link between the Manifest Image and the Scientific Image as a major advantage of the nonrelativistic Bohm theory over monist interpretations, he acknowledges that the Manifest Image is not sacrosanct:

\begin{quote}
\begin{singlespace}
There is no requirement that all, or even most, of the Manifest Image be vindicated by the Scientific Image: we might conclude on the basis of our physical theory that some of the most central aspects of our pre-theoretical picture of the world are false. So the demand for isomorphism seems to entail rather little \cite{MaudlinDescrying}.
\end{singlespace}
\end{quote}

\noindent But presumably, the fallibility of the Manifest Image should include the possibility that, contrary to immediate appearances, the world around us is not constructed out of fermionic matter, but instead may be built out of something like the electromagnetic field - assuming that is, that the behavior of this field, together with the links that the minimalist model implicitly takes to hold between field beable configurations and the world of appearances, suffice to save the appearances. Presumably, one such link would entail that there is a charged particle whenever the field beable takes the configuration of a field emanating from a moving charge (e.g. something akin to the field associated with a classical Lienard-Wiechert potential). Clearly, for the minimalist model to succeed, such correspondences need to fleshed out in more detail. Once this is done, the question as to whether the theory is empirically adequate becomes a theoretical rather than a metaphysical one. And this theoretical question seems at the moment to be an open one. 

While the link between the Scientific Image and the Manifest Image that would be required on a minimalist model with only EM field beables would be much less direct than the link presupposed by nonrelativistic particle models, Maudlin's analysis seems to suggest that we should not dismiss a theory simply because it undercuts our Manifest Image of the world. To be sure, a picture in which everything around us is ultimately constructed out of different configurations of an electromagnetic field does just this; but, as Maudlin himself should admit, this alone does not consitute sufficient reason for dismissing the theory.  

Nevertheless, it is also important to recognise that minimalist models may, if we like, incorporate fermionic beables as well. These beables, although they will not possess any independent dynamics of their own - their values and evolution will depend entirely on the value of the electromagnetic field beables and on the quantum state - can be expected to behave very much in accord with the way matter behaves on in the Manifest Image, at least insofar as distributions of fermionic matter can be expected to be localised in the appropriate sorts of situations. In such a case, the appropriate isomorphism between the Scientific and Manifest Images may be simpler and more direct than in the model with only electromagnetic beables, since this model will, as in the case on non-relativistic Bohm theory, simply associate cat-like distributions of fermionic matter in the Scientific Image with the cats of the Manifest Image. The only potentially salient difference, in this case, with the nonrelativistic particle model is that in this case the fermionic degrees of freedom do not possess their own dynamics but instead `piggyback' on the dynamics of the electromagnetic field beables. Whether the lack of any independent dynamics on the part the fermionic beables somehow ought to disqualify them as a legitimate substrate of the Manifest Image I leave as a subject for another discussion.

\section{Summary}

In this chapter, I have provided templates for the DS reduction of two models of the nonrelativistic quantum mechanics of a charged spin-1/2 particle - the nonrelativistic 2-particle Coulomb model and the 1-particle Pauli model - and discussed the connection between Bell's nonrelativistic Bohmian model of a spin-1/2 particle and Struyve and Westman's minimalist QED model.

\chapter{Conclusion: DS Reduction and Nagelian Reduction}
\label{ch6}

In Chapter 1, I highlighted a number of parallels between GNS reduction and DS reduction, and suggested that these parallels were strong enough to justify the claim that DS reduction simply is GNS reduction applied within the context of a semantic, dynamical systems view of physical theories. Presently, I argue that in the narrowed context of physical, dynamical systems reduction, in which the various concepts of GNS reduction can be given more precise meanings than they are given in formulations that attempt to encompass reduction across all of the sciences, many of the usual criticisms of GNS reduction can be straighforwardly addressed. I consider these criticisms one-by-one, and in order. For the reader's convenience, I reproduce the summary of these criticisms here, as quoted from \cite{dizadji2010s}:

\begin{itemize}
\begin{singlespace}
\item \textit{Problem 1: The syntactic view of theories}. Nagel formulated his theory in the framework of the so-called syntactic view of theories, which regards the- ories as axiomatic systems formulated in first-order logic whose non-logical vocabulary is bifurcated into observational and theoretical terms. This view is deemed untenable for many reasons, one of them being that first-order logic is too weak to adequately formalise theories and that the distinction between observational and theoretical terms is unsustainable. This, so one often hears, renders Nagelian reduction untenable.
\item  \textit{Problem 2: The content of bridge laws}. There is a question about what kind of statements bridge laws are. Nagel considers three options (1961, 354-355): they can be claims of meaning equivalence, conventional stipulations, or assertions about matters of fact. The third option can be broken down further, since a statement connecting two quantities could assert the identity of two properties, the presence of a (merely) \textit{de facto} correlation between them, or the existence of a nomic connection. Although the issue of the content bridge laws is not per se an objection, it is a question that has often been discussed in ways that gave rise to various objections, in particular in connection with multiple realisability, to which we turn now.
\item  \textit{Problem 3: Bridge laws and multiple realisability}. The issue of multiple real- isability (MR) is omnipresent in discussions of reduction. A $T_{P}$ -property is multiply realisable if it corresponds to more than one different $T_{F}$ -properties. The standard example of a multiply realisable property is that of pain: Pain can be realised by different physical states, for instance in a humanÕs and in a dogÕs brain. The issue also seems to arise in SM because, as Sklar points out, temperature is multiply realisable. MR is commonly considered to undermine reduction. ... [One] argument from MR is that, in order to reduce $T_{P}$ -phenomena to $T_{F}$ -phenomena, $T_{P}$ -properties must be shown to be Ônothing over and aboveÕ $T_{F}$ -properties. That is, it must be shown that $T_{P}$ -properties do not exist as something extra or in addition to $T_{F}$ -properties: There is only one group of entities, $T_{F}$ -properties. Showing this requires the identification of $T_{P}$ - properties with $T_{F}$ -properties. But a multiply realisable $T_{P}$ -property is not identifiable with a $T_{F}$ -property. This undercuts reduction.
\item \textit{Problem 4: The Epistemology of Bridge Laws}. How are bridge laws established? Nagel points out that this is a difficult issue since we cannot test bridge laws independently. The kinetic theory of gases can be put to test only after we have adopted Equation 5 as a bridge law, but then we can only test the entire ÔpackageÕ of the kinetic theory and the bridge law, while it is impossible to subject the bridge law to independent tests. While this is not a problem if one sees bridge laws as analytical statements or mere conventions, it is an issue for those who see bridge laws as making factual claims.
\item \textit{Problem 5: Strong analogy}. Strong analogy is essential to GNS. This raises three issues. The first is that the notion of strong analogy is too vague and hard to pin down to do serious work in a reduction. It is a commonplace that everything is similar to everything else, and hence saying that one theory is analogous to another one is a vacuous claim.  
 \cite{dizadji2010s}
 \end{singlespace}
\end{itemize}

\noindent

\section{Problem 1: The Syntactic View of Theories}

DS reduction is formulated specifically within the semantic view of theories, and concerns the reduction of models to other models. For this reason, it is not subject to arguments against the syntactic view of theories. As the authors of \cite{dizadji2010s} note, although Nagel was a proponent of the syntactic view of theories, there is no textual evidence to the effect that he saw it as an essential prerequisite for Nagelian reduction.

Independently of what Nagel or Schaffner may originally have intended, though, I have explicitly demonstrated in the preceding chapters how the core elements of Nagelian reduction (as intepreted by the GNS account) apply within a semantic, dynamical systems view of theories, and have thereby shown that the syntactic view is not a prerequisite for the application of Nagel and Schaffner's insights. Of course, the semantic view adopted here, like the syntactic view, is subject to its own criticisms, though to address these would be to go beyond the scope of this thesis (see, for instance, \cite{frigg2009scientific} for some of these criticisms).

\section{Problem 2: The Content of Bridge Laws} 

In \underbar{The Structure of Science}, Nagel raises the question as to the logical status of the connecting assumptions employed in his account of reduction:

\begin{quote}
\begin{singlespace}
There appear to be just three possibilities as to the nature of the linkages postulated by these additional assumptions [i.e., the bridge rules]: (1) The first is that the links are \textit{logical connections} between established meanings of expressions. The assumptions then assert `A' to be logically related (presumably by synonymy or by some form of one-way analytical entailment) to a theoretical expression `B' in the primary science. On this alternative, the meaning of `A' as fixed by the rules or habits of usage of the secondary science must be explicable in terms of the established meanings of theoretical primitives in the primary discipline. (2) The second possibility is that the linkages are \textit{conventions}, created by deliberate fiat. The assumptions are then coordinating definitions, which institute a correspondence between `A' and a certain theoretical primitive, or some construct formed out of the theoretical primitives, of the primary science. On this alternative, unlike the preceding one, the meaning of `A' is not being explicated or analyzed in terms of the meanings of theoretical primitives. On the contrary, if `A' is an observation term of the secondary science, the assumptions in this case \textit{assign} an experimental significance to a certain theoretical expression of the primary science, consistent with other such assignments that may have been previously made. (3) The third possibility is that the linkages are \textit{factual} or \textit{material}. The assumptions then are physical hypotheses, asserting that the occurrence of the state of affairs signified by a certain theoretical expression `B' in the primary science is a sufficient (or necessary and sufficient) condition for the state of affairs designated by `A.' It will be evident that in this case independent evidence must in principle be obtainable for the occurrence of each of the two states of affairs, so that the expressions designating the two states must have identifiably different meanings. On this alternative, therefore, the meaning of `A' is not related analytically to the meaning of `B.' Accordingly, the additional assumptions cannot be certified as true by logical analysis alone, and the hypothesis they formulate must be supported by empirical evidence. 
\end{singlespace}
\end{quote}

\noindent Nagel is occupied largely with the question of whether bridge rules are analytic (possibilities (1) and (2)) or synthetic (possibility (3)) in nature, and if they are analytic, what the precise nature of their analyticity is - that is, whether their analyticity is more aptly characterized by proposal (1) or proposal (2). 

In the context of DS reduction, concepts such as bridge maps, bridge rules, image and analogue models, and `strong analogy' have precise mathematical definitions. The question as to their status as analytic or synthetic claims seems beside the point, to the extent that it has any meaning in a semantic, dynamical systems context. Unlike most of the core elements of Nagelian reduction, Nagel's comments regarding the logical status of bridge laws do seem more appropriate to the syntactic view of theories than they do to the semantic view. In the semantic view, a model as a whole represents some domain of reality; it is difficult to see how one might isolate individual propositions within the model and identify them as analytically or synthetically true claims. Of course, the claim that it is difficult to see how to apply the analytic/synthetic distinction within a semnatic, dynamical systems framework doesn't mean it can't be done. For this reason, I present what I believe is the most natural attempt to carry over Nagel's inquiry as to the logical status of laws into the context of DS reduction, although, insofar as Nagel's three possibilities can be given any reasonable construal in a dynamical systems framework, I reach a different conclusion from Nagel on this matter. Specifically, Nagel concludes that the bridging assumptions cannot be logical connections and that they must be either conventions or empirical claims. That is, he denies (1) as a possibility and claims that a mixture of (2) and (3) serve to characteise the nature of the bridging assumptions in a reduction. On the other hand, I claim that within the context of DS reduction, it is (3) that can be ruled out, and that there are senses in which both (1) and (2) serve to characterise the nature of the bridging assumptions. 

%As we have seen in the preceding chapters, on a dynamical systems approach, the bridging assumptions that translate between the mathematical frameworks of two models from two different theories take the form of a function from the state space of the low-level model into the state space of the high-level model, furnished by the bridge map, together with a variable assignment to the value of the function furnished by the bridge rules. The fact that the image of some element $x^{l}_{0}$ under the function $ B_{l}^{h}$ exhibits  approximately the same evolution under the dynamics induced by the low-level theory as the corresponding element $ x^{h}_{0}$ in the high level state does space under the dynamics of $M_{h}$ is then a fact deducible purely mathematically. 

\subsection{Do Bridge Maps Reflect Synthetic Claims?}

I begin by addressing Nagel's third possibility, that the bridging assumptions have the status of physical hypotheses - that is, that they are synthetic propositions. Given a high-level model $(S_{h}, D_{h})$ and a low-level model $(S_{l}, D_{l})$, the question as to which bridge maps, if any, satisfy the conditions for DS reduction, and in which domains of $S_{l}$ and over what timescales and to what margins of approximation, is a question purely of \textit{mathematics}. It is \textit{not} a question requiring empirical investigation for its resolution. Given a high- and a low- level model that describe the same system, I do not need to perform any experiments to determine whether any bridge maps, and if so which ones, exist that satisfy the conditions of DS reduction; in principle, this can be decide purely by theoretical means, with a pen and paper. The specification of the two models, along with the stipulation that they describe the same physical system, very severely constrain what the bridge map can be, since they specify the dynamical maps and the symmetries with which the bridge map (which, recall, must be time-independent) must be compatible. Once we have specified the high- and low- level models and stipulated that they describe the same system, It is simply not the case that prior to further empirical investigation the bridge map could conceivably be anything and that it is up to us to find out through experiment what it is. The constraints of DS reduction very severely limit the set of possibilities, even if they do not single out a unique one.

To elaborate, bridge maps $B_{l}^{h}$ are functions from $S_{l}$ to $S_{h}$, and given the dynamical maps $D_{l}$ and $D_{h}$ of the high- and low- level models, as well as a particular approximation margin $\delta$ and a reduction timescale $\tau$, it is possible to decide entirely by mathematical means whether the DSR condition, $\bigg|  B_{l}^{h} \big(D_{l}(x^{l}_{0})\big) - D_{h}\big(B_{l}^{h}(x^{l}_{0})\big) \bigg|_{h} < \delta$ for $0 \leq t \leq \tau$, as well as the condition 2) ensuring compatibility with the dynamical symmetries of the models, is satisfied for a particular choice of $B$, without ever performing a single experiment. The only assumptions of an empirical nature that enter into the analysis are that the high- and low- level models both approximately describe the same phenomena. But this empirical assumption is one concerning the relationship between the models and the physical world, not concerning the relationship \textit{between the models}, as bridge maps and bridge rules do. Once the models are specified and stipulated to describe the same system, the question of what bridge maps may exist that connect the two within the constraints of the DSR condition is, again, solely a mathematical one. For this reason, we should discard Nagel's third possibility, namely that the linkages posited by the bridging assumptions are physical hypotheses, at least within the context of physical reductions that fit within the framework of DS reduction.

To this line of argument, one might object that bridge maps nevertheless reflect synthetic claims with empirical content because the assertion that the state $x^{l}$ of $M_{l}$ subvenes or instantiates the state $x'^{h}=B(x^{l})$ of $M_{h}$ is one with empirical content that goes beyond that of $M_{l}$ and $M_{h}$ alone. But to say that such a claim of instantiation has independent empirical content is to say that it asserts a contingent link between properties, or more appropriately in the context of DS reduction, a contingent link between the states, posited by the two theories - a link that, logically speaking, need not have been so. This, in turn, entails that given the empirical assumptions that the high- and low- level models both describe some system, nature, so to speak, still has complete discretion in deciding what the linkages between the states of the models should be, so that these linkages reflect empirical facts that are logically independent of the empirical claims that the high- and low- level models describe the same system. But of course this is not the case, because the allowable linkages are tighly constrained by the DSR conditions, and therefore not independent of the empirical claims that the high- and low- level model describe a given system (since the allowable linkages depend strongly on what the models are). 

The only recourse, it seems, for those who wish to retain the notion that bridge maps possess empirical content over and above the empirical claims that the high- and low- level models describe a particular system, is the potential non-uniqueness of the bridge map that satisfies the DSR conditions. Perhaps, in this case, there is still a range of possibilities compatible with the DSR constraints for nature to choose from, so that while many linkages are possible that satisfy the constraints, as a matter of empirical fact only one reflects the correct linkage between the states of the high- and low- level models. 

There are a number of responses one can give to this line of thinking. The first is to ask, in cases where a bridge map is already known to satisfy the DSR conditions to within a certain margin of error and within a particular timescale and on a particular domain of states, for a concrete example of an alternative bridge map that satisfies these conditions to within the same margins before one begins to seriously worry about non-uniqueness of the bridge map. The question of a bridge map's uniqueness given these constraints is one that cannot be addressed in general but must be assessed on a case-by-case basis, unless one constructs a general recipe for constructing alternative bridge maps that satisfy the DSR conditions to with the given specifications. Pending such a general recipe, or proposals of alternative bridge maps in specific cases, one can simply choose to be relaxed about the possible non-uniqueness of bridge maps, and to worry about it when it has more concretely been shown to be an issue and not merely a hypothetical possibility. Ideally, though, a proof of uniqueness of the bridge map in all cases would be the best and most conclusive way to confront to the line of thinking expounded in the previous paragraph. 

If a general recipe for constructing alternative bridge maps does exist, then this may suggest further conditions that need to be added to those already imposed by the framework of DS reduction. 

On the other hand, one also might respond to the worry about the non-uniqueness of the bridge map by questioning whether it is in fact a problem to begin with. Any bridge map identifies some mathematical structure that can be defined within the low-level model; if more than one structure exists that satisfies the necessary conditions to within the specified margins, then it is simply inappropriate to question which is the `real' structure and which are those that nature chose not to make use of. For, given that the low-level model applies, \textit{all} of these structures are defined within the low-level model and therefore exist within the model and are in some sense equally `real;' they simply represent different functions of the low-level state. Of course, in such a case there still remains the question of which among these various structures corresponds to the physical system in question. In such a case (again, if any such cases exist) there may be genuine underdetermination as to which of the various structures in the low-level theory represents the physical system in question. In such a case, it will indeed turn out to be a matter of fact, independent of the claims that the high- and low- level theories apply, as to which of the many alternative low-level structures (all satisfying the DSR conditions to within the same specifications) represents the system in quesiton. 

Having listed these possible responses to the non-uniqueness worry, I choose here to adopt a relaxed attitude to this possibility until a concrete example of non-uniqueness or a general recipe for constructing alternative bridge maps has been put forward. 

Finally, as a caveat, it should be noted here that if $B$ is a bridge map satisying the DSR conditions, and $T_{l}$ is a symmetry of the low-level dynamics, then $B\circ T_{l}$ is also a bridge map satisfying these conditions. Likewise, if $T_{h}$ is a high-level symmetry, then $T_{h} \circ B$ will also be bridge map satisfying the DSR conditions. Thus, uniqueness here should be understood as uniqueness \textit{up to} symmetries of the high and low-level models.

\subsection{The Analytic Nature of Bridge Maps and Bridge Rules}

If they are not characterised by Nagel's third possibility, might the bridge maps and bridge rules of DS reduction be characterised by one of his first two possibilities? Insofar as bridge maps are required to satisfy the DSR condition with respect to the dynamical maps of the two models, as indeed any bridge map in a DS reduction would, they reflect the non-empirical fact of a certain correspondence, defined by the DSR condition, between the mathematical structures of the high and low level models. It would be incorrect to describe the claim of the existence of such a map as a definition or a convention, as per Nagel's second possibility, because the claim that between two models there exists such a map is a mathematical fact, not a reflection of some arbitrary choices on our part. 

However, there is a sense in which not bridge maps but bridge rules, as defined in Chapter 1, are simply coordinating definitions or conventions as per Nagel's second possibility, in that bridge rules simply assign a label familiar to the mathematical language of the high level model to a term of the same mathematical form constructed from the mathematical ingredients of the low level model. That is, they coordinate some element of the high level model with some construction within the low level model that is constrained to be same kind of mathematical object (e.g., a vector in a particular Hilbert space, or point on a 6n-dimensional symplectic manifold, etc.). Beyond the constraint that it represent a variable of the same mathematical form, the choice of what label to assign the image under the bridge map is, of course, a matter of arbitrary choice; so bridge rules (again, as differentiated from bridge maps) are conventional in the trivial sense that we may assign whatever letter or name we like to the relevant quantity without altering its significance. 

What about Nagel's first possibility, on which `the meaning of ``A'' as fixed by the rules or habits of usage of the secondary science [in a DS context, the high level model] must be explicable in terms of the established meanings of theoretical primitives in the primary discipline [in a DS context, the low level model]' \cite{NagelSS}? There is an approximate sense in which bridge maps and bridge rules, as understood in the context of DS reduction, fit this description. In DS reduction, the laws of the low-level model (i.e., the dynamical map of the low level model, along possibly with other constraints on the state evolution), combined with an appropriate domain restriction within the low-level state space, logically entail that the quantity defined by the bridge map satisfies the laws of the image model, which reflect in an approximate sense the behavior characteristic of the high-level model. Insofar as the meaning of a term in the high level model is established by the mathematical behavior that it exhibits within that model, and insofar as the meaning of a term constructed within the low-level model is established by its behavior in the low-level model, the meaning of the term construed according to the bridge map in the low-level theory entails, within a restricted domain of the low-level theory, that it exhibits approximately the same behaviour, and in this sense holds approximately the same meaning, as the corresponding term in the high-level theory. The bridge rule and strong analogy relations that together connect a term in the high-level model to a corresponding image term in the low-level model in this sense reflect a logical entailment from the meaning of a term in the low-level theory to the (approximate) meaning of the corresponding term in the high-level theory.

%the meaning of the bridge map term logically entails (though in an approximate sense) the meaning of the corresponding term in the high-level model. To summarise: the fact that a certain term defined within the low-level model exhibits approximately the same behavior as a certain term in the high-level model is a deductive consequence of the laws of the low-level model and the definition of the term within that model;  insofar as the meaning of these two terms is determined by the way they behave in their respective models, the meaning of the term in high-level model is approximately entailed by the meaning of the corresponding term in the low-level theory. By `approximate entailment' I simply mean that the behavior of term in the low-level theory is approximately the same as that of the corresponding 

\section{Problem 3: Bridge Laws and Multiple Realisability}

One major criticism of Nagelian reduction states that because multiple realisability entails that certain high-level properties cannot be identified with low-level properties, and because the bridge laws of Nagelian reduction - or so the line of argument goes - must be statements of \textit{identity} between high and low level properties, the Nagelian account fails. While Nagel himself did not require bridge laws to be identities in his original account of reduction, a number of authors have argued that these laws must be identities because reductions require inclusion of the domain of one theory into that of another, and such inclusion requires the identification of properties in the high-level theory with those in the low-level theory (see, for instance, \cite{Causey1977}, Ch. 4). As Sklar has argued in \cite{Sklar1967}, bridge laws must be identities in order to distinguish reductions from mere correlations between different sets of laws; he cites the example of the Wiedemann-Franz law, which expresses a correlation between thermal and electrical conductivity in a metal, and, in particular, permits the derivation of certain laws of electrical conductivity from certain laws of thermal conductivity. Yet we would not wish to say that the laws of electrical conductivity have thereby been reduced to those of thermal conductivity, even though one may be deduced from them. The reason is that the nature of the correlations established by the Wiedemann-Franz law are not identities but `mere' correlations; the domain of the theory of electrical theory has not been subsumed into that of the theory of thermal conductivity as it ought to be in a reduction. 

If reduction rests on bridge laws that are identities, and multiple realisation of some high-level properties by some low-level ones is incompatible precludes identification of each high-level property with some unique low-level property, then - or so the antireductionist argument from multiple realisability goes - multiple realisability precludes reduction. A number of lines of response to this antireductionist argument have been developed in the literature. While there is not space here to discuss them all in detail, I will briefly describe a few. The first denies, \textit{contra} Causey and Sklar, that bridge rules need to be identities in order to effect genuine reductions, but argues instead that they need only be one-way conditionals (see \cite{Richardson1979} and  \cite{Bickle2003}). The second upholds the requirement that bridge laws be identities, but instead argues that a given high-level property  
realised by various low-level properties should in fact be divided into several distinct high-level properties, each of which can be identified with a particular low-level property; for example, the property of pain, rather than being multiply realised in humans, dogs and Martians, in fact should be conceived of as three distinct properties: pain-in-humans, pain-in-dogs, pain-in-Martians. The third line of argument is to regard the disjunction of all low-level properties that realise a particular high-level property as a property in itself, and to identify this disjunction with the high-level property (again, this approach upholds the requirement that bridge laws be identities). 

In the context of DS reduction, where we deal with the reduction of models rather than of laws or theories, it is indeed the case that we expect the domain of a high-level model to be subsumed into that of a low-level model, \textit{for a given set of applications of that high-level model}. For a different set of applications of the same high-level model, it is entirely possible that an altogether different low-level model from the first will be required to effect a reduction. For example, a particlar classical model of the simple harmonic oscillator might reduce to one quantum model in applications where the physical system in question is a bob on a spring; on the other hand, the very same classical model may be reduced to an altogether different quantum model in the case where the physical system being described is a charge moving along an axis bored through a uniformly charged sphere (which will also produce a linear restoring force on the charge). Certainly, the quantum mechanical models underlying these two applications of the classical simple harmonic oscillator model will be very different - for example, as regards the origin of the restoring force, since in the case of the spring the restoring force is compounded from the individual forces between the atoms inside the spring, and in the case of the electric charge it is simply the result of a classical background electrostatic field. We should not expect that the domain of the high-level model, without further specification as to the class of applications, will be subsumed into the domain of any single low-level model, for it may be the case that different low-level models are needed to reduce the high-level model in the context of different applications of the high-level model. In this sense, a uniform, systematic reduction of the high-level model will not be possible, and the reduction must proceed on a more piecemeal basis (though it will often be possible to retain significantly more generality than to proceed with the reductions on a system-by-system basis; rather, \textit{classes} of systems can be treated all at once; for example, all low-level systems that can be modeled with the same form, but different parameter values, for their quantum Hamiltonian might constitute one class of models to which a particular class of classical high-level models reduces). Given that a particular high-level model may have a number of distinct low-level models to which it reduces depending on the particular application in question, it would be inappropriate to require that the relation between the components of the high-level model and those of the various low-level models to which it reduces (in the DS sense) be one of identity. 

Yet even specialising to a particular class of applications of a high-level model, and requiring its domain relative to this range of applications to be subsumed within that of some low-level model, there are reasons to doubt that the bridge maps of DS reduction should be required to be identifies, even though the terms linked by the bridge map will both refer to the same physical state of affairs. The term `identity' in the mathematical context of DS reduction suggests that the bridge maps should be required to be one-to-one; certainly, this is not the case with any of the bridge maps we have seen in any of the examples considered, as there are often numerous states in the low-level theory that will emulate the behavior of a given high-level state given some timescale and margin of error. The best we can do - within the level of precision characterising the high-level model's success - is to say that the high-level state used to describe a particular system is instantiated or realised by some particular one or other of the many numerous low-level states that emulate its behavior under the bridge map. But the correspondence certainly is not one-to-one, and for this reason also it would be a mistake to require that bridge maps be identities. 

In short, we have seen that DS reduction serves to describe a wide range of inter-theory (or rather inter-model) relations in physics, but that its ability to do so rests on an understanding of bridge maps that permits them to be many-one rather than one-one. If we were to require that bridge maps of DS reduction be one-one, the account would likely no longer successfully characterise the inter-theory relations discussed here. We can either choose to abandon the condition of identity on bridge maps, and incorporate many intertheory relations into our account of reduction; or we could continue to insist that `reduction' requires identity and simply call the relationship that does hold among these models, and characterised by the conditions set out in the first chapter, something other than reduction. I have opted for the first option here, given that DS reduction does exhibit many of the other characteristics typically associated with the reduction relation. 

But if the links entailed by bridge maps are not required to be identities, how do we avoid Sklar's concern that they simply reflect correlations? Instead of interpreting these links as identities, we should understand them links as claims of \textit{co-reference}: that is, within a given physical context, for instance, both the classical phase space point $(X,P)$ and a state $| \psi \rangle$ such that  $\langle \psi| \hat{X} | \psi \rangle,  \langle \psi | \hat{P} | \psi \rangle$ refer to the same condition of the physical system in question; however, because the low-level model describes this system in greater detail and with greater accuracy, it is possible that multiple states $| \psi \rangle$ are compatible with the condition (or rather set of possible conditions) represented by $(X,P)$.

Since the anti-reductionist's multiple realisability argument loses traction once the links between high- and low-level descriptions are no longer required to be identities, the fact of multuple realisability simply occurs as a characteristic of DS reduction, rather than a reason to doubt its applicability. I will now survey the different possible senses in which multiple realisability can occur in DS reduciton.

\vspace{5mm}

%Yet nothing in Nagel's original account required that the connecting assumptions which were later referred to as bridge laws be identity claims. By simply abandoning - or rather, by simply not adopting in the first place - the requirement that bridge laws be identities between high- and low- level properties, the Nagelian account is immune to this particular criticism from multiple realisability. The DS account certainly does not require that bridge maps and bridge rules signify identities between high- and low- level models; presumably, if it did, this requirement would likely entail that the bridge map be one-to-one, that the bridge map be unique, and that the low level model be the only model of the low-level theory to which the high level model reduced. As I explain in further detail, multiple realisability is simply a characterisitic of DS reduction, not an obstacle for the DS account to overcome.  

In the context of DS reduction, multiple realisability can be attributed to a variety of potential characteristics of the reduction relation between two models. First, it can be attributed to the possibility of there being more than one low-level model to which a given high-level model reduces. For example, the classical model of the simple harmonic oscillator is multiply instantiated, even within classical theory, as a bob on a spring or as a charge moving through the hollowed-out axis of a uniformly charged sphere. Certainly, then, the quantum mechanical models underlying these two instantiations of the simple harmonic oscillator model will be very different - for example, as regards the origin of the restoring force, since in the case of the spring the restoring force is compounded from the individual forces between the atoms inside the spring, and in the case of the electric charge it is simply the result of a classical background electrostatic field. 

Multiple realisability also can be associated with the existence of more than one bridge map between a  single high level and a single low level model, and with the distinct domains of approximate $M_{h}$-behavior in the low-level state space that are associated with these distinct bridge maps. For example, a classical model of two planets orbiting each other can be instantiated in two different ways by a classical model of two pairs of oribting planets in which the pairs are widely separated in space so as not to affect each other; one bridge map will associate the pair of planets in the first model with one pair in the second model, while a different bridge map will associate the pair of planets in the first model with the other pair in the second model. 
%For example, a classical model of the motion of $n$ planets, functioning as a high-level model, can be doubly realised by another classical model containing $2n$ planets, where each realisation is associated with a different set of $n$ degrees of freedom. More generally, a high-level model containing relatively few degrees of freedom may be multiply realised in a low-level model containing many more degrees of freedom, where each distinct realisation is associated with a different set of degrees of freedom in the low-level model. 

Finally, multiple realisability could be associated with the fact that the bridge map is typically many-one, even when restricted to the domain of states that exhibit approximate $M_{h}$ behavior under the bridge map. For example, in the reduction of the single-particle model of classical mechanics to the single-particle model of quantum mechanics discussed in the first chapter, there may be a range of narrow wave packets, with varying widths and slightly varying shapes, all of which instantiate the same classical phase space point, and the same classical dynamics, when expectation values are taken. 

Still, while acknowledging that DS reduction incorporates these various forms of mutliple realisation, one could object that precisely for this reason it should not be considered as a form of reduction at all, for reduction ought to require that the corresponding elements of the high and low-level models be identified. To some extent, this objection is simply one of what a legitimate application of the term `reduction' is, and of whether one must include identifications as a necessary component of its proper usage. What I hope I've demonstrated at least somewhat convincingly in the preceding chapters is that the DS account succeeds in characterising a number of inter-theory relations in physics, irrespective of whether these particular inter-theory relations conform to some previously conceived notion of reduction. Moreover, as I have argued in Chapter 1, the DS account characterises these inter-theory relations in a manner that conforms with at least some uses of the term reduction that are flexible enough to accommodate multiple realisation - in particular, the usage of the term associated with the GNS account.

\section{Problem 4: The Epistemology of Bridge Laws}

How are bridge laws estabished on the GNS account? As argued above, they are not established empirically, but mathematically. Given a model of a high-level theory and a model of a low-level theory, both of which describe the same physical system, it is a purely mathematical fact whether there exist bridge laws and corresponding domains within the low-level state space such that the DSR condition is satisfied. If one considers a particular function from the low-level state space to the high-level state space, and the fixed dynamical maps of the high- and low- level models, as well as a particular approximation margin and a reduction timescale, it is possible to decide entirely by mathematical means whether the DSR condition is satisfied, without ever performing a single experiment.

%I have argued that on the DS account of reduction, the bridging assumptions consist of a bridge map and of bridge rules. These are stipulated be fiat, so there is no question of how they are established. It is a question of mathematics, not of empirical fact, whether the laws of the low level model, combined with the birdge map and bridge rules approximate, in the sense `strong analogy' that I discuss in the next section, the dynamics of the high level theory. If they do, then we may say only after the fact that the bridge map and bridge rules establish some relationship of coreference between ; but again, this claim of coreference is not an independent statement of empirical fact, but rather a recongnition that two terms defined in the context of different models occupy similar structural roles within some limited domain of the low-level model's state space, in the sense that there is some function from the low level state space into the high level state space such that the DSR condition is satisfied over suitably long timescales. For example,  in the single-particle models of quantum and classical mechanics, expectation values of position and momentum on the one hand and classical position and momentum on the other, exhibit appromimately the same dynamics over a certain timescale in the domain of narrow wave packet state. That they do so is a \textit{mathematical} fact, not an empirical one. 

\section{Problem 5: Strong Analogy}

The relation of `strong analogy' between an image and analogue theory is supposed to signify some approximate agreement between the two, at least on the GNS model 
\footnote{It is possible that this construal of the concept of `strong analogy' is more narrow than Nagel or Schaffner intended; here I follow the usage of this term employed in \cite{dizadji2010s}, which takes it simply to signify approximate agreement between the analogue and high-level theories; on the DS approach, approximate agreement between analogue and high-level models is, as discussed in Chapter 1, determined according to the norm on the high-level state space. Insofar as any ambiguities in the meaning of `strong analogy' are resolved by the DS approach, I only argue that this is the case with respect to the construal of this term adopted \cite{dizadji2010s}.}. 
Yet what this approximate agreement consists in is left open on the GNS account. Given the more specialised context of DS reduction, we can make the sense of `approximate agreement' and therefore of `strong analogy,' exact. Specifically, as we have seen, on the DS approach the condition of strong analogy between a high- and a low- level model requires that 

\begin{equation}
\big| x'^{h}(t) - x^{h}(t) \big| < \delta \ \forall \ 0 \leq t \leq \tau,
\end{equation}
 
\noindent where $\tau$ again is the reduction timescale and $\delta$ some margin of error. 

Immediately we can see that this notion is much more precise than the simple requirement of approximate agreement between theories. The only potential ambiguities, it seems, with this notion of `strong analogy' concern the choice of norm on the high level state space, and the arbitrariness in the choices of margin of error. Although from a mathematical point of view it is usually possible to define any number of norms on the high level state space, in practice there is usually one that presents itself as the natural norm to use on the space; indeed, the DSR condition in all cases that we have considered is satisfied with respect to the most obvious norm on the high-level space. One may ask what uniquely qualifies these norms as the `natural' ones to use, or whether there is likely to be a uniquely obvious norm to use in every case of dynamical systems reduction. I will not attempt to address these concerns here, except to note that in the particular examples that I have considered, it was possible to demonstrate DS reduction according to the most obvious choice of norm on the high level space in question, and that there was never any cause to consider other possible norms.   

As regards the potential concern about the arbitrariness in the margin of error characterising the approximation, this is no more serious a concern than it is in the most straightforward cases of approximation: for which values of $x$ does the first-order Taylor expansion, $1 + x$, of $f(x)=e^{x}$ count as a good approximation to this function? Clearly, there is no unique answer, for the answer depends on how close an agreement one desires between the value of the first-order expansion and the value of the function for the approximation to be `good'; narrower agreement will require more stringent restrictions on $x$. Likewise, in the context of the DSR condition, narrower restrictions on the margin of error in the approximation typically will reduce the timescale over which the approximation can be expected to hold, while allowing for larger margins of error usually will increase this timescale.

\section{Summary}

In the preceding chapters, I have exibited what I believe to be the correct methodology for the reduction of physical theories by building upon an insight that has received too little attention in the literature on reduction - namely, the DSR condition. I also claim that beyond its methdological virtues, DS reduction provides the correct account of the relationship between different models of physical theories that describe the same phenomena. Unlike much of the philosophical literature on the subject of reduction, my analysis has not endeavoured to include the whole of science or the whole of reduction, but rather has specialised to particular set of reductions within physics. However, I hope that the reader is convinced that what we pay for with this loss of generality in our analysis is a significant gain in the precision with which we may discuss various notions relating to the reduction of physical theories. By specialising in this manner, a number of important insights about reduction from the general philosophy of science literature can be given more exact formulations. In particular, I showed that many of the insights of Nagelian reduction, construed according to the GNS account - can be carried over into the semantic framework of dynamical systems reduction and much of the ambgiuity surrounding them eliminated in the process.

\bibliographystyle{plain}
\bibliography{Thesis-corrections}

\end{document}